\documentclass[journal]{IEEEtran}
\usepackage{cite}

\ifCLASSINFOpdf
\else
   \usepackage[dvips]{graphicx}
\fi
\usepackage{amsmath}
\usepackage{amssymb}
\usepackage{algorithm}
\usepackage{algorithmicx}
\usepackage{algpseudocode}
\usepackage{enumitem}

\newtheorem{definition}{Definition}
\newtheorem{assumption}{Assumption}
\newtheorem{theorem}{Theorem}
\newtheorem{remark}{Remark}
\newtheorem{proposition}{Proposition}
\newtheorem{lemma}{Lemma}
\newtheorem{corollary}{Corollary}

\newcommand{\ato}{\overset{\mathrm{a.s.}}{\to}}
\newcommand{\aeq}{\overset{\mathrm{a.s.}}{=}}
\newcommand{\ag}{\overset{\mathrm{a.s.}}{>}}
\newcommand{\A}{\mathrm{A}}
\newcommand{\B}{\mathrm{B}}
\newcommand{\arginf}{\mathop{\mathrm{arginf}}\limits}

\begin{document}
%
\title{Orthogonal Approximate Message-Passing\\ for Spatially Coupled Linear Models}
%
%
%

\author{Keigo~Takeuchi,~\IEEEmembership{Member,~IEEE}
\thanks{K.~Takeuchi is with the Department of Electrical and Electronic Information Engineering, Toyohashi University of Technology, Toyohashi 441-8580, Japan (e-mail: takeuchi@ee.tut.ac.jp).}
\thanks{
The author was in part supported by the Grant-in-Aid 
for Scientific Research~(B) (JSPS KAKENHI Grant Numbers 21H01326), Japan. 
The material in this paper will be presented in part at 2023 IEEE 
International Conference on Acoustics, Speech and Signal Processing.
}
}

%
%

\markboth{IEEE transactions on information theory}%
{Takeuchi: Orthogonal Approximate Message-Passing for Spatially Coupled Linear Models}
%

\IEEEpubid{0000--0000/00\$00.00~\copyright~2015 IEEE}


\maketitle

\begin{abstract}
Orthogonal approximate message-passing (OAMP) is proposed for signal 
recovery from right-orthogonally invariant linear measurements with spatial 
coupling. Conventional state evolution is generalized to a unified framework 
of state evolution for the spatial coupling and long-memory case. The unified 
framework is used to formulate the so-called Onsager correction in OAMP for 
spatially coupled systems. The state evolution recursion of Bayes-optimal 
OAMP is proved to converge for spatially coupled systems via Bayes-optimal 
long-memory OAMP and its state evolution. This paper proves the 
information-theoretic optimality of Bayes-optimal OAMP for noiseless spatially 
coupled systems with right-orthogonally invariant sensing matrices.  
\end{abstract}


%
\IEEEpeerreviewmaketitle

\section{Introduction}
\subsection{Compressed Sensing with Zero-Mean i.i.d.\ Matrices} 
\IEEEPARstart{C}{ompressed} sensing~\cite{Donoho06,Candes061} is a 
technique to reconstruct unknown sparse signals from compressed measurements. 
When the signals are independent and identically distributed (i.i.d.), 
the R\'enyi information dimension~\cite{Renyi59} of each signal characterizes  
the information-theoretic compression limit~\cite{Wu10} for noiseless 
measurements. For instance, the Bernoulli-Gaussian (BG) signal with signal 
density $\rho\in[0, 1]$---the occurrence probability of non-zero signals---has 
the information dimension $\rho$. A goal in compressed sensing 
is to establish a reconstruction algorithm that achieves the 
information-theoretic compression limit. 

Approximate message-passing (AMP)~\cite{Donoho09,Rangan11} is a 
low-complexity and powerful algorithm for compressed sensing. 
AMP reconstructs unknown signals via message passing (MP) between the matched 
filter (MF) for interference suppression and a separable denoiser.  
AMP using the Bayes-optimal denoiser---called Bayes-optimal AMP---can be 
regarded as an asymptotically exact approximation of loopy belief 
propagation (BP)~\cite{Kabashima03}. 

The performance of AMP was rigorously analyzed via state 
evolution~\cite{Bayati11,Bayati15}, which was motivated by Bolthausen's 
conditioning technique~\cite{Bolthausen14}. The asymptotic dynamics of AMP 
is characterized via one-dimensional (1D) discrete systems called state 
evolution recursions. When the compression rate is 
larger than a value called BP threshold~\cite{Takeuchi15} in this paper, 
Bayes-optimal AMP 
was proved to achieve the Bayes-optimal performance asymptotically for 
zero-mean i.i.d.\ sub-Gaussian sensing matrices~\cite{Bayati11,Bayati15}. 
However, there is a gap between the BP threshold and the 
information-theoretic limit if the state evolution recursions have 
multiple fixed points. 

Spatial coupling~\cite{Kudekar11} was proposed to improve the BP performance 
of low-density parity-check (LDPC) codes toward the Bayes-optimal performance. 
This improvement was referred to as threshold saturation via spatial coupling 
in \cite{Kudekar11}. 
Spatial coupling is a universal technique to improve the performance of 
iterative algorithms in the other problems~\cite{Hassani12,Takeuchi12}. 
In particular, AMP~\cite{Krzakala12,Donoho13,Takeuchi15} was shown 
to achieve\footnote{
Achievability in \cite{Donoho13} is different from that in \cite{Wu10}. 
A compression rate was said to be achievable in 
\cite{Donoho13} if the mean-square error averaged over all sections converges 
almost surely to zero. On the other hand, \cite{Wu10} defined achievability 
in terms of block error probability. This paper uses the achievability 
in \cite{Donoho13}.  
} the information-theoretic compression limit for spatially coupled 
zero-mean i.i.d.\ Gaussian sensing matrices. More precisely, state 
evolution~\cite{Javanmard13} for spatially coupled dense systems was utilized 
in \cite{Donoho13} to obtain rigorous state evolution recursions. On 
the other hand, spatially coupled sparse systems were used in 
\cite{Takeuchi15} to bypass the technical difficulty in dense systems. 
The two systems result in the same state evolution recursions as each other, 
so that they can achieve the information-theoretic compression limit. 

\IEEEpubidadjcol

Spatial coupling was also applied to AMP decoding~\cite{Barbier17,Rush21} for 
sparse superposition codes~\cite{Joseph12} over the additive white Gaussian 
noise (AWGN) channel. Decoding in sparse superposition codes is equivalent 
to the reconstruction of signals having section-wise sparsity. 
Spatially coupled sparse superposition codes were proved to achieve the 
capacity of the AWGN channel via state evolution~\cite{Rush21}.  

Threshold saturation via spatial coupling can be analyzed with a potential 
function~\cite{Yedla14,Takeuchi15} that is defined from state evolution 
recursions without spatial coupling. This potential-function approach was 
originally motivated by \cite{Hassani12,Takeuchi12} and simplifies the proof 
of threshold saturation conducted by \cite{Donoho13}.  

The potential function defined in \cite{Takeuchi15} is equivalent to a 
replica-symmetric potential used in characterizing the Bayes-optimal 
performance for systems without spatial coupling~\cite{Tanaka02,Guo051}. 
The potential was originally derived with the replica method under the 
replica-symmetry assumption---non-rigorous tool in statistical 
physics~\cite{Mezard87,Nishimori01}---and rigorously 
justified in \cite{Reeves19,Barbier20}. It is possible to simplify the proof 
of threshold saturation in \cite{Donoho13} by using a relationship between 
the R\'enyi information dimension and the mutual 
information~\cite[Theorem~6]{Wu11}. 

\subsection{Beyond Zero-Mean i.i.d.\ Sensing Matrices} 
We have so far discussed zero-mean i.i.d.\ sensing matrices. 
Beyond zero-mean i.i.d.\ matrices, the replica-symmetric potential  
in \cite{Tanaka02,Guo051} was generalized to the case of right-orthogonally 
invariant sensing matrices via the replica method~\cite{Takeda06,Tulino13}. 
Right-orthogonal invariance implies that the right-singular vectors of the 
sensing matrix are orthonormal and Haar-distributed~\cite{Hiai00,Tulino04}. 
Zero-mean i.i.d.\ Gaussian matrices are included in the class of 
right-orthogonally invariant matrices. See \cite{Barbier18,Li22} 
for a theoretical progress to prove the replica-symmetric potential 
in \cite{Takeda06,Tulino13} rigorously. 

There are general ensembles beyond zero-mean i.i.d.\ sensing matrices
such that AMP fails to converge~\cite{Caltagirone14,Rangan191}. 
To solve this convergence issue in AMP, several MP 
algorithms~\cite{Kabashima14,Vila15,Manoel15,Rangan17,Ma17,Rangan192,Yuan21} 
were proposed. The most promising algorithm is orthogonal AMP 
(OAMP)~\cite{Ma17} or equivalently vector AMP (VAMP)~\cite{Rangan192}. 
In this paper, they are called OAMP.  

A prototype of OAMP was originally proposed in \cite[Appendix~D]{Opper05} 
as a single-loop algorithm to solve a fixed point of the 
expectation-consistent (EC) 
free energy. Bayes-optimal OAMP can be regarded as an asymptotically exact 
approximation~\cite{Cespedes14,Takeuchi201} of expectation propagation 
(EP)~\cite{Minka01}. Bayes-optimal OAMP~\cite{Ma17} can solve reconstruction 
problems beyond zero-mean i.i.d.\ sensing matrices while it needs the 
high-complexity linear minimum mean-square error (LMMSE) filter.

State evolution was generalized to the case of right-orthogonally invariant 
sensing matrices~\cite{Rangan192,Takeuchi201} to prove a 
conjecture~\cite{Ma17} for state evolution. The conjecture was 
resolved positively:  
State evolution recursions for Bayes-optimal OAMP was proved to have the same 
fixed points as the replica-symmetric potential derived in 
\cite{Takeda06,Tulino13,Barbier18,Li22}. 
See \cite{Takeuchi22,Takeuchi213,Liu221} 
for the convergence of the state evolution recursions to a fixed point. 

The purpose of this paper is to prove that, via spatial coupling, 
Bayes-optimal OAMP achieves the information-theoretic compression limit for 
right-orthogonally invariant sensing matrices. There is a gap between 
the OAMP performance and information-theoretic limit when the state evolution 
recursions have multiple fixed points. 
Spatial coupling is utilized to fill this gap 
and prove the information-theoretic optimality of OAMP.  

This paper proposes both OAMP and long-memory (LM) OAMP 
(LM-OAMP)~\cite{Takeuchi213,Takeuchi22} for 
spatially coupled and right-orthogonally invariant systems. 
LM-OAMP should be regarded as a proof strategy to guarantee the convergence 
of Bayes-optimal OAMP to a fixed point. For systems without spatial coupling, 
Bayes-optimal LM-OAMP was proved in \cite{Takeuchi213,Takeuchi22} to converge 
and to be asymptotically equivalent to Bayes-optimal OAMP. This paper 
generalizes these results to the spatial coupling case. 

LM-OAMP is an instance of LM-MP, which utilizes messages in all preceding 
iterations to update the 
current message while conventional MP uses messages only in the latest 
iteration. An instance of LM-MP was originally proposed in \cite{Opper16} 
via non-rigorous dynamical functional theory. On the basis of rigorous state 
evolution in this direction, LM-MP---called AMP for rotationally invariant 
matrices---was proposed in \cite{Fan22,Venkataramanan21}.  

Another LM-MP was proposed via state evolution in 
\cite{Takeuchi211}, which is a generalization of \cite{Takeuchi201} to 
the LM case. Convolutional AMP (CAMP)~\cite{Takeuchi202,Takeuchi211,
Takeuchi221} achieves the Bayes-optimal performance for right-orthogonally 
invariant sensing matrices with low-to-moderate condition numbers. 
Memory AMP (MAMP)~\cite{Liu222} improves the convergence property 
of CAMP for high condition numbers. See \cite{Skuratovs221,Skuratovs222} 
for the other instance of LM-MP, inspired by \cite{Takeuchi172}. 
Since the main purpose of these algorithms is a complexity reduction of OAMP, 
this type of LM-MP is out of the scope of this paper. 

\subsection{Contributions}
The main contributions of this paper are fourfold: 
A first contribution is a connection between the information-theoretic 
compression limit~\cite{Wu10} and the replica-symmetric potential for 
right-orthogonally invariant matrices~\cite{Takeda06,Tulino13,Barbier18,Li22} 
in the noiseless limit (Corollary~\ref{corollary2}). This connection 
bridges a gap between the information-theoretic limit and the 
potential-function approach for analyzing spatial 
coupling~\cite{Takeuchi15,Yedla14}. As a by-product, the proof in 
\cite{Donoho13} can be simplified for the case of uniform spatial coupling: 
For spatially coupled zero-mean i.i.d.\ 
Gaussian sensing matrices, the information-theoretic 
optimality of AMP follows immediately from the state evolution 
in \cite{Javanmard13} and the replica-symmetric potential 
in \cite{Takeuchi15} to characterize the AMP performance. 

A second contribution is a generalization of conventional state evolution 
in \cite{Takeuchi201,Takeuchi211} to the case of spatial coupling, 
right-orthogonal invariance, and LM-MP (Theorem~\ref{theorem_SE_tech}). 
As special cases, state evolution recursions for both OAMP and LM-OAMP are 
proved in Theorems~\ref{theorem_SE} and \ref{theorem_SE_LM}, respectively. 
Bayes-optimal LM-OAMP is used to justify the convergence of state evolution 
recursions of Bayes-optimal OAMP for spatially coupled systems 
(Theorem~\ref{theorem_SE_Bayes}), by proving the convergence of the state 
evolution recursions for LM-OAMP and the equivalence between Bayes-optimal 
LM-OAMP and Bayes-optimal OAMP, as proved in \cite{Takeuchi213,Takeuchi22}. 
As a by-product, the state evolution result can be utilized to evaluate 
the asymptotic performance of LM-MP that aims to reduce the computational 
complexity of OAMP for spatially coupled systems. However, research 
for complexity reduction is left as future work. 

From a technical point of view, a third contribution is an asymptotically 
exact approximation of state evolution recursions for Bayes-optimal OAMP 
(Lemma~\ref{lemma_approximation}). The state evolution recursions are 
approximated so that they are included in the class of spatially coupled 
systems considered in \cite{Takeuchi15}. As a result, we can utilize 
the potential-function approach~\cite{Takeuchi15} to analyze the properties 
of the state evolution recursions for Bayes-optimal OAMP. By proving that 
a potential function defined in the approach~\cite{Takeuchi15} is equivalent 
to the replica-symmetric potential for right-orthogonally invariant 
matrices~\cite{Takeda06,Tulino13,Barbier18,Li22}, we arrive at the 
information-theoretic optimality of Bayes-optimal OAMP for spatially coupled 
systems (Theorem~\ref{theorem_optimality}).  

The last contribution is numerical results. Bayes-optimal OAMP for spatially 
coupled systems is shown to be superior to that for conventional systems 
without spatial coupling in the so-called waterfall region. For spatially 
coupled sensing matrices with orthogonal rows, Bayes-optimal OAMP is a 
low-complexity alternative of Bayes-optimal AMP for zero-mean 
i.i.d.\ Gaussian sensing matrices with spatial coupling.  

The second and last contributions were presented in part in a conference 
paper~\cite{Takeuchi23}. 

\subsection{Organization}
The remainder of this paper is organized as follows: After summarizing the 
notation used in this paper, Section~\ref{sec2} reviews compressed sensing 
for conventional measurements without spatial coupling. Conditional mutual 
information is selected as performance measure in signal reconstruction and 
connected to the replica-symmetric potential for right-orthogonally invariant 
sensing matrices, which is defined with the 
R-transform~\cite{Tulino04} of the sensing matrix. After presenting basic 
properties of the R-transform used throughout this paper, we prove a 
relationship between the information-theoretic compression limit and 
the replica-symmetric potential in Theorem~\ref{theorem_potential}. 

Section~\ref{sec3} presents compressed sensing from spatially coupled 
measurements. In Section~\ref{sec4}, we propose OAMP for signal recovery 
from the spatially coupled measurements. 

The two main theorems---Theorems~\ref{theorem_SE} and 
\ref{theorem_optimality}---are presented in Section~\ref{sec5}. 
State evolution recursions for OAMP are proved in Theorem~\ref{theorem_SE}. 
This paper also proves the convergence of the state evolution recursions 
of Bayes-optimal OAMP for the spatial coupling case in the same theorem. 
Theorem~\ref{theorem_optimality} claims 
the information-theoretic optimality of Bayes-optimal OAMP for spatially 
coupled systems in the noiseless case. 

The proof of Theorem~\ref{theorem_SE} is presented in 
Section~\ref{proof_theorem_SE}. To prove the convergence of the 
state evolution recursions for Bayes-optimal OAMP, we follow 
\cite{Takeuchi213,Takeuchi22} to formulate LM-OAMP for the spatially 
coupled system. State evolution recursions for LM-OAMP are proved in 
Theorem~\ref{theorem_SE_LM}---a special case of Theorem~\ref{theorem_SE_tech} 
presented in Appendix~\ref{proof_theorem_SE_LM}, claiming state 
evolution results for the case of spatial coupling, right-orthogonal 
invariance, and LM-MP. Theorem~\ref{theorem_SE} is obtained by proving 
the convergence of the state evolution recursions for Bayes-optimal LM-OAMP in 
Theorem~\ref{theorem_SE_Bayes}, as well as the equivalence between 
Bayes-optimal OAMP and Bayes-optimal LM-OAMP. 

The proof of Theorem~\ref{theorem_optimality} is presented in 
Section~\ref{sec_proof_optimality}. 
After numerical results are presented in Section~\ref{sec8}, this paper 
is concluded in Section~\ref{sec9}. 

\subsection{Notation}
Throughout this paper, the transpose and trace of a matrix $\boldsymbol{M}$ 
are denoted by $\boldsymbol{M}^{\mathrm{T}}$ and $\mathrm{Tr}(\boldsymbol{M})$, 
respectively. The vector $\boldsymbol{e}_{n}$ represents the $n$th column 
of the identity matrix $\boldsymbol{I}$ while $\boldsymbol{1}$ is a vector 
of which the elements are all one. The notation $\boldsymbol{O}$ represents 
an all-zero matrix. For $\{\boldsymbol{M}_{i}\}_{i=1}^{n}$, 
the notation $\mathrm{diag}\{\boldsymbol{M}_{1},\ldots,\boldsymbol{M}_{n}\}$ 
denotes the block diagonal matrix having the $i$th diagonal block 
$\boldsymbol{M}_{i}$. The norm $\|\cdot\|$ represents the Euclidean norm. 
For a symmetric matrix $\boldsymbol{S}$, the minimum eigenvalue of 
$\boldsymbol{S}$ is written as $\lambda_{\mathrm{min}}(\boldsymbol{S})$.  

For a vector $\boldsymbol{v}_{\mathcal{I}}$ with a set of indices $\mathcal{I}$, 
the $n$th element $[\boldsymbol{v}_{\mathcal{I}}]_{n}$ of 
$\boldsymbol{v}_{\mathcal{I}}$ is written as $v_{n,\mathcal{I}}$. Similarly, 
the $t$th column of a matrix $\boldsymbol{M}_{\mathcal{I}}$ is represented as
$\boldsymbol{m}_{t,\mathcal{I}}$. 

The notation $\boldsymbol{x}\sim\mathcal{N}(\boldsymbol{\mu},
\boldsymbol{\Sigma})$ 
means that a random vector $\boldsymbol{x}$ follows the Gaussian distribution 
with mean $\boldsymbol{\mu}$ and covariance $\boldsymbol{\Sigma}$. 
The almost sure convergence and equivalence are denoted by $\ato$ and 
$\aeq$, respectively. The notation $\ag$ is defined in a similar manner. 

For a scalar function $f:\mathbb{R}\to\mathbb{R}$ and a vector 
$\boldsymbol{x}\in\mathbb{R}^{n}$, the notation $f(\boldsymbol{x})$ means 
the element-wise application of $f$ to $\boldsymbol{x}$, i.e.\ 
$[f(\boldsymbol{x})]_{i}=f(x_{i})$. The arithmetic mean of 
$\boldsymbol{x}\in\mathbb{R}^{n}$ is written as 
$\langle\boldsymbol{x}\rangle=n^{-1}\sum_{i=1}^{n}x_{i}$. For a multi-variate 
function $f:\mathbb{R}^{t}\to\mathbb{R}$, the notation $\partial_{i}$ represents 
the partial derivative of $f$ with respect to the $i$th variable. 

The space of all possible $N\times N$ orthogonal matrices is denoted by 
$\mathcal{O}_{N}$. The notation 
$\boldsymbol{M}^{\dagger}=(\boldsymbol{M}^{\mathrm{T}}\boldsymbol{M})^{-1}
\boldsymbol{M}^{\mathrm{T}}$ represents the pseudo-inverse of a full-rank matrix 
$\boldsymbol{M}\in\mathbb{R}^{m\times n}$ satisfying $m\geq n$. 
The singular-value decomposition (SVD) of $\boldsymbol{M}$ is written as
$\boldsymbol{M}=\boldsymbol{\Phi}_{\boldsymbol{M}}\boldsymbol{\Sigma}_{\boldsymbol{M}}
\boldsymbol{\Psi}_{\boldsymbol{M}}^{\mathrm{T}}$ with
$\boldsymbol{\Phi}_{\boldsymbol{M}}\in\mathcal{O}_{m}$ and 
$\boldsymbol{\Psi}_{\boldsymbol{M}}\in\mathcal{O}_{n}$.
The matrix $\boldsymbol{P}_{M}^{\perp}=\boldsymbol{I}
-\boldsymbol{M}(\boldsymbol{M}^{\mathrm{T}}\boldsymbol{M})^{-1}
\boldsymbol{M}^{\mathrm{T}}$ is the projection onto the orthogonal complement 
of the space spanned by the columns of $\boldsymbol{M}$. 

\section{Conventional Compressed Sensing} \label{sec2}
\subsection{System Model}
This section reviews compressed sensing from linear measurements 
without spatial coupling. Let $M\in\mathbb{N}$ and $N\in\mathbb{N}$ denote 
the dimensions of measurement and signal vectors, respectively. The 
measurement vector $\boldsymbol{y}\in\mathbb{R}^{M}$ is given by 
\begin{equation} \label{uncoupled_model}
\boldsymbol{y} = \boldsymbol{A}\boldsymbol{x} + \boldsymbol{n}.
\end{equation}
In (\ref{uncoupled_model}), $\boldsymbol{A}\in\mathbb{R}^{M\times N}$ represents 
a known sensing matrix. The vectors $\boldsymbol{x}\in\mathbb{R}^{N}$ and 
$\boldsymbol{n}\in\mathbb{R}^{M}$ denote sparse signal and noise vectors, 
respectively. The random variables in the triple 
$\{\boldsymbol{A}, \boldsymbol{x}, \boldsymbol{n}\}$ are independent. 
The purpose of compressed sensing is to reconstruct the unknown 
sparse signal vector $\boldsymbol{x}$ from the knowledge on the sensing 
matrix $\boldsymbol{A}$ and the compressed measurement vector 
$\boldsymbol{y}$ with $M\leq N$. 

The sensing matrix $\boldsymbol{A}$ is assumed to be sampled from the ensemble 
of right-orthogonally invariant matrices uniformly and randomly. 
\begin{definition} \label{def1}
A matrix $\boldsymbol{A}$ is said to be right-orthogonally invariant if 
the SVD $\boldsymbol{A}
=\boldsymbol{U}\boldsymbol{\Sigma}\boldsymbol{V}^{\mathrm{T}}$ satisfies the 
following conditions:
\begin{itemize}
\item The orthogonal matrix $\boldsymbol{V}\in\mathcal{O}_{N}$ is independent 
of $\boldsymbol{U}\boldsymbol{\Sigma}$ and Haar-distributed on the space 
of all possible $N\times N$ orthogonal matrices. 
\item The empirical eigenvalue distribution of $\boldsymbol{A}^{\mathrm{T}}
\boldsymbol{A}$ converges almost surely to a compactly supported deterministic 
distribution with unit mean in the large system limit, where both $M$ and $N$ 
tend to infinity with the compression rate $\delta=M/N\in(0, 1]$ 
kept constant.
\end{itemize}
\end{definition}

The unit-mean assumption implies the almost sure convergence 
$N^{-1}\mathrm{Tr}(\boldsymbol{A}^{\mathrm{T}}\boldsymbol{A})\ato1$ in the 
large system limit. 

The original definition of the right-orthogonal invariance only includes the 
former assumption. Nonetheless, this paper includes the latter assumption 
in the definitions of right-orthogonal invariant matrices. The 
latter assumption is needed to perform rigorous state evolution analysis. 

We refer to the ensemble of right-orthogonally invariant matrices as 
$(M, N)$-ensemble. Note that $(M, N)$-ensemble is defined for fixed 
statistical properties of $\boldsymbol{U}\in\mathcal{O}_{M}$ and diagonal 
$\boldsymbol{\Sigma}\in\mathbb{R}^{M\times N}$. In other words, 
$(M, N)$-ensemble depends on the joint distribution of $\boldsymbol{U}$ and 
$\boldsymbol{\Sigma}$ while it is not written explicitly. 

It is practically important to relax the right-orthogonal invariance to 
weaker assumptions, including 
discrete cosine transform (DCT) or Hadamard matrices with random 
permutation~\cite{Candes062}. See \cite{Anderson14,Male20,Dudeja22} 
for theoretical progress in this direction. 

\subsection{Conditional Mutual Information}
We measure the optimal reconstruction performance with the conditional 
mutual information $I(\boldsymbol{x}; \boldsymbol{y} | \boldsymbol{A})$ 
in nats between the signal vector $\boldsymbol{x}$ and the measurement vector 
$\boldsymbol{y}$ given $\boldsymbol{A}$. It might be standard to use the 
minimum mean-square error (MMSE) $\mathbb{E}[\|\boldsymbol{x}
-\mathbb{E}[\boldsymbol{x}|\boldsymbol{y}, \boldsymbol{A}]\|^{2}]$ in 
compressed sensing. Nonetheless, the conditional mutual information is useful 
to define a potential that characterizes the MMSE performance.  

This paper focuses on a rigorous result~\cite{Barbier18} on the conditional 
mutual information since another rigorous result requires a strong 
assumption~\cite[Assumption 1.4]{Li22} on the condition number of the 
sensing matrix. To present the former rigorous result~\cite{Barbier18} , 
we need the following assumptions: 
\begin{assumption} \label{assumption0}
\begin{itemize}
\item The signal vector $\boldsymbol{x}$ has i.i.d.\ bounded elements with 
zero mean and unit variance. 
\item The noise vector $\boldsymbol{n}\sim\mathcal{N}(\boldsymbol{0},
\sigma^{2}\boldsymbol{I}_{M})$ has independent Gaussian elements with 
zero mean and variance $\sigma^{2}>0$. 
\item The sensing matrix is represented as the product $\boldsymbol{A}
=\boldsymbol{D}\boldsymbol{W}$ of 
two independent matrices $\boldsymbol{D}\in\mathbb{R}^{M\times M}$ and 
$\boldsymbol{W}\in\mathbb{R}^{M\times N}$. The matrix $\boldsymbol{W}$ has 
independent zero-mean Gaussian elements with variance $1/M$ while 
$\boldsymbol{D}$ is the product of a finite number of independent matrices 
having i.i.d.\ bounded or Gaussian elements. Furthermore, $\boldsymbol{D}$ 
satisfies the almost sure convergence 
$M^{-1}\mathrm{Tr}(\boldsymbol{D}^{\mathrm{T}}\boldsymbol{D})\ato1$. 
\end{itemize}
\end{assumption}

The boundedness of the signal vector excludes the BG prior. The ensemble of 
sensing matrices $\boldsymbol{A}=\boldsymbol{D}\boldsymbol{W}$ is a subclass 
of $(M, N)$-ensemble since the Gaussian matrix $\boldsymbol{W}$ is 
right-orthogonally invariant. The assumption 
$M^{-1}\mathrm{Tr}(\boldsymbol{D}^{\mathrm{T}}\boldsymbol{D})\ato1$ 
should be regarded as a normalization to include 
$\boldsymbol{A}=\boldsymbol{D}\boldsymbol{W}$ in $(M, N)$-ensemble. 
This structure $\boldsymbol{A}=\boldsymbol{D}\boldsymbol{W}$ was not required 
in the replica conjecture~\cite{Takeda06,Tulino13}, as well as 
the boundedness of the signal vector. Thus, it is still open to relax 
these assumptions. 

\begin{remark}
Without loss of generality, we can transform $\boldsymbol{D}$ into a 
diagonal matrix. Consider the SVD $\boldsymbol{D}
=\boldsymbol{\Phi}_{\boldsymbol{D}}\boldsymbol{\Sigma}_{\boldsymbol{D}}
\boldsymbol{\Psi}_{\boldsymbol{D}}^{\mathrm{T}}$. Left-multiplying 
(\ref{uncoupled_model}) by $\boldsymbol{\Phi}_{\boldsymbol{D}}^{\mathrm{T}}$ yields 
\begin{equation}
\boldsymbol{\Phi}_{\boldsymbol{D}}^{\mathrm{T}}\boldsymbol{y} 
= \boldsymbol{\Sigma}_{\boldsymbol{D}}
\boldsymbol{\Psi}_{\boldsymbol{D}}^{\mathrm{T}}\boldsymbol{W}\boldsymbol{x} 
+ \boldsymbol{\Phi}_{\boldsymbol{D}}^{\mathrm{T}}\boldsymbol{n}
\sim \boldsymbol{\Sigma}_{\boldsymbol{D}}\boldsymbol{W}\boldsymbol{x} 
+ \boldsymbol{n}, 
\end{equation}
where the last statistical equivalence follows from the left-orthogonal 
invariance of $\boldsymbol{W}$ and the orthogonal invariance of 
$\boldsymbol{n}\sim\mathcal{N}(\boldsymbol{0},\sigma^{2}\boldsymbol{I}_{M})$. 
Thus, $\boldsymbol{D}$ can be transformed into the diagonal matrix 
$\boldsymbol{\Sigma}_{\boldsymbol{D}}$. Since $\boldsymbol{D}$ is the product 
of a finite number of independent matrices having i.i.d.\ bounded or 
Gaussian elements, $\boldsymbol{\Sigma}_{\boldsymbol{D}}$ is in a subclass of 
general diagonal matrices. 
\end{remark}

The conditional mutual information can be described with a replica-symmetric 
potential, which was derived in \cite{Takeda06,Tulino13} via the replica 
method under the replica symmetry assumption. 
Let $f_{\mathrm{RS}}:[0, 1]\times[0,\infty)\to\mathbb{R}$ denote 
the replica-symmetric potential, given by 
\begin{equation} \label{RS_potential}
f_{\mathrm{RS}}(E,s) 
= I(s) + \frac{1}{2}\int_{0}^{E/\sigma^{2}}
R_{\boldsymbol{A}^{\mathrm{T}}\boldsymbol{A}}(-z)dz - \frac{sE}{2}. 
\end{equation}
In (\ref{RS_potential}), $R_{\boldsymbol{A}^{\mathrm{T}}\boldsymbol{A}}$ denotes 
the R-transform of the empirical eigenvalue distribution 
of $\boldsymbol{A}^{\mathrm{T}}\boldsymbol{A}$ in the large system 
limit~\cite{Tulino04}, defined shortly. 
The first term $I(s)=I(x_{1}; \sqrt{s}x_{1}+z_{1})$ 
denotes the mutual information in nats between the first element $x_{1}$ 
of the signal vector and a virtual AWGN measurement $\sqrt{s}x_{1}+z_{1}$ 
with signal-to-noise ratio (SNR) $s\geq0$ for $z_{1}\sim\mathcal{N}(0,1)$ 
independent of~$x_{1}$. The variable $s$ is regarded as the asymptotic 
signal-to-interference-plus-noise ratio (SINR). In this sense, $z_{1}$ 
corresponds to the effective interference plus noise. 

The variable $E$ will be evaluated at 
$E=\mathrm{MMSE}(s)$, given by 
\begin{equation} \label{MMSE_func}
\mathrm{MMSE}(s)
=\mathbb{E}\left[
 (x_{1}-\mathbb{E}[x_{1}|\sqrt{s}x_{1}+z_{1}])^{2}
\right]
\end{equation}  
We use the optimality of the posterior mean estimator to obtain the 
following trivial upper bound:
\begin{equation} \label{MMSE_upper_bound} 
\mathrm{MMSE}(s)\leq\mathbb{E}[x_{1}^{2}],   
\end{equation}
considering the estimator 
$\hat{x}_{1}=0$ for $x_{1}$ given any observation $\sqrt{s}x_{1}+z_{1}$. 
When $\mathbb{E}[x_{1}^{2}]=1$ holds, thus, the finite interval $[0, 1]$ is 
considered as the domain of $f_{\mathrm{RS}}$ for $E$ while $[0,\infty)$ is 
considered for $s$.  

\begin{theorem}[\cite{Barbier18}]
Suppose that Assumption~\ref{assumption0} holds. 
Then, the normalized conditional mutual information is given by 
\begin{equation} \label{asymptotic_mutual_information}
\frac{1}{N}I(\boldsymbol{x}; \boldsymbol{y}|\boldsymbol{A}) 
\to \inf_{s\geq0}\sup_{E\in[0, 1]}f_{\mathrm{RS}}(E,s) 
\end{equation} 
in the large system limit, where the replica-symmetric potential 
$f_{\mathrm{RS}}$ is defined as (\ref{RS_potential}). 
\end{theorem}

The optimizer $(E_{\mathrm{opt}}, s_{\mathrm{opt}})$ for the inf-sup 
problem~(\ref{asymptotic_mutual_information}) characterizes the asymptotic 
MMSE and SINR for the Bayes-optimal reconstruction of the signal vector 
based on the measurement model~(\ref{uncoupled_model}). More precisely, 
the MMSE $N^{-1}\mathbb{E}[\|\boldsymbol{x}-\mathbb{E}[\boldsymbol{x}|
\boldsymbol{A},\boldsymbol{y}]\|^{2}]$ converges to $E_{\mathrm{opt}}$ in the 
large system limit. See \cite{Barbier18} for the details. 

\subsection{Optimizer}
The goal of this section is to investigate properties of the optimizer 
in the inf-sup problem~(\ref{asymptotic_mutual_information}) as a technical 
step to characterize the information-theoretic compression limit.  
We first investigate properties of the optimizer in a general inf-sup problem 
including the inf-sup problem~(\ref{asymptotic_mutual_information}). 

\begin{lemma} \label{lemma_infsup_general}
Let $f:[E_{\mathrm{min}}, E_{\mathrm{max}}]\to[-\infty,\infty)$ denote a function 
satisfying $f(E)>-\infty$ for some $E\in[E_{\mathrm{min}}, E_{\mathrm{max}}]$. 
Suppose that $g:[0,\infty)\to[-\infty,\infty)$ is an upper semicontinuous and 
concave function such that $g(s)>-\infty$ holds for some $s\geq0$, 
that the infimum of $sE-g(s)$ over $s\geq0$ is attained at $s=0$, 
an interior point $s>0$, and $s=\infty$ for all $E=E_{\mathrm{max}}$, 
$E\in(E_{\mathrm{min}}, E_{\mathrm{max}})$, and $E=E_{\mathrm{min}}$, respectively.
The function $\psi(E,s)=f(E) + g(s) - sE$ satisfies the following properties: 
\begin{itemize}
\item Suppose that $f$ is upper semicontinuous and concave, and that 
there are some $s_{\mathrm{min}}\geq0$ and $s_{\mathrm{max}}>s_{\mathrm{min}}$ 
such that the infimum of $sE-f(E)$ over $E\in[E_{\mathrm{min}}, E_{\mathrm{max}}]$ 
is attained at $E=E_{\mathrm{min}}$, an interior point 
$E\in(E_{\mathrm{min}}, E_{\mathrm{max}})$, and $E=E_{\mathrm{max}}$ 
for all $s= s_{\mathrm{max}}$, $s\in(s_{\mathrm{min}},s_{\mathrm{max}})$, and 
$s= s_{\mathrm{min}}$, respectively. Then, we have 
\begin{equation} \label{exchange} 
\inf_{s\geq0}\sup_{E\in [E_{\mathrm{min}},E_{\mathrm{max}}]}\psi(E,s) 
= \inf_{E\in[E_{\mathrm{min}},E_{\mathrm{max}}]}\sup_{s\geq0}\psi(E,s). 
\end{equation}

\item Suppose that $g$ is differentiable and strictly concave, 
and that $f$ is differentiable and non-decreasing. Then, we have   
\begin{equation} 
\inf_{E\in[E_{\mathrm{min}},E_{\mathrm{max}}]}\sup_{s\geq0}\psi(E,s) 
= \inf_{(E, s)}\psi(E,s),
\label{infsup_general}
\end{equation}
where the infimum on the right-hand side (RHS) is over 
\begin{equation} \label{set}
\left\{
 (E, s)\in[E_{\mathrm{min}}, E_{\mathrm{max}}]\times[0, \infty]: 
 E = g'(s), s=f'(E) 
\right\}. 
\end{equation}
\end{itemize}
\end{lemma}
\begin{IEEEproof}
The former and latter parts in Lemma~\ref{lemma_infsup_general} correspond to 
\cite[Corollary~7]{Barbier191} and \cite[Lemma~23]{Barbier191}, respectively. 
We prove Lemma~\ref{lemma_infsup_general} under weaker conditions than in 
\cite{Barbier191} by removing unnecessary conditions in \cite{Barbier191}. 
See Appendix~\ref{proof_lemma_infsup_general} for the details. 
\end{IEEEproof}

To use Lemma~\ref{lemma_infsup_general} for evaluation of the optimizer 
$(E_{\mathrm{opt}}, s_{\mathrm{opt}})$ in the inf-sup 
problem~(\ref{asymptotic_mutual_information}), we need to confirm that 
the first two terms in the replica-symmetric potential~(\ref{RS_potential}) 
satisfy the assumptions in Lemma~\ref{lemma_infsup_general}. For that purpose, 
we start with the definition of the $\eta$-transform for the empirical 
eigenvalue distribution of $\boldsymbol{A}^{\mathrm{T}}\boldsymbol{A}$ 
in the large system limit, which is used to define the R-transform. 

Let $\eta_{\boldsymbol{A}^{\mathrm{T}}\boldsymbol{A}}$ denote 
the $\eta$-transform of the empirical eigenvalue distribution of 
$\boldsymbol{A}^{\mathrm{T}}\boldsymbol{A}$ in the large system 
limit~\cite[Definition 2.11]{Tulino04}, given by\footnote{
The $\eta$-transform is also defined as 
$\tilde{\eta}_{\boldsymbol{A}^{\mathrm{T}}\boldsymbol{A}}(z)
=1-1/\eta_{\boldsymbol{A}^{\mathrm{T}}\boldsymbol{A}}(-z)$ in random matrix 
theory~\cite[Eq.~(10.8)]{Mingo17}. 
} 
\begin{equation} \label{eta_transform} 
\eta_{\boldsymbol{A}^{\mathrm{T}}\boldsymbol{A}}(z) 
= \lim_{M=\delta N\to\infty}\frac{1}{N}\mathrm{Tr}\left\{
 (\boldsymbol{I}_{N} + z\boldsymbol{A}^{\mathrm{T}}\boldsymbol{A})^{-1}
\right\}
\end{equation}
for all $z\geq0$. Let $r$ denote the rank of 
$\boldsymbol{A}^{\mathrm{T}}\boldsymbol{A}$. 
From the definition of the $\eta$-transform in (\ref{eta_transform}) 
we have 
\begin{equation} \label{eta_transform_tmp}
\eta_{\boldsymbol{A}^{\mathrm{T}}\boldsymbol{A}}(z)
= \lim_{M=\delta N\to\infty}\left(
 \frac{1}{N}\sum_{n=1}^{r}\frac{1}{1+\lambda_{n}z} + 1-\frac{r}{N}
\right),
\end{equation}
where $\{\lambda_{n}>0\}$ are strictly positive eigenvalues of 
$\boldsymbol{A}^{\mathrm{T}}\boldsymbol{A}$. 

We define the $k$th moment of the asymptotic eigenvalue distribution of 
$\boldsymbol{A}^{\mathrm{T}}\boldsymbol{A}$ as 
\begin{equation}
\mu_{k}=\lim_{M=\delta N\to\infty}\frac{1}{N}\mathrm{Tr}\{
(\boldsymbol{A}^{\mathrm{T}}\boldsymbol{A})^{k}\}. 
\end{equation}

The $\eta$-transform has the following basic properties, which are 
trivial from the definition of the $\eta$-transform. 

\begin{lemma} \label{lemma_eta_transform}
The $\eta$-transform~(\ref{eta_transform}) satisfies the following properties: 
\begin{itemize}
\item $z\eta_{\boldsymbol{A}^{\mathrm{T}}\boldsymbol{A}}(z)$ is strictly increasing 
for all $z\geq0$.
\item If all moments $\{\mu_{k}\}$ exist, 
the $\eta$-transform~(\ref{eta_transform}) is infinitely 
continuously-differentiable for all $z\geq0$. 
\end{itemize}
\end{lemma}
\begin{IEEEproof}
See Appendix~\ref{proof_lemma_eta_transform}. 
\end{IEEEproof}

In the subsequent sections we consider sensing matrices with bounded $\mu_{k}$ 
for all $k$ while the boundedness of $\mu_{k}$ is explicitly assumed in 
Section~\ref{sec2}. Thus, we need not investigate the differentiability of the 
$\eta$-transform in the subsequent sections. 
When all moments $\{\mu_{k}\}$ exist, 
the $\eta$-transform has the following series-expansion: 
\begin{equation} \label{eta_series} 
\eta_{\boldsymbol{A}^{\mathrm{T}}\boldsymbol{A}}(z)  
= \sum_{k=0}^{\infty}\mu_{k}(-z)^{k} 
\end{equation}
if (\ref{eta_series}) is bounded in a neighborhood of $z=0$. 

We next consider the R-transform $R_{\boldsymbol{A}^{\mathrm{T}}\boldsymbol{A}}$, 
which is implicitly defined as 
\begin{equation} \label{R_transform}
\eta_{\boldsymbol{A}^{\mathrm{T}}\boldsymbol{A}}(z)
= \frac{1}{1 + zR_{\boldsymbol{A}^{\mathrm{T}}\boldsymbol{A}}
(-z\eta_{\boldsymbol{A}^{\mathrm{T}}\boldsymbol{A}}(z))}, 
\end{equation}
with $R_{\boldsymbol{A}^{\mathrm{T}}\boldsymbol{A}}(0)
=\lim_{z\uparrow0}R_{\boldsymbol{A}^{\mathrm{T}}\boldsymbol{A}}(z)$. 
Solving the definition of the R-transform in (\ref{R_transform}) 
with respect to $R_{\boldsymbol{A}^{\mathrm{T}}\boldsymbol{A}}$, we have 
\begin{equation} \label{R_transform_tmp}
R_{\boldsymbol{A}^{\mathrm{T}}\boldsymbol{A}}
(-z\eta_{\boldsymbol{A}^{\mathrm{T}}\boldsymbol{A}}(z))
= \frac{1-\eta_{\boldsymbol{A}^{\mathrm{T}}\boldsymbol{A}}(z)}
{z\eta_{\boldsymbol{A}^{\mathrm{T}}\boldsymbol{A}}(z)}.  
\end{equation} 
Lemma~\ref{lemma_eta_transform} implies that the domain of the R-transform 
is equal to the interval $(z_{\mathrm{min}},0]$ with 
$z_{\mathrm{min}}=-\lim_{z\to\infty}z
\eta_{\boldsymbol{A}^{\mathrm{T}}\boldsymbol{A}}(z)$. 

The R-transform satisfies the following basic properties, which are 
trivial when the R-transform is defined with an equivalent definition: 
a formal power series with the coefficients equal to free 
cumulants~\cite[Eq.~(2.84)]{Tulino04}. 

\begin{lemma} \label{lemma_R_transform_0}
If all moments $\{\mu_{k}\}$ exist, the R-transform~(\ref{R_transform}) 
is infinitely continuously-differentiable for all $z\in(z_{\mathrm{min}},0]$. 
In particular, we have 
\begin{equation} \label{R_transform_0}
R_{\boldsymbol{A}^{\mathrm{T}}\boldsymbol{A}}(0) = \mu_{1}, 
\end{equation}
\begin{equation} \label{R_transform_deriv_0}
R_{\boldsymbol{A}^{\mathrm{T}}\boldsymbol{A}}'(0)=\mu_{2} - \mu_{1}^{2}. 
\end{equation}
\end{lemma}
\begin{IEEEproof}
See Appendix~\ref{proof_lemma_R_transform_0}. 
\end{IEEEproof}

Lemma~\ref{lemma_R_transform_0} allows us to use differentiability and 
continuity of the R-transform freely since bounded $\{\mu_{k}\}$ are 
considered in this paper. Nonetheless, we assume explicitly regularity 
conditions for the R-transform in Section~\ref{sec2} to clarify 
what conditions we use in their proofs. 

In this paper, we consider the R-transform satisfying the following 
conditions:
\begin{definition} \label{definition_R_proper}
The R-transform is said to be proper for all $z\leq0$ 
if $z_{\mathrm{min}}=-\lim_{z\to\infty}z\eta_{\boldsymbol{A}^{\mathrm{T}}\boldsymbol{A}}(z)
=-\infty$ holds and if $\mu_{2}>\mu_{1}^{2}$ holds.  
\end{definition}

The former condition $z_{\mathrm{min}}=-\infty$ in 
Definition~\ref{definition_R_proper} implies that the domain of 
the R-transform is the interval $(-\infty, 0]$. As a result, the 
replica-symmetric potential~(\ref{RS_potential}) is well defined 
in the limit $\sigma^{2}\downarrow0$. The latter condition 
$\mu_{2}>\mu_{1}^{2}$ excludes the constant R-transform 
$R_{\boldsymbol{A}^{\mathrm{T}}\boldsymbol{A}}(z)=\mu_{1}$ for all $z$, because of 
Lemma~\ref{lemma_R_transform_0}. The zero variance $\mu_{2}-\mu_{1}^{2}=0$ 
holds when the empirical eigenvalue distribution of 
$\boldsymbol{A}^{\mathrm{T}}\boldsymbol{A}$ converges almost surely to 
the Dirac distribution that takes $\mu_{1}$ with probability~$1$. 
This convergence occurs when the sensing matrix is square and has identical 
singular values with the exception of $o(N)$ singular values. Since the 
R-transform cannot distinguish this sensing matrix from the identity matrix, 
the constant R-transform is excluded as a trivial case. 

The following two lemmas present sufficient conditions for technical 
assumptions required in proving the information-theoretic optimality 
in compressed sensing. 

\begin{lemma} \label{lemma_R_transform_positive}
Suppose that all moments $\{\mu_{k}\}$ exist. Then, 
\begin{itemize}
\item The R-transform~(\ref{R_transform}) is non-negative  
for all $z\in(z_{\mathrm{min}}, 0]$. In particular, it is positive 
for all $z\in(z_{\mathrm{min}}, 0]$ if $\mu_{1}>0$ holds. 
\item The R-transform~(\ref{R_transform}) is non-decreasing for all 
$z\in(z_{\mathrm{min}}, 0]$. In particular, it is strictly increasing 
for all $z\in(z_{\mathrm{min}}, 0]$ if $\mu_{2}>\mu_{1}^{2}$ holds. 
\end{itemize}
\end{lemma}
\begin{IEEEproof}
See Appendix~\ref{proof_lemma_R_transform_positive}. 
\end{IEEEproof}

\begin{lemma} \label{lemma_R_transform_delta}
Let $r$ denote the rank of $\boldsymbol{A}$ and
suppose that the ratio $r/N$ tends to 
$\delta$ in the large system limit. If $\delta<1$ or the following condition 
for $\delta=1$ is satisfied:  
\begin{equation}  \label{diverging_condition}
\lim_{M=\delta N\to\infty}
\frac{1}{N}\sum_{n=1}^{N}\frac{1}{\lambda_{n}}
\to \infty,   
\end{equation}
then $\lim_{z\to\infty}z\eta_{\boldsymbol{A}^{\mathrm{T}}\boldsymbol{A}}(z)=\infty$ and 
$\lim_{z\to\infty}zR_{\boldsymbol{A}^{\mathrm{T}}\boldsymbol{A}}(-z)=\delta$ hold. 
\end{lemma}
\begin{IEEEproof}
We first prove $\lim_{z\to\infty}\eta_{\boldsymbol{A}^{\mathrm{T}}\boldsymbol{A}}(z)
=1-\delta$ and $\lim_{z\to\infty}z\eta_{\boldsymbol{A}^{\mathrm{T}}\boldsymbol{A}}(z)
=\infty$. From the representation of 
the $\eta$-transform in (\ref{eta_transform_tmp}), we have 
$\lim_{z\to\infty}\eta_{\boldsymbol{A}^{\mathrm{T}}\boldsymbol{A}}(z)
=1-\delta$. For $\delta<1$, 
$\lim_{z\to\infty}z\eta_{\boldsymbol{A}^{\mathrm{T}}\boldsymbol{A}}(z)=\infty$ 
is trivial from (\ref{eta_transform_tmp}). On the other hand, for $\delta=1$ 
we use the assumption~(\ref{diverging_condition}) to prove 
\begin{IEEEeqnarray}{rl}
\lim_{z\to\infty}z\eta_{\boldsymbol{A}^{\mathrm{T}}\boldsymbol{A}}(z)
=& \lim_{z\to\infty}\lim_{M=\delta N\to\infty}
\frac{1}{N}\sum_{n=1}^{N}\frac{z}{1+\lambda_{n}z} 
\nonumber \\
>& \lim_{M=\delta N\to\infty}
\frac{1}{N}\sum_{n=1}^{N}\frac{1}{\lambda_{n}} \to \infty. 
\end{IEEEeqnarray}

We next prove $\lim_{z\to\infty}zR_{\boldsymbol{A}^{\mathrm{T}}\boldsymbol{A}}(-z)=\delta$. 
Using the limit $z\eta_{\boldsymbol{A}^{\mathrm{T}}\boldsymbol{A}}(z)\to\infty$ 
and the definition of the R-transform in (\ref{R_transform_tmp}),   
we obtain 
\begin{IEEEeqnarray}{rl} \label{R_transform_inf}
\lim_{z\to\infty}zR_{\boldsymbol{A}^{\mathrm{T}}\boldsymbol{A}}(-z)
=& \lim_{z\to\infty}z\eta_{\boldsymbol{A}^{\mathrm{T}}\boldsymbol{A}}(z)
R_{\boldsymbol{A}^{\mathrm{T}}\boldsymbol{A}}
(-z\eta_{\boldsymbol{A}^{\mathrm{T}}\boldsymbol{A}}(z)) 
\nonumber \\
=& 1 - \lim_{z\to\infty}\eta_{\boldsymbol{A}^{\mathrm{T}}\boldsymbol{A}}(z)
= \delta,
\end{IEEEeqnarray}
where the last follows from 
$\eta_{\boldsymbol{A}^{\mathrm{T}}\boldsymbol{A}}(z)\to1-\delta$. 
\end{IEEEproof}

We are ready to investigate properties of the second term in the 
replica-symmetric potential~(\ref{RS_potential}).  
To investigate properties of the first term, 
we use a general formula between mutual information and 
MMSE~\cite{Guo052} and the smoothness of 
MMSE~\cite[Propositions~7 and 9]{Guo11}. 
 
\begin{proposition}[\cite{Guo052}]
\begin{equation} \label{IM_relationship} 
\frac{d}{ds}I(s)
= \frac{1}{2}\mathrm{MMSE}(s). 
\end{equation} 
\end{proposition}
\begin{proposition}[\cite{Guo11}] \label{proposition_MSE}
The function $\mathrm{MMSE}(s)$ is infinitely continuously-differentiable 
for all $s\geq0$. In particular, we have 
\begin{equation}
\frac{d}{ds}\mathrm{MMSE}(s) 
= - \mathbb{E}\left[
 \left\{
  \mathbb{E}[(x_{1} - \mathbb{E}[x_{1}|u_{1}])^{2} | u_{1}]
 \right\}^{2}
\right],
\end{equation}
with $u_{1}=\sqrt{s}x_{1}+z_{1}$ and $z_{1}\sim\mathcal{N}(0,1)$ independent 
of $x_{1}$. 
\end{proposition}

The following lemma implies that it is sufficient to consider 
extremizers in solving the inf-sup 
problem~(\ref{asymptotic_mutual_information}):  
 
\begin{lemma} \label{lemma_infsup}
Suppose $\mathbb{E}[x_{1}^{2}]=1$ and assume that 
the R-transform $R_{\boldsymbol{A}^{\mathrm{T}}\boldsymbol{A}}(z)$ is 
proper, continuous, non-decreasing, and non-negative for all $z\leq0$. 
Then, the following identity holds: 
\begin{equation} \label{infsup}
\inf_{s\geq0}\sup_{E\in[0, 1]}f_{\mathrm{RS}}(E,s)
= \inf_{(E, s)\in\mathcal{S}}f_{\mathrm{RS}}(E,s), 
\end{equation}
where $\mathcal{S}\subset[0,1]\times[0,\infty]$ denotes the set of 
extremizers 
\begin{IEEEeqnarray}{r}
\mathcal{S} 
= \{(E,s)\in[0,1]\times[0,\infty] : 
E = \mathrm{MMSE}(s), \nonumber \\
s = R_{\boldsymbol{A}^{\mathrm{T}}\boldsymbol{A}}(-E/\sigma^{2})/\sigma^{2}\}. 
\label{extremizer}
\end{IEEEeqnarray}
\end{lemma}

\begin{IEEEproof}[Proof of Lemma~\ref{lemma_infsup}]
We utilize Lemma~\ref{lemma_infsup_general} to prove 
the identity~(\ref{infsup}). Let $g(s)=2I(s)$ and 
\begin{equation}
f(E) = \int_{0}^{E/\sigma^{2}}R_{\boldsymbol{A}^{\mathrm{T}}\boldsymbol{A}}(-z)dz. 
\end{equation}
Then, we have $\psi(E,s)=2f_{\mathrm{RS}}(E,s)$. 

We confirm that $g(s)=2I(s)$ satisfies all conditions in 
Lemma~\ref{lemma_infsup_general}. 
From the general formula~(\ref{IM_relationship}) between mutual information 
and MMSE, we have $g'(s)=\mathrm{MMSE}(s)$, which implies  
$g'(0)=1$ and $\lim_{s\to\infty}g'(s)=0$. Since $g'$ is continuous and 
strictly decreasing for $s\geq0$ from Proposition~\ref{proposition_MSE}, 
$g(s)$ is continuously differentiable and strictly concave for all $s\geq0$. 
Furthermore, the infimum of $sE-g(s)$ over $s\geq0$ is attained at $s=0$, 
the unique solution $s=s^{*}>0$ to $E=g'(s^{*})$, and $s=\infty$ for 
$E=E_{\mathrm{max}}$, $E\in(E_{\mathrm{min}}, E_{\mathrm{max}})$, and 
$E=E_{\mathrm{min}}$, respectively, with $E_{\mathrm{min}}=0$ and 
$E_{\mathrm{max}}=1$. Thus, all conditions for $g$ 
in Lemma~\ref{lemma_infsup_general} are satisfied.  

We next investigate properties of $f(E)$, which has the derivative 
$f'(E)=R_{\boldsymbol{A}^{\mathrm{T}}\boldsymbol{A}}(-E/\sigma^{2})/\sigma^{2}$. 
Since the R-transform $R_{\boldsymbol{A}^{\mathrm{T}}\boldsymbol{A}}(z)$ has been 
assumed to be proper, continuous, non-decreasing, and non-negative 
for all $z\leq0$, $f(E)$ is a continuously differentiable, 
concave, and non-decreasing function of $E\in[E_{\mathrm{min}},E_{\mathrm{max}}]$. 
Furthermore, we find that the infimum of $sE-f(E)$ over 
$E\in[E_{\mathrm{min}},E_{\mathrm{max}}]$ is attained at $E=E_{\mathrm{min}}$, 
an interior solution $E=E^{*}\in(E_{\mathrm{min}}, E_{\mathrm{max}})$ 
to $s=R_{\boldsymbol{A}^{\mathrm{T}}\boldsymbol{A}}(-E^{*}/\sigma^{2})/\sigma^{2}$, 
and $E=E_{\mathrm{max}}$ for $s=s_{\mathrm{max}}$, 
$s\in(s_{\mathrm{min}},s_{\mathrm{max}})$, and $s=s_{\mathrm{min}}$, respectively, 
with $s_{\mathrm{min}}=R_{\boldsymbol{A}^{\mathrm{T}}\boldsymbol{A}}
(-E_{\mathrm{max}}/\sigma^{2})/\sigma^{2}\geq0$ and 
$s_{\mathrm{max}}=R_{\boldsymbol{A}^{\mathrm{T}}\boldsymbol{A}}
(-E_{\mathrm{min}}/\sigma^{2})/\sigma^{2}
=R_{\boldsymbol{A}^{\mathrm{T}}\boldsymbol{A}}(0)/\sigma^{2}> s_{\mathrm{min}}$, because 
of Lemma~\ref{lemma_R_transform_0} and the assumption $\mu_{2}>\mu_{1}^{2}$. 
Thus, $f(E)$ satisfies all conditions in Lemma~\ref{lemma_infsup_general}, 
so that we can use Lemma~\ref{lemma_infsup_general} to arrive at 
the identity~(\ref{infsup}). 
\end{IEEEproof}

Lemma~\ref{lemma_infsup} implies that the optimizer $(E, s)$ 
for the inf-sup problem~(\ref{infsup}) satisfies $E=\mathrm{MMSE}(s)$. 
This allows us to regard the variable $E$ as the MMSE. 

\subsection{Information-Theoretic Compression Limit}
We consider the noiseless limit $\sigma^{2}\downarrow0$. 
The R\'enyi information dimension is useful to characterize the performance 
of the Bayes-optimal reconstruction in the system~(\ref{uncoupled_model}) 
without spatial coupling, as well as the performance 
of Bayes-optimal OAMP for spatially coupled systems. 

\begin{definition}[\cite{Renyi59}]
For a random variable $X\in\mathbb{R}$, let $X_{n}=\lfloor nX \rfloor/n$ 
denote a discrete random variable rounded down with the floor operation for 
$n\in\mathbb{N}$. 
The random variable $X$ is said to have the R\'enyi information 
dimension $d_{\mathrm{I}}$ if the normalized entropy 
$-\mathbb{E}[\log\mathrm{Pr}(X_n)]/\log n$ converges to $d_{\mathrm{I}}$ as 
$n\to\infty$. 
\end{definition}
\begin{theorem} \label{theorem_potential}
Suppose that all moments $\{\mu_{k}\}$ are bounded and 
assume the following conditions: 
\begin{itemize}
\item The signal $x_{1}$ has $\mathbb{E}[x_{1}^{2}]=1$ and 
the R\'enyi information dimension $d_{\mathrm{I}}$.
\item The R-transform $R_{\boldsymbol{A}^{\mathrm{T}}\boldsymbol{A}}(z)$ is proper, 
continuous, non-decreasing, and positive for all $z\leq0$. 
\item $\lim_{z\to\infty}zR_{\boldsymbol{A}^{\mathrm{T}}\boldsymbol{A}}(-z)=\delta$ holds. 
\end{itemize}
If and only if the compression rate $\delta$ is larger than $d_{\mathrm{I}}$, 
the optimizer $(E_{\mathrm{opt}}, s_{\mathrm{opt}})$ for the inf-sup 
problem~(\ref{asymptotic_mutual_information}) is noise-limited: 
$E_{\mathrm{opt}}\downarrow0$ and $s_{\mathrm{opt}}\to\infty$ hold 
as $\sigma^{2}\downarrow0$. In particular, 
the optimizer $(E_{\mathrm{opt}}, s_{\mathrm{opt}})$ is unique 
as $\sigma^{2}\downarrow0$ if $\delta$ is larger than $d_{\mathrm{I}}$.  
\end{theorem}
\begin{IEEEproof}
See Appendix~\ref{proof_theorem_potential}. 
\end{IEEEproof}

Theorem~\ref{theorem_potential} implies that error-free reconstruction 
$E_{\mathrm{opt}}\downarrow0$ is possible as long as the compression rate  
$\delta$ is larger than the information-theoretic compression limit 
$d_{\mathrm{I}}$~\cite{Wu10}. The last two assumptions in 
Theorem~\ref{theorem_potential} provide sufficient conditions which 
the sensing matrix should satisfy.  

Theorem~\ref{theorem_potential} reproduces a known result on 
the optimality of zero-mean i.i.d.\ Gaussian sensing 
matrices~\cite{Donoho13,Takeuchi15}. 

\begin{corollary} \label{corollary1}
Assume the following conditions: 
\begin{itemize}
\item The signal $x_{1}$ has $\mathbb{E}[x_{1}^{2}]=1$ and 
the R\'enyi information dimension $d_{\mathrm{I}}$. 
\item The sensing matrix $\boldsymbol{A}$ has independent zero-mean 
Gaussian elements with variance $1/M$. 
\end{itemize}
If and only if $\delta>d_{\mathrm{I}}$ holds, the optimizer 
$(E_{\mathrm{opt}},s_{\mathrm{opt}})$ satisfies $E_{\mathrm{opt}}\downarrow0$ and 
$s_{\mathrm{opt}}\to\infty$ as $\sigma^{2}\downarrow0$. In particular, 
the optimizer $(E_{\mathrm{opt}}, s_{\mathrm{opt}})$ is unique 
as $\sigma^{2}\downarrow0$ if $\delta$ is larger than $d_{\mathrm{I}}$.   
\end{corollary}
\begin{IEEEproof}
We know that the R-transform for zero-mean i.i.d.\ Gaussian sensing matrices 
is given by $R_{\boldsymbol{A}^{\mathrm{T}}\boldsymbol{A}}(z)=\delta/(\delta-z)$ 
for all $z\leq0$~\cite[Section 2.4.2]{Tulino04}, 
which satisfies the last two assumptions in Theorem~\ref{theorem_potential}. 
Thus, Theorem~\ref{theorem_potential} implies Corollary~\ref{corollary1}. 
\end{IEEEproof}

The replica symmetric potential~(\ref{RS_potential}) was used to characterize 
the asymptotic performance of AMP for spatially coupled zero-mean i.i.d.\ 
Gaussian sensing matrices in \cite{Takeuchi15}. More precisely, state evolution 
recursions proved in \cite{Javanmard13} are equivalent to density evolution 
recursions considered in \cite{Takeuchi15}. Thus, properties of the state 
evolution recursions are also characterized with the replica symmetric 
potential. To prove the information-theoretic optimality of AMP, we can use 
Corollary~\ref{corollary1} instead of a proof in \cite{Donoho13}.   
 
It is an interesting issue to specify the class of sensing matrices 
that satisfy the last two assumptions for the R-transform 
in Theorem~\ref{theorem_potential}. 
We have the following corollary: 

\begin{corollary} \label{corollary2}
Suppose that all moments $\{\mu_{k}\}$ are bounded and that $\mu_{1}>0$ and 
$\mu_{2}>\mu_{1}^{2}$ hold. 
Let $r$ denote the rank of $\boldsymbol{A}$ and assume that the ratio $r/N$ 
tends to $\delta$ in the large system limit. Furthermore, postulate the 
following conditions: 
\begin{itemize}
\item The signal $x_{1}$ has $\mathbb{E}[x_{1}^{2}]=1$ and 
the R\'enyi information dimension $d_{\mathrm{I}}$.
\item $\delta<1$ or the condition~(\ref{diverging_condition}) for $\delta=1$ 
holds. 
\end{itemize}
If and only if $\delta>d_{\mathrm{I}}$ holds, the optimizer 
$(E_{\mathrm{opt}},s_{\mathrm{opt}})$ satisfies $E_{\mathrm{opt}}\downarrow0$ and 
$s_{\mathrm{opt}}\to\infty$ as $\sigma^{2}\downarrow0$. In particular, 
the optimizer $(E_{\mathrm{opt}}, s_{\mathrm{opt}})$ is unique 
as $\sigma^{2}\downarrow0$ if $\delta$ is larger than $d_{\mathrm{I}}$.   
\end{corollary}
\begin{IEEEproof}
Lemmas~\ref{lemma_R_transform_0} and \ref{lemma_R_transform_positive} imply 
the continuity, non-decreasing, and positivity properties of the 
R-transform $R_{\boldsymbol{A}^{\mathrm{T}}\boldsymbol{A}}(z)$. Furthermore, 
we use Lemma~\ref{lemma_R_transform_delta} to find 
$\lim_{z\to\infty}zR_{\boldsymbol{A}^{\mathrm{T}}\boldsymbol{A}}(-z)=\delta$ and 
$\lim_{z\to\infty}z\eta_{\boldsymbol{A}^{\mathrm{T}}\boldsymbol{A}}(z)=\infty$. 
The latter property and the assumption $\mu_{2}>\mu_{1}^{2}$ indicate that 
the R-transform~(\ref{R_transform}) is proper for all $z\leq0$. 
Thus, Theorem~\ref{theorem_potential} implies Corollary~\ref{corollary2}. 
\end{IEEEproof}

Corollary~\ref{corollary2} implies that the information-theoretic compression 
limit is achievable for the non-trivial case $\delta<1$ 
when the sensing matrix $\boldsymbol{A}$ is picked up 
from $(M, N)$-ensemble with full rank. 

\section{Spatial Coupling} \label{sec3}
We extend the conventional measurement model in (\ref{uncoupled_model}) to 
a spatially coupled model with the number of sections $L$ and coupling 
width $W<L$, defined via a deterministic base matrix 
$\boldsymbol{B}_{\boldsymbol{\Gamma}}\in\mathbb{R}^{(L+W)\times L}$. 
The $(\ell,l)$ element $\gamma[\ell][l]$ of 
$\boldsymbol{B}_{\boldsymbol{\Gamma}}$ is 
non-zero for all $\ell\in\mathcal{L}_{W}=\{0,\ldots,L+W-1\}$ and 
$l\in\mathcal{L}_{0}=\{0,\ldots,L-1\}$ satisfying $\ell-l\in\{0,\ldots,W\}$. 
Otherwise, $\gamma[\ell][l]=0$ holds. Furthermore,  
we impose the power normalization 
\begin{equation}
\frac{1}{L}\sum_{l=0}^{L-1}\sum_{w=0}^{W}\gamma^{2}[l+w][l] = 1. 
\end{equation} 
In particular, $W=0$ implies no spatial coupling. 

This paper does not postulate special structures of the base matrix 
$\boldsymbol{B}_{\boldsymbol{\Gamma}}$ in state evolution analysis. 
In proving the information-theoretic optimality, as well as numerical 
evaluation, we focus on the following base matrix with 
uniform coupling weights: 
\begin{equation} \label{uniform_coupling}
\boldsymbol{B}_{\boldsymbol{\Gamma}} 
= \frac{1}{\sqrt{W+1}}
\begin{bmatrix}
1 &  & \boldsymbol{O} \\
\vdots & \ddots & \\
1 & & 1\\
& \ddots & \vdots \\
\boldsymbol{O} & & 1 
\end{bmatrix}\in\mathbb{R}^{(L+W)\times L}. 
\end{equation}

For a given base matrix $\boldsymbol{B}_{\boldsymbol{\Gamma}}$, 
the $M[\ell]$-dimensional 
measurement vector $\boldsymbol{y}[\ell]\in\mathbb{R}^{M[\ell]}$ for 
section~$\ell\in\mathcal{L}_{W}$ in a spatially coupled system  
is defined as 
\begin{equation} \label{spatial_coupling}
\boldsymbol{y}[\ell] 
= \sum_{w=0}^{W}\gamma[\ell][\ell-w]\boldsymbol{A}[\ell][\ell-w]
\boldsymbol{x}[\ell-w] + \boldsymbol{n}[\ell].
\end{equation}
In (\ref{spatial_coupling}), $\boldsymbol{n}[\ell]\in\mathbb{R}^{M[\ell]}$ 
denotes an additive noise vector in section~$\ell$ and satisfies 
$\lim_{M[\ell]\to\infty}M^{-1}[\ell]
\mathbb{E}[\|\boldsymbol{n}[\ell]\|^{2}]=\sigma^{2}$ for variance $\sigma^{2}>0$. 
The $N[l]$-dimensional vector $\boldsymbol{x}[l]\in\mathbb{R}^{N[l]}$ 
represents unknown sparse signals in section~$l\in\mathcal{L}_{0}$ and 
satisfies the power normalization 
$\lim_{N[l]\to\infty}N^{-1}[l]\mathbb{E}[\|\boldsymbol{x}[l]\|^{2}]=1$. 
The matrix $\boldsymbol{A}[\ell][l]\in\mathbb{R}^{M[\ell]\times N[l]}$ is 
a sensing matrix in section $(\ell, l)$.  
The random variables in the triple 
$\{\{\boldsymbol{A}[\ell][l]\}, \{\boldsymbol{x}[l]\}, 
\{\boldsymbol{n}[\ell]\}\}$ are independent. 
For notational convenience, we introduce 
$\gamma[\ell][l]=0$ for all 
$(\ell,l)\notin\mathcal{L}_{W}\times\mathcal{L}_{0}$, 
$\boldsymbol{x}[l]=\boldsymbol{0}$, and $N[l]=0$ for $l\notin\mathcal{L}_{0}$.  
Since the system~(\ref{spatial_coupling}) is independent of 
$\boldsymbol{A}[\ell][l]$ in all positions~$(\ell, l)$ satisfying 
$\gamma[\ell][l]=0$, we assume $\boldsymbol{A}[\ell][l]=\boldsymbol{O}$ in 
the positions. 

\begin{figure}[t]
\begin{center}
\includegraphics[width=\hsize]{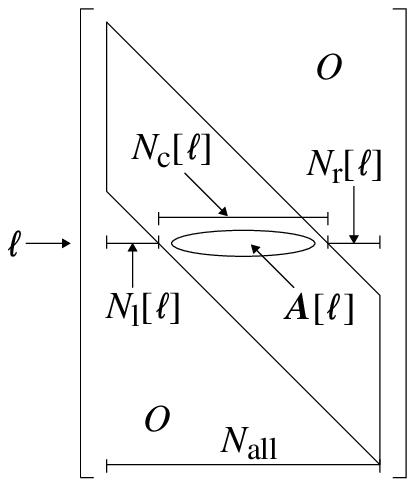}
\caption{
Band structure of the overall sensing matrix. 
}
\label{fig1} 
\end{center}
\end{figure}

We rewrite the spatially coupled system~(\ref{spatial_coupling}) in a vector 
form. Focus on the summation in the spatially coupled 
system~(\ref{spatial_coupling}) for the bulk 
region $\ell\in\{W,\ldots,L-1\}$. The summation may be regarded as 
the multiplication of the block matrix 
$(\gamma[\ell][\ell-W]\boldsymbol{A}[\ell][\ell-W],\ldots, 
\gamma[\ell][\ell]\boldsymbol{A}[\ell][\ell])$ by the block vector 
$(\boldsymbol{x}^{\mathrm{T}}[\ell-W], \ldots,
\boldsymbol{x}^{\mathrm{T}}[\ell])^{\mathrm{T}}$. In the proposed OAMP, however, 
the summation should be regarded as the multiplication of 
$\boldsymbol{A}[\ell]=(W+1)^{-1/2}
(\boldsymbol{A}[\ell][\ell-W],\ldots, \boldsymbol{A}[\ell][\ell])$ 
by $\vec{\boldsymbol{x}}[\ell]
=(W+1)^{1/2}(\gamma[\ell][\ell-W]\boldsymbol{x}^{\mathrm{T}}[\ell-W],\ldots, 
\gamma[\ell][\ell]\boldsymbol{x}^{\mathrm{T}}[\ell])^{\mathrm{T}}$. In other words, 
the coupling coefficients $\{\gamma[\ell][l]\}$ are moved 
from the sensing-matrix side to the signal-vector side.  
The prefactor $(W+1)^{-1/2}$ in $\boldsymbol{A}[\ell]$ normalizes 
the eigenvalues of $\boldsymbol{A}^{\mathrm{T}}[\ell]\boldsymbol{A}[\ell]$ 
in the limit $W\to\infty$. 
The prefactor $(W+1)^{1/2}$ in $\vec{\boldsymbol{x}}[\ell]$ 
compensates for the normalization in $\boldsymbol{A}[\ell]$. Owing to this 
prefactor, each element in $\vec{\boldsymbol{x}}[\ell]$ is kept ${\cal O}(1)$ 
for the uniform coupling weights~(\ref{uniform_coupling}).

To present appropriate definitions in the boundary region, 
we first introduce several notations. As shown in Fig.~\ref{fig1}, 
let $N_{\mathrm{c}}[\ell]=\sum_{l=\ell-W}^{\ell}N[l]$ denote 
the number of non-zero columns in the $\ell$th row section of the sensing 
matrices. We write 
the numbers of zero columns on the left and right sides of the non-zero 
columns in row section~$\ell$ as $N_{\mathrm{l}}[\ell]=\sum_{l=0}^{\ell-W-1}N[l]$ 
and $N_{\mathrm{r}}[\ell]=\sum_{l=\ell+1}^{L-1}N[l]$, respectively. 
The notation $N_{\mathrm{all}}=\sum_{l=0}^{L-1}N[l]$ represents the number of 
columns in section~$\ell$. 

Define the normalized non-zero blocks 
$\boldsymbol{A}[\ell]\in\mathbb{R}^{M[\ell]\times N_{\mathrm{c}}[\ell]}$ in row 
section~$\ell\in\mathcal{L}_{W}$ as 
\begin{equation} \label{A_row}
\boldsymbol{A}[\ell]=\left\{
 |\mathcal{W}[\ell]|^{-1/2}\boldsymbol{A}[\ell][\ell-w]: w\in\mathcal{W}[\ell]
\right\}, 
\end{equation}
where the set of indices $\mathcal{W}[\ell]$ is given by 
\begin{equation} \label{set_W}
\mathcal{W}[\ell]=\{w_{\mathrm{min}}[\ell],\ldots,w_{\mathrm{max}}[\ell]\}, 
\end{equation}
with $w_{\mathrm{min}}[\ell] = \max\{\ell-(L-1),0\}$ and 
$w_{\mathrm{max}}[\ell] =  \min\{W,\ell\}$. 
The set $\mathcal{W}[\ell]$ 
reduces to $\mathcal{W}[\ell]=\{0,\ldots,\ell\}$, 
$\mathcal{W}[\ell]=\{0,\ldots,W\}$, and $\mathcal{W}[\ell]
=\{\ell-(L-1),\ldots,W\}$ for $\ell<W$, $W\leq\ell<L$, and $L\leq\ell$, 
respectively. 
Using these notations, we find that the spatially coupled 
system~(\ref{spatial_coupling}) reduces to 
\begin{equation} \label{vector_model}
\boldsymbol{y}[\ell] 
= \boldsymbol{A}[\ell]\vec{\boldsymbol{x}}[\ell]
+ \boldsymbol{n}[\ell], 
\end{equation}
with 
\begin{equation}
\vec{\boldsymbol{x}}[\ell]
= \sqrt{|\mathcal{W}[\ell]|}\begin{bmatrix}
\gamma[\ell][\ell - w_{\mathrm{max}}[\ell]]
\boldsymbol{x}[\ell - w_{\mathrm{max}}[\ell]] \\
\vdots \\
\gamma[\ell][\ell - w_{\mathrm{min}}[\ell]]
\boldsymbol{x}[\ell - w_{\mathrm{min}}[\ell]]
\end{bmatrix}.  
\end{equation}

To represent $\vec{\boldsymbol{x}}[\ell]\in\mathbb{R}^{N_{\mathrm{c}}[\ell]}$ 
with the overall signal vector 
$\boldsymbol{x}=(\boldsymbol{x}^{\mathrm{T}}[0],\ldots,
\boldsymbol{x}^{\mathrm{T}}[L-1])^{\mathrm{T}}\in\mathbb{R}^{N_{\mathrm{all}}}$, 
we define 
\begin{equation} \label{Gamma_l} 
\boldsymbol{\Gamma}[\ell]
= \left(
 \boldsymbol{O}_{N_{\mathrm{c}}[\ell]\times N_{\mathrm{l}}[\ell]},  
 \tilde{\boldsymbol{\Gamma}}[\ell], 
 \boldsymbol{O}_{N_{\mathrm{c}}[\ell]\times N_{\mathrm{r}}[\ell]}
\right)\in\mathbb{R}^{N_{\mathrm{c}}[\ell]\times N_{\mathrm{all}}}, 
\end{equation}
with $\tilde{\boldsymbol{\Gamma}}[\ell]
=\mathrm{diag}\{\gamma[\ell][\ell-w]\boldsymbol{I}_{N[\ell-w]}: 
w\in\mathcal{W}[\ell]\}$. 
The product $\boldsymbol{\Gamma}[\ell]\boldsymbol{x}
\in\mathbb{R}^{N_{\mathrm{c}}[\ell]}$ is a vector with $|\mathcal{W}[\ell]|$ 
sections. In particular, for $w\in\mathcal{W}[\ell]$ we have the $w$th section 
$\{\boldsymbol{\Gamma}[\ell]\boldsymbol{x}\}_{w}=
\gamma[\ell][\ell-w]\boldsymbol{x}[\ell-w]\in\mathbb{R}^{N[\ell-w]}$.  
Using these notations, we can rewrite the signal vector 
$\vec{\boldsymbol{x}}[\ell]$ as  
\begin{equation} \label{signal_representation}  
\vec{\boldsymbol{x}}[\ell]
=\sqrt{|\mathcal{W}[\ell]|}\boldsymbol{\Gamma}[\ell]\boldsymbol{x}. 
\end{equation}

\begin{figure}[t]
\begin{algorithm}[H]
\caption{Orthogonal AMP with $T$ iterations} 
\label{alg1}
\begin{algorithmic}[1]
\State 
For all $\ell\in\mathcal{L}_{W}$,
let $\vec{\boldsymbol{x}}_{\B\to\A,0}[\ell]=\boldsymbol{0}$ and 
$v_{\B\to\A,0}[\ell]=|\mathcal{W}[\ell]|N_{\mathrm{c}}^{-1}[\ell]
\sum_{w\in\mathcal{W}[\ell]}N[\ell-w]\gamma^{2}[\ell][\ell-w]$.    
\For{$t=0,\ldots,T-1$}
\For{$\ell=0,\ldots,L+W-1$} 

\State 
\hspace{-2em}$\vec{\boldsymbol{x}}_{\A,t}^{\mathrm{post}}[\ell] 
= \vec{\boldsymbol{x}}_{\B\to\A,t}[\ell] 
+ \boldsymbol{F}_{t}^{\mathrm{T}}[\ell]
(\boldsymbol{y}[\ell] - \boldsymbol{A}[\ell]
\vec{\boldsymbol{x}}_{\B\to\A,t}[\ell])$.

\State
\hspace{-2em}$v_{\A,t}^{\mathrm{post}}[\ell] 
= \frac{\sigma^{2}}{N_{\mathrm{c}}[\ell]}\mathrm{Tr}\left(
 \boldsymbol{F}_{t}^{\mathrm{T}}[\ell]\boldsymbol{F}_{t}[\ell]
\right)$ 

$+ \frac{v_{\B\to\A,t}[\ell]}{N_{\mathrm{c}}[\ell]}\mathrm{Tr}\left\{
 (\boldsymbol{I} - \boldsymbol{F}_{t}^{\mathrm{T}}[\ell]\boldsymbol{A}[\ell])
 (\boldsymbol{I} - \boldsymbol{F}_{t}^{\mathrm{T}}[\ell]
 \boldsymbol{A}[\ell])^{\mathrm{T}}
\right\}$. 

\State 
$\eta_{\A,t}[\ell]= N_{\mathrm{c}}^{-1}[\ell]\mathrm{Tr}
(\boldsymbol{I} - \boldsymbol{F}_{t}^{\mathrm{T}}[\ell]\boldsymbol{A}[\ell])$.

\State 
$\vec{\boldsymbol{x}}_{\A\to\B,t}[\ell]
= \frac{\vec{\boldsymbol{x}}_{\A,t}^{\mathrm{post}}[\ell] 
- \eta_{\A,t}[\ell]\vec{\boldsymbol{x}}_{\B\to\A,t}[\ell]}
{\sqrt{|\mathcal{W}[\ell]|}(1 - \eta_{\A,t}[\ell])}$. 

\State 
$v_{\A\to\B,t}[\ell] 
= \frac{v_{\A,t}^{\mathrm{post}}[\ell] - \eta_{\A,t}^{2}[\ell]v_{\B\to\A,t}[\ell]}
{|\mathcal{W}[\ell]|(1-\eta_{\A,t}[\ell])^{2}}$. 
\EndFor 

\For{$l=0,\ldots,L-1$} 
\State 
$v_{\A\to \B,t}^{\mathrm{suf}}[l] 
= \left(
 \sum_{w=0}^{W}\frac{\gamma^{2}[l+w][l]}{v_{\A\to \B,t}[l+w]}
\right)^{-1}$.

\State 
Let $\vec{\boldsymbol{x}}_{\A\to\B,t}[l+w][w]\in\mathbb{R}^{N}$ denote 
the $w$th section in $\vec{\boldsymbol{x}}_{\A\to\B,t}[l+w]$ 
for $w\in\mathcal{W}[l+w]$ 
and compute $\boldsymbol{x}_{\A\to \B,t}^{\mathrm{suf}}[l]
= v_{\A\to \B,t}^{\mathrm{suf}}[l]\sum_{w=0}^{W}\gamma[l+w][l]
\frac{\vec{\boldsymbol{x}}_{\A\to\B,t}[l+w][w]}{v_{\A\to \B,t}[l+w]}$.  

\State 
$\boldsymbol{x}_{\B,t+1}^{\mathrm{post}}[l]
=f_{t}[l](\boldsymbol{x}_{\A\to \B,t}^{\mathrm{suf}}[l])$. 

\State 
Let $v_{\B,t+1}^{\mathrm{post}}[l]$ be a consistent estimator of 
$N^{-1}[l]\mathbb{E}[\|\boldsymbol{x}[l]
-\boldsymbol{x}_{\B,t+1}^{\mathrm{post}}[l]\|^{2}]$~\cite[Eq.~(23)]{Takeuchi213}. 
\EndFor

\For{$\ell=0,\ldots,L+W-1$}
\State 
$\vec{\boldsymbol{x}}_{\B,t+1}^{\mathrm{post}}[\ell]
=\sqrt{|\mathcal{W}[\ell]}\mathrm{vec}\{
\gamma[\ell][\ell-w]\boldsymbol{x}_{\B,t+1}^{\mathrm{post}}[\ell-w]:$

\hspace{6em}$w\in\mathcal{W}[\ell]\}$. 

\State 
$\eta_{\B,t}[\ell][w] 
= \frac{|\mathcal{W}[\ell]|\gamma^{2}[\ell][\ell-w]
v_{\A\to \B,t}^{\mathrm{suf}}[\ell-w]}
{v_{\A\to\B,t}[\ell]}$

\hspace{6em}$\cdot\langle f'_{t}[\ell-w]
(\boldsymbol{x}_{\A\to\B,t}^{\mathrm{suf}}[\ell-w])\rangle$.  

\State 
$\eta_{\B,t}[\ell]
= \sum_{w\in\mathcal{W}[\ell]}\frac{N[\ell-w]}{N_{\mathrm{c}}[\ell]}
\eta_{\B,t}[\ell][w]$.  

\State 
$\vec{\boldsymbol{x}}_{\B\to\A,t+1}[\ell] 
= (1 - \eta_{\B,t}[\ell]/|\mathcal{W}[\ell]|)^{-1}$

\hspace{2em}
$\cdot\left\{\vec{\boldsymbol{x}}_{\B,t+1}^{\mathrm{post}}[\ell]
- |\mathcal{W}[\ell]|^{-1/2}\eta_{\B,t}[\ell]
\vec{\boldsymbol{x}}_{\A\to\B,t}[\ell]\right\}$. 

\State 
$v_{\B\to\A,t+1}[\ell]=(1 - \eta_{\B,t}[\ell]/|\mathcal{W}[\ell]|)^{-2}$
 
$\cdot\Bigl\{\sum_{w\in\mathcal{W}[\ell]}\frac{N[\ell-w]}{N_{\mathrm{c}}[\ell]}
|\mathcal{W}[\ell]|\gamma^{2}[\ell][\ell-w]v_{\B,t+1}^{\mathrm{post}}[\ell-w]$ 

$-|\mathcal{W}[\ell]|^{-1}\eta_{\B,t}^{2}[\ell]v_{\A\to\B,t}[\ell]\Bigr\}$.  

\EndFor 
\EndFor
\State Output $\boldsymbol{x}_{\mathrm{B},T}^{\mathrm{post}}[l]$ as an estimator of 
$\boldsymbol{x}[l]$ for all $l\in\mathcal{L}_{0}$. 
\end{algorithmic}
\end{algorithm}
\end{figure}

\section{Orthogonal AMP} \label{sec4}
\subsection{Overview}
For the spatially coupled system~(\ref{vector_model}), this paper proposes 
OAMP in Algorithm~\ref{alg1}, 
which consists of the two modules---called modules A and B. 
Module~A uses a linear filter to compute posterior messages 
while module~B utilizes a separable nonlinear 
denoiser to refine the messages in module~A. To realize asymptotic 
Gaussianity for the estimation errors, each module computes extrinsic 
messages via the so-called Onsager correction of the posterior messages. 

In module~A for the spatial coupling case, the signal vectors 
$\{\vec{\boldsymbol{x}}[\ell]\}$ in the extended space are estimated 
in parallel for all~$\ell$. 
Message computation for each $\ell$ is equivalent 
to that for conventional OAMP~\cite{Ma17,Rangan192}, with the exception 
of normalization due to spatial coupling. In module~A, 
the signal vectors $\{\vec{\boldsymbol{x}}[\ell]\}$ for 
all row section $\ell\in\mathcal{L}_{W}$ are regarded as independent Gaussian 
random vectors with i.i.d.\ elements. Since each element in $\boldsymbol{x}$ is 
broadcast over adjacent signal vectors $\{\vec{\boldsymbol{x}}[\ell]\}$ 
via (\ref{signal_representation}), 
the dependencies between $\{\vec{\boldsymbol{x}}[\ell]\}$ through 
the original signal vector $\boldsymbol{x}$ are not taken into account. 

Module~B for the spatial coupling case takes the dependencies into account, 
as well as the signal prior distributions. In other words,  
module~B operates in the original space $\mathbb{R}^{N_{\mathrm{all}}}$ while 
module~A operates in $|\mathcal{L}_{W}|$ extended signal 
spaces $\{\mathbb{R}^{N_{\mathrm{c}}[\ell]}\}_{\ell\in\mathcal{L}_{W}}$. 
The vector $\vec{\boldsymbol{x}}[\ell]$ in (\ref{signal_representation}) 
corresponds 
to the signal vector in the $\ell$th extended space. Thus, the original signal 
vector $\boldsymbol{x}$ can be reconstructed from the extended spaces via the 
pseudo-inverse $|\mathcal{W}[\ell]|^{-1/2}
\boldsymbol{\Gamma}^{\dagger}\vec{\boldsymbol{x}}$, 
with $\vec{\boldsymbol{x}}=(\vec{\boldsymbol{x}}^{\mathrm{T}}[0],\ldots,
\vec{\boldsymbol{x}}^{\mathrm{T}}[L+W-1])^{\mathrm{T}}$. This reconstruction of 
the original signal vector is performed in module~B. 

The main novelty in module~B is in the Onsager correction to realize 
asymptotic Gaussianity for the estimation errors of messages 
fed back to module~A. The asymptotic Gaussianity is realized in the extended 
signal space where module~A operates, rather than 
in the original signal space where module~B operates. To design 
this Onsager correction appropriately, this paper establishes 
a unified framework of state evolution for the spatial coupling case. 

\subsection{Module~A (Linear Estimation)}
Module~A consists of two steps: A first step is computation of posterior 
messages based on linear filters. The second step is the Onsager correction 
of the posterior messages to realize asymptotic Gaussianity in module~B. 

Let $\vec{\boldsymbol{x}}_{\B\to \A,t}[\ell]\in\mathbb{R}^{N_{\mathrm{c}}[\ell]}$ and 
$v_{\B\to \A,t}[\ell]>0$ denote the mean and variance messages of 
$\vec{\boldsymbol{x}}[\ell]$ in (\ref{signal_representation}) passed from 
module~B to module~A in iteration~$t$, respectively. 
The variance message $v_{\B\to \A,t}[\ell]$ 
corresponds to a consistent estimator of the the mean-square error (MSE) 
$N_{\mathrm{c}}^{-1}[\ell]\mathbb{E}[\|\vec{\boldsymbol{x}}[\ell]
-\vec{\boldsymbol{x}}_{\B\to \A,t}[\ell]\|^{2}]$ in the large system limit. 

Similarly, we write the mean and variance messages of 
$|\mathcal{W}[\ell]|^{-1/2}\vec{\boldsymbol{x}}[\ell]
=\boldsymbol{\Gamma}[\ell]\boldsymbol{x}$ passed in the opposite direction as 
$\vec{\boldsymbol{x}}_{\A\to \B,t}[\ell]\in\mathbb{R}^{N_{\mathrm{c}}[\ell]}$ and 
$v_{\A\to \B,t}[\ell]>0$. Owing to the prefactor in 
$|\mathcal{W}[\ell]|^{-1/2}\vec{\boldsymbol{x}}[\ell]$, the scaled MSE 
$N_{\mathrm{c}}^{-1}[\ell]\mathbb{E}[
\||\mathcal{W}[\ell]|^{-1/2}\vec{\boldsymbol{x}}[\ell]
-\vec{\boldsymbol{x}}_{\A\to \B,t}[\ell]\|^{2}]$ is 
kept ${\cal O}(1)$ in the limit $W\to\infty$. 

In a first step of iteration~$t$, module~A computes the posterior mean and 
variance of $\vec{\boldsymbol{x}}[\ell]$ based on a linear filter 
$\boldsymbol{F}_{t}[\ell]\in\mathbb{R}^{M[\ell]\times N_{\mathrm{c}}[\ell]}$ 
for each $\ell\in\mathcal{L}_{W}$. 
The posterior mean $\vec{\boldsymbol{x}}_{\A,t}^{\mathrm{post}}[\ell]
\in\mathbb{R}^{N_{\mathrm{c}}[\ell]}$ in the $\ell$th extended space is defined as 
\begin{equation} \label{mean_post_A}
\vec{\boldsymbol{x}}_{\A,t}^{\mathrm{post}}[\ell] 
= \vec{\boldsymbol{x}}_{\B\to\A,t}[\ell]
+ \boldsymbol{F}_{t}^{\mathrm{T}}[\ell]
(\boldsymbol{y}[\ell] - \boldsymbol{A}[\ell]
\vec{\boldsymbol{x}}_{\B\to\A,t}[\ell]) 
\end{equation}
for section~$\ell\in\mathcal{L}_{W}$. The corresponding posterior variance 
$v_{\A,t}^{\mathrm{post}}[\ell]$ is given by 
\begin{IEEEeqnarray}{l}
v_{\A,t}^{\mathrm{post}}[\ell] 
= \frac{\sigma^{2}}{N_{\mathrm{c}}[\ell]}\mathrm{Tr}\left(
 \boldsymbol{F}_{t}^{\mathrm{T}}[\ell]\boldsymbol{F}_{t}[\ell]
\right) \nonumber \\
+ \frac{v_{\B\to\A,t}[\ell]}{N_{\mathrm{c}}[\ell]}\mathrm{Tr}\left\{
 (\boldsymbol{I} - \boldsymbol{F}_{t}^{\mathrm{T}}[\ell]\boldsymbol{A}[\ell])
 (\boldsymbol{I} - \boldsymbol{F}_{t}^{\mathrm{T}}[\ell]
 \boldsymbol{A}[\ell])^{\mathrm{T}}
\right\}. 
\nonumber \\
\label{var_post_A}
\end{IEEEeqnarray}

For $t=0$, we use the initial conditions 
$\vec{\boldsymbol{x}}_{\B\to\A,0}[\ell]=\boldsymbol{0}$ 
and $v_{\B\to\A,0}[\ell]=|\mathcal{W}[\ell]|N_{\mathrm{c}}^{-1}[\ell]
\sum_{w\in\mathcal{W}[\ell]}N[\ell-w]\gamma^{2}[\ell][\ell-w]$, of which the latter 
is equal to $|\mathcal{W}[\ell]|N_{\mathrm{c}}^{-1}[\ell]\mathrm{Tr}
(\boldsymbol{\Gamma}[\ell]\boldsymbol{\Gamma}^{\mathrm{T}}[\ell])$. 

Consider the LMMSE filter 
\begin{equation} \label{LMMSE}
\boldsymbol{F}_{t}[\ell]
=v_{\B\to\A,t}[\ell]\boldsymbol{\Xi}_{t}^{-1}[\ell]\boldsymbol{A}[\ell],
\end{equation}
with 
\begin{equation} 
\boldsymbol{\Xi}_{t}[\ell]
= \sigma^{2}\boldsymbol{I}_{M[\ell]} 
+ v_{\B\to\A,t}[\ell]\boldsymbol{A}[\ell]\boldsymbol{A}^{\mathrm{T}}[\ell]. 
\end{equation}
Substituting (\ref{LMMSE}) into the definition of $v_{\A,t}^{\mathrm{post}}[\ell]$ 
in (\ref{var_post_A}), we find that $v_{\A,t}^{\mathrm{post}}[\ell]$ reduces to 
\begin{equation} \label{var_post_A_LMMSE} 
v_{\A,t}^{\mathrm{post}}[\ell]
= \eta_{\A,t}[\ell]v_{\B\to\A,t}[\ell],
\end{equation}
with
\begin{equation} \label{eta_A}
\eta_{\A,t}[\ell]
= \frac{1}{N_{\mathrm{c}}[\ell]}\mathrm{Tr}\left(
 \boldsymbol{I}_{N_{\mathrm{c}}[\ell]} 
 - \boldsymbol{F}_{t}^{\mathrm{T}}[\ell]\boldsymbol{A}[\ell]
\right).
\end{equation}

The second step is the Onsager correction to realize 
asymptotic Gaussianity in module~B. Module~A computes 
the extrinsic mean $\vec{\boldsymbol{x}}_{\A\to\B,t}[\ell]$ and variance 
$v_{\A\to\B,t}[\ell]>0$ of 
$|\mathcal{W}[\ell]|^{-1/2}\vec{\boldsymbol{x}}[\ell]$ in the extended space as 
\begin{equation} \label{mean_AB}
\vec{\boldsymbol{x}}_{\A\to\B,t}[\ell]
= |\mathcal{W}[\ell]|^{-1/2}\frac{\vec{\boldsymbol{x}}_{\A,t}^{\mathrm{post}}[\ell] 
- \eta_{\A,t}[\ell]\vec{\boldsymbol{x}}_{\B\to\A,t}[\ell]}{1 - \eta_{\A,t}[\ell]},
\end{equation}
\begin{equation} \label{var_AB} 
v_{\A\to\B,t}[\ell]  
= \frac{1}{|\mathcal{W}[\ell]|}
\frac{v_{\A,t}^{\mathrm{post}}[\ell] - \eta_{\A,t}^{2}[\ell]v_{\B\to\A,t}[\ell]}
{(1-\eta_{\A,t}[\ell])^{2}}, 
\end{equation}
with $\eta_{\A,t}[\ell]$ defined in (\ref{eta_A}). 
In particular, for the LMMSE filter~(\ref{LMMSE}) we substitute the posterior 
variance~(\ref{var_post_A_LMMSE}) into the definition of $v_{\A\to\B,t}[\ell]$ 
in (\ref{var_AB}) to obtain 
\begin{equation} \label{var_AB_LMMSE}
v_{\A\to\B,t}[\ell] 
= \frac{1}{|\mathcal{W}[\ell]|}
\frac{\eta_{\A,t}[\ell]v_{\B\to\A,t}[\ell]}{1-\eta_{\A,t}[\ell]}. 
\end{equation}

The normalization in $v_{\A\to\B,t}[\ell]$ can be understood as follows: 
As $W\to\infty$, the vector $\boldsymbol{y}[\ell]$ in 
(\ref{vector_model}) is an extremely compressed measurement of 
$\vec{\boldsymbol{x}}[\ell]$. Thus, signal reconstruction 
based on the LMMSE filter~(\ref{LMMSE}) results in poor performance. 
As proved in state evolution analysis, we have $1-\eta_{A,t}[\ell]
={\cal O}(|\mathcal{W}[\ell]|^{-1})$ for 
$\ell\in\{W,\ldots,L-1\}$ as $W\to\infty$, which implies 
$\eta_{A,t}[\ell]\to 1$. As a result, 
the extrinsic variance $\bar{v}_{\A\to\B,t}[\ell]$ in (\ref{var_AB_LMMSE}) 
is kept ${\cal O}(1)$ as $W\to\infty$. 

The discussion mentioned above is for individual variance messages 
$\{v_{\B\to\A,t}[\ell]\}$ and does not necessarily imply that module~A cannot 
refine the messages passed from module~B at all as $W\to\infty$. Since 
$|\mathcal{W}[\ell]|$ extensive messages contribute to estimation of 
each signal element in $\boldsymbol{x}$, module~A can still provide an 
impact of ${\cal O}(1)$ on estimation performance of module~B as $W\to\infty$.

\subsection{Module~B (Nonlinear Estimation)}
Module~B consists of four steps: A first step is the extraction of messages 
in the original space $\mathbb{R}^{N}$ from the messages in the extended 
spaces $\{\mathbb{R}^{N_{\mathrm{c}}[\ell]}\}_{\ell\in\mathcal{L}_{W}}$. A second step 
is computation of a sufficient statistic for estimation of the 
signal vector $\boldsymbol{x}\in\mathbb{R}^{N}$ given the extracted messages. 
A third step is evaluation of posterior messages based on the sufficient 
statistic. These two steps are equivalent to direct computation of the 
posterior messages given the extracted messages in the original space. 
In the last step, the posterior messages in the original space are transformed 
into those in the extended spaces. Then, the transformed messages are 
Onsager-corrected to realize asymptotic Gaussianity in module~A. 

In the first step of iteration~$t$, module~B extracts information required for 
estimation of $\boldsymbol{x}[l]$ from 
$\{\vec{\boldsymbol{x}}_{\A\to\B,t}[\ell]\}$ in the extended space. 
Using the definitions of $\vec{\boldsymbol{x}}[\ell]$ and  
$\boldsymbol{\Gamma}[\ell]$ in (\ref{signal_representation}) and 
(\ref{Gamma_l}), respectively, 
we find that $\boldsymbol{x}[l]$ is contained only in 
$\{\vec{\boldsymbol{x}}[l+w][w]: w\in\{0,\ldots,W\}\}$, 
with $\vec{\boldsymbol{x}}[\ell][w]\in\mathbb{R}^{N[\ell-w]}$ denoting the $w$th 
section in $\vec{\boldsymbol{x}}[\ell]\in\mathbb{R}^{N_{\mathrm{c}}[\ell]}$ for 
$w\in\mathcal{W}[\ell]$. More precisely, we have 
\begin{equation} \label{extraction}
|\mathcal{W}[l+w]|^{-1/2}\vec{\boldsymbol{x}}[l+w][w]
= \gamma[l+w][l]\boldsymbol{x}[l]   
\end{equation}
for $w\in\mathcal{W}[\ell]$. 
Since $\vec{\boldsymbol{x}}_{\A\to\B,t}[\ell]$ is the extrinsic mean for 
$|\mathcal{W}[\ell]|^{-1/2}\vec{\boldsymbol{x}}[\ell]$, the extracted message  
$\boldsymbol{x}_{\A\to\B,t}[l][w]\in\mathbb{R}^{N[l]}$ in the original space 
is defined as 
\begin{equation}  \label{mean_AB_projection}
\boldsymbol{x}_{\A\to\B,t}[l][w]
=\vec{\boldsymbol{x}}_{\A\to\B,t}[l+w][w] 
\end{equation} 
for $w\in\{0,\ldots,W\}$, 
where $\vec{\boldsymbol{x}}_{\A\to\B,t}[\ell][w]\in\mathbb{R}^{N[\ell-w]}$ denotes 
the $w$th section in $\vec{\boldsymbol{x}}_{\A\to\B,t}[\ell]
\in\mathbb{R}^{N_{\mathrm{c}}[\ell]}$ for $w\in\mathcal{W}[\ell]$. 

The second step is computation of a sufficient statistic 
$\boldsymbol{x}_{\A\to \B,t}^{\mathrm{suf}}[l]\in\mathbb{R}^{N[l]}$ 
for estimation of $\boldsymbol{x}[l]$ and the corresponding variance 
$v_{\A\to \B,t}^{\mathrm{suf}}[l]>0$. As derived in Appendix~\ref{appen_suf}, 
the mean message $\boldsymbol{x}_{\A\to \B,t}^{\mathrm{suf}}[l]$ in section~$l$ and 
corresponding variance $v_{\A\to \B,t}^{\mathrm{suf}}[l]$ are computed as 
\begin{equation} 
\boldsymbol{x}_{\A\to \B,t}^{\mathrm{suf}}[l]
= v_{\A\to \B,t}^{\mathrm{suf}}[l]\sum_{w=0}^{W}\gamma[l+w][l]
\frac{\boldsymbol{x}_{\A\to \B,t}[l][w]}{v_{\A\to \B,t}[l+w]},
\label{mean_suf_B}
\end{equation}
\begin{equation} \label{var_suf_B}
v_{\A\to \B,t}^{\mathrm{suf}}[l] 
= \left(
 \sum_{w=0}^{W}\frac{\gamma^{2}[l+w][l]}{v_{\A\to \B,t}[l+w]}
\right)^{-1}.  
\end{equation}

The third step is computation of posterior messages 
$\boldsymbol{x}_{\B,t+1}^{\mathrm{post}}
=[(\boldsymbol{x}_{\B,t+1}^{\mathrm{post}}[0])^{\mathrm{T}},\ldots,
(\boldsymbol{x}_{\B,t+1}^{\mathrm{post}}[L-1])^{\mathrm{T}}]^{\mathrm{T}}$ 
based on separable denoisers $\{f_{t}[l]\}$, given by 
\begin{equation} \label{mean_post_B}
\boldsymbol{x}_{\B,t+1}^{\mathrm{post}}[l] 
= f_{t}[l](\boldsymbol{x}_{\A\to\B,t}^{\mathrm{suf}}[l]). 
\end{equation}  
The corresponding variance $v_{\B,t+1}^{\mathrm{post}}[l]$ needs to be 
a consistent estimator of $N^{-1}[l]\mathbb{E}[\|\boldsymbol{x}[l]
-\boldsymbol{x}_{\B,t+1}^{\mathrm{post}}[l]\|^{2}]$ in the large system limit. 
See \cite[Eq.~(23)]{Takeuchi213} for the details. 

For the signal vector $\boldsymbol{x}[l]$ with i.i.d.\ elements, 
consider the Bayes-optimal denoiser $f_{t}[l](u)=
f_{\mathrm{opt}}(u;v_{\A\to \B,t}^{\mathrm{suf}}[l])$, with 
\begin{equation} \label{opt_denoiser}
f_{\mathrm{opt}}(u;v_{\A\to \B,t}^{\mathrm{suf}}[l])
=\mathbb{E}[x_{1}[l] | u=x_{1}[l]+z_{t}[l]],
\end{equation} 
where $z_{t}[l]\sim\mathcal{N}(0,v_{\A\to \B,t}^{\mathrm{suf}}[l])$ denotes 
a zero-mean Gaussian random variable with variance 
$v_{\A\to \B,t}^{\mathrm{suf}}[l]$ and independent of $x_{1}[l]$. 
This definition is justified via state evolution. 
In this case, the following posterior variance should be used:  
\begin{IEEEeqnarray}{rl} 
v_{\B,t+1}^{\mathrm{post}}[l]=\frac{1}{N[l]}\mathbb{E}&\left[
 \left\|
  \boldsymbol{x}[l] 
  - f_{\mathrm{opt}}(\boldsymbol{x}_{\A\to\B,t}^{\mathrm{suf}}[l]
  ; v_{\A\to \B,t}^{\mathrm{suf}}[l])
 \right\|^{2}
\right. \nonumber \\
& \Bigl| 
\boldsymbol{x}_{\A\to\B,t}^{\mathrm{suf}}[l], v_{\A\to \B,t}^{\mathrm{suf}}[l]
\Bigr]. 
\label{var_post_B} 
\end{IEEEeqnarray}

The last step is the Onsager correction of the posterior mean 
$|\mathcal{W}[\ell]|^{1/2}\boldsymbol{\Gamma}[\ell]
\boldsymbol{x}_{\B,t+1}^{\mathrm{post}}\in\mathbb{R}^{N_{\mathrm{c}}[\ell]}$ 
in the extended space to realize asymptotic Gaussianity in module~A. Let
\begin{equation} \label{eta_B}
\eta_{\B,t}[\ell]
= \sum_{w\in\mathcal{W}[\ell]}\frac{N[\ell-w]}{N_{\mathrm{c}}[\ell]}
\eta_{\B,t}[\ell][w],   
\end{equation} 
with 
\begin{IEEEeqnarray}{rl} 
\eta_{\B,t}[\ell][w] 
=& \frac{|\mathcal{W}[\ell]|\gamma^{2}[\ell][\ell-w]
v_{\A\to \B,t}^{\mathrm{suf}}[\ell-w]}
{v_{\A\to\B,t}[\ell]}
\nonumber \\
&\cdot\langle f'_{t}[\ell-w]
(\boldsymbol{x}_{\A\to\B,t}^{\mathrm{suf}}[\ell-w])\rangle.
\label{eta_B_w}
\end{IEEEeqnarray}
The normalized message $|\mathcal{W}[\ell]|^{-1/2}\eta_{\B,t}[\ell][w]$ is 
the average of the partial derivative 
of the $n$th element in the $w$th section of $|\mathcal{W}[\ell]|^{1/2}
\boldsymbol{\Gamma}[\ell]\boldsymbol{x}_{\B,t+1}^{\mathrm{post}}$ 
with respect to $\vec{x}_{\A\to\B,n,t}[\ell][w]$ over all $n\in N[\ell-w]$. 
The extrinsic mean $\vec{\boldsymbol{x}}_{\B\to\A,t+1}[\ell]
\in\mathbb{R}^{N_{\mathrm{c}}[\ell]}$ and variance $v_{\B\to\A,t+1}[\ell]>0$ of 
$\vec{\boldsymbol{x}}[\ell]$ in the extended spaces are computed as   
\begin{IEEEeqnarray}{rl} 
&(1 - \eta_{\B,t}[\ell]/|\mathcal{W}[\ell]|)
\vec{\boldsymbol{x}}_{\B\to\A,t+1}[\ell] 
\nonumber \\
=& |\mathcal{W}[\ell]|^{1/2}\boldsymbol{\Gamma}[\ell]
\boldsymbol{x}_{\B,t+1}^{\mathrm{post}} 
- |\mathcal{W}[\ell]|^{-1/2}\eta_{\B,t}[\ell]
\vec{\boldsymbol{x}}_{\A\to\B,t}[\ell], 
\label{mean_BA}
\end{IEEEeqnarray}
\begin{IEEEeqnarray}{rl}
&(1 - \eta_{\B,t}[\ell]/|\mathcal{W}[\ell]|)^{2}v_{\B\to\A,t+1}[\ell] 
\nonumber \\
=& \sum_{w\in\mathcal{W}[\ell]}\frac{N[\ell-w]}{N_{\mathrm{c}}[\ell]}
|\mathcal{W}[\ell]|\gamma^{2}[\ell][\ell-w]v_{\B,t+1}^{\mathrm{post}}[\ell-w] 
\nonumber \\
&- \frac{1}{|\mathcal{W}[\ell]|}\eta_{\B,t}^{2}[\ell]v_{\A\to\B,t}[\ell].  
\label{var_BA} 
\end{IEEEeqnarray}

These non-trivial messages have been designed so as to realize asymptotic 
Gaussianity via state evolution. To confirm why the Onsager correction 
realizes asymptotic Gaussianity, one needs to understand a general error model 
proposed in Appendix~\ref{proof_theorem_SE_LM}. The Onsager correction is a 
natural definition in terms of the general error model. 

For the Bayes-optimal denoiser 
$f_{\mathrm{opt}}(u;v_{\A\to \B,t}^{\mathrm{suf}}[l])$ in (\ref{opt_denoiser}), 
we use (\ref{eta_B}), (\ref{eta_B_w}), 
and the well-known identity $\langle \{f_{\mathrm{opt}}
(\boldsymbol{x}_{\A\to\B,t}^{\mathrm{suf}}[l];v_{\A\to \B,t}^{\mathrm{suf}}[l])\}'\rangle
=v_{\B,t+1}^{\mathrm{post}}[l]/v_{\A\to\B,t}^{\mathrm{suf}}[l]$, with 
$v_{\B,t+1}^{\mathrm{post}}[l]$ given in (\ref{var_post_B}), to find that 
the extrinsic variance~(\ref{var_BA}) reduces to 
\begin{equation} \label{var_BA_opt}
v_{\B\to\A,t+1}[\ell] 
= \frac{\eta_{\B,t}[\ell]v_{\A\to\B,t}[\ell]}
{1 - \eta_{\B,t}[\ell]/|\mathcal{W}[\ell]|}, 
\end{equation}
with 
\begin{equation}
\eta_{\B,t}[\ell]
= \sum_{w\in\mathcal{W}[\ell]}\frac{N[\ell-w]}{N_{\mathrm{c}}[\ell]}
\frac{|\mathcal{W}[\ell]|\gamma^{2}[\ell][\ell-w]
v_{\B,t+1}^{\mathrm{post}}[\ell-w]}{v_{\A\to\B,t}[\ell]}.  
\end{equation}

When the uniform coupling weights~(\ref{uniform_coupling}) are considered, 
$\eta_{\B,t}[\ell]$ is ${\cal O}(1)$ as $W\to\infty$. Thus, the extrinsic 
variance~(\ref{var_BA_opt}) tends to $v_{\B\to\A,t+1}[\ell] 
\to\eta_{\B,t}[\ell]v_{\A\to\B,t}[\ell]$ as $W\to\infty$. This convergence and 
$\eta_{\A,t}[\ell]\to1$ are key properties to prove the information-theoretic 
optimality of Bayes-optimal OAMP.

\section{Main Results} \label{sec5}
State evolution analysis is presented for OAMP in the spatially 
coupled system~(\ref{vector_model}). This paper extends a unified 
framework~\cite{Takeuchi211} of state evolution to that for the spatially 
coupled system. A general error model for the spatially coupled system is 
proposed and analyzed in the large system limit---$M[\ell]$ and $N[l]$ 
tend to infinity for all $\ell$ and $l$ while the ratio 
$\alpha[\ell][l]=N[l]/M[\ell]$ kept constant. By proving that the 
proposed general error model contains the error model for OAMP 
in the spatially coupled system, we derive state evolution recursions for 
OAMP. 

In state evolution analysis, we postulate the following assumptions: 
\begin{assumption} \label{assumption_x} 
For some $\epsilon>0$, the signal vector 
$\boldsymbol{x}\in\mathbb{R}^{N_{\mathrm{all}}}$ 
in (\ref{signal_representation}) has i.i.d.\ elements with zero mean, 
unit variance, and a bounded $(2+\epsilon)$th moment.  
\end{assumption}

Assumption~\ref{assumption_x} simplifies state evolution analysis. 
Non-separable denoising~\cite{Berthier19,Ma19,Fletcher19} might be needed 
if dependent signal elements were considered. 

\begin{assumption} \label{assumption_A}
The rescaled row section $|\mathcal{W}[\ell]|^{1/2}\boldsymbol{A}[\ell]$ 
given via (\ref{A_row}) is picked up from 
$(M[\ell],N_{\mathrm{c}}[\ell])$-ensemble 
in Definition~\ref{def1} uniformly and randomly.  
\end{assumption}

An important point in Assumption~\ref{assumption_A} is that the 
right-orthogonal invariance in Definition~\ref{def1} is assumed not for 
each section $\boldsymbol{A}[\ell][l]$ but each row section 
$\boldsymbol{A}[\ell]$ in (\ref{A_row}). This assumption allows us to 
analyze the dynamics of OAMP via rigorous state evolution. 

Another important point is that the empirical eigenvalue distribution of 
$\boldsymbol{A}^{\mathrm{T}}[\ell]\boldsymbol{A}[\ell]$ converges almost surely 
to a compactly supported distribution 
in $(M[\ell],N_{\mathrm{c}}[\ell])$-ensemble. As a result, all moments of the 
asymptotic eigenvalue distribution are bounded. Thus, we need not investigate 
the boundedness of the moments to use technical results in Section~\ref{sec2}.

\begin{assumption} \label{assumption_w}
The noise vectors $\{\boldsymbol{n}[\ell]: \ell\in\mathcal{L}_{W}\}$ in 
(\ref{vector_model}) are independent vectors. 
Each vector $\boldsymbol{n}[\ell]$ satisfies orthogonal invariance, 
$\lim_{M[\ell]\to\infty}M^{-1}[\ell]\|\boldsymbol{n}[\ell]\|^{2}\aeq\sigma^{2}>0$, 
and bounded $(2+\epsilon)$th moments for some $\epsilon>0$.    
\end{assumption}

The AWGN vector $\boldsymbol{n}[\ell]\sim\mathcal{N}(\boldsymbol{0},
\sigma^{2}\boldsymbol{I}_{M[\ell]})$ with variance $\sigma^{2}$ satisfies 
Assumption~\ref{assumption_w}. The orthogonal invariance in 
$\boldsymbol{n}[\ell]$ may be induced via the left-orthogonal invariance of 
the row section $\boldsymbol{A}[\ell]$. 

\begin{assumption} \label{assumption_filter}
The linear filter $\boldsymbol{F}_{t}[\ell]$ in module~A has the same SVD 
structure $\boldsymbol{F}_{t}[\ell]=\boldsymbol{U}[\ell]
\boldsymbol{\Sigma}_{\boldsymbol{F}_{t}[\ell]}\boldsymbol{V}^{\mathrm{T}}[\ell]$ as 
the SVD $\boldsymbol{A}[\ell]=\boldsymbol{U}[\ell]\boldsymbol{\Sigma}[\ell]
\boldsymbol{V}^{\mathrm{T}}[\ell]$, in which 
$\boldsymbol{\Sigma}_{\boldsymbol{F}_{t}[\ell]}^{\mathrm{T}}
\boldsymbol{\Sigma}_{\boldsymbol{F}_{t}[\ell]}$ is in the space spanned by  
$\{(\boldsymbol{\Sigma}^{\mathrm{T}}[\ell]
\boldsymbol{\Sigma}[\ell])^{j}\}_{j=0}^{\infty}$. 
\end{assumption}

Assumption~\ref{assumption_filter} contains practical linear filters, such as 
the LMMSE filter~(\ref{LMMSE_LM}), the MF 
$\boldsymbol{F}_{t}[\ell]=\boldsymbol{A}[\ell]$, and the zero-forcing (ZF) 
filter $\boldsymbol{F}_{t}[\ell]=\boldsymbol{A}^{\dagger}[\ell]$.  

\begin{assumption} \label{assumption_denoiser}
The scalar denoiser $f_{t}[l]$ in module~B is Lipschitz-continuous and 
nonlinear. 
\end{assumption}

The Lipschitz-continuity is the standard assumption in state evolution 
analysis. The nonlinearity is required to prevent module~B from outputting 
$\vec{\boldsymbol{x}}_{\B\to\A,t}[\ell]=\boldsymbol{0}$. This situation occurs 
in Bayes-optimal OAMP for Gaussian signaling since module~B in the initial 
iteration can compute the Bayes-optimal estimator of the signal vector. Thus, 
the nonlinearity should be regarded as an assumption to exclude the trivial 
case in which no iterations are needed. 

We first define state evolution recursions for OAMP 
in the spatially coupled system~(\ref{vector_model}), derived via 
state evolution. 
Let $\alpha_{\mathrm{c}}[\ell]=N_{\mathrm{c}}[\ell]/M[\ell]=\sum_{l=\ell-W}^{\ell}
\alpha[\ell][l]$. State evolution recursions for module~A 
with the initial condition $\bar{v}_{\B\to\A,0}[\ell]
=|\mathcal{W}[\ell]|\alpha_{\mathrm{c}}^{-1}[\ell]
\sum_{w\in\mathcal{W}[\ell]}\alpha[\ell][\ell-w]
\gamma^{2}[\ell][\ell-w]$ are given by 
\begin{IEEEeqnarray}{l}
\bar{v}_{\A,t}^{\mathrm{post}}[\ell] 
= \lim_{\{M[\ell], N[l]\}\to\infty}
\frac{\sigma^{2}}{N_{\mathrm{c}}[\ell]}\mathrm{Tr}\left(
 \boldsymbol{F}_{t}^{\mathrm{T}}[\ell]\boldsymbol{F}_{t}[\ell]
\right) \nonumber \\
+ \frac{\bar{v}_{\B\to\A,t}[\ell]}{N_{\mathrm{c}}[\ell]}\mathrm{Tr}\left\{
 (\boldsymbol{I} - \boldsymbol{F}_{t}^{\mathrm{T}}[\ell]\boldsymbol{A}[\ell])
 (\boldsymbol{I} - \boldsymbol{F}_{t}^{\mathrm{T}}[\ell]
 \boldsymbol{A}[\ell])^{\mathrm{T}}
\right\},
\nonumber \\
\label{var_post_A_SE}
\end{IEEEeqnarray}
\begin{equation} \label{var_AB_SE}
\bar{v}_{\A\to\B,t}[\ell]  
= \frac{1}{|\mathcal{W}[\ell]|}
\frac{\bar{v}_{\A,t}^{\mathrm{post}}[\ell] 
- \bar{\eta}_{\A,t}^{2}[\ell]\bar{v}_{\B\to\A,t}[\ell]}
{(1-\bar{\eta}_{\A,t}[\ell])^{2}}, 
\end{equation}
where the limit in (\ref{var_post_A_SE}) represents the large system limit, 
with 
\begin{equation} \label{eta_A_SE}
\bar{\eta}_{\A,t}[\ell]
= \lim_{\{M[\ell], N[l]\}\to\infty}\frac{1}{N_{\mathrm{c}}[\ell]}\mathrm{Tr}\left(
 \boldsymbol{I}_{N_{\mathrm{c}}[\ell]} 
 - \boldsymbol{F}_{t}^{\mathrm{T}}[\ell]\boldsymbol{A}[\ell]
\right). 
\end{equation}
The variables $\bar{v}_{\A,t}^{\mathrm{post}}[\ell]$ and 
$\bar{\eta}_{\A,t}[\ell]$ converge almost surely to deterministic quantities  
from Assumptions~\ref{assumption_A} and \ref{assumption_filter}. 
They can be evaluated in closed form if the asymptotic eigenvalue 
distribution of $\boldsymbol{A}^{\mathrm{T}}[\ell]\boldsymbol{A}[\ell]$ has 
a closed-form expression. 

State evolution recursions for module~B are given by 
\begin{equation}  \label{var_suf_B_SE}
\bar{v}_{\A\to \B,t}^{\mathrm{suf}}[l] 
= \left(
 \sum_{w=0}^{W}\frac{\gamma^{2}[l+w][l]}
 {\bar{v}_{\A\to \B,t}[l+w]}
\right)^{-1},  
\end{equation}
\begin{equation} \label{var_post_B_SE}
\bar{v}_{\B,t+1}^{\mathrm{post}}[l] 
= \mathbb{E}\left[
 \{x_{1}[l] - f_{t}[l](x_{1}[l] + z_{t}[l])\}^{2}
\right], 
\end{equation}
\begin{IEEEeqnarray}{rl}
&(1 - \bar{\eta}_{\B,t}[\ell]/|\mathcal{W}[\ell]|)^{2}\bar{v}_{\B\to\A,t+1}[\ell] 
\nonumber \\
=& \sum_{w\in\mathcal{W}[\ell]}\frac{\alpha[\ell][\ell-w]}{\alpha_{\mathrm{c}}[\ell]}
|\mathcal{W}[\ell]|\gamma^{2}[\ell][\ell-w]\bar{v}_{\B,t+1}^{\mathrm{post}}[\ell-w] 
\nonumber \\
&- \frac{1}{|\mathcal{W}[\ell]|}\bar{\eta}_{\B,t}^{2}[\ell]
\bar{v}_{\A\to\B,t}[\ell],
\label{var_BA_SE} 
\end{IEEEeqnarray}
with 
\begin{equation} \label{eta_B_SE}
\bar{\eta}_{\B,t}[\ell]
= \sum_{w\in\mathcal{W}[\ell]}\frac{\alpha[\ell][\ell-w]}{\alpha_{\mathrm{c}}[\ell]}
\bar{\eta}_{\B,t}[\ell][w],   
\end{equation} 
\begin{IEEEeqnarray}{rl} 
\bar{\eta}_{\B,t}[\ell][w] 
=& \frac{|\mathcal{W}[\ell]|\gamma^{2}[\ell][\ell-w]
\bar{v}_{\A\to \B,t}^{\mathrm{suf}}[\ell-w]}{\bar{v}_{\A\to\B,t}[\ell]}
\nonumber \\
&\cdot\mathbb{E}[f'_{t}[\ell-w](x_{1}[\ell-w] + z_{t}[\ell-w])].
\label{eta_B_w_SE}
\end{IEEEeqnarray}
In these expressions, $z_{t}[l]$ denotes a zero-mean Gaussian random 
variable with variance $\bar{v}_{\A\to \B,t}^{\mathrm{suf}}[l]$, 
independent of $x_{1}[l]$. 

In particular, for Bayes-optimal OAMP with the LMMSE filter~(\ref{LMMSE}) 
and the Bayes-optimal denoiser 
$f_{t}[l](u)=f_{\mathrm{opt}}(u; \bar{v}_{\A\to \B,t}^{\mathrm{suf}}[l])$ in 
(\ref{opt_denoiser}), (\ref{var_post_A_SE}) and (\ref{var_AB_SE}) 
reduce to  
\begin{equation} \label{var_AB_SE_Bayes}
\bar{v}_{\A\to\B,t}[\ell] 
= \frac{1}{|\mathcal{W}[\ell]|}
\frac{\bar{\eta}_{\A,t}[\ell]\bar{v}_{\B\to\A,t}[\ell]}
{1-\bar{\eta}_{\A,t}[\ell]},
\end{equation}
where $\bar{\eta}_{\A,t}[\ell]$ is defined in (\ref{eta_A_SE}), with
\begin{equation}  \label{LMMSE_SE_OAMP} 
\boldsymbol{F}_{t}[\ell]
=\bar{v}_{\B\to\A,t}[\ell]\boldsymbol{\Xi}_{t}^{-1}[\ell]\boldsymbol{A}[\ell],
\end{equation}
\begin{equation} 
\boldsymbol{\Xi}_{t}[\ell]
= \sigma^{2}\boldsymbol{I}_{M[\ell]} 
+ \bar{v}_{\B\to\A,t}[\ell]\boldsymbol{A}[\ell]\boldsymbol{A}^{\mathrm{T}}[\ell]. 
\end{equation}
Furthermore, (\ref{eta_B_SE}) and (\ref{var_BA_SE}) reduce to  
\begin{equation} \label{eta_B_SE_Bayes}
\bar{\eta}_{\B,t}[\ell]
= \sum_{w\in\mathcal{W}[\ell]}\frac{\alpha[\ell][\ell-w]}{\alpha_{\mathrm{c}}[\ell]}
\frac{|\mathcal{W}[\ell]|\gamma^{2}[\ell][\ell-w]
\bar{v}_{\B,t+1}^{\mathrm{post}}[\ell-w]}
{\bar{v}_{\A\to\B,t}[\ell]},
\end{equation}
\begin{equation} \label{var_BA_SE_Bayes}
\bar{v}_{\B\to\A,t+1}[\ell] 
= \frac{\bar{\eta}_{\B,t}[\ell]\bar{v}_{\A\to\B,t}[\ell]}
{1 - \bar{\eta}_{\B,t}[\ell]/|\mathcal{W}[\ell]|}, 
\end{equation}
with $\bar{v}_{\B,t+1}^{\mathrm{post}}[l]$ defined in (\ref{var_post_B_SE}). 

The state evolution recursions are the asymptotic counterpart to 
the variance messages in OAMP. Rather, the variance messages have been 
designed such that they become consistent estimators of the variables in the 
state evolution recursions. 

\begin{theorem} \label{theorem_SE}
Suppose that Assumptions~\ref{assumption_x}--\ref{assumption_denoiser} hold. 
\begin{itemize}
\item The MSE $N^{-1}[l]\|\boldsymbol{x}[l]
-\boldsymbol{x}_{\B,t}[l]\|^{2}$ for OAMP converges almost surely to 
$\bar{v}_{\B,t}^{\mathrm{post}}[l]$ in the large system limit, in which 
$\bar{v}_{\B,t}^{\mathrm{post}}[l]$ is given via the state evolution 
recursions~(\ref{var_post_A_SE})--(\ref{var_BA_SE}). 
\item Consider the LMMSE filter~(\ref{LMMSE}) and the Bayes-optimal 
denoiser~(\ref{opt_denoiser}). Then, the state evolution 
recursions~(\ref{var_AB_SE_Bayes})--(\ref{var_BA_SE_Bayes}) 
for Bayes-optimal OAMP converge to a fixed point as $t\to\infty$. 
\end{itemize}  
\end{theorem}
\begin{IEEEproof}
See Section~\ref{proof_theorem_SE}. 
\end{IEEEproof}

Theorem~\ref{theorem_SE} implies asymptotic Gaussianity for the estimation 
error $\boldsymbol{x}[l]-\boldsymbol{x}_{\B,t}[l]$: The MSE 
$N^{-1}[l]\|\boldsymbol{x}[l]-\boldsymbol{x}_{\B,t}[l]\|^{2}$ converges 
almost surely to $\bar{v}_{\B,t}^{\mathrm{post}}[l]$ in 
(\ref{var_post_B_SE})---given via the Gaussian random variable $z_{t}[l]$. 
The asymptotic Gaussianity implies that the Bayes-optimal denoiser 
$f_{\mathrm{opt}}(u;v_{\A\to \B,t}^{\mathrm{suf}}[l])$ given in (\ref{opt_denoiser}) 
minimizes the asymptotic MSE $\bar{v}_{\B,t+1}^{\mathrm{post}}[l]$. 
In this sense, OAMP with the Bayes-optimal denoiser, as well as 
the LMMSE filter, is called Bayes-optimal OAMP. 

We next prove the information-theoretic optimality of 
Bayes-optimal OAMP in terms of the R\'enyi information dimension. 
To use existing results~\cite{Yedla14,Takeuchi15} on spatial coupling, 
we assume $M[\ell]=M$, $N[l]=N$, and 
the uniform coupling weights~(\ref{uniform_coupling}). In this case, 
$\bar{v}_{\A\to \B,t}^{\mathrm{suf}}[l]$ and 
$\bar{\eta}_{\B,t}[\ell]$ given in (\ref{var_suf_B_SE}) and 
(\ref{eta_B_SE_Bayes}) for Bayes-optimal OAMP reduce to 
\begin{equation}  \label{var_suf_B_SE_uniform}
\bar{v}_{\A\to \B,t}^{\mathrm{suf}}[l] 
= \left(
 \frac{1}{W+1}\sum_{w=0}^{W}\frac{1}{\bar{v}_{\A\to \B,t}[l+w]}
\right)^{-1},  
\end{equation}
\begin{equation} \label{eta_B_SE_Bayes_uniform}
\bar{\eta}_{\B,t}[\ell]
= \frac{1}{W+1}\sum_{w\in\mathcal{W}[\ell]}
\frac{\bar{v}_{\B,t+1}^{\mathrm{post}}[\ell-w]}
{\bar{v}_{\A\to\B,t}[\ell]},   
\end{equation}
respectively. 

The overall compression rate 
$N_{\mathrm{all}}^{-1}\sum_{\ell\in\mathcal{L}_{W}}M[\ell]$ tends to 
$(1+\Delta)\delta$ in the continuum limit $L, W\to\infty$, 
with $\Delta=W/L$ kept constant, after taking the large system limit $M, 
N\to\infty$ with $\delta=M/N$ kept constant. Thus, the overall compression 
rate converges to $\delta$ as $\Delta\downarrow0$. 

The continuum limit was originally considered in \cite{Donoho13,Takeuchi15} 
to obtain an exact continuum approximation of state evolution recursions for 
AMP. This paper takes the same limit to utilize an existing result in 
\cite{Takeuchi15}. 

To prove the information-theoretic optimality of Bayes-optimal OAMP, 
we need the following results: 
\begin{proposition} \label{proposition_W}
Let $\{a_{W}>0\}_{W=1}^{\infty}$ denote a positive and diverging sequence 
at a sublinear speed in $W$: $\lim_{W\to\infty}a_{W}=\infty$ and 
$\lim_{W\to\infty}a_{W}/W=0$. Then, $|\mathcal{W}[\ell]|=W+1$ and 
$|\mathcal{W}[\ell]|\geq1 + W/a_{W}$ hold 
for all $\ell\in\{W,\ldots,L-1\}$ 
and $\ell\in\{\lceil W/a_{W}\rceil,\ldots,W-1\}
\cup\{L,\ldots,L+W-1-\lceil W/a_{W}\rceil\}$ 
in the continuum limit, respectively. 
\end{proposition}
\begin{IEEEproof}
For $\ell\in\{W,\ldots,L-1\}$, 
we use the definition of $\mathcal{W}[\ell]$ in (\ref{set_W}) to have 
$|\mathcal{W}[\ell]|=W+1$. For $\ell\in\{\lceil W/a_{W}\rceil,\ldots,
W-1\}$, similarly, we obtain $|\mathcal{W}[\ell]|=\ell+1\geq W/a_{W} + 1$. 
For $\ell\in\{L,\ldots,L+W-1-\lceil W/a_{W}\rceil\}$, we have 
$|\mathcal{W}[\ell]|=L+W-\ell\geq1+W/a_{W}$. 
\end{IEEEproof}

Proposition~\ref{proposition_W} implies 
$|\mathcal{W}[\ell]|\geq W/a_{W}\to\infty$ for all 
$\ell\in\{\lceil W/a_{W}\rceil, \ldots, L+W-1-\lceil W/a_{W}\rceil\}$. 
It is used to control approximation errors of the state evolution 
recursions in the boundaries.  

\begin{theorem} \label{theorem_optimality} 
Consider $M[\ell]=M$, $N[\ell]=N$, and the uniform coupling 
weights~(\ref{uniform_coupling}). Suppose that Assumptions~\ref{assumption_x} 
and \ref{assumption_A} hold. Let $\{a_{W}>0\}_{W=1}^{\infty}$ 
denote a positive and diverging sequence at a sublinear speed in $W$: 
$\lim_{W\to\infty}a_{W}=\infty$ and $\lim_{W\to\infty}a_{W}/W=0$. 
Suppose that there is some function $R(z)$ such that 
the R-transform of $\boldsymbol{G}[\ell]=|\mathcal{W}[\ell]|
\boldsymbol{A}^{\mathrm{T}}[\ell]\boldsymbol{A}[\ell]$ satisfies 
\begin{equation} \label{R_transform_limit} 
\lim_{W=\Delta L\to\infty}a_{W}\left|
 R_{\boldsymbol{G}[\ell]}\left(
  \frac{z}{|\mathcal{W}[\ell]|}
 \right)
 - R(z)
\right| <\infty 
\end{equation}
for all $\ell\in\{\lceil W/a_{W}\rceil,\ldots,L+W-1-\lceil W/a_{W}\rceil\}$. 
Furthermore, assume the following conditions: 
\begin{itemize}
\item The signal $x_{1}$ has the R\'enyi information dimension $d_{\mathrm{I}}$. 
\item $R(z)$ is proper, twice continuously differentiable, strictly increasing, 
and positive for all $z\leq0$. 
\item $\lim_{z\to\infty}zR(-z)=\delta$ holds. 
\end{itemize}
Let $E_{\mathrm{opt}}>0$ denotes the global minimizer of the replica-symmetric 
potential $f_{\mathrm{RS}}(E,s)$ in (\ref{RS_potential}) with 
$R_{\boldsymbol{A}^{\mathrm{T}}\boldsymbol{A}}(z)=R(z)$ and 
$s=R(-E/\sigma^{2})/\sigma^{2}$. If $E_{\mathrm{opt}}$ is unique, 
then, the state evolution recursions for Bayes-optimal OAMP satisfies 
\begin{equation} \label{zero_MSE}
\lim_{\Delta\downarrow0}\lim_{t\to\infty}
\lim_{W=\Delta L\to\infty}\frac{1}{|\mathcal{L}_{0}|}\sum_{l\in\mathcal{L}_{0}}
\bar{v}_{\B,t}^{\mathrm{post}}[l]
\leq E_{\mathrm{opt}}.
\end{equation}
In particular, $E_{\mathrm{opt}}$ is unique and tends to zero 
as $\sigma^{2}\downarrow0$ if the ratio $\delta$ is larger than $d_{\mathrm{I}}$. 
\end{theorem}
\begin{IEEEproof}
See Section~\ref{sec_proof_optimality}. 
\end{IEEEproof}

Theorem~\ref{theorem_optimality} implies that Bayes-optimal OAMP for the 
spatially coupled system~(\ref{vector_model}) can achieve the 
Bayes-optimal MSE performance $E_{\mathrm{opt}}$ for the uncoupled 
system~(\ref{uncoupled_model}) with the R-transform of the sensing matrix 
given by $R(z)$. Furthermore, Bayes-optimal OAMP for the spatially coupled 
system can achieve the information-theoretic compression limit $d_{\mathrm{I}}$. 

Let us investigate the relationship between the original R-transform 
$R_{\boldsymbol{G}[\ell]}$ and $R$ for characterizing the performance of 
Bayes-optimal OAMP. 
The ensemble of zero-mean i.i.d.\ Gaussian matrices satisfies 
the conditions in Theorem~\ref{theorem_optimality}. 

\begin{corollary} \label{corollary3}
Consider $M[\ell]=M$, $N[\ell]=N$, and the uniform coupling 
weights~(\ref{uniform_coupling}). Suppose that Assumption~\ref{assumption_x} 
holds and that $|\mathcal{W}[\ell]|^{1/2}
\boldsymbol{A}[\ell]$ is picked up from the ensemble of 
zero-mean i.i.d.\ Gaussian matrices with variance $1/M$. 
Furthermore, assume that 
the signal $x_{1}$ has the R\'enyi information dimension $d_{\mathrm{I}}$. 
Then, the condition~(\ref{R_transform_limit}) in 
Theorem~\ref{theorem_optimality} holds for $R(z)=\delta/(\delta-z)$. 
Furthermore, the state evolution recursions for Bayes-optimal OAMP satisfies 
(\ref{zero_MSE}) if $E_{\mathrm{opt}}$ is unique. In particular, 
$E_{\mathrm{opt}}$ is unique and tends to zero as $\sigma^{2}\downarrow0$ 
if the ratio $\delta$ is larger than $d_{\mathrm{I}}$. 
\end{corollary}
\begin{IEEEproof}
As shown in the proof of Corollary~\ref{corollary1}, we have 
\begin{equation}
R_{\boldsymbol{G}[\ell]}(z)
= \frac{|\mathcal{W}[\ell]|^{-1}\delta}{|\mathcal{W}[\ell]|^{-1}\delta-z},
\end{equation}
which implies that the R-transform 
$R(z)=\delta/(\delta-z)$ satisfies the 
condition~(\ref{R_transform_limit}) in Theorem~\ref{theorem_optimality}. 
It is straightforward to confirm that $R(z)$ satisfies all conditions 
in Theorem~\ref{theorem_optimality}. Thus, Corollary~\ref{corollary3} holds. 
\end{IEEEproof}

Corollary~\ref{corollary3} implies that $R(z)$ is equal to the R-transform 
of zero-mean i.i.d.\ Gaussian sensing matrices without spatial coupling. 
Thus, Bayes-optimal OAMP for the spatially 
coupled system~(\ref{vector_model}) achieves the Bayes-optimal performance 
for the underlying system~(\ref{uncoupled_model}) without spatial coupling when 
zero-mean i.i.d.\ Gaussian sensing matrices are considered. 

We next show that $R(z)$ does not necessarily coincide with 
$R_{\boldsymbol{G}[\ell]}(z)$ in general. To present examples in which they are 
different, the following result is useful when the $\eta$-transform of 
$\boldsymbol{G}[\ell]$ can be evaluated explicitly: 

\begin{corollary} \label{corollary4}
Consider $M[\ell]=M$, $N[\ell]=N$, and the uniform coupling 
weights~(\ref{uniform_coupling}). Suppose that Assumptions~\ref{assumption_x} 
and \ref{assumption_A} hold. Let $r[\ell]$ denote the rank of 
$\boldsymbol{G}[\ell]$ and suppose that the ratio $r[\ell]/N$ tends to 
$\delta$ in the large system limit. Let $\{a_{W}>0\}_{W=1}^{\infty}$ 
denote a positive and diverging sequence at a sublinear speed in $W$: 
$\lim_{W\to\infty}a_{W}=\infty$ and $\lim_{W\to\infty}a_{W}/W=0$. 
Assume that there is some bounded function $\eta(z)$ such that 
the positive eigenvalues $\{\lambda_{n}[\ell]>0\}$ of $\boldsymbol{G}[\ell]$ 
satisfies 
\begin{IEEEeqnarray}{r} 
\lim_{W=\Delta L\to\infty}a_{W}\left|
 \lim_{M=\delta N\to\infty}\frac{1}{N}\sum_{n=1}^{r[\ell]}
 \frac{1}{1+\lambda_{n}[\ell]z/|\mathcal{W}[\ell]|}
\right. \nonumber \\
- \{\eta(z) - 1 + \delta\}
\Biggr| 
<\infty
\label{eta_transform_limit}
\end{IEEEeqnarray}
for all $\ell\in\{\lceil W/a_{W}\rceil,\ldots,L+W-1-\lceil W/a_{W}\rceil\}$. 
Furthermore, let $R(z)=\{\eta(-z)-1\}/z$ for $z<0$, with $R(0)=1$, and 
assume the following conditions: 
\begin{itemize}
\item The signal $x_{1}$ has the R\'enyi information dimension $d_{\mathrm{I}}$. 
\item $R(z)$ is proper, twice continuously 
differentiable, strictly increasing, and positive for all $z\leq0$. 
\end{itemize}
Then, $R(z)$ satisfies the condition~(\ref{R_transform_limit}) in 
Theorem~\ref{theorem_optimality}. 
Furthermore, the state evolution recursions for Bayes-optimal OAMP satisfies 
(\ref{zero_MSE}) if $E_{\mathrm{opt}}$ is unique. In particular, 
$E_{\mathrm{opt}}$ is unique and tends to zero as $\sigma^{2}\downarrow0$ 
if the ratio $\delta$ is larger than $d_{\mathrm{I}}$.
\end{corollary}
\begin{IEEEproof}
See Appendix~\ref{proof_corollary4}
\end{IEEEproof}

From Corollary~\ref{corollary4} we obtain two corollaries, of which the 
former is for sensing matrices with orthogonal rows while the latter considers 
sensing matrices with condition number larger than $1$.  

\begin{corollary} \label{corollary5} 
Consider $M[\ell]=M$, $N[\ell]=N$, and the uniform coupling 
weights~(\ref{uniform_coupling}). Suppose that Assumption~\ref{assumption_x} 
holds and 
that $|\mathcal{W}[\ell]|^{1/2}\boldsymbol{A}[\ell]\in
\mathbb{R}^{M\times |\mathcal{W}[\ell]|N}$ is right-orthogonally invariant and 
has the singular value $\sqrt{|\mathcal{W}[\ell]|/\delta}$ 
with multiplicity $M$, i.e.\ $\boldsymbol{A}[\ell]$ has orthogonal rows. 
Assume that the signal $x_{1}$ has the R\'enyi information dimension 
$d_{\mathrm{I}}$. Then, the condition~(\ref{R_transform_limit}) in 
Theorem~\ref{theorem_optimality} holds for $R(z)=\delta/(\delta-z)$. 
Furthermore, the state evolution recursions for Bayes-optimal OAMP satisfies 
(\ref{zero_MSE}) if $E_{\mathrm{opt}}$ is unique. In particular, $E_{\mathrm{opt}}$ 
is unique and tends to zero as $\sigma^{2}\downarrow0$ 
if $\delta$ is larger than $d_{\mathrm{I}}$. 
\end{corollary}
\begin{IEEEproof}
We prove the condition~(\ref{eta_transform_limit}) in 
Corollary~\ref{corollary4}. Using $r[\ell]/N\to\delta$ and 
$\lambda_{n}=|\mathcal{W}[\ell]|/\delta$ yields 
\begin{equation}
\lim_{M=\delta N\to\infty}\frac{1}{N}\sum_{n=1}^{r[\ell]}
 \frac{1}{1+\lambda_{n}[\ell]z/|\mathcal{W}[\ell]|}
= \frac{\delta}{1+\delta^{-1}z}.
\end{equation}
Since the condition~(\ref{eta_transform_limit}) in 
Corollary~\ref{corollary4} holds 
for $\eta(z)-1=-\delta z/(\delta + z)$, we arrive at 
\begin{equation}
R(z) = \frac{\eta(-z) - 1}{z} = \frac{\delta }{\delta - z}, 
\end{equation}
which satisfies the assumptions on $R(z)$ in Corollary~\ref{corollary4}. 
Thus, Corollary~\ref{corollary4} implies Corollary~\ref{corollary5}. 
\end{IEEEproof}

Corollary~\ref{corollary5} implies that $R(z)$ is not the R-transform 
of the underlying uncoupled sensing matrix with orthogonal rows but that of 
zero-mean i.i.d.\ Gaussian matrices. Thus, Bayes-optimal OAMP for spatially 
coupled sensing matrices with orthogonal rows achieves the same 
performance as that for spatially coupled zero-mean i.i.d.\ Gaussian sensing 
matrices as long as the continuum limit is considered.

\begin{corollary} \label{corollary6}
Consider $M[\ell]=M$, $N[\ell]=N$, and the uniform coupling 
weights~(\ref{uniform_coupling}). Suppose that Assumption~\ref{assumption_x} 
holds and that 
$|\mathcal{W}[\ell]^{1/2}\boldsymbol{A}[\ell]\in
\mathbb{R}^{M\times |\mathcal{W}[\ell]|N}$ is right-orthogonally invariant and 
has non-zero 
singular values $\sigma_{0}\geq\cdots\geq\sigma_{M-1}>0$ satisfying condition 
number $\kappa=\sigma_{0}/\sigma_{M-1}>1$, 
$\sigma_{m}/\sigma_{m-1}=\kappa^{-1/(M-1)}$, 
and $\sigma_{0}^{2}=|\mathcal{W}[\ell]|N(1-\kappa^{-2/(M-1)})
/(1-\kappa^{-2M/(M-1)})$. Furthermore, assume that 
the signal $x_{1}$ has the R\'enyi information dimension~$d_{\mathrm{I}}$. 
Then, the condition~(\ref{R_transform_limit}) in 
Theorem~\ref{theorem_optimality} holds for 
\begin{equation} \label{R_transform_geometric}
R(z) = \int_{1}^{\kappa^{2}}\frac{dy}{\kappa^{2}-1-Czy},
\end{equation}
with $C=2\delta^{-1}\ln\kappa$. 
The state evolution recursions for Bayes-optimal OAMP satisfies 
(\ref{zero_MSE}) if $E_{\mathrm{opt}}$ is unique. In particular, 
$E_{\mathrm{opt}}$ is unique and tends to zero as $\sigma^{2}\downarrow0$ 
if the ratio $\delta$ is larger than $d_{\mathrm{I}}$.
\end{corollary}
\begin{IEEEproof}
See Appendix~\ref{proof_corollary6}. 
\end{IEEEproof}

Corollary~\ref{corollary6} implies that $R(z)$ in (\ref{R_transform_geometric}) 
is different from the R-transform of the underlying uncoupled sensing matrix 
with condition number $\kappa>1$, which cannot be evaluated explicitly. 

\section{Proof of Theorem~\ref{theorem_SE}} 
\label{proof_theorem_SE}
\subsection{Proof Strategy}
This paper follow \cite{Takeuchi213,Takeuchi22} to prove the latter part in 
Theorem~\ref{theorem_SE}. LM-OAMP for the spatially coupled 
system~(\ref{vector_model}) is proposed as a tool to prove the 
convergence of the state evolution recursions for Bayes-optimal OAMP. 
The convergence of the state evolution recursions for Bayes-optimal OAMP is 
guaranteed by proving their convergence for Bayes-optimal LM-OAMP and 
the reduction of Bayes-optimal LM-OAMP to Bayes-optimal OAMP. 

The former part in Theorem~\ref{theorem_SE} is proved by generalizing 
existing state evolution in \cite{Takeuchi211} to the spatial coupling case. 
To derive state evolution recursions for both OAMP and LM-OAMP, this paper 
establishes a unified framework of state evolution for the spatial coupling 
case. A conventional general error model in \cite{Takeuchi211} is extended 
to that for the spatial coupling case. The proposed general error model 
contains both OAMP and LM-OAMP as instances. The former part in 
Theorem~\ref{theorem_SE} is obtained by proving the asymptotic Gaussianity 
for the general error model via state evolution. 

\subsection{Long-Memory Orthogonal AMP}
\subsubsection{Overview}
We start with the definition of LM-OAMP. 
The main difference between OAMP and LM-OAMP is in the second and last steps 
of module~B: In computing a sufficient statistic for estimation of 
$\boldsymbol{x}$, 
LM-OAMP utilizes all messages in the preceding iterations while OAMP only uses 
the messages in the latest iteration. As a result, the Onsager correction 
in the last step depends on all preceding messages. 
The LM processing in the second step guarantees that the MSE for 
LM-OAMP is monotonically decreasing as the iteration proceeds. 

LM-OAMP requires the covariance between estimation errors for different 
iterations in computing a sufficient statistic for estimation of 
$\boldsymbol{x}$ given all preceding messages. As a result, LM-OAMP computes 
mean and covariance messages in all steps 
while OAMP uses the mean and variance messages. 

For notational convenience, we use the same notation for LM-OAMP as that for 
OAMP. This paper proves that Bayes-optimal LM-OAMP is equivalent to 
Bayes-optimal OAMP in the large system limit. Thus, we do not need to 
distinguish the two algorithms as long as the LMMSE filter and Bayes-optimal 
denoiser are considered. 

Let $\vec{\boldsymbol{x}}_{\B\to \A,t}[\ell]\in\mathbb{R}^{N_{\mathrm{c}}[\ell]}$ and 
$\boldsymbol{V}_{\B\to \A,t}[\ell]\in\mathbb{R}^{(t+1)\times (t+1)}$ denote the mean 
and covariance messages of $\vec{\boldsymbol{x}}[\ell]$ passed from module~B 
to module~A in iteration~$t$, respectively. We write the mean and 
covariance messages of $|\mathcal{W}[\ell]|^{-1/2}\vec{\boldsymbol{x}}[\ell]$ 
passed in the opposite direction as 
$\vec{\boldsymbol{x}}_{\A\to \B,t}[\ell]\in\mathbb{R}^{N_{\mathrm{c}}[\ell]}$ and 
$\boldsymbol{V}_{\A\to \B,t}[\ell]\in\mathbb{R}^{(t+1)\times (t+1)}$.  

The covariance matrix $\boldsymbol{V}_{\B\to \A,t}[\ell]$ has the 
$(\tau',\tau)$ element $v_{\B\to \A,\tau',\tau}[\ell]$, which corresponds to 
a consistent estimator of the error covariance 
$N_{\mathrm{c}}^{-1}[\ell]\mathbb{E}[(\vec{\boldsymbol{x}}[\ell]
-\vec{\boldsymbol{x}}_{\B\to \A,\tau'}[\ell])^{\mathrm{T}}
(\vec{\boldsymbol{x}}[\ell]-\vec{\boldsymbol{x}}_{\B\to \A,\tau}[\ell])]$. 
Similarly, the covariance matrix $\boldsymbol{V}_{\A\to \B,t}[\ell]$ has 
the $(\tau',\tau)$ element $v_{\A\to \B,\tau',\tau}[\ell]$. 

\subsubsection{Module~A (Linear Estimation)}
Module~A in LM-OAMP consists of two steps similar to those in OAMP. The main 
difference is in computing covariance messages, instead of the variance 
messages in OAMP. 

In iteration~$t$, module~A first computes the posterior mean and covariance 
of $\vec{\boldsymbol{x}}[\ell]$ based on a linear filter 
$\boldsymbol{F}_{t}[\ell]\in\mathbb{R}^{M[\ell]\times N_{\mathrm{c}}[\ell]}$ 
for each $\ell\in\mathcal{L}_{W}$. 
The posterior mean $\vec{\boldsymbol{x}}_{\A,t}^{\mathrm{post}}[\ell]
\in\mathbb{R}^{N_{\mathrm{c}}[\ell]}$ in the extended space is defined as 
\begin{equation} \label{mean_post_A_LM}
\vec{\boldsymbol{x}}_{\A,t}^{\mathrm{post}}[\ell] 
= \vec{\boldsymbol{x}}_{\B\to\A,t}[\ell]
+ \boldsymbol{F}_{t}^{\mathrm{T}}[\ell]
(\boldsymbol{y}[\ell] - \boldsymbol{A}[\ell]
\vec{\boldsymbol{x}}_{\B\to\A,t}[\ell]) 
\end{equation}
for $\ell\in\mathcal{L}_{W}$, which is exactly the same as 
(\ref{mean_post_A}) in OAMP. The corresponding posterior covariance 
$v_{\A,t',t}^{\mathrm{post}}[\ell]$ is given by 
\begin{IEEEeqnarray}{l}
v_{\A,t',t}^{\mathrm{post}}[\ell] 
= \frac{\sigma^{2}}{N_{\mathrm{c}}[\ell]}\mathrm{Tr}\left(
 \boldsymbol{F}_{t}^{\mathrm{T}}[\ell]\boldsymbol{F}_{t'}[\ell]
\right) \nonumber \\
+ \frac{v_{\B\to\A,t',t}[\ell]}{N_{\mathrm{c}}[\ell]}\mathrm{Tr}\left\{
 (\boldsymbol{I} - \boldsymbol{F}_{t}^{\mathrm{T}}[\ell]\boldsymbol{A}[\ell])
 (\boldsymbol{I} - \boldsymbol{F}_{t'}^{\mathrm{T}}[\ell]
 \boldsymbol{A}[\ell])^{\mathrm{T}}
\right\} 
\nonumber \\
\label{cov_post_A} 
\end{IEEEeqnarray}
for section~$\ell\in\mathcal{L}_{W}$, which is justified via state evolution. 
If $v_{\B\to\A,t,t}[\ell]$ in LM-OAMP is equal to $v_{\B\to\A,t}[\ell]$ in OAMP, 
we have the identity $v_{\A,t,t}^{\mathrm{post}}[\ell]=v_{\A,t}^{\mathrm{post}}[\ell]$ 
for (\ref{var_post_A}) in OAMP. As is the case in OAMP, the LMMSE filter is 
defined as  
\begin{equation} \label{LMMSE_LM}
\boldsymbol{F}_{t}[\ell]
=v_{\B\to\A,t,t}[\ell]\boldsymbol{\Xi}_{t}^{-1}[\ell]\boldsymbol{A}[\ell],
\end{equation}
with 
\begin{equation} 
\boldsymbol{\Xi}_{t}[\ell]
= \sigma^{2}\boldsymbol{I}_{M[\ell]} 
+ v_{\B\to\A,t,t}[\ell]\boldsymbol{A}[\ell]\boldsymbol{A}^{\mathrm{T}}[\ell]. 
\end{equation}

For $t=0$, 
the initial conditions $\vec{\boldsymbol{x}}_{\B\to\A,0}[\ell]=\boldsymbol{0}$ 
and $v_{\B\to\A,0,0}[\ell]=|\mathcal{W}[\ell]|N_{\mathrm{c}}^{-1}[\ell]
\sum_{w\in\mathcal{W}[\ell]}N[\ell-w]\gamma^{2}[\ell][\ell-w]$ are used. 

The second step is the Onsager correction to realize 
asymptotic Gaussianity in module~B. Module~A computes 
the extrinsic mean $\vec{\boldsymbol{x}}_{\A\to\B,t}[\ell]$ and covariance 
$v_{\A\to\B,t',t}[\ell]$ of 
$|\mathcal{W}[\ell]|^{-1/2}\vec{\boldsymbol{x}}[\ell]$ in the extended space as 
\begin{equation} \label{mean_AB_LM}
\vec{\boldsymbol{x}}_{\A\to\B,t}[\ell] 
= |\mathcal{W}[\ell]^{-1/2}
\frac{\vec{\boldsymbol{x}}_{\A,t}^{\mathrm{post}}[\ell] 
- \eta_{\A,t}[\ell]\vec{\boldsymbol{x}}_{\B\to\A,t}[\ell]}{1 - \eta_{\A,t}[\ell]},
\end{equation}
\begin{equation} \label{cov_AB}
v_{\A\to\B,t',t}[\ell] 
= \frac{1}{|\mathcal{W}[\ell]|}
\frac{v_{\A,t',t}^{\mathrm{post}}[\ell] - \eta_{\A,t'}[\ell]\eta_{\A,t}[\ell]
v_{\B\to\A,t',t}[\ell]}
{(1-\eta_{\A,t'}[\ell])(1-\eta_{\A,t}[\ell])}, 
\end{equation}
with $\eta_{\A,t}[\ell]$ defined in (\ref{eta_A}). 

\subsubsection{Module~B (Nonlinear Estimation)}
Module~B in LM-OAMP consists of four steps similar to those in OAMP. 
The first step is exactly the same as that in OAMP. 
A difference between OAMP and LM-OAMP is in the second step, i.e.\ 
computation of a sufficient statistic based on all preceding messages. 
In the last two steps, LM-OAMP computes posterior/extrinsic mean and 
{\em co}variance messages while OAMP computes posterior/extrinsic mean and 
variance messages. In particular, the Onsager correction in the last step 
uses all preceding messages. 

In iteration~$t$, module~B first extracts information required for 
estimation of $\boldsymbol{x}[l]$ from 
$\{\vec{\boldsymbol{x}}_{\A\to\B,t}[\ell]\}$ in the extended space. This 
computation is the same as (\ref{mean_AB_projection}) in OAMP. 

The second step is computation of a sufficient statistic 
$\boldsymbol{x}_{\A\to \B,t}^{\mathrm{suf}}[l]\in\mathbb{R}^{N[l]}$ 
for estimation of $\boldsymbol{x}[l]$ given all preceding messages 
and the corresponding covariance $\{v_{\A\to \B,t',t}^{\mathrm{suf}}[l]\}$. 
We write the mean messages for section~$l$ passed from module~A to 
module~B in all preceding iterations up to $t$ as 
$\boldsymbol{X}_{\A\to \B,t+1}[l][w]
=(\boldsymbol{x}_{\A\to \B,0}[l][w],\ldots,\boldsymbol{x}_{\A\to \B,t}[l][w])
\in\mathbb{R}^{N[l]\times(t+1)}$, with $\boldsymbol{x}_{\A\to \B,t}[l][w]$ 
given in (\ref{mean_AB_projection}). For all $l\in\mathcal{L}_{0}$, 
the mean message $\boldsymbol{x}_{\A\to \B,t}^{\mathrm{suf}}[l]$ in section~$l$ and 
corresponding covariance $v_{\A\to \B,t',t}^{\mathrm{suf}}[l]$ are given by 
\begin{IEEEeqnarray}{rl} 
\boldsymbol{x}_{\A\to \B,t}^{\mathrm{suf}}[l] 
=& v_{\A\to \B,t,t}^{\mathrm{suf}}[l]\sum_{w=0}^{W}\gamma[l+w][l]
\nonumber \\
&\cdot\boldsymbol{X}_{\A\to \B,t+1}[l][w]
\boldsymbol{V}_{\A\to \B,t}^{-1}[l+w]\boldsymbol{1},
\label{mean_suf_B_LM}
\end{IEEEeqnarray}
\begin{equation} 
\frac{1}{v_{\A\to \B,t',t}^{\mathrm{suf}}[l]} 
= \sum_{w=0}^{W}\gamma^{2}[l+w][l]\boldsymbol{1}^{\mathrm{T}}
\boldsymbol{V}_{\A\to \B,t}^{-1}[l+w]\boldsymbol{1}
\label{cov_suf_B}
\end{equation}
for all $t'\in\{0,\ldots,t\}$. See Appendix~\ref{appen_suf_LM} for the 
derivation of (\ref{mean_suf_B_LM}) and (\ref{cov_suf_B}). 

The third step is computation of posterior messages 
$\boldsymbol{x}_{\B,t+1}^{\mathrm{post}}
=[(\boldsymbol{x}_{\B,t+1}^{\mathrm{post}}[0])^{\mathrm{T}},\ldots,
(\boldsymbol{x}_{\B,t+1}^{\mathrm{post}}[L-1])^{\mathrm{T}}]^{\mathrm{T}}$ 
based on separable denoisers $\{f_{t}[l]\}$, given by 
\begin{equation} \label{mean_post_B_LM}
\boldsymbol{x}_{\B,t+1}^{\mathrm{post}}[l] 
= f_{t}[l](\boldsymbol{x}_{\A\to\B,t}^{\mathrm{suf}}[l]),
\end{equation}  
which is the same as (\ref{mean_post_B}) in OAMP. 
The corresponding covariance $v_{\B,t',t+1}^{\mathrm{post}}[l]$ needs to be 
a consistent estimator of $N^{-1}[l]\mathbb{E}[(\boldsymbol{x}[l]
-\boldsymbol{x}_{\B,t'}^{\mathrm{post}}[l])^{\mathrm{T}}
(\boldsymbol{x}[l]-\boldsymbol{x}_{\B,t+1}^{\mathrm{post}}[l])]$. 
See \cite[Eq.~(23)]{Takeuchi213} for the details. 

For the signal vector $\boldsymbol{x}[l]$ with i.i.d.\ elements, 
consider the Bayes-optimal denoiser $f_{t}[l](u)
=f_{\mathrm{opt}}(u;v_{\A\to \B,t,t}^{\mathrm{suf}}[l])$. 
In this case, the following posterior covariance should be used:  
\begin{IEEEeqnarray}{rl} 
v_{\B,t'+1,t+1}^{\mathrm{post}}[l]&
\nonumber \\
=\frac{1}{N[l]}\mathbb{E}\Biggl[
\bigl\{
\boldsymbol{x}[l]& 
- f_{\mathrm{opt}}(\boldsymbol{x}_{\A\to\B,t'}^{\mathrm{suf}}[l]
;v_{\A\to \B,t',t'}^{\mathrm{suf}}[l])
\bigr\}^{\mathrm{T}} \nonumber \\
\cdot\bigl\{
 \boldsymbol{x}[l]& 
 - f_{\mathrm{opt}}(\boldsymbol{x}_{\A\to\B,t}^{\mathrm{suf}}[l] 
 ;v_{\A\to \B,t,t}^{\mathrm{suf}}[l])
\bigr\}
\nonumber \\
\Biggr| 
\boldsymbol{x}_{\A\to\B,t'}^{\mathrm{suf}}[l],&
\boldsymbol{x}_{\A\to\B,t}^{\mathrm{suf}}[l], 
v_{\A\to \B,t',t'}^{\mathrm{suf}}[l], v_{\A\to \B,t,t}^{\mathrm{suf}}[l]
\Biggr], \label{cov_post_B} 
\end{IEEEeqnarray}
where conditioning with respect to $v_{\A\to\B,t',t}^{\mathrm{suf}}[l]$ is omitted, 
because of $v_{\A\to\B,t',t}^{\mathrm{suf}}[l]=v_{\A\to\B,t,t}^{\mathrm{suf}}[l]$ for 
$t'\leq t$. 

The last step is the Onsager correction of the posterior message 
$|\mathcal{W}[\ell]|^{1/2}
\boldsymbol{\Gamma}[\ell]\boldsymbol{x}_{\B,t+1}^{\mathrm{post}}$ in the extended 
space to realize asymptotic Gaussianity in module~A. 
Let $|\mathcal{W}[\ell]|^{-1/2}\eta_{\B,\tau,t}[\ell][w]$ denote the average of 
the partial derivative of 
the $n$th element in the $w$th section of $|\mathcal{W}[\ell]|^{1/2}
\boldsymbol{\Gamma}[\ell]\boldsymbol{x}_{\B,t+1}^{\mathrm{post}}$ 
with respect to $\vec{x}_{\A\to\B,n,\tau}[\ell][w]$ over all $n\in N[\ell-w]$, 
\begin{equation} 
\eta_{\B,\tau,t}[\ell][w] 
= \eta_{\B,t}[\ell][w] 
\frac{\boldsymbol{e}_{\tau}^{\mathrm{T}}\boldsymbol{V}_{\A\to\B,t}^{-1}[\ell]
\boldsymbol{1}}
{\boldsymbol{1}^{\mathrm{T}}\boldsymbol{V}_{\A\to\B,t}^{-1}[\ell]
\boldsymbol{1}},
\label{zeta_B}
\end{equation} 
with
\begin{IEEEeqnarray}{rl} 
\eta_{\B,t}[\ell][w] 
&= |\mathcal{W}[\ell]|\gamma^{2}[\ell][\ell-w]\boldsymbol{1}^{\mathrm{T}}
\boldsymbol{V}_{\A\to\B,t}^{-1}[\ell]\boldsymbol{1}
\nonumber \\
\cdot& v_{\A\to \B,t,t}^{\mathrm{suf}}[\ell-w]
\langle f'_{t}[\ell-w]
(\boldsymbol{x}_{\A\to\B,t}^{\mathrm{suf}}[\ell-w])\rangle.
\label{eta_B_w_LM}
\end{IEEEeqnarray}
Furthermore, we define 
\begin{equation} \label{eta_B_ell_LM}
\eta_{\B,\tau,t}[\ell]
= \sum_{w\in\mathcal{W}[\ell]}\frac{N[\ell-w]}{N_{\mathrm{c}}[\ell]}
\eta_{\B,\tau,t}[\ell][w]. 
\end{equation}
The extrinsic mean $\vec{\boldsymbol{x}}_{\B\to\A,t+1}[\ell]
\in\mathbb{R}^{N_{\mathrm{c}}[\ell]}$ and covariance $v_{\B\to\A,t'+1,t+1}[\ell]$ 
of $\vec{\boldsymbol{x}}[\ell]$ are computed as   
\begin{IEEEeqnarray}{rl} 
(1 - \eta_{\B,t}[\ell]/|\mathcal{W}[\ell]|)&
\vec{\boldsymbol{x}}_{\B\to\A,t+1}[\ell] 
= |\mathcal{W}[\ell]|^{1/2}\boldsymbol{\Gamma}[\ell]
\boldsymbol{x}_{\B,t+1}^{\mathrm{post}} 
\nonumber \\
- |\mathcal{W}[\ell]|^{-1/2}&\sum_{\tau=0}^{t}\eta_{\B,\tau,t}[\ell]
\vec{\boldsymbol{x}}_{\A\to\B,\tau}[\ell], 
\label{mean_BA_LM}
\end{IEEEeqnarray}
\begin{IEEEeqnarray}{rl}
&(1 - \eta_{\B,t'}[\ell]/|\mathcal{W}[\ell]|)
(1 - \eta_{\B,t}[\ell]/|\mathcal{W}[\ell]|)
v_{\B\to\A,t'+1,t+1}[\ell] 
\nonumber \\
=& \sum_{w\in\mathcal{W}[\ell]}\frac{N[\ell-w]}{N_{\mathrm{c}}[\ell]}
|\mathcal{W}[\ell]|\gamma^{2}[\ell][\ell-w]v_{\B,t'+1,t+1}^{\mathrm{post}}[\ell-w]
\nonumber \\
&- \frac{1}{|\mathcal{W}[\ell]|}\frac{\eta_{\B,t'}[\ell]\eta_{\B,t}[\ell]}
{\boldsymbol{1}^{\mathrm{T}}\boldsymbol{V}_{\A\to\B,t}^{-1}[\ell]\boldsymbol{1}},
\label{cov_BA} 
\end{IEEEeqnarray}
with $\eta_{\B,t}[\ell]$ given in (\ref{eta_B}). 

Module~A in LM-OAMP requires the covariance message $v_{\B\to\A,0,t+1}[\ell]$, 
which is obtained by letting $f_{-1}[l]=0$ in (\ref{cov_BA}),  
\begin{IEEEeqnarray}{rl}
&(1 - \eta_{\B,t}[\ell]/|\mathcal{W}[\ell]|)
v_{\B\to\A,0,t+1}[\ell] 
\nonumber \\
=& \sum_{w\in\mathcal{W}[\ell]}\frac{N[\ell-w]}{N_{\mathrm{c}}[\ell]}
|\mathcal{W}[\ell]|\gamma^{2}[\ell][\ell-w]v_{\B,0,t+1}^{\mathrm{post}}[\ell-w], 
\label{cov_BA0}
\end{IEEEeqnarray}
where the posterior covariance $v_{\B,0,t+1}^{\mathrm{post}}[l]$ needs to be 
a consistent estimator of $N^{-1}[l]\mathbb{E}[\boldsymbol{x}^{\mathrm{T}}[l]
(\boldsymbol{x}[l]-\boldsymbol{x}_{\B,t+1}^{\mathrm{post}}[l])]$. 

For the Bayes-optimal denoiser, we use 
\begin{IEEEeqnarray}{rl}
v_{\B,0,t+1}^{\mathrm{post}}[l]
=\frac{1}{N[l]}\mathbb{E}&\left[
 \left\|
  \boldsymbol{x}[l] 
  - f_{\mathrm{opt}}(\boldsymbol{x}_{\A\to\B,t}^{\mathrm{suf}}[l]
  ; v_{\A\to \B,t}^{\mathrm{suf}}[l])
 \right\|^{2}
\right. \nonumber \\
& \Bigl| 
\boldsymbol{x}_{\A\to\B,t}^{\mathrm{suf}}[l], v_{\A\to \B,t}^{\mathrm{suf}}[l]
\Bigr], \label{cov_post_B0} 
\end{IEEEeqnarray}
which is justified via the fact that the posterior mean estimator 
$f_{\mathrm{opt}}$ is uncorrelated with its estimation error.

As expected from \cite{Takeuchi213,Takeuchi22}, 
the following proposition implies the equivalence between Bayes-optimal OAMP 
and Bayes-optimal LM-OAMP.   

\begin{proposition} \label{proposition_equivalence} 
Consider the signal vector $\boldsymbol{x}[l]$ with i.i.d.\ elements, 
the LMMSE filter, and the Bayes-optimal denoiser. If the covariance matrix 
$\boldsymbol{V}_{\A\to\B,t}[\ell]$ is positive definite, then LM-OAMP is 
equivalent to OAMP: The messages $\vec{\boldsymbol{x}}_{\A\to\B,t}[\ell]$, 
$v_{\A\to\B,t',t}[\ell]$, $\vec{\boldsymbol{x}}_{\B\to\A,t+1}[\ell]$, and 
$v_{\B\to\A,t'+1,t+1}[\ell]$ in (\ref{mean_AB_LM}), (\ref{cov_AB}), 
(\ref{mean_BA_LM}), and (\ref{cov_BA}) are respectively equal to 
(\ref{mean_AB}), (\ref{var_AB_LMMSE}), (\ref{mean_BA}), and 
(\ref{var_BA_opt}) in OAMP for all $t$ and $t'\in\{0,\ldots,t\}$. 
Furthermore, $v_{\B\to\A,0,t+1}[\ell]$ in (\ref{cov_BA0}) is equal to 
(\ref{var_BA_opt}) in OAMP for all $t$.  
\end{proposition}
\begin{IEEEproof}
See Appendix~\ref{proof_proposition_equivalence}.   
\end{IEEEproof}

The positive definiteness of $\boldsymbol{V}_{\A\to\B,t}[\ell]$ is justified 
in the large system limit via state evolution. 
Proposition~\ref{proposition_equivalence} allows us to analyze the asymptotic 
dynamics of Bayes-optimal OAMP and its convergence property via those for 
Bayes-optimal LM-OAMP. In particular, the convergence property of LM-OAMP 
can be analyzed more straightforwardly than that of OAMP. In this sense, 
LM-OAMP is regarded as a technical tool to prove the convergence of state 
evolution recursions for Bayes-optimal OAMP.

\subsection{State Evolution}
State evolution analysis is presented for LM-OAMP in the spatially 
coupled system~(\ref{vector_model}). This paper extends a unified 
framework~\cite{Takeuchi211} of state evolution to that for the spatially 
coupled system. A general error model for the spatially coupled system is 
proposed and analyzed in the large system limit. By proving that the 
proposed general error model contains the error model for LM-OAMP 
in the spatially coupled system, we derive state evolution recursions for 
LM-OAMP. 

We first define state evolution recursions for LM-OAMP. 
State evolution recursions for module~A with the initial 
condition $\bar{v}_{\B\to\A,0,0}[\ell]=|\mathcal{W}[\ell]|
\alpha_{\mathrm{c}}^{-1}[\ell]\sum_{w\in\mathcal{W}[\ell]}\alpha[\ell][\ell-w]
\gamma^{2}[\ell][\ell-w]$ are given by 
\begin{IEEEeqnarray}{l}
\bar{v}_{\A,t',t}^{\mathrm{post}}[\ell] 
= \lim_{\{M[\ell], N[l]\}\to\infty}\left\{
 \frac{\sigma^{2}}{N_{\mathrm{c}}[\ell]}\mathrm{Tr}\left(
  \boldsymbol{F}_{t}^{\mathrm{T}}[\ell]\boldsymbol{F}_{t'}[\ell]
 \right)
\right. \nonumber \\
+ \left.
 \frac{\bar{v}_{\B\to\A,t',t}[\ell]}{N_{\mathrm{c}}[\ell]}\mathrm{Tr}\left\{
  (\boldsymbol{I} - \boldsymbol{F}_{t}^{\mathrm{T}}[\ell]\boldsymbol{A}[\ell])
  (\boldsymbol{I} - \boldsymbol{F}_{t'}^{\mathrm{T}}[\ell]
  \boldsymbol{A}[\ell])^{\mathrm{T}}
 \right\}
\right\}, \nonumber \\
\label{cov_post_A_SE}
\end{IEEEeqnarray}
\begin{equation} \label{cov_AB_SE}
\bar{v}_{\A\to\B,t',t}[\ell] 
= \frac{1}{|\mathcal{W}[\ell]|}\frac{\bar{v}_{\A,t',t}^{\mathrm{post}}[\ell] 
- \bar{\eta}_{\A,t'}[\ell]\bar{\eta}_{\A,t}[\ell]\bar{v}_{\B\to\A,t',t}[\ell]}
{(1-\bar{\eta}_{\A,t'}[\ell])(1-\bar{\eta}_{\A,t}[\ell])}, 
\end{equation}
where $\bar{\eta}_{\A,t}[\ell]$ is the same as (\ref{eta_A_SE}) for OAMP. 

Let $\bar{\boldsymbol{V}}_{\A\to \B,t}[\ell]\in\mathbb{R}^{(t+1)\times(t+1)}$ denote 
the covariance matrix with the $(\tau',\tau)$ element 
$[\bar{\boldsymbol{V}}_{\A\to \B,t}[\ell]]_{\tau',\tau}
=\bar{v}_{\A\to\B,\tau',\tau}[\ell]$. 
State evolution recursions for module~B are given by 
\begin{equation} \label{cov_suf_B_SE}
\frac{1}{\bar{v}_{\A\to \B,t',t}^{\mathrm{suf}}[l]} 
= \sum_{w=0}^{W}\gamma^{2}[l+w][l]
\boldsymbol{1}^{\mathrm{T}}\bar{\boldsymbol{V}}_{\A\to \B,t}^{-1}[l+w]\boldsymbol{1}, 
\end{equation}
\begin{IEEEeqnarray}{r}
\bar{v}_{\B,t'+1,t+1}^{\mathrm{post}}[l] 
= \mathbb{E}[\{x_{1}[l] - f_{t'}[l](x_{1}[l] + z_{t'}[l])\}
\nonumber \\
\cdot\{x_{1}[l] - f_{t}[l](x_{1}[l] + z_{t}[l])\}], 
\label{cov_post_B_SE}
\end{IEEEeqnarray}
\begin{IEEEeqnarray}{rl}
&(1 - \bar{\eta}_{\B,t'}[\ell]/|\mathcal{W}[\ell]|)
(1 - \bar{\eta}_{\B,t}[\ell]/|\mathcal{W}[\ell]|)
\bar{v}_{\B\to\A,t'+1,t+1}[\ell] 
\nonumber \\
=& \sum_{w\in\mathcal{W}[\ell]}\frac{\alpha[\ell][\ell-w]}{\alpha_{\mathrm{c}}[\ell]}
|\mathcal{W}[\ell]|\gamma^{2}[\ell][\ell-w]
\bar{v}_{\B,t'+1,t+1}^{\mathrm{post}}[\ell-w]
\nonumber \\
&- \frac{1}{|\mathcal{W}[\ell]|}
\frac{\bar{\eta}_{\B,t'}[\ell]\bar{\eta}_{\B,t}[\ell]}
{\boldsymbol{1}^{\mathrm{T}}\bar{\boldsymbol{V}}_{\A\to\B,t}^{-1}[\ell]
\boldsymbol{1}},
\label{cov_BA_SE} 
\end{IEEEeqnarray}
with 
\begin{equation} \label{eta_B_SE_LM}
\bar{\eta}_{\B,t}[\ell]=\sum_{w\in\mathcal{W}[\ell]}
\frac{\alpha[\ell][\ell-w]}{\alpha_{\mathrm{c}}[\ell]}\bar{\eta}_{\B,t}[\ell][w],
\end{equation}
\begin{IEEEeqnarray}{l} 
\bar{\eta}_{\B,t}[\ell][w] 
= |\mathcal{W}[\ell]|\gamma^{2}[\ell][\ell-w]\boldsymbol{1}^{\mathrm{T}}
\bar{\boldsymbol{V}}_{\A\to\B,t}^{-1}[\ell]\boldsymbol{1}
\nonumber \\
\cdot\bar{v}_{\A\to \B,t,t}^{\mathrm{suf}}[\ell-w]
\mathbb{E}\left[
 f'_{t}[\ell-w](x_{1}[\ell-w] + z_{t}[\ell-w])
\right]. \label{zeta_B_SE} 
\end{IEEEeqnarray}
In these expressions, $\{z_{t}[l]\}$ denote zero-mean Gaussian random 
variables with covariance $\mathbb{E}[z_{t'}[l]z_{t}[l]]
=\bar{v}_{\A\to \B,t',t}^{\mathrm{suf}}[l]$, independent of $x_{1}[l]$. 

For $t'=-1$, we use 
\begin{IEEEeqnarray}{rl}
&(1 - \bar{\eta}_{\B,t}[\ell]/|\mathcal{W}[\ell]|)
\bar{v}_{\B\to\A,0,t+1}[\ell] 
\nonumber \\
=& \sum_{w\in\mathcal{W}[\ell]}\frac{\alpha[\ell][\ell-w]}{\alpha_{\mathrm{c}}[\ell]}
|\mathcal{W}[\ell]|\gamma^{2}[\ell][\ell-w]
\bar{v}_{\B,0,t+1}^{\mathrm{post}}[\ell-w] 
\label{cov_BA_SE_0}
\end{IEEEeqnarray}
instead of (\ref{cov_BA_SE}), with 
\begin{equation}
\bar{v}_{\B,0,t+1}^{\mathrm{post}}[l] 
= \mathbb{E}[x_{1}[l]
\{x_{1}[l] - f_{t}[l](x_{1}[l] + z_{t}[l])\}].  
\end{equation}

The following theorem implies that the former part in Theorem~\ref{theorem_SE} 
is correct. 

\begin{theorem} \label{theorem_SE_LM}
Suppose that Assumptions~\ref{assumption_x}--\ref{assumption_denoiser} 
hold. Then, the MSE $N^{-1}[l]\|\boldsymbol{x}[l]
-\boldsymbol{x}_{\B,t}[l]\|^{2}$ for OAMP and the error covariance 
$N^{-1}[l](\boldsymbol{x}[l]-\boldsymbol{x}_{\B,t'}[l])^{\mathrm{T}}
(\boldsymbol{x}[l]-\boldsymbol{x}_{\B,t}[l])$ for LM-OAMP converge almost 
surely to $\bar{v}_{\B,t}^{\mathrm{post}}[l]$ and 
$\bar{v}_{\B,t',t}^{\mathrm{post}}[l]$ in the large system limit, respectively. 
The asymptotic MSE $\bar{v}_{\B,t}^{\mathrm{post}}[l]$ is given via the state 
evolution recursions~(\ref{var_post_A_SE})--(\ref{var_BA_SE}) for OAMP 
while $\bar{v}_{\B,t',t}^{\mathrm{post}}[l]$ is given via the state evolution 
recursions~(\ref{cov_post_A_SE})--(\ref{cov_BA_SE}) for LM-OAMP. Furthermore, 
the covariance matrix $\bar{\boldsymbol{V}}_{\A\to \B,t}[\ell]$ is positive 
definite.  
\end{theorem}
\begin{IEEEproof}
See Appendix~\ref{proof_theorem_SE_LM}. 
\end{IEEEproof}

Theorem~\ref{theorem_SE_LM} implies asymptotic Gaussianity for the estimation 
error $\boldsymbol{x}[l]-\boldsymbol{x}_{\B,t}[l]$: The error covariance 
$N^{-1}[l](\boldsymbol{x}[l]-\boldsymbol{x}_{\B,t'}[l])^{\mathrm{T}}
(\boldsymbol{x}[l]-\boldsymbol{x}_{\B,t}[l])$ converges almost surely to 
$\bar{v}_{\B,t',t}^{\mathrm{post}}[l]$ in (\ref{cov_post_B_SE})---given via the 
Gaussian random variables $\{z_{t'}[l], z_{t}[l]\}$. 

As a conclusion of Theorem~\ref{theorem_SE_LM} and 
Proposition~\ref{proposition_equivalence}, we arrive at the equivalence 
between Bayes-optimal OAMP and Bayes-optimal LM-OAMP in the large system 
limit. The latter part in Theorem~\ref{theorem_SE} is obtained by proving 
the convergence of state evolution recursions for Bayes-optimal OAMP. 

\begin{theorem} \label{theorem_SE_Bayes} 
Consider the LMMSE filter~(\ref{LMMSE}) and the Bayes-optimal 
denoiser~(\ref{opt_denoiser}). Suppose that 
Assumptions~\ref{assumption_x}--\ref{assumption_denoiser} hold. 
\begin{itemize}
\item LM-OAMP is asymptotically equivalent to OAMP: 
The messages $\vec{\boldsymbol{x}}_{\A\to\B,t}[\ell]$, 
$v_{\A\to\B,t',t}[\ell]$, $\vec{\boldsymbol{x}}_{\B\to\A,t+1}[\ell]$, and 
$v_{\B\to\A,t'+1,t+1}[\ell]$ in (\ref{mean_AB_LM}), (\ref{cov_AB}), 
(\ref{mean_BA_LM}), and (\ref{cov_BA}) are respectively equal to 
(\ref{mean_AB}), (\ref{var_AB_LMMSE}), (\ref{mean_BA}), and 
(\ref{var_BA_opt}) in OAMP for all $t$ and $t'\in\{0,\ldots,t\}$ 
in the large system limit.  
Furthermore, $v_{\B\to\A,0,t+1}[\ell]$ in (\ref{cov_BA0}) is equal to 
(\ref{var_BA_opt}) in OAMP for all $t$.  

\item The error covariance 
$N^{-1}[l](\boldsymbol{x}[l]-\boldsymbol{x}_{\B,t'}[l])^{\mathrm{T}}
(\boldsymbol{x}[l]-\boldsymbol{x}_{\B,t}[l])$ for Bayes-optimal LM-OAMP 
converges almost surely to $\bar{v}_{\B,t}^{\mathrm{post}}[l]$ for all 
$t'\in\{0,\ldots,t\}$ in the large system limit, in which 
$\bar{v}_{\B,t}^{\mathrm{post}}[l]$ is given via the state evolution 
recursions~(\ref{var_AB_SE_Bayes})--(\ref{var_BA_SE_Bayes}) for 
Bayes-optimal OAMP. 

\item The state evolution 
recursions~(\ref{var_AB_SE_Bayes})--(\ref{var_BA_SE_Bayes}) 
for Bayes-optimal OAMP converge to a fixed point as $t\to\infty$. 
\end{itemize}
\end{theorem}
\begin{IEEEproof}
The first statement follows from Proposition~\ref{proposition_equivalence} 
and the positive definiteness of $\bar{\boldsymbol{V}}_{\A\to\B,t}[\ell]$ in 
Theorem~\ref{theorem_SE_LM}. The second statement can be proved by repeating 
the proof of Proposition~\ref{proposition_equivalence} for the state 
evolution recursions~(\ref{cov_post_A_SE})--(\ref{cov_BA_SE}), which 
describe the asymptotic dynamics of the error covariance 
$N^{-1}[l](\boldsymbol{x}[l]-\boldsymbol{x}_{\B,t'}[l])^{\mathrm{T}}
(\boldsymbol{x}[l]-\boldsymbol{x}_{\B,t}[l])$ from Theorem~\ref{theorem_SE_LM}.  

We prove the last statement. The first two statements in 
Theorem~\ref{theorem_SE_Bayes} imply that 
it is sufficient to prove the convergence of 
the state evolution recursions for Bayes-optimal LM-OAMP, 
because of the equivalence between Bayes-optimal OAMP 
and Bayes-optimal LM-OAMP. 

Let us prove the convergence. We use \cite[Lemma~2]{Takeuchi213} to find that 
the asymptotic MSE $\bar{v}_{\B,t+1,t+1}^{\mathrm{post}}[l]$ in 
(\ref{cov_post_B_SE}) for Bayes-optimal LM-OAMP is 
monotonically non-increasing as $t$ increases. As a result, there exists 
$\lim_{t\to\infty}\bar{v}_{\B,t+1,t+1}^{\mathrm{post}}[l]$. Since the second statement 
in Theorem~\ref{theorem_SE_Bayes} implies $\bar{v}_{\B,t',t+1}^{\mathrm{post}}[l]
=\bar{v}_{\B,t+1,t+1}^{\mathrm{post}}[l]$ for all $t'\in\{0,\ldots,t+1\}$, 
the covariance $\bar{v}_{\B,t',t+1}^{\mathrm{post}}[l]$ converges for all 
$t'\in\{0,\ldots,t+1\}$ as $t\to\infty$. Thus, the state evolution 
recursions for Bayes-optimal LM-OAMP converge to a fixed point. 
\end{IEEEproof}

Theorem~\ref{theorem_SE_Bayes} implies that LM-OAMP is equivalent to OAMP 
in the large system limit, as long as the LMMSE filter and the Bayes-optimal 
denoiser are considered. Theorem~\ref{theorem_SE} follows from 
Theorem~\ref{theorem_SE_LM} and the last statement in 
Theorem~\ref{theorem_SE_Bayes}. 

\section{Proof of Theorem~\ref{theorem_optimality}}
\label{sec_proof_optimality} 
\subsection{Overview}
The proof of Theorem~\ref{theorem_optimality} consists of three steps: 
In a first step, via the change of variables, the state evolution recursions 
for Bayes-optimal OAMP are connected to the replica-symmetric 
potential~(\ref{RS_potential}). Unfortunately, the state evolution recursions 
obtained via the change of variables are not included in the class of 
spatially coupled systems analyzed in \cite{Yedla14,Takeuchi15}.  

In a second step, we approximate the state evolution recursions so that 
the obtained recursions are included in the class of 
spatially coupled systems in \cite{Yedla14,Takeuchi15}. We prove that 
the approximate state evolution recursions are an exact approximation of 
the original state evolution recursions in the continuum limit.  

The last step is evaluation of the fixed point of the approximate state 
evolution recursions via existing results~\cite{Yedla14,Takeuchi15}. 
By proving that a potential for characterizing the fixed point is connected 
to the replica-symmetric potential~(\ref{RS_potential}), 
we arrive at Theorem~\ref{theorem_optimality}.

\subsection{Change of Variables}
To connect the state evolution recursions for Bayes-optimal OAMP with 
the replica symmetric potential~(\ref{RS_potential}), we consider 
the change of variables $s_{t}[l]=1/\bar{v}_{\A\to\B,t}^{\mathrm{suf}}[l]$ and 
\begin{IEEEeqnarray}{rl}
E_{t+1}[\ell] 
=& \frac{1}{W+1}\sum_{w\in\mathcal{W}[\ell]}\bar{v}_{\B,t+1}^{\mathrm{post}}[\ell-w]
\nonumber \\
=& \frac{1}{W+1}\sum_{w\in\mathcal{W}[\ell]}\mathrm{MMSE}(s_{t}[\ell-w]),  
\label{E_t}
\end{IEEEeqnarray}
where we have represented the MMSE $\bar{v}_{\B,t+1}^{\mathrm{post}}[l]$ in 
(\ref{var_post_B_SE}) with $\mathrm{MMSE}(\cdot)$ given in (\ref{MMSE_func}). 
For $t=0$, we use the initial condition 
$s_{0}[l]=1/\bar{v}_{\A\to\B,0}^{\mathrm{suf}}[l]$, computed via the state evolution 
recursions for Bayes-optimal OAMP. 

The goal in the first step is to derive the following state evolution 
recursion: 
\begin{equation} \label{s_t}
s_{t}[l] = \frac{1}{W+1}\sum_{\ell=l}^{l+W}
g[\ell]\left(
 \frac{\bar{\eta}_{\A,t}[\ell]E_{t}[\ell]}
 {1 - \bar{\eta}_{\B,t-1}[\ell]/|\mathcal{W}[\ell]|}
\right),
\end{equation}
with 
\begin{equation} \label{g_l_func}
g[\ell](z) = \frac{1}{\sigma^{2}}
R_{\boldsymbol{G}[\ell]}
\left(
 - \frac{z}{|\mathcal{W}[\ell]|\sigma^{2}}
\right). 
\end{equation}
The functions $g[\ell](z)$ in (\ref{g_l_func}) and $\mathrm{MMSE}(s)$ are 
connected to the second term in the replica-symmetric 
potential~(\ref{RS_potential}) and the derivative of the first term, 
respectively, via the relationship~(\ref{IM_relationship}) between 
mutual information and MMSE.  

We first represent the state evolution recursion~(\ref{var_AB_SE_Bayes}) for 
module~A with the R-transform. Repeating the derivation of 
(\ref{eta_transform_tmp}) for $\bar{\eta}_{\A,t}[\ell]$ in (\ref{eta_A_SE}) 
with the LMMSE filter~(\ref{LMMSE_SE_OAMP}), 
we obtain 
\begin{equation} 
\bar{\eta}_{\A,t}[\ell] = \eta_{\boldsymbol{A}^{\mathrm{T}}[\ell]\boldsymbol{A}[\ell]}
\left(
 \frac{\bar{v}_{\B\to\A,t}[\ell]}{\sigma^{2}} 
\right),
\end{equation} 
where $\eta_{\boldsymbol{A}^{\mathrm{T}}[\ell]\boldsymbol{A}[\ell]}$ denotes the 
$\eta$-transform~(\ref{eta_transform}) of 
$\boldsymbol{A}^{\mathrm{T}}[\ell]\boldsymbol{A}[\ell]$ 
in the large system limit. 
Using the identity $\eta_{\boldsymbol{A}^{\mathrm{T}}[\ell]\boldsymbol{A}[\ell]}
(z)=\eta_{\boldsymbol{G}[\ell]}(z/|\mathcal{W}[\ell]|)$ obtained from 
the definition of the $\eta$-transform in (\ref{eta_transform}) yields 
\begin{equation} \label{eta_A_SE_eta}
\bar{\eta}_{\A,t}[\ell] = \eta_{\boldsymbol{G}[\ell]}
\left(
 \frac{\bar{v}_{\B\to\A,t}[\ell]}{|\mathcal{W}[\ell]|\sigma^{2}} 
\right). 
\end{equation} 
Thus, from the representation of the R-transform in (\ref{R_transform_tmp}) 
at $z=\bar{v}_{\B\to\A,t}[\ell]/(\sigma^{2}|\mathcal{W}[\ell]|)$ we have  
\begin{equation}
\frac{1}{\sigma^{2}}R_{\boldsymbol{G}[\ell]}
\left(
 - \frac{\bar{\eta}_{\A,t}[\ell]\bar{v}_{\B\to\A,t}[\ell]}
 {\sigma^{2}|\mathcal{W}[\ell]|}
\right)
= \frac{|\mathcal{W}[\ell]|(1 - \bar{\eta}_{\A,t}[\ell])}
{\bar{\eta}_{\A,t}[\ell]\bar{v}_{\B\to\A,t}[\ell]}.   
\end{equation}
Applying this expression to the definition of $\bar{v}_{\A\to\B,t}[\ell]$ in 
(\ref{var_AB_SE_Bayes}), we arrive at 
\begin{equation}\label{var_AB_SE_R}
\frac{1}{\bar{v}_{\A\to\B,t}[\ell]} 
=  \frac{1}{\sigma^{2}}
R_{\boldsymbol{G}[\ell]}
\left(
 - \frac{\bar{\eta}_{\A,t}[\ell]\bar{v}_{\B\to\A,t}[\ell]}
 {|\mathcal{W}[\ell]|\sigma^{2}}
\right). 
\end{equation}

We next derive the state evolution recursion~(\ref{s_t}). 
Substituting the expression of $\bar{v}_{\A\to\B,t}[\ell]$ in 
(\ref{var_AB_SE_R}) into the definition of $\bar{v}_{\A\to\B,t}^{\mathrm{suf}}[l]$ 
in (\ref{var_suf_B_SE_uniform}) for Bayes-optimal OAMP, we have 
\begin{equation} \label{s_t_tmp}
s_{t}[l] 
= \frac{1}{\bar{v}_{\A\to\B,t}^{\mathrm{suf}}[l]} 
= \frac{1}{W+1}\sum_{\ell=l}^{l+W}
g[\ell](\bar{\eta}_{\A,t}[\ell]\bar{v}_{\B\to\A,t}[\ell]), 
\end{equation}
with $g[\ell](z)$ defined in (\ref{g_l_func}). 
Using the identity $E_{t+1}[\ell]=\bar{\eta}_{\B,t}[\ell]\bar{v}_{\A\to\B,t}[\ell]$ 
obtained from $E_{t+1}[\ell]$ in (\ref{E_t}) and 
$\bar{\eta}_{\B,t}[\ell]$ in (\ref{eta_B_SE_Bayes_uniform}), 
we find that (\ref{var_BA_SE_Bayes}) reduces to  
\begin{equation}
\bar{v}_{\B\to\A,t+1}[\ell]
= \frac{E_{t+1}[\ell]}{1 - \bar{\eta}_{\B,t}[\ell]/|\mathcal{W}[\ell]|}. 
\end{equation}
Substituting this expression into (\ref{s_t_tmp}), we arrive at 
the state evolution recursion~(\ref{s_t}). 

\subsection{Approximate State Evolution Recursions} 
The state evolution recursions~(\ref{E_t}) and (\ref{s_t}) with respect to 
$s_{t}[l]$ and $E_{t}[\ell]$ are not included in the class of spatially 
coupled dynamical systems in \cite{Yedla14,Takeuchi15}. Thus, 
we consider approximate state evolution recursions with 
the initial condition $\tilde{s}_{0}[l]=s_{0}[l]$, given by 
\begin{equation} \label{E_t_tilde}
\tilde{E}_{t+1}[\ell] 
= \frac{1}{W+1}\sum_{w\in\mathcal{W}[\ell]}\mathrm{MMSE}(\tilde{s}_{t}[\ell-w]),  
\end{equation}
\begin{equation} \label{s_t_tilde}
\tilde{s}_{t}[l] = \frac{1}{W+1}\sum_{\ell=l}^{l+W}g(\tilde{E}_{t}[\ell]), 
\end{equation}
with 
\begin{equation} \label{g_func}
g(z)= \frac{1}{\sigma^{2}}R\left(
 - \frac{z}{\sigma^{2}}, 
\right), 
\end{equation}
where $R(z)$ is defined in the 
assumption~(\ref{R_transform_limit}) in Theorem~\ref{theorem_optimality}. 
The approximate state evolution recursions are included in 
the class of spatially coupled systems in \cite{Yedla14,Takeuchi15}. 

The former state evolution recursion~(\ref{E_t_tilde}) is equivalent to 
the original recursion~(\ref{E_t}). On the other hand, the 
latter~(\ref{s_t_tilde}) is obtained by letting $\bar{\eta}_{\A,t}[\ell]=1$, 
$\bar{\eta}_{\B,t-1}[\ell]/|\mathcal{W}[\ell]|=0$, and 
$R_{\boldsymbol{G}[\ell]}(z/|\mathcal{W}[\ell]|)=R(z)$ in the original 
recursion~(\ref{s_t}), of which the last is motivated by the 
assumption~(\ref{R_transform_limit}) in Theorem~\ref{theorem_optimality}.  
The first two replacements are due to the following lemma:  

\begin{lemma} \label{lemma_eta}
Let $\{a_{W}>0\}_{W=1}^{\infty}$ denote a positive and diverging sequence 
at a sublinear speed in $W$: $\lim_{W\to\infty}a_{W}=\infty$ and 
$\lim_{W\to\infty}a_{W}/W=0$. Suppose $\mathbb{E}[x_{1}^{2}[l]]=1$ holds and 
that there is some function $R(z)$ such 
that the R-transform of $\boldsymbol{G}[\ell]$ satisfies 
the assumption~(\ref{R_transform_limit}) in Theorem~\ref{theorem_optimality} 
for all $\ell\in\{\lceil W/a_{W}\rceil,\ldots,L+W-1-\lceil W/a_{W}\rceil\}$ 
and  $\lim_{z\to\infty}zR(-z)=\delta$ holds. Then, we have 
\begin{equation} \label{eta_A_SE_limit}
\lim_{W=\Delta L\to\infty}\frac{W}{a_{W}}|\bar{\eta}_{\A,t}[\ell] - 1|<\infty,
\end{equation} 
\begin{equation} \label{eta_B_SE_limit}
\lim_{W=\Delta L\to\infty}\frac{W}{a_{W}}\frac{\bar{\eta}_{\B,t}[\ell]}
{|\mathcal{W}[\ell]|}<\infty 
\end{equation}
in the continuum limit, for all $\ell\in\{\lceil W/a_{W}\rceil,\ldots,
L+W-1-\lceil W/a_{W}\rceil\}$.
\end{lemma}
\begin{IEEEproof}
We first prove the former bound~(\ref{eta_A_SE_limit}). Let 
$y=v\eta_{\boldsymbol{G}[\ell]}(v/|\mathcal{W}[\ell]|)$ with 
$v=\bar{v}_{\B\to\A,t}[\ell]/\sigma^{2}$. Evaluating the representation of 
the R-transform $R_{\boldsymbol{G}[\ell]}$ in (\ref{R_transform_tmp}) at 
$z=v/|\mathcal{W}[\ell]|$ yields 
\begin{equation}
R_{\boldsymbol{G}[\ell]}\left(
 - \frac{y}{|\mathcal{W}[\ell]|}
\right) 
= |\mathcal{W}[\ell]|\frac{1-\eta_{\boldsymbol{G}[\ell]}(v/|\mathcal{W}[\ell]))}
{y}. 
\end{equation}
We use the definition of $\bar{\eta}_{\A,t}[\ell]$ in (\ref{eta_A_SE_eta}) 
and the assumption~(\ref{R_transform_limit}) in 
Theorem~\ref{theorem_optimality} to obtain 
\begin{equation}
|\mathcal{W}[\ell]||\bar{\eta}_{\A,t}[\ell] - 1|
= |yR(-y)| + {\cal O}(a_{W}^{-1}). 
\end{equation}
The assumption $\lim_{y\to\infty}yR(-y)=\delta$ in Lemma~\ref{lemma_eta} implies 
the boundedness of the RHS. 
Since Proposition~\ref{proposition_W} implies 
$|\mathcal{W}[\ell]|\geq W/a_{W}$, 
we arrive at the former bound~(\ref{eta_A_SE_limit}).  

We next prove the latter bound~(\ref{eta_B_SE_limit}). 
It is sufficient to prove the 
boundedness of $\bar{\eta}_{\B,t}[\ell]$ in (\ref{eta_B_SE_Bayes_uniform}). 
As shown in (\ref{MMSE_upper_bound}), 
the MMSE $\bar{v}_{\B,t+1}^{\mathrm{post}}[l]$ in the numerator is bounded from 
above by the prior variance $\mathbb{E}[x_{1}^{2}[l]]=1$. 
The variance $\bar{v}_{\A\to\B,t}[\ell]$ in the denominator has to be positive 
since $\bar{\boldsymbol{V}}_{\A\to\B,t}[\ell]$ has been proved to be positive 
definite in Theorem~\ref{theorem_SE_LM}. Thus, 
$\bar{\eta}_{\B,t}[\ell]$ in (\ref{eta_B_SE_Bayes_uniform}) is bounded. 
\end{IEEEproof}

The goal in the second step is to prove that the approximate state evolution 
recursions~(\ref{E_t_tilde}) and (\ref{s_t_tilde}) are an exact 
approximation of the original state evolution recursions  
in the continuum limit. 

\begin{lemma} \label{lemma_approximation} 
Suppose that $\mathbb{E}[x_{1}^{2}[l]]=1$ holds and 
that all moments for the asymptotic eigenvalue distribution of 
$\boldsymbol{G}[\ell]$ are bounded. 
Let $\{a_{W}>0\}_{W=1}^{\infty}$ denote a positive and diverging sequence 
at a sublinear speed in $W$: $\lim_{W\to\infty}a_{W}=\infty$ and 
$\lim_{W\to\infty}a_{W}/W=0$. Assume the following conditions:
\begin{itemize}
\item There is some function $R(z)$ such that the R-transform of 
$\boldsymbol{G}[\ell]$ satisfies the assumption~(\ref{R_transform_limit}) in 
Theorem~\ref{theorem_optimality} 
for all $\ell\in\{\lceil W/a_{W}\rceil,\ldots,L+W-1-\lceil W/a_{W}\rceil\}$.
\item $R(z)$ is proper, continuously differentiable, and 
non-decreasing for all $z\leq0$. 
\item $\lim_{z\to\infty}zR(-z)=\delta$ holds.
\end{itemize} 
Then, for all $\ell\in\mathcal{L}_{W}$, $l\in\mathcal{L}_{0}$, 
and iteration~$\tau$, 
\begin{equation} \label{E_t_dif}
\lim_{W=\Delta L\to\infty}
a_{W}|\tilde{E}_{\tau+1}[\ell] - E_{\tau+1}[\ell]|<\infty,
\end{equation}
\begin{equation} \label{s_t_dif}
\lim_{W=\Delta L\to\infty}a_{W}|\tilde{s}_{\tau}[l] - s_{\tau}[l]|<\infty. 
\end{equation}
\end{lemma}
\begin{IEEEproof}
See Appendix~\ref{proof_lemma_approximation}. 
\end{IEEEproof}

Lemma~\ref{lemma_approximation} allows us to evaluate the dynamics of 
the original state evolution recursions for Bayes-optimal OAMP via the 
approximate state evolution~(\ref{E_t_tilde}) and (\ref{s_t_tilde}) 
as long as the continuum limit is considered. 

\subsection{Analysis via Potential} 
The fixed point of the approximate state evolution recursions~(\ref{E_t_tilde}) 
and (\ref{s_t_tilde}) was analyzed 
in \cite{Yedla14,Takeuchi15}. The existing results are different in terms of 
the order of limits: \cite{Takeuchi15} considered 
$\lim_{\Delta\downarrow0}\lim_{t\to\infty}\lim_{W=\Delta L\to\infty}$ 
while \cite{Yedla14} took 
$\lim_{W\to\infty}\lim_{L\to\infty}\lim_{t\to\infty}$. In the proof of 
Theorem~\ref{theorem_optimality}, we focus on the former limit to 
use Lemma~\ref{lemma_approximation}.  

Define a potential function $F:[0,1]\to\mathbb{R}$ as 
\begin{equation} \label{potential_func}
F(\tilde{E}) 
= \int_{0}^{g(\tilde{E})}\mathrm{MMSE}(s)ds + \int_{0}^{\tilde{E}}g(z)dz 
- \tilde{E}g(\tilde{E}), 
\end{equation}
where $g(z)$ is given in (\ref{g_func}).  
Note that $F(\tilde{E})$ depends on $\delta$ through 
$R(z)$ in $g(z)$. 

\begin{theorem}[\cite{Takeuchi15}] \label{theorem_threshold}
Suppose that $R(z)$ in (\ref{g_func}) is strictly increasing and twice 
continuously differentiable for all $z\leq0$.
Let $\tilde{E}_{\mathrm{opt}}$ denote the global minimizer of 
the potential function~(\ref{potential_func}). If $\tilde{E}_{\mathrm{opt}}$ 
is unique, then    
\begin{equation}
\lim_{\Delta\downarrow0}\lim_{t\to\infty}\lim_{W=\Delta L\to\infty}
\frac{1}{|\mathcal{L}_{W}|}\sum_{\ell\in\mathcal{L}_{W}}\tilde{E}_{t}[\ell]
\leq \tilde{E}_{\mathrm{opt}}.
\end{equation}
\end{theorem}
\begin{IEEEproof}
Consider $v_{\ell}(t)=-\tilde{E}_{t}[\ell]$ and $u_{l}(t)=s_{t}[l]$ 
with $\phi(s)=-\mathrm{MMSE}(s)$, $\psi(v)=g(-v)$, and $\beta=1$  
in \cite[Eqs.~(41) and (42)]{Takeuchi15}. 
From Proposition~\ref{proposition_MSE} and the assumption on $R(z)$ in 
Theorem~\ref{theorem_threshold}, the two functions $\phi$ and $\psi$ 
are strictly increasing and twice continuously differentiable. A potential 
function $V(\psi(-\tilde{E}))$ in \cite[Eq.~(58)]{Takeuchi15} reduces to 
(\ref{potential_func}). Thus, from 
\cite[Theorem~5 and Corollary~1]{Takeuchi15} there is some function 
$\tilde{E}:[0, 1]\to[0, 1]$ such that the upper bound 
$\tilde{E}(x)\leq \tilde{E}_{\mathrm{opt}}$ holds for all $x$ and
\begin{equation}
\lim_{\Delta\downarrow0}\lim_{t\to\infty}\lim_{W=\Delta L\to\infty}
\left|
 \frac{1}{|\mathcal{L}_{W}|}\sum_{\ell\in\mathcal{L}_{W}}
 \left\{
  \tilde{E}_{t}[\ell]
  - \tilde{E}\left(
   \frac{\ell}{|\mathcal{L}_{W}|}
  \right)
 \right\}
\right| =0. 
\end{equation}
Applying the upper bound $\tilde{E}(x)\leq \tilde{E}_{\mathrm{opt}}$ 
to this expression, we arrive at Theorem~\ref{theorem_threshold}. 
\end{IEEEproof}

\begin{remark}
The other existing result~\cite[Theorem~1]{Yedla14} implies 
\begin{equation}
\lim_{W\to\infty}\lim_{L\to\infty}
\lim_{t\to\infty}\max_{\ell\in\mathcal{L}_{W}}\tilde{E}_{t}[\ell]
\leq \tilde{E}_{\mathrm{opt}} 
\end{equation}
under slightly weaker assumptions than those in 
Theorem~\ref{theorem_threshold}, where $R(z)$ is assumed to be non-decreasing 
and continuously differentiable for all $z\leq0$. However, this upper bound 
is not matched with Lemma~\ref{lemma_approximation} since the limit 
$t\to\infty$ is taken before the continuum limit. 
\end{remark}

\begin{remark}
The so-called BP threshold~\cite{Takeuchi15} was 
defined as the infimum $\delta_{\mathrm{BP}}$ of $\delta$ such that the potential 
function~(\ref{potential_func}) has a unique minimizer for all 
$\delta\in(\delta_{\mathrm{BP}},1]$. Furthermore, define  
the potential threshold $\delta_{\mathrm{opt}}$ as the infimum of 
$\delta$ such that the global minimizer $\tilde{E}_{\mathrm{opt}}$ of 
the potential function~(\ref{potential_func}) is unique and equal to 
the smallest local minimizer of (\ref{potential_func}).  
When $\delta_{\mathrm{opt}}<\delta_{\mathrm{BP}}$ 
holds, the potential function~(\ref{potential_func}) has multiple minimizers 
for all $\delta\in[\delta_{\mathrm{opt}}, \delta_{\mathrm{BP}})$. In this case, 
the state evolution recursions for Bayes-optimal OAMP have multiple 
fixed points for all $\delta\in[\delta_{\mathrm{opt}}, \delta_{\mathrm{BP}})$. 
Spatial coupling is a general technique to guarantee that the state evolution 
recursions converge to the best fixed point as long as $\delta$ is larger 
than the potential threshold $\delta_{\mathrm{opt}}$. 
\end{remark}

We are ready to prove Theorem~\ref{theorem_optimality}. 
\begin{IEEEproof}[Proof of Theorem~\ref{theorem_optimality}]
We first prove the upper bound,  
\begin{equation} \label{target_upper_bound}
\lim_{\Delta\downarrow0}\lim_{t\to\infty}\lim_{W=\Delta L\to\infty}
\frac{1}{|\mathcal{L}_{W}|}\sum_{\ell\in\mathcal{L}_{W}}E_{t}[\ell]
\leq \tilde{E}_{\mathrm{opt}} 
\end{equation}
if $\tilde{E}_{\mathrm{opt}}$ is unique. 
Using the triangle inequality and Lemma~\ref{lemma_approximation} yields 
\begin{IEEEeqnarray}{rl}
&\frac{1}{|\mathcal{L}_{W}|}\sum_{\ell\in\mathcal{L}_{W}}E_{t}[\ell]
\nonumber \\
\leq& \frac{1}{|\mathcal{L}_{W}|}\sum_{\ell\in\mathcal{L}_{W}}
|E_{t}[\ell] - \tilde{E}_{t}[\ell]| 
+ \frac{1}{|\mathcal{L}_{W}|}\sum_{\ell\in\mathcal{L}_{W}}\tilde{E}_{t}[\ell]
\nonumber \\
\to& \lim_{W=\Delta L\to\infty}
\frac{1}{|\mathcal{L}_{W}|}\sum_{\ell\in\mathcal{L}_{W}}\tilde{E}_{t}[\ell]
\end{IEEEeqnarray}
in the continuum limit. Applying Theorem~\ref{theorem_threshold}, 
we arrive at the upper bound~(\ref{target_upper_bound}) 
if $\tilde{E}_{\mathrm{opt}}$ is unique. 

We next prove the identity $\tilde{E}_{\mathrm{opt}}=E_{\mathrm{opt}}$. 
Applying $g(z)$ given in (\ref{g_func}) and 
the general formula~(\ref{IM_relationship}) between mutual information 
and MMSE to the potential function~(\ref{potential_func}), we have 
\begin{equation}
F(\tilde{E}) 
= 2I(s) + \int_{0}^{\tilde{E}}\frac{1}{\sigma^{2}}R\left(
 - \frac{z}{\sigma^{2}}
\right)dz - s\tilde{E}, 
\end{equation} 
with $s=R(-\tilde{E}/\sigma^{2})/\sigma^{2}$. Using the change of 
variables $\tilde{z}=z/\sigma^{2}$ for the second term, we find that 
$F(\tilde{E})/2$ is equal to the replica-symmetric potential 
$f_{\mathrm{RS}}(\tilde{E},s)$ in (\ref{RS_potential}) 
with $R_{\boldsymbol{A}^{\mathrm{T}}\boldsymbol{A}}(z)=R(z)$ and 
$s=R(-\tilde{E}/\sigma^{2})/\sigma^{2}$. Lemma~\ref{lemma_infsup} for 
$R_{\boldsymbol{A}^{\mathrm{T}}\boldsymbol{A}}(z)=R(z)$ implies that the global 
minimizer $\tilde{E}_{\mathrm{opt}}$ of the potential 
function~(\ref{potential_func}) is equal to $E_{\mathrm{opt}}$ in the global 
optimizer $(E_{\mathrm{opt}}, s_{\mathrm{opt}})$ of the optimization 
problem~(\ref{infsup}). 

We prove the main statement~(\ref{zero_MSE}) in 
Theorem~\ref{theorem_optimality}.  
Using the following identity obtained from the definition of $E_{t}[\ell]$ 
in (\ref{E_t}): 
\begin{equation}
\sum_{\ell\in\mathcal{L}_{W}}E_{t}[\ell]
=\sum_{l\in\mathcal{L}_{0}}\bar{v}_{\B,t}^{\mathrm{post}}[l], 
\end{equation}
we have  
\begin{IEEEeqnarray}{rl} 
&\lim_{\Delta\downarrow0}\lim_{t\to\infty}
\lim_{W=\Delta L\to\infty}\frac{1}{|\mathcal{L}_{0}|}\sum_{l\in\mathcal{L}_{0}}
\bar{v}_{\B,t}^{\mathrm{post}}[l]
\nonumber \\
=& \lim_{\Delta\downarrow0}(1+\Delta)\lim_{t\to\infty}
\lim_{W=\Delta L\to\infty}\frac{1}{|\mathcal{L}_{W}|}
\sum_{\ell\in\mathcal{L}_{W}}E_{t}[\ell]
\leq E_{\mathrm{opt}},
\nonumber \\
\end{IEEEeqnarray}
where the inequality follows from $\tilde{E}_{\mathrm{opt}}=E_{\mathrm{opt}}$,  
the uniqueness assumption of $E_{\mathrm{opt}}$, and the upper 
bound~(\ref{target_upper_bound}). 
Thus, we arrive at the main statement~(\ref{zero_MSE}). 

Finally, we evaluate $E_{\mathrm{opt}}$. Since $E_{\mathrm{opt}}$ is the global 
optimizer of the optimization problem~(\ref{infsup}), 
Theorem~\ref{theorem_potential} for 
$R_{\boldsymbol{A}^{\mathrm{T}}\boldsymbol{A}}(z)=R(z)$ implies that $E_{\mathrm{opt}}$ 
is unique and tends to zero as $\sigma^{2}\downarrow0$ 
for all $\delta>d_{\mathrm{I}}$. Thus, Theorem~\ref{theorem_optimality} holds. 
\end{IEEEproof}

\section{Numerical Results} \label{sec8}
\subsection{Numerical Conditions}
In all numerical results, the BG prior with signal density $\rho\in[0, 1]$ 
was assumed: The signal vector $\boldsymbol{x}$ has independent elements 
that take $0$ with probability $1-\rho$ and are sampled from the 
Gaussian distribution $\mathcal{N}(0,1/\rho)$ with probability $\rho$.
This signal vector satisfies Assumption~\ref{assumption_x}. 
 
For spatially coupled systems, we assumed $M[\ell]=M$, $N[l]=N$, and 
the uniform coupling weights~(\ref{uniform_coupling}).
Two kinds of sensing matrices were considered: 
One is i.i.d.\ Gaussian sensing matrices 
$|\mathcal{W}[\ell]|^{1/2}\boldsymbol{A}[\ell]$ given via (\ref{A_row}) 
that have independent zero-mean Gaussian elements with variance $1/M$. 
The other is artificial ill-conditioned sensing matrices~\cite{Takeuchi211}. 
The SVD structure $|\mathcal{W}[\ell]|^{1/2}\boldsymbol{A}[\ell]
=\boldsymbol{\Sigma}[\ell]\boldsymbol{H}[\ell]$ is considered. The 
singular values 
in $\boldsymbol{\Sigma}[\ell]\in\mathbb{R}^{M\times |\mathcal{W}[\ell]|N}$ are 
defined in Corollary~\ref{corollary6} with condition number $\kappa>1$. 
The unit condition number $\kappa=1$ indicates that the sensing matrix 
has orthogonal rows, as considered in Corollary~\ref{corollary5}. 
The orthogonal matrix $\boldsymbol{H}[\ell]\in\mathcal{O}_{|\mathcal{W}[\ell]|N}$ 
denotes the Hadamard matrix with random row permutation, which can be 
regarded as a practical alternative of Haar-distributed orthogonal 
matrices~\cite{Dudeja22}. 

We considered damped OAMP with the LMMSE filter and the Bayes-optimal 
denoiser---called Bayes-optimal OAMP. Damping~\cite{Rangan192} was employed 
in module~B: The original messages passed from module~B to module~A 
for $t>0$ were replaced by  
\begin{equation}
\vec{\boldsymbol{x}}_{\B\to\A,t+1}[\ell]
:= \zeta\vec{\boldsymbol{x}}_{\B\to\A,t+1}[\ell]
+ (1-\zeta)\vec{\boldsymbol{x}}_{\B\to\A,t}[\ell], 
\end{equation} 
\begin{equation}
v_{\B\to\A,t+1}[\ell]
:= \zeta v_{\B\to\A,t+1}[\ell]
+ (1-\zeta)v_{\B\to\A,t}[\ell],
\end{equation}
with damping factor $\zeta\in[0,1]$, where 
$\vec{\boldsymbol{x}}_{\B\to\A,t+1}[\ell]$ and $v_{\B\to\A,t+1}[\ell]$ 
on the RHSs are given in (\ref{mean_BA}) and (\ref{var_BA}), 
respectively. The damping technique is empirically known to improve 
the convergence property of OAMP for finite-sized systems.  

For comparison, damped AMP with the Bayes-optimal 
denoiser~\cite{Donoho09,Donoho13}---called Bayes-optimal AMP---was 
considered. Damping~\cite{Rangan191} was applied to mean and variance 
messages just before denoising since damping after denoising was not 
effective. 

For spatially coupled systems, different MSEs are achieved in different 
sections. We focus on the largest MSE among all sections. 
$10^{4}$ independent numerical trials were simulated for spatially coupled 
systems while $10^{5}$ independent trials were simulated for conventional 
systems without spatial coupling. 

\subsection{State Evolution}
\begin{figure}[t]
\begin{center}
\includegraphics[width=\hsize]{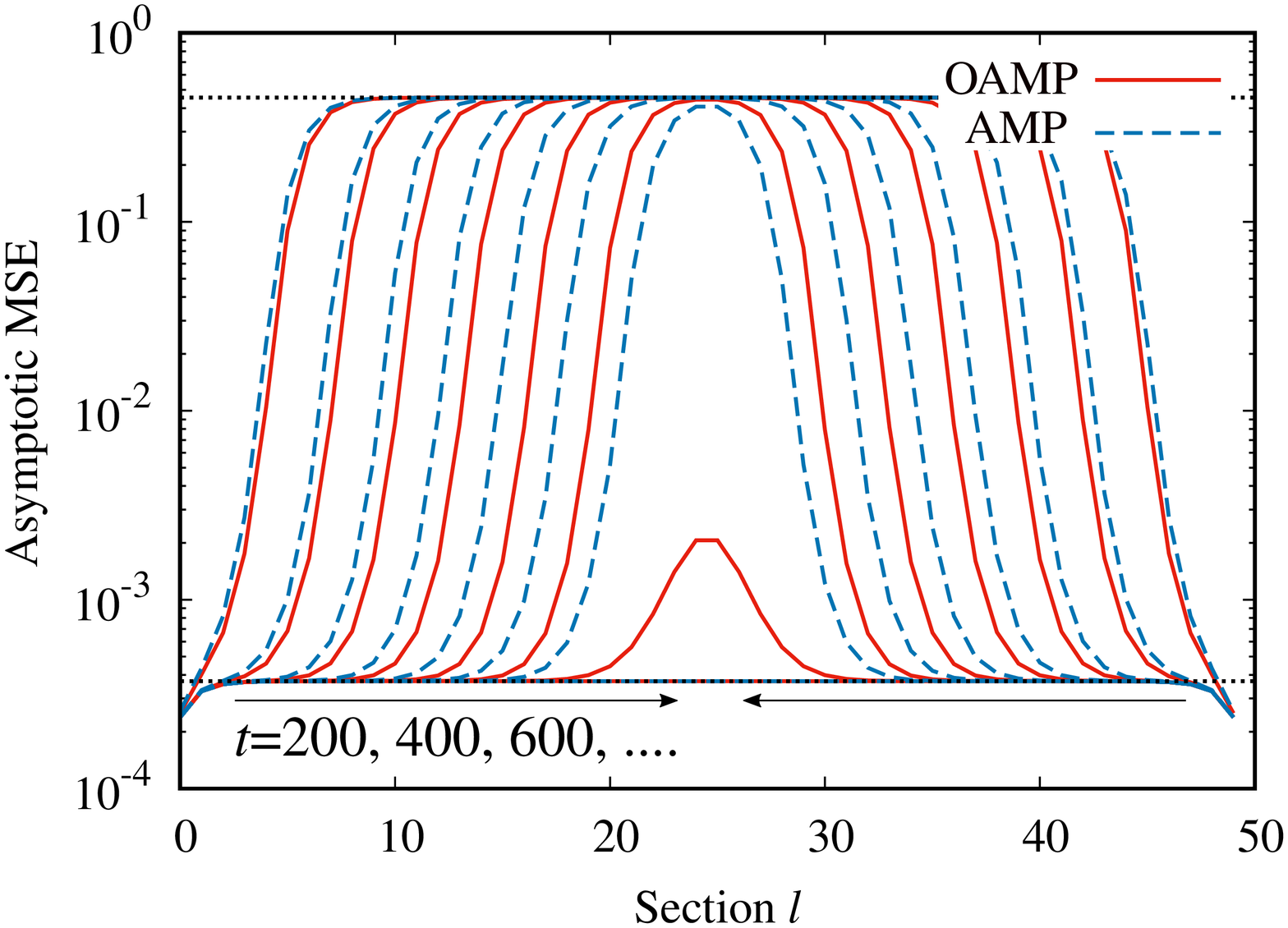}
\caption{
Asymptotic MSE versus section~$l$ for zero-mean i.i.d.\ Gaussian sensing 
matrices, $L=50$, $W=1$, $M/N=0.18$, $\rho=0.1$, and $1/\sigma^{2}=30$~dB. 
The top and bottom horizontal lines are the asymptotic MSEs to which the 
state evolution recursions without spatial coupling converge 
when the original initialization and 
the artificial initialization $\bar{v}_{\B\to\A,0}[\ell]=10^{-6}$ are used, 
respectively. 
}
\label{fig2} 
\end{center}
\end{figure}

The asymptotic dynamics of OAMP is investigated via state evolution. 
As shown in Fig.~\ref{fig2}, the MSEs at both ends decrease to a small value 
in the early stage. Then, the small MSEs propagate toward the center sections. 
Eventually, the MSEs in all sections converge to the MSE shown as the bottom 
horizontal line, which is equal to the Bayes-optimal MSE 
for $M/N=0.18$~\cite{Takeda06,Tulino13,Barbier18}, while the MSE of OAMP 
without spatial coupling converges to the top horizontal line.   

In terms of the comparison between OAMP and AMP, OAMP converges to the 
Bayes-optimal MSE faster than AMP. When the SVDs of $\boldsymbol{A}[\ell]$ 
in (\ref{A_row}) are pre-computed, the per-iteration complexity of OAMP is 
the same as that of AMP. Since the SVD pre-computation is of course dominant 
in the complexity, a reduction in the complexity of OAMP is an important 
future direction. 

We next focus on thresholds~\cite{Takeuchi15}. 
For systems without spatial coupling, a threshold $\delta^{*}$ is defined 
as the infimum of $\delta=M/N$ such that the state evolution recursions 
converge to a unique fixed point, which corresponds to the Bayes-optimal 
performance. Thus, OAMP is Bayes-optimal for $\delta>\delta^{*}$ 
while it is not for $\delta<\delta^{*}$. 

For spatially coupled systems, a threshold $\delta_{\mathrm{SC}}^{*}$ is defined 
as the infimum of $\delta$ such that the state evolution recursions converge to 
the Bayes-optimal MSE. In general, the threshold $\delta_{\mathrm{SC}}^{*}$ is 
smaller than $\delta^{*}$ when the state evolution recursions have multiple 
fixed points for systems without spatial coupling. 

\begin{figure}[t]
\begin{center}
\includegraphics[width=\hsize]{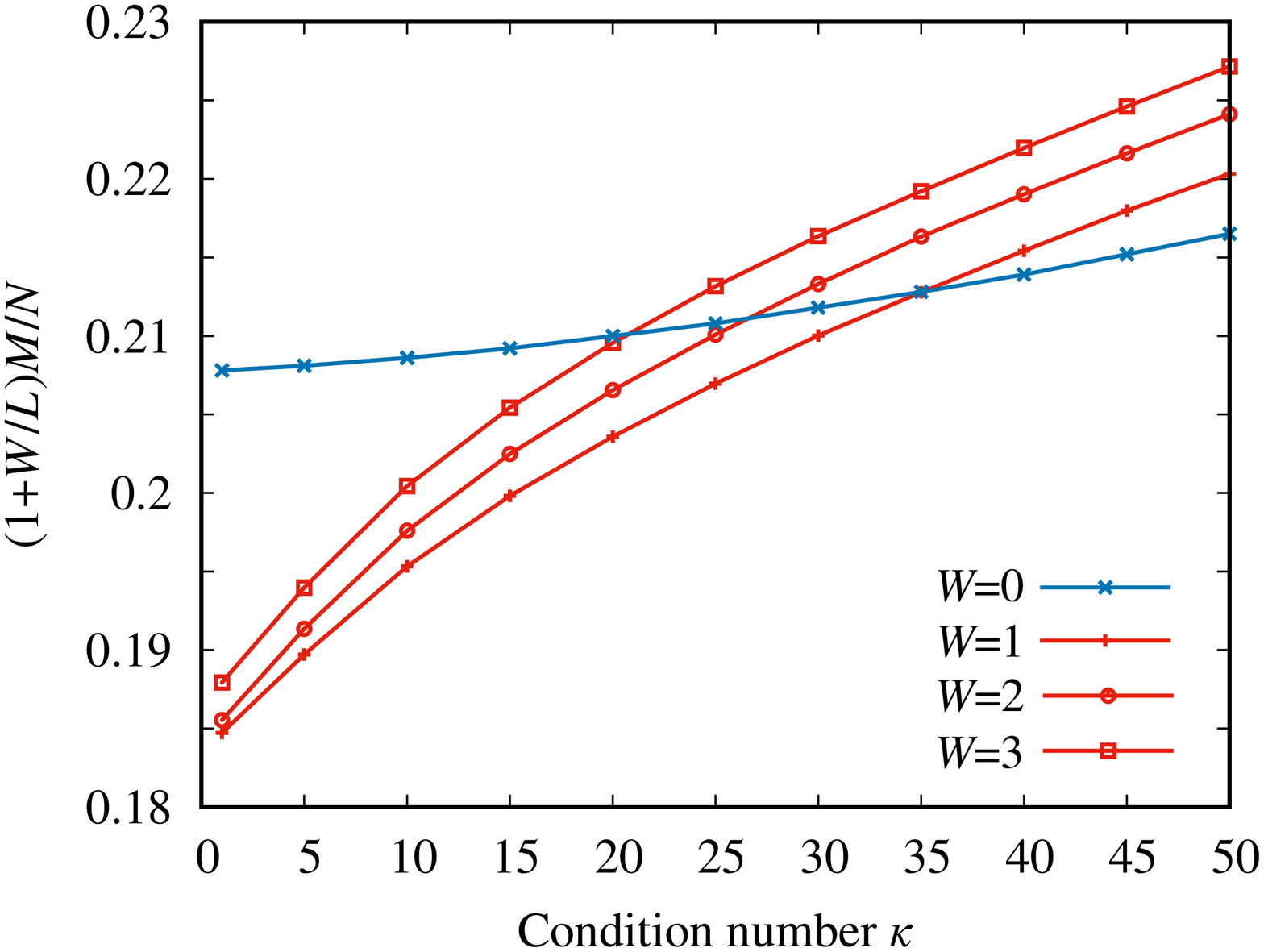}
\caption{
Threshold of OAMP for ill-conditioned sensing matrices, $L=50$, $\rho=0.1$, 
$1/\sigma^{2}=30$~dB, and $1000$~iterations. 
}
\label{fig3} 
\end{center}
\end{figure}

Figure~\ref{fig3} shows the thresholds for systems without spatial coupling 
($W=0$) and spatially coupled systems ($W\geq1$). A a fair comparison, 
$(1+W/L)\delta_{\mathrm{SC}}^{*}$ is plotted for spatially coupled systems while 
$\delta^{*}$ is shown for $W=0$. The thresholds were numerically 
estimated via $1000$ iterations of the state evolution recursions. 

We first focus on the spatially coupled systems with $W\geq1$. As shown 
in Fig.~\ref{fig3}, $(1+W/L)\delta_{\mathrm{SC}}^{*}$ degrades as $W$ increases 
while numerical evaluation showed that the threshold 
$\delta_{\mathrm{SC}}^{*}$ itself improves as $W$ grows. This observation 
implies that a loss in the compression rate is more dominant than an 
improvement in the threshold for $L=50$ and $W\geq1$. We need to consider 
larger $L$ to reduce the loss factor $(1+W/L)$ in the compression rate.  

We next compare $W=0$ and $W=1$. For $W=1$, as implied in 
Corollary~\ref{corollary5}, the threshold $\delta_{\mathrm{SC}}^{*}$ of OAMP 
for $\kappa=1$ is equal to that of AMP for zero-mean i.i.d.\ Gaussian sensing 
matrices with spatial coupling.  
As the condition number $\kappa$ increases, the gap between $\delta^{*}$ and 
$\delta_{\mathrm{SC}}^{*}$ shrinks. As a result, $W=1$ is better than $W=0$ only 
for low-to-moderate condition number when a loss in the compression rate is 
taken into account. Thus, spatial coupling is effective 
for small-to-moderate $\kappa$. 
Note that this conclusion depends heavily on the signal prior: 
It depends heavily on the signal prior whether the state evolution 
recursions have a unique fixed point for high condition number.      

\subsection{Numerical Simulations}
\begin{figure}[t]
\begin{center}
\includegraphics[width=\hsize]{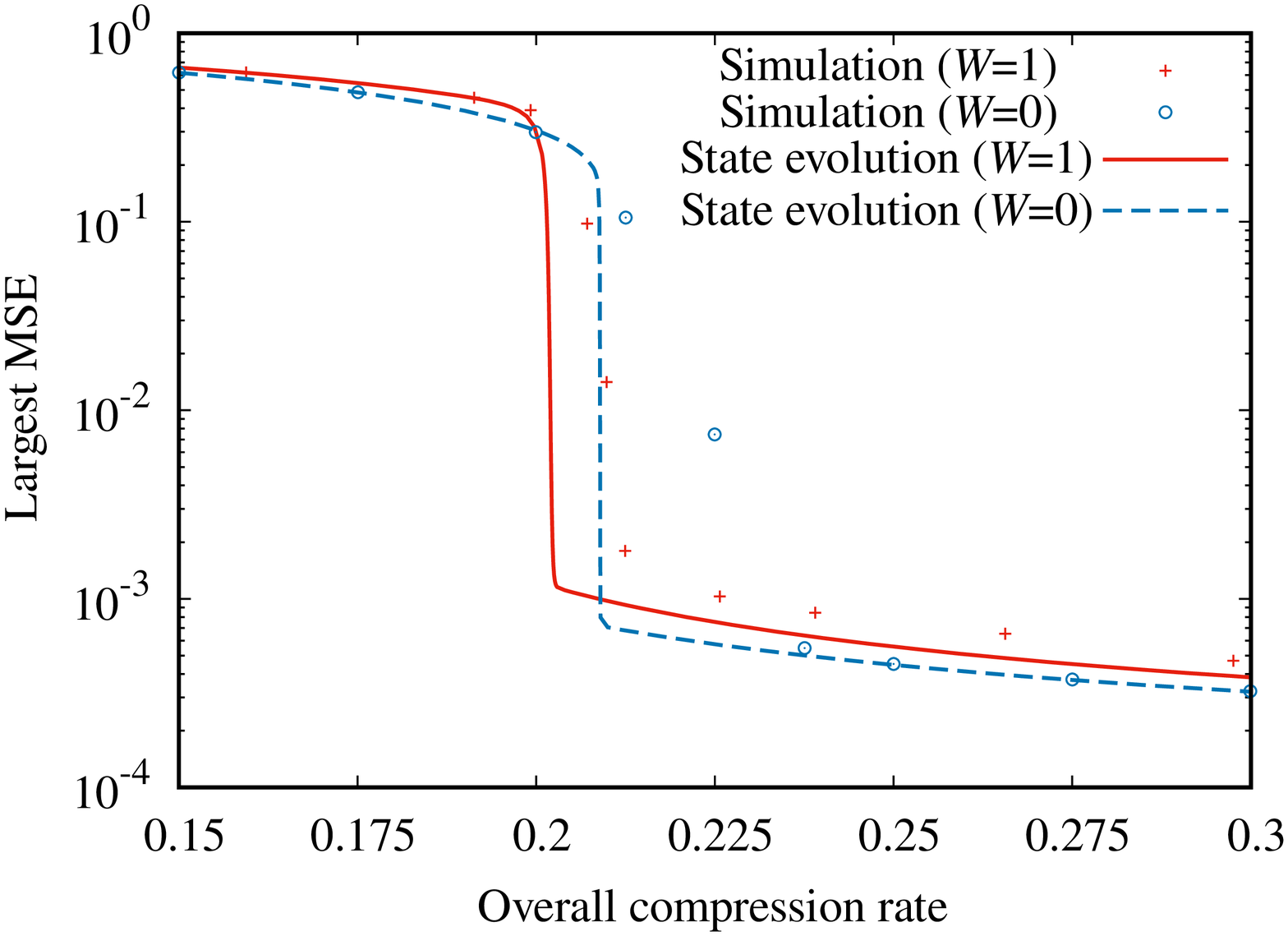}
\caption{
Largest MSE versus the overall compression rate $(1+W/L)\delta$ for 
OAMP, ill-conditioned sensing matrices with condition number $\kappa=10$, 
$L=16$, $N=2^{13}$, $\rho=0.1$, $1/\sigma^{2}=30$~dB, and $200$~iterations. 
}
\label{fig4} 
\end{center}
\end{figure}

OAMP for spatially coupled systems is compared to that for systems without 
spatial coupling. Figure~\ref{fig4} shows the largest MSE among all sections 
for sensing matrices with condition number $\kappa=10$. The used damping 
factors are presented in Table~\ref{table1}. Since the threshold improves 
via spatial coupling, OAMP for spatially coupled systems with $W=1$ achieves 
small MSEs for smaller compression rates than that for $W=0$. 

A disadvantage of spatial coupling is in the region of large compression rates. 
The MSEs for $W=1$ are slightly larger than those for $W=0$ in that region, 
because the $W$-dependency of the R-transform for the sensing 
matrices~(\ref{A_row}) and the rate loss via spatial coupling. 
While the rate loss decreases as $L$ grows, the change 
of the R-transform degrades the performance as $W$ increases. Thus, small 
$W$ should be used in spatially coupled systems. 

\begin{figure}[t]
\begin{center}
\includegraphics[width=\hsize]{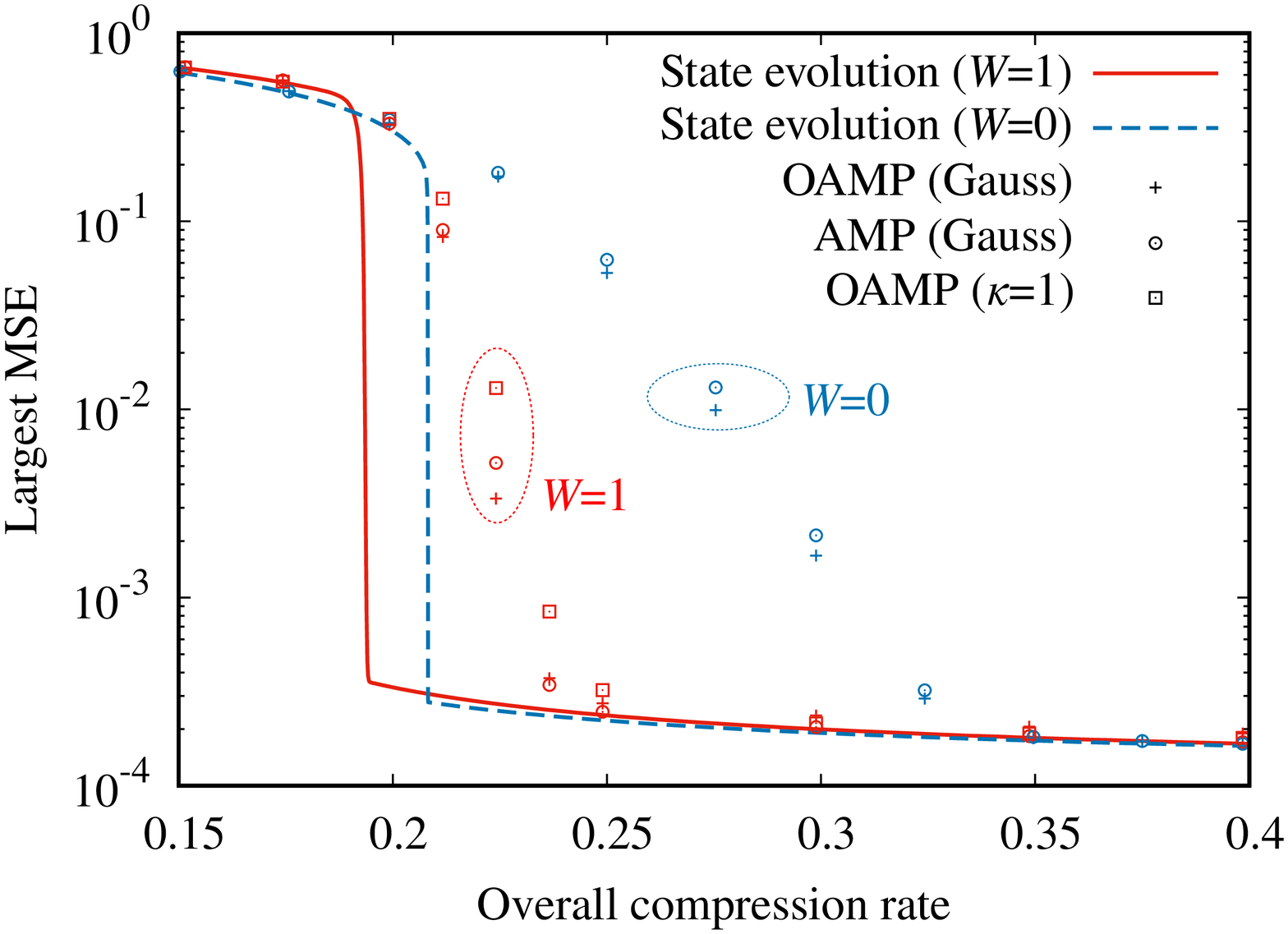}
\caption{
Largest MSE versus the overall compression rate $(1+W/L)\delta$ for 
OAMP, zero-mean i.i.d.\ Gaussian sensing matrices, 
$L=16$, $N=512$, $\rho=0.1$, $1/\sigma^{2}=30$~dB, and $200$~iterations. 
OAMP was also simulated for spatially coupled sensing matrices with 
condition number $\kappa=1$. 
}
\label{fig5} 
\end{center}
\end{figure}

OAMP is compared to AMP for zero-mean i.i.d.\ Gaussian sensing matrices. 
As shown in Fig.~\ref{fig5}, the two algorithms are comparable to each other 
for both $W=0$ and $W=1$. Furthermore, OAMP for the unit condition number 
$\kappa=1$ is also comparable to that for zero-mean i.i.d.\ Gaussian sensing 
matrices for $W=1$. The latter result is consistent with 
Corollary~\ref{corollary5}, claiming that the R-transform for the unit 
condition number $\kappa=1$ reduces to that for zero-mean i.i.d.\ Gaussian 
matrices in the limit $W\to\infty$. These results imply that sensing matrices 
with the unit condition number are a low-complexity alternative of zero-mean 
i.i.d.\ Gaussian sensing matrices for OAMP to achieve the 
information-theoretic compression limit via spatial coupling.

\section{Conclusions} \label{sec9}
This paper has established the unified framework of state evolution 
for LM-MP to reconstruct the signal vectors from right-orthogonally 
invariant linear measurements with spatial coupling. The unified framework 
has been utilized to propose OAMP and LM-OAMP for spatially coupled systems, 
of which the latter is regarded as a tool for proving the convergence 
of the state evolution recursions for Bayes-optimal OAMP. For the noiseless 
case, Bayes-optimal OAMP has been proved to achieve the information-theoretic 
compression limit for right-orthogonally invariant matrices with 
spatial coupling. 

\begin{table}[t]
\begin{center}
\caption{
Damping factors used in Figs.~\ref{fig4}. 
}
\label{table1}
\begin{tabular}{|l|}
\hline
$(M, \zeta)$ for $W=1$ \\
\hline
$(1229, 0.7)$, $(1475, 0.6)$, $(1536, 0.7)$, $(1597, 0.7)$, 
$(1618, 0.7)$, \\
$(1638, 0.7)$, $(1740, 0.7)$, $(1843, 0.7)$, $(2048, 0.7)$, 
$(2294, 0.65)$  \\
\hline 
\hline
$(M, \zeta)$ for $W=0$ \\
\hline
$(1229, 1)$, $(1434, 1)$, $(1638, 1)$, $(1741, 0.95)$, 
$(1843, 0.95)$,  \\
\hline  $(1946, 0.95)$, $(2048, 1)$, $(2253, 1)$, $(2457, 1)$ \\
\hline
\end{tabular}
\end{center}
\end{table}

\begin{table}[t]
\begin{center}
\caption{
Damping factors used in Figs.~\ref{fig5}. 
}
\label{table2}
\begin{tabular}{|l|}
\hline
$(M, \zeta)$ for $W=1$ \\
\hline
$(1229, 0.7)$, $(1475, 0.6)$, $(1536, 0.7)$, $(1597, 0.7)$, 
$(1618, 0.7)$, \\
$(1638, 0.7)$, $(1740, 0.7)$, $(1843, 0.7)$, $(2048, 0.7)$, 
$(2294, 0.65)$  \\
\hline 
\hline
$(M, \zeta)$ for $W=0$ \\
\hline
$(1229, 1)$, $(1434, 1)$, $(1638, 1)$, $(1741, 0.95)$, 
$(1843, 0.95)$,  \\
\hline  $(1946, 0.95)$, $(2048, 1)$, $(2253, 1)$, $(2457, 1)$ \\
\hline
\end{tabular}
\end{center}
\end{table}
 
Possible directions for future research are twofold: One direction is a 
construction of low-complexity LM-MP that achieves the information-theoretic 
compression limit. We need to generalize existing LM-MP, such as 
MAMP~\cite{Liu222} or VAMP with warm-started conjugate gradient 
(WS-CG)~\cite{Skuratovs221,Skuratovs222}, to the spatial coupling 
case. 

The other direction is challenging. This paper has assumed the joint 
right-orthogonal invariance of all non-zero sensing matrices in each row 
section. As a result, 
the R-transform of the sensing matrices depends on the coupling width~$W$. 
Numerical results have shown that the $W$-dependency of the R-transform 
degrades the performance of OAMP in the large-compression-rate regime compared 
to $W=0$. To circumvent this disadvantage, we need to assume the 
right-orthogonal invariance of {\em each} non-zero sensing matrix in each 
row section. For spatially coupled systems under such an assumption, however, 
it is challenging to establish a unified framework of state evolution. 

\appendices 

\section{Proof of Lemma~\ref{lemma_infsup_general}}
\label{proof_lemma_infsup_general}
We follow an existing proof strategy~\cite[Appendix~D]{Barbier191} 
to prove Lemma~\ref{lemma_infsup_general}. If the two functions $f$ and $g$ 
were defined on $\mathbb{R}$, the former part in 
Lemma~\ref{lemma_infsup_general} could be proved straightforwardly by using 
the Legendre-Fenchel transform and its inverse transform. However, 
we need a technical result (i.e.\ Lemma~\ref{lemma_restriction}) 
to restrict the domains of $f$ and $g$ 
to $[E_{\mathrm{min}}, E_{\mathrm{max}}]$ and $[0, \infty)$, respectively. 

We start with the definition of the Legendre-Fenchel transform for concave 
functions.
\begin{definition}
Let $\phi:[x_{\mathrm{min}}, x_{\mathrm{max}}]\to[-\infty,\infty)$ denote a 
concave function such that $\phi(x)>-\infty$ holds for some 
$x\in[x_{\mathrm{min}}, x_{\mathrm{max}}]$. The Legendre-Fenchel transform 
$\phi^{*}$ of $\phi$ is defined as 
\begin{equation} \label{original_Legendre_transform}
\phi^{*}(y) = \inf_{x\in [x_{\mathrm{min}}, x_{\mathrm{max}}]}\{xy - \phi(x)\}
\end{equation} 
for all $y\in\mathbb{R}$. 
\end{definition}

Extend the domain of $\phi$ from $[x_{\mathrm{min}}, x_{\mathrm{max}}]$ to 
$\mathbb{R}$, by letting $\phi(x)=-\infty$ for all 
$x\notin[x_{\mathrm{min}}, x_{\mathrm{max}}]$. If $\phi$ is upper semicontinuous 
and concave on the closed interval $[x_{\mathrm{min}}, x_{\mathrm{max}}]$, 
the extended function $\phi$ is also upper semicontinuous and concave 
on $\mathbb{R}$. Since the infimum in (\ref{original_Legendre_transform}) is 
not attained at $x\notin[x_{\mathrm{min}}, x_{\mathrm{max}}]$, we have the 
following proposition:

\begin{proposition} \label{proposition_extension}
Suppose that $\phi:\mathbb{R}\to[-\infty,\infty)$ is a 
concave function such that $\phi(x)>-\infty$ and $\phi(x')=-\infty$ hold 
for some $x\in[x_{\mathrm{min}}, x_{\mathrm{max}}]$ and all 
$x'\notin[x_{\mathrm{min}}, x_{\mathrm{max}}]$. Then, 
\begin{equation}
\inf_{x\in\mathbb{R}}\{xy - \phi(x)\} = \inf_{x\in[x_{\mathrm{min}}, x_{\mathrm{max}}]}
\{xy - \phi(x)\}
\end{equation}
for all $y\in\mathbb{R}$. 
\end{proposition}

From Proposition~\ref{proposition_extension}, without loss of generality, 
we can extend the domain of $\phi$ to $\mathbb{R}$. Similarly, we can extend 
the domain of another function on $[0,\infty)$ to $\mathbb{R}$. In proving 
Lemma~\ref{lemma_infsup_general}, we use the Fenchel-Moreau 
theorem~\cite[Theorem VI.5.3(e)]{Ellis06}, i.e.\ the invertibility of the 
Legendre-Fenchel transform. 

\begin{lemma}[Fenchel-Moreau Theorem] \label{lemma_Fenchel_Moreau}
Suppose that $\phi:\mathbb{R}\to[-\infty,\infty)$ is an upper semicontinuous 
and concave function such that $\phi(x)>-\infty$ holds for some 
$x\in\mathbb{R}$. Then, $(\phi^{*})^{*}(x)=\phi(x)$ holds for 
all $x\in\mathbb{R}$. 
\end{lemma}

Note that the lower semicontinuity is required in the Fenchel-Moreau theorem 
for convex functions. 

The following result is useful in considering the inverse transform 
for the Legendre-Fenchel transform of $\phi$ with the domain 
$[x_{\mathrm{min}}, x_{\mathrm{max}}]$ extended to $\mathbb{R}$. 

\begin{lemma} \label{lemma_restriction}
Suppose that $\phi:\mathbb{R}\to[-\infty,\infty)$ is a concave function 
such that $\phi(x)>-\infty$ and $\phi(x')=-\infty$ hold 
for some $x\in[x_{\mathrm{min}}, x_{\mathrm{max}}]$ and all 
$x'\notin[x_{\mathrm{min}}, x_{\mathrm{max}}]$. 
Define $\mathcal{Y}\subset\mathbb{R}$ denote the set of $y\in\mathbb{R}$ such 
that the infimum in the Legendre-Fenchel 
transform~(\ref{original_Legendre_transform}) is attained in the 
interior of $[x_{\mathrm{min}}, x_{\mathrm{max}}]$, i.e.\ 
\begin{equation}
\mathcal{Y}=\left\{
 y\in\mathbb{R}: \arginf_{x\in [x_{\mathrm{min}}, x_{\mathrm{max}}]}
 \{xy - \phi(x)\}\in(x_{\mathrm{min}}, x_{\mathrm{max}})
\right\}. 
\end{equation}
Let $y_{0}\in\mathbb{R}$ denote the unique intersection of two 
affine functions: $x_{\mathrm{min}}y - \phi(x_{\mathrm{min}})
=x_{\mathrm{max}}y - \phi(x_{\mathrm{max}})$ at $y=y_{0}$.  
If the infimum in (\ref{original_Legendre_transform}) is attained at 
$x=x_{\mathrm{min}}$ and $x=x_{\mathrm{max}}$ for $y=\sup\mathcal{Y}$ and 
$y=\inf\mathcal{Y}$, respectively, and if $y_{0}\notin\mathcal{Y}_{1}
=\{y\in(\inf\mathcal{Y},\sup\mathcal{Y}): 
y\notin\mathcal{Y}\}$ holds, then 
the Legendre-Fenchel transform $\phi^{*}$ satisfies 
\begin{equation}
\inf_{y\in\mathbb{R}}\{xy - \phi^{*}(y)\} 
= \inf_{y\in\bar{\mathcal{Y}}}\{xy - \phi^{*}(y)\} 
\end{equation}
for all $x\in[x_{\mathrm{min}}, x_{\mathrm{max}}]$, with 
$\bar{\mathcal{Y}}=\mathcal{Y}\cup\{\inf\mathcal{Y}, \sup\mathcal{Y}\}$. 
\end{lemma}
\begin{IEEEproof}
Since $\inf_{y\in\mathbb{R}}\{xy - \phi^{*}(y)\} 
\leq \inf_{y\in\bar{\mathcal{Y}}}\{xy - \phi^{*}(y)\}$ is trivial, 
we prove $\inf_{y\in\mathbb{R}}\{xy - \phi^{*}(y)\} 
\geq \inf_{y\in\bar{\mathcal{Y}}}\{xy - \phi^{*}(y)\}$ 
for all $x\in[x_{\mathrm{min}}, x_{\mathrm{max}}]$. 

We first focus on $y\geq\sup\mathcal{Y}$. 
The definition of $\mathcal{Y}$ implies that the infimum 
in (\ref{original_Legendre_transform}) is attained 
at $x=x_{\mathrm{min}}$ or $x=x_{\mathrm{max}}$.  
For all $y\geq\sup\mathcal{Y}$ we use $x_{\mathrm{max}}\geq x_{\mathrm{min}}$ 
to have 
\begin{align}
&x_{\mathrm{max}}y - \phi(x_{\mathrm{max}}) 
- \{x_{\mathrm{min}}y - \phi(x_{\mathrm{min}})\} \nonumber \\
&\geq x_{\mathrm{max}}\sup\mathcal{Y} - \phi(x_{\mathrm{max}}) 
- \{x_{\mathrm{min}}\sup\mathcal{Y} - \phi(x_{\mathrm{min}})\} \geq 0, 
\end{align}
where the last inequality follows from the optimality assumption of 
$x=x_{\mathrm{min}}$ at $y=\sup\mathcal{Y}$. Thus, we obtain 
\begin{equation}
\phi^{*}(y) = x_{\mathrm{min}}y - \phi(x_{\mathrm{min}}) 
\end{equation}
for all $y\geq\sup\mathcal{Y}$. Using this expression, 
for all $x\geq x_{\mathrm{min}}$ we arrive at 
\begin{align}
&\inf_{y\geq\sup\mathcal{Y}}\{xy - \phi^{*}(y)\} 
= \inf_{y\geq\sup\mathcal{Y}}\{(x-x_{\mathrm{min}})y + \phi(x_{\mathrm{min}})\}
\nonumber \\
&= x\sup\mathcal{Y} - \phi^{*}(\sup\mathcal{Y})  
\geq \inf_{y\in\bar{\mathcal{Y}}}\{xy - \phi^{*}(y)\}. 
\end{align}

We next consider $y\leq\inf\mathcal{Y}$. 
Repeating the same argument with the optimality assumption of 
$x=x_{\mathrm{max}}$ at $y=\inf\mathcal{Y}$, for all $x\leq x_{\mathrm{max}}$ 
we have 
\begin{equation}
\inf_{y\leq\inf\mathcal{Y}}\{xy - \phi^{*}(y)\} 
\geq \inf_{y\in\bar{\mathcal{Y}}}\{xy - \phi^{*}(y)\}.
\end{equation}

Finally, we focus on the remaining set 
$\mathcal{Y}_{1}=\{y\in(\inf\mathcal{Y},\sup\mathcal{Y}): 
y\notin\mathcal{Y}\}$ if $\mathcal{Y}_{1}\neq\emptyset$. 
Since the intersection $y_{0}\notin\mathcal{Y}_{1}$ has been assumed, 
$\phi^{*}(y)=\min\{x_{\mathrm{min}}y - \phi(x_{\mathrm{min}}), 
x_{\mathrm{max}}y - \phi(x_{\mathrm{max}})\}$ reduces to 
$\phi^{*}(y)=x_{\mathrm{max}}y - \phi(x_{\mathrm{max}})$ 
and $\phi^{*}(y)=x_{\mathrm{min}}y - \phi(x_{\mathrm{min}})$ for all 
$y\in\{y\in\mathcal{Y}_{1}: y<y_{0}\}$ and 
$y\in\{y\in\mathcal{Y}_{1}: y>y_{0}\}$, respectively.  
Thus, the infimum of $xy-\phi^{*}(y)$ over 
$y\in\mathcal{Y}_{1}$ is attained at an endpoint for $\mathcal{Y}_{1}$, 
which is also an endpoint for $\bar{\mathcal{Y}}$. These observations imply
\begin{equation}
\inf_{y\in\mathcal{Y}_{1}}\{xy - \phi^{*}(y)\} 
\geq \inf_{y\in\bar{\mathcal{Y}}}\{xy - \phi^{*}(y)\}.
\end{equation}
Since $(-\infty,\inf\mathcal{Y}]\cup\mathcal{Y}_{1}\cup\mathcal{Y}\cup
[\sup\mathcal{Y},\infty)=\mathbb{R}$ holds, we combine these inequalities 
to arrive at $\inf_{y\in\mathbb{R}}\{xy - \phi^{*}(y)\} 
\geq \inf_{y\in\mathcal{Y}}\{xy - \phi^{*}(y)\}$ 
for all $x\in[x_{\mathrm{min}}, x_{\mathrm{max}}]$. 
\end{IEEEproof}

Lemma~\ref{lemma_restriction} is useful for restricting the domain of 
the Legendre-Fenchel transform. When the intersection $y_{0}$ is included 
in $\mathcal{Y}_{1}$, the infimum of $xy-\phi^{*}(y)$ over $y\in\mathcal{Y}_{1}$ 
is attained at $y=y_{0}\notin\mathcal{Y}$. Thus, the condition 
$y_{0}\notin\mathcal{Y}_{1}$ is necessary for Lemma~\ref{lemma_restriction}. 

We are ready for proving the former part in Lemma~\ref{lemma_infsup_general}. 

\begin{IEEEproof}[Proof of (\ref{exchange})]
Using $\psi(E,s)=f(E) + g(s) - sE$ and the definition of 
the Legendre-Fenchel transform, we have 
\begin{equation} \label{infsup_tmp1}
\inf_{s\geq0}\sup_{E\in [E_{\mathrm{min}},E_{\mathrm{max}}]}\psi(E,s) 
= \inf_{s\geq0}\{g(s) - f^{*}(s)\}, 
\end{equation}
where $f^{*}$ is the Legendre-Fenchel transform of $f$ on $\mathbb{R}$, which 
satisfies $f(E)=-\infty$ for all $E\notin[E_{\mathrm{min}},E_{\mathrm{max}}]$.  
See Proposition~\ref{proposition_extension}. 

Extend the domain of $g$ from $[0,\infty)$ to $\mathbb{R}$ by letting 
$g(s)=-\infty$ for all $s<0$. Proposition~\ref{proposition_extension} implies 
that the Legendre-Fenchel transform $g^{*}$ of $g$ on $\mathbb{R}$ is equal 
to $g^{*}(E)=\inf_{s\geq0}\{sE - g(s)\}$. Since $g$ is upper semicontinuous and 
concave on $\mathbb{R}$, Lemma~\ref{lemma_Fenchel_Moreau} implies 
$g(s)=\inf_{E\in\mathbb{R}}\{sE - g^{*}(E)\}$. From the assumption for 
the Legendre-Fenchel transform of $g$ in Lemma~\ref{lemma_infsup_general}, 
we use Lemma~\ref{lemma_restriction} with $\mathcal{Y}=(E_{\mathrm{min}}, 
E_{\mathrm{max}})$ and $\mathcal{Y}_{1}=\emptyset$ to obtain 
$g(s)=\inf_{E\in[E_{\mathrm{min}},E_{\mathrm{max}}]}\{sE - g^{*}(E)\}$ for all 
$s\geq0$. Substituting this expression 
into the RHS of (\ref{infsup_tmp1}) yields 
\begin{align}
&\inf_{s\geq0}\sup_{E\in [E_{\mathrm{min}},E_{\mathrm{max}}]}\psi(E,s) 
\nonumber \\
&= \inf_{s\geq0}\left\{
 \inf_{E\in[E_{\mathrm{min}},E_{\mathrm{max}}]}\{sE - g^{*}(E)\} - f^{*}(s)
\right\} \nonumber \\
&= \inf_{E\in[E_{\mathrm{min}},E_{\mathrm{max}}]}\left\{
 \inf_{s\geq0}\{sE - f^{*}(s)\} - g^{*}(E) 
\right\}. \label{infsup_tmp2}
\end{align}

To evaluate $\inf_{s\geq0}\{sE - f^{*}(s)\}$, we use the assumption for the 
Legendre-Fenchel transform of $f$ in Lemma~\ref{lemma_infsup_general}. 
Lemma~\ref{lemma_restriction} with $\mathcal{Y}=(s_{\mathrm{min}}, s_{\mathrm{max}})$ 
and $\mathcal{Y}_{1}=\emptyset$ implies 
$\inf_{s\in[s_{\mathrm{min}},s_{\mathrm{max}}]}\{sE - f^{*}(s)\}
=\inf_{s\in\mathbb{R}}\{sE - f^{*}(s)\}$. Since we have 
$\inf_{s\geq0}\{sE - f^{*}(s)\}\leq
\inf_{s\in[s_{\mathrm{min}},s_{\mathrm{max}}]}\{sE - f^{*}(s)\}$
from the inclusion  
$[s_{\mathrm{min}},s_{\mathrm{max}}]\subset[0,\infty)$, we obtain the inequality 
$\inf_{s\geq0}\{sE - f^{*}(s)\}\leq\inf_{s\in\mathbb{R}}\{sE - f^{*}(s)\}$. 
Using the trivial inequality 
$\inf_{s\geq0}\{sE - f^{*}(s)\}\geq\inf_{s\in\mathbb{R}}\{sE - f^{*}(s)\}$, 
we have $\inf_{s\geq0}\{sE - f^{*}(s)\}
=\inf_{s\in\mathbb{R}}\{sE - f^{*}(s)\}=f(E)$, in which the last follows 
from Lemma~\ref{lemma_Fenchel_Moreau}.  
Substituting this expression into (\ref{infsup_tmp2}) yields 
\begin{align}
&\inf_{s\geq0}\sup_{E\in [E_{\mathrm{min}},E_{\mathrm{max}}]}\psi(E,s) 
= \inf_{E\in[E_{\mathrm{min}},E_{\mathrm{max}}]}\left\{
 f(E) - g^{*}(E)
\right\}
\nonumber \\
&= \inf_{E\in[E_{\mathrm{min}},E_{\mathrm{max}}]}\left\{
 f(E) - \inf_{s\geq0}\{sE - g(s)\} 
\right\} \nonumber \\
&= \inf_{E\in [E_{\mathrm{min}},E_{\mathrm{max}}]}\sup_{s\geq0}\psi(E,s). 
\end{align}
Thus, (\ref{exchange}) is correct. 
\end{IEEEproof}

To prove the latter part in Lemma~\ref{lemma_infsup_general}, we present known 
properties of a differentiable and strictly concave function $\phi$ on 
$[x_{\mathrm{min}}, x_{\mathrm{max}}]$. 

\begin{lemma}  \label{lemma_Legendre}
Suppose that $\phi:[x_{\mathrm{min}}, x_{\mathrm{max}}]\to[-\infty,\infty)$ is 
differentiable and strictly concave. Let $\mathcal{Y}^{*}\subset\mathbb{R}$ 
denote the set of $y\in\mathbb{R}$ such that $y=\phi'(x)$ has a 
unique solution $x=x^{*}\in[x_{\mathrm{min}}, x_{\mathrm{max}}]$. 
Then, the Legendre-Fenchel 
transform $\phi^{*}$ in (\ref{original_Legendre_transform}) reduces to 
\begin{equation} \label{Legendre_transform}
\phi^{*}(y) = x^{*}y - \phi(x^{*})
\end{equation}
for all $y\in\mathcal{Y}^{*}$. 
Furthermore, $\phi^{*}$ is differentiable. In particular, we have 
\begin{equation} \label{Legendre_derivative}
(\phi^{*})'(y) = x^{*}
\end{equation}
for all $y\in\mathcal{Y}^{*}$. 
\end{lemma}
\begin{IEEEproof}
The expression~(\ref{Legendre_transform}) is trivial. 
The differentiability of $\phi^{*}$ follows 
from the strict concavity of $\phi$~\cite[Theorem VI.5.6]{Ellis06}. 
The expression~(\ref{Legendre_derivative}) is due to 
\cite[Theorem VI.5.3(d)]{Ellis06}.  
\end{IEEEproof}

We are ready for proving the latter part in Lemma~\ref{lemma_infsup_general}. 

\begin{IEEEproof}[Proof of (\ref{infsup_general})] 
Let $g^{*}(E)=\inf_{s\geq0}\{sE - g(s)\}$ denote the Legendre-Fenchel transform 
of $g$. We use $\psi(E,s)=f(E) + g(s) - sE$ to have 
$\inf_{E\in[E_{\mathrm{min}},E_{\mathrm{max}}]}\sup_{s\geq0}\psi(E,s)
= \inf_{E\in[E_{\mathrm{min}},E_{\mathrm{max}}]}\{f(E) - g^{*}(E)\}$.
The derivative $g'$ is strictly decreasing since $g$ is differentiable 
and strictly concave. From Lemma~\ref{lemma_Legendre} and 
the assumption for the Legendre-Fenchel transform 
of $g$ in Lemma~\ref{lemma_infsup_general}, we obtain 
$g^{*}(E)=s^{*}E-g(s^{*})$ for all $E\in[E_{\mathrm{min}}, E_{\mathrm{max}}]$, 
with $E_{\mathrm{min}}=\lim_{s\to\infty}g'(s)$ and $E_{\mathrm{max}}=g'(0)$, in which 
$s^{*}\geq0$ is the unique solution to $E=g'(s^{*})$. 

Let $E^{*}\in[E_{\mathrm{min}}, E_{\mathrm{max}}]$ denote a solution to the 
minimization problem $\inf_{E\in[E_{\mathrm{min}},E_{\mathrm{max}}]}\{f(E)-g^{*}(E)\}$. 
Since Lemma~\ref{lemma_Legendre} implies the differentiability of $g^{*}$, 
we find that the first-order optimality condition for the minimization 
problem is equal to 
\begin{align}
f'(E^{*}) - (g^{*})'(E^{*}) - \mu_{\mathrm{min}} + \mu_{\mathrm{max}} = 0,
\nonumber \\
\mu_{\mathrm{min}}(E_{\mathrm{min}}-E^{*})=0, \quad  
\mu_{\mathrm{max}}(E^{*}- E_{\mathrm{max}})=0, \label{KKT}
\end{align}
with some Lagrange multipliers $\mu_{\mathrm{min}}, \mu_{\mathrm{max}}\geq0$. 

It is sufficient to confirm $f'(E^{*})=s^{*}$ since $E^{*}=g'(s^{*})$ holds 
for $E^{*}\in[E_{\mathrm{min}},E_{\mathrm{max}}]$. 
In the case $E^{*}\in(E_{\mathrm{min}},E_{\mathrm{max}})$, i.e.\ $\mu_{\mathrm{min}}
=\mu_{\mathrm{max}}=0$, we use (\ref{Legendre_derivative}) for $\phi=g$ and 
$\mathcal{Y}^{*}=[E_{\mathrm{min}}, E_{\mathrm{max}}]$ to find that 
the first-order optimality condition~(\ref{KKT}) reduces to 
$f'(E^{*})=(g^{*})'(E^{*})=s^{*}$. Thus, 
the solution $(E^{*}, s^{*})$ is included in the set~(\ref{set}).   

Consider the case $E^{*}=E_{\mathrm{min}}$, i.e.\ $\mu_{\max}=0$. 
The definition $E_{\mathrm{min}}=\lim_{s\to\infty}g'(s)$ implies the optimizer  
$(E^{*}, s^{*})=(E_{\mathrm{min}},\infty)$. 
We use (\ref{Legendre_derivative}) for $\phi=g$ and 
$\mathcal{Y}^{*}=[E_{\mathrm{min}}, E_{\mathrm{max}}]$ to have 
$(g^{*})'(E^{*})=s^{*}=\infty$. Thus, 
the first-order optimality condition~(\ref{KKT}) implies 
$f'(E^{*})=\infty=s^{*}$. Thus, $(E^{*},s^{*})$ is included 
in the set~(\ref{set}). 

Finally, consider $E^{*}=E_{\mathrm{max}}$, i.e.\ $\mu_{\mathrm{min}}=0$. 
The definition $E_{\mathrm{max}}=g'(0)$ implies the optimizer 
$(E^{*}, s^{*})=(E_{\mathrm{max}},0)$. 
From the first-order optimality condition~(\ref{KKT}) we have 
$f'(E^{*})-(g^{*})'(E^{*})+\mu_{\mathrm{max}}=0$. Since $(g^{*})'(E^{*})=s^{*}=0$ 
holds from (\ref{Legendre_derivative}) for $\phi=g$ and 
$\mathcal{Y}^{*}=[E_{\mathrm{min}}, E_{\mathrm{max}}]$, we have 
$f'(E^{*})=-\mu_{\mathrm{max}}\leq0$. Using the non-decreasing assumption 
$f'(E^{*})\geq0$, we arrive at $f'(E^{*})=0=s^{*}$. 
Thus, $(E^{*},s^{*})$ is included in the set~(\ref{set}). 

We have proved the lower bound 
\begin{equation}
\inf_{E\in[E_{\mathrm{min}},E_{\mathrm{max}}]}\sup_{s\geq0}\psi(E,s) 
\geq \inf_{(E, s)}\psi(E,s),
\end{equation}
where the infimum on the RHS is over the set~(\ref{set}). 
To prove the converse inequality, we assume that $(E^{*},s^{*})$ is included 
in the set~(\ref{set}). We use the definition of $g^{*}$ to obtain the lower 
bound $\psi(E^{*},s^{*})=f(E^{*}) - \inf_{s\geq0}\{sE^{*} - g(s)\}
=\sup_{s\geq0}\psi(E^{*},s)\geq \inf_{E\in[E_{\mathrm{min}},E_{\mathrm{max}}]}
\sup_{s\geq0}\psi(E,s)$, which implies   
\begin{equation}
\inf_{(E, s)}\psi(E,s)
\geq \inf_{E\in[E_{\mathrm{min}},E_{\mathrm{max}}]}\sup_{s\geq0}\psi(E,s).
\end{equation}
Combining the two bounds, we arrive at (\ref{infsup_general}). 
\end{IEEEproof}

\section{Properties of $\eta$-Transform and R-Transform}
\subsection{Proof of Lemma~\ref{lemma_eta_transform}}
\label{proof_lemma_eta_transform}
We first prove the former property. 
Repeating the derivation of (\ref{eta_transform_tmp}) yields
\begin{equation}
z\eta_{\boldsymbol{A}^{\mathrm{T}}\boldsymbol{A}}(z)
= \lim_{M=\delta N\to\infty}\left\{
 \frac{1}{N}\sum_{n=1}^{r}\frac{z}{1+\lambda_{n}z}
 + \left(
 1-\frac{r}{N}
 \right)z
\right\},
\label{eta_transform_inf}
\end{equation} 
which is strictly increasing for all $z\geq0$ since (\ref{eta_transform_inf}) 
is the sum of strictly increasing functions of $z\geq0$. 

We next prove the latter property. 
Let $f(z;\lambda_{n})=(1+\lambda_{n}z)^{-1}$ denote the $n$th term 
in the summation on the RHS of (\ref{eta_transform_tmp}).  
Evaluating the $k$th derivative $f^{(k)}$ of $f$ with respect to $z$ 
for any $k\in\mathbb{N}$ yields 
\begin{equation}
f^{(k)}(z;\lambda_{n})
= \frac{(-1)^{k}k!\lambda_{n}^{k}}{(1+\lambda_{n}z)^{k+1}}, 
\end{equation}
which has the $z$-independent upper bound 
$|f^{(k)}(z;\lambda_{n})|\leq k!\lambda_{n}^{k}$ for all $z\geq0$. Furthermore, 
the assumption $\mu_{k}<\infty$ is equivalent to the boundedness of 
$N^{-1}\sum_{n=1}^{r}\lambda_{n}^{k}$ in the large system limit. 
Thus, we can interchange the $k$th derivative and the large system limit 
in (\ref{eta_transform_tmp}) to obtain 
\begin{equation} \label{eta_transform_deriv}
\eta_{\boldsymbol{A}^{\mathrm{T}}\boldsymbol{A}}^{(k)}(z)
= \lim_{M=\delta N\to\infty}\frac{1}{N}\sum_{n=1}^{r}f^{(k)}(z;\lambda_{n}),  
\end{equation}
which is bounded. Thus, the latter property in 
Lemma~\ref{lemma_eta_transform} holds.

\subsection{Proof of Lemma~\ref{lemma_R_transform_0}}
\label{proof_lemma_R_transform_0}
We first prove that the R-transform~(\ref{R_transform}) is infinitely 
continuously-differentiable for all $z\in(z_{\mathrm{min}},0)$. 
From Lemma~\ref{lemma_eta_transform} and the implicit function 
theorem, it is sufficient to confirm that the image of  
$z\eta_{\boldsymbol{A}^{\mathrm{T}}\boldsymbol{A}}(z)$ for $z>0$ is 
$(0, -z_{\mathrm{min}})$, which follows from Lemma~\ref{lemma_eta_transform}.  

We next prove the differentiability of the R-transform at $z=0$. Since 
all moments $\{\mu_{k}\}$ are assumed to be bounded, we have the 
series-expansion~\cite[Eq.~(2.84)]{Tulino04} in a neighborhood of $z=0$, 
\begin{equation}
R_{\boldsymbol{A}^{\mathrm{T}}\boldsymbol{A}}(z) 
= \sum_{k=1}^{\infty}c_{k}z^{k-1}, 
\end{equation}  
where $c_{k}$ denotes the so-called $k$th free cumulant of the asymptotic 
eigenvalue distribution of $\boldsymbol{A}^{\mathrm{T}}\boldsymbol{A}$. 
Thus, the R-transform is infinitely continuously-differentiable at $z=0$. 

Finally, we confirm the well-known results $c_{1}=\mu_{1}$ and 
$c_{2}=\mu_{2}-\mu_{1}^{2}$~\cite[p.~48]{Tulino04} to prove 
(\ref{R_transform_0}) and (\ref{R_transform_deriv_0}).  
We use Taylor's theorem to expand the $\eta$-transform in 
(\ref{eta_transform}) around $z=0$ up to the second order, 
\begin{equation}
\eta_{\boldsymbol{A}^{\mathrm{T}}\boldsymbol{A}}(-z)
= 1 +\mu_{1}z + \mu_{2}z^{2} + o(1) 
\end{equation}
as $z\to0$. 
Evaluating the series-expansion of the RHS
in the R-transform~(\ref{R_transform_tmp}) yields   
\begin{equation}
\frac{1-\eta_{\boldsymbol{A}^{\mathrm{T}}\boldsymbol{A}}(-z)}
{-z\eta_{\boldsymbol{A}^{\mathrm{T}}\boldsymbol{A}}(-z)}
= \mu_{1} + (\mu_{2}-\mu_{1}^{2})z + o(1)
\end{equation}
as $z\to0$. Since $\eta_{\boldsymbol{A}^{\mathrm{T}}\boldsymbol{A}}(0)=1$ holds, 
we take the limit $z\to0$ in (\ref{R_transform_tmp}) to obtain 
$R_{\boldsymbol{A}^{\mathrm{T}}\boldsymbol{A}}(0)
=\lim_{z\uparrow0}R_{\boldsymbol{A}^{\mathrm{T}}\boldsymbol{A}}
(z\eta_{\boldsymbol{A}^{\mathrm{T}}\boldsymbol{A}}(-z))=\mu_{1}$. 
Similarly, we have 
\begin{equation}
R_{\boldsymbol{A}^{\mathrm{T}}\boldsymbol{A}}'(0)
= \lim_{z\uparrow0}\frac{R_{\boldsymbol{A}^{\mathrm{T}}\boldsymbol{A}}
(z\eta_{\boldsymbol{A}^{\mathrm{T}}\boldsymbol{A}}(-z)) 
- \mu_{1}}{z\eta_{\boldsymbol{A}^{\mathrm{T}}\boldsymbol{A}}(-z)}
= \mu_{2} - \mu_{1}^{2}. 
\end{equation}
Thus, Lemma~\ref{lemma_R_transform_0} holds. 

\subsection{Proof of Lemma~\ref{lemma_R_transform_positive}}
\label{proof_lemma_R_transform_positive}
We first confirm $\eta_{\boldsymbol{A}^{\mathrm{T}}\boldsymbol{A}}(z)\in(0, 1)$ for 
all $z>0$. Since the positivity follows from the definition of the 
$\eta$-transform in (\ref{eta_transform}), we need to prove 
$\eta_{\boldsymbol{A}^{\mathrm{T}}\boldsymbol{A}}(z)<1$. 
Applying the strict upper bound $1/(1+\lambda_{n}z)<1$ due to $\lambda_{n}z>0$ 
to (\ref{eta_transform_tmp}), we have 
\begin{equation}
\eta_{\boldsymbol{A}^{\mathrm{T}}\boldsymbol{A}}(z)
< \lim_{M=\delta N\to\infty}\left(
 \frac{r}{N} + 1-\frac{r}{N}
\right)=1
\end{equation}
for all $z>0$. 

We next prove the former properties. The properties for $z=0$ follow from 
Lemma~\ref{lemma_R_transform_0} and the assumption $\mu_{1}>0$. 
For all $z\in(z_{\mathrm{min}},0)$, on the other hand, we use 
$\eta_{\boldsymbol{A}^{\mathrm{T}}\boldsymbol{A}}(z)\in(0,1)$ for all $z>0$ 
to find that the R-transform $R_{\boldsymbol{A}^{\mathrm{T}}\boldsymbol{A}}(z)$ 
in (\ref{R_transform_tmp}) is positive for 
all $z\in(z_{\mathrm{min}},0)$. Thus, we arrive at the former properties.

Finally, we prove the latter properties. The properties for $z=0$ follow 
from Lemma~\ref{lemma_R_transform_0} and the assumption $\mu_{2}>\mu_{1}^{2}$. 
Thus, we focus on the open interval $(z_{\mathrm{min}}, 0)$. 
Differentiating both sides in (\ref{R_transform_tmp}) with respect to $z$ 
yields 
\begin{equation} \label{R_transform_deriv}
R_{\boldsymbol{A}^{\mathrm{T}}\boldsymbol{A}}'
(-z\eta_{\boldsymbol{A}^{\mathrm{T}}\boldsymbol{A}})
= \frac{z\eta_{\boldsymbol{A}^{\mathrm{T}}\boldsymbol{A}}
\eta_{\boldsymbol{A}^{\mathrm{T}}\boldsymbol{A}}'
+ (1-\eta_{\boldsymbol{A}^{\mathrm{T}}\boldsymbol{A}})
(z\eta_{\boldsymbol{A}^{\mathrm{T}}\boldsymbol{A}})'}
{(z\eta_{\boldsymbol{A}^{\mathrm{T}}\boldsymbol{A}})^{2}
(z\eta_{\boldsymbol{A}^{\mathrm{T}}\boldsymbol{A}})'}. 
\end{equation}
Lemma~\ref{lemma_eta_transform} implies that 
$-z\eta_{\boldsymbol{A}^{\mathrm{T}}\boldsymbol{A}}(z)$ is one-to-one mapping from 
$(0, \infty)$ onto $(z_{\mathrm{min}}, 0)$. Thus, it is sufficient to prove 
that the numerator in (\ref{R_transform_deriv}) is positive for all $z>0$. 

To prove the positivity of the numerator, we use definition of 
the $\eta$-transform in (\ref{eta_transform_tmp}) to evaluate 
$(z\eta_{\boldsymbol{A}^{\mathrm{T}}\boldsymbol{A}})'$ as 
\begin{IEEEeqnarray}{rl}
(z\eta_{\boldsymbol{A}^{\mathrm{T}}\boldsymbol{A}})'
=& \lim_{M=\delta N\to\infty}\left\{
 \frac{1}{N}\sum_{n=1}^{r}\frac{1}{(1+\lambda_{n}z)^{2}}
 + 1-\frac{r}{N}
\right\} \nonumber \\
=& \lim_{M=\delta N\to\infty}\frac{1}{N}\mathrm{Tr}\left\{
 (\boldsymbol{I}_{N} + z\boldsymbol{A}^{\mathrm{T}}\boldsymbol{A})^{-2}
\right\}. 
\end{IEEEeqnarray}
Using this identity and the definition of the $\eta$-transform in 
(\ref{eta_transform}), we find that the numerator 
in (\ref{R_transform_deriv}) reduces to 
\begin{IEEEeqnarray}{l}
z\eta_{\boldsymbol{A}^{\mathrm{T}}\boldsymbol{A}}
\eta_{\boldsymbol{A}^{\mathrm{T}}\boldsymbol{A}}'
+ (1-\eta_{\boldsymbol{A}^{\mathrm{T}}\boldsymbol{A}})
(z\eta_{\boldsymbol{A}^{\mathrm{T}}\boldsymbol{A}})' 
= (z\eta_{\boldsymbol{A}^{\mathrm{T}}\boldsymbol{A}})' 
-\eta_{\boldsymbol{A}^{\mathrm{T}}\boldsymbol{A}}^{2}
\nonumber \\
= \lim_{M=\delta N\to\infty}\left\{
 \frac{1}{N}\mathrm{Tr}\left(
  \boldsymbol{S}_{z}^{2}
 \right)
 - \left[
  \frac{1}{N}\mathrm{Tr}\left(
   \boldsymbol{S}_{z}
  \right)
 \right]^{2} 
\right\}\geq0
\end{IEEEeqnarray}
for all $z>0$, with $\boldsymbol{S}_{z}=(\boldsymbol{I}_{N} 
+ z\boldsymbol{A}^{\mathrm{T}}\boldsymbol{A})^{-1}$, where the last inequality 
follows from Jensen's inequality. In particular, the equality holds only when 
all eigenvalues of $\boldsymbol{S}_{z}$ are identical, so that we have the 
strict inequality under the assumption $\mu_{2}>\mu_{1}^{2}$. 
Thus, the latter properties hold.

\section{Proof of Theorem~\ref{theorem_potential}}
\label{proof_theorem_potential}
\subsection{Overview}
The converse theorem is due to \cite{Wu10}. Thus, we only prove the 
achievability: The optimizer $(E_{\mathrm{opt}}, s_{\mathrm{opt}})$ is unique and 
noise-limited if $\delta>d_{\mathrm{I}}$ holds. 
We use Lemma~\ref{lemma_infsup} to focus on the extremizers 
$(E, s)\in\mathcal{S}$ in (\ref{extremizer}) in solving the inf-sup 
problem~(\ref{infsup}).  

We classify the extremizers~(\ref{extremizer}) into two classes: A first 
class contains noise-limited extremizers satisfying $\sigma^{2}s>0$ 
in the limit $\sigma^{2}\downarrow0$ while the other class includes 
interference-limited extremizers $\sigma^{2}s\downarrow0$ 
as $\sigma^{2}\downarrow0$. The noise-limited extremizers require diverging 
SINR $s\to\infty$ in the noiseless limit $\sigma^{2}\downarrow0$ while 
the interference-limited extremizers include extremizers 
with bounded SINR $s<\infty$ in the noiseless limit.

We prove the achievability by showing 
\begin{equation} \label{former_limit}
\lim_{\sigma^{2}\downarrow0}\frac{1}{\ln\sigma^{-2}}\left\{
 f_{\mathrm{RS}}(E,s) + \frac{d_{\mathrm{I}}}{2}\ln\sigma^{2} 
\right\}
= 0
\end{equation}
for the former class of extremizers $(s,E)\in\mathcal{S}$ that 
satisfy $\sigma^{2}s>0$ in the limit $\sigma^{2}\downarrow0$. Otherwise, 
for any $\epsilon>0$ and $\delta\neq d_{\mathrm{I}}$ 
\begin{equation} \label{latter_limit}
\liminf_{\sigma^{2}\downarrow0}\frac{1}{\ln\sigma^{-2}}\left\{
 f_{\mathrm{RS}}(s,E) + \frac{d_{\mathrm{I}}}{2}\ln\sigma^{2} 
\right\}
\geq \frac{\delta-d_{\mathrm{I}}-\epsilon}{2} 
\end{equation}
for the latter class of extremizers $(s,E)\in\mathcal{S}$ satisfying 
$\sigma^{2}s\downarrow0$ as $\sigma^{2}\downarrow0$. 
The proofs of (\ref{former_limit}) and (\ref{latter_limit}) are given 
in Appendices~\ref{proof_former_limit} and \ref{proof_latter_limit}, 
respectively. 

We confirm that extremizers in the former class are the solution to the 
minimization problem~(\ref{infsup}) for $\delta>d_{\mathrm{I}}$. 
Let $\epsilon=(\delta-d_{\mathrm{I}})/2>0$ in (\ref{latter_limit}). 
Since the lower bound $(\delta-d_{\mathrm{I}}-\epsilon)/2$ in 
(\ref{latter_limit}) is positive, 
the left-hand side (LHS) of (\ref{latter_limit}) is larger than 
that of (\ref{former_limit}). 
This observation implies that the 
former extremizers satisfying $\sigma^{2}s>0$ are the solution to the 
minimization problem~(\ref{infsup}) of the replica-symmetric 
potential~(\ref{RS_potential}) for $\delta>d_{\mathrm{I}}$. 

Extremizers in the former class satisfy $s\to\infty$ as 
$\sigma^{2}\downarrow0$, so that $E=\mathrm{MMSE}(s)$ must converge to zero. 
These observations imply that $(E_{\mathrm{opt}}, s_{\mathrm{opt}})=(0, \infty)$ 
is the unique and global optimizer of (\ref{infsup}) as 
$\sigma^{2}\downarrow0$. Thus, Theorem~\ref{theorem_potential} holds. 

\subsection{Proof of (\ref{former_limit})}
\label{proof_former_limit}
We first prove the following result on the R-transform: 
\begin{lemma} \label{lemma_R_transform_bound}
Suppose that all moments $\{\mu_{k}\}$ are bounded and that 
$R_{\boldsymbol{A}^{\mathrm{T}}\boldsymbol{A}}(z)$ is non-decreasing and non-negative 
for all $z\leq0$. Then, the R-transform 
$R_{\boldsymbol{A}^{\mathrm{T}}\boldsymbol{A}}(z)$ is bounded for all $z\leq0$.  
\end{lemma}
\begin{IEEEproof}
The boundedness of $R_{\boldsymbol{A}^{\mathrm{T}}\boldsymbol{A}}(z)$ for $z\leq0$ 
follows from $R_{\boldsymbol{A}^{\mathrm{T}}\boldsymbol{A}}(0)=\mu_{1}<\infty$,   
obtained from Lemma~\ref{lemma_R_transform_0}, and 
the assumptions for $R_{\boldsymbol{A}^{\mathrm{T}}\boldsymbol{A}}(z)$.
\end{IEEEproof}

We next prove the boundedness $\sigma^{2}s<\infty$ and 
$E/\sigma^{2}<\infty$ as $\sigma^{2}\downarrow0$ 
for the noise-limited extremizer $(E,s)\in\mathcal{S}$ 
satisfying $\sigma^{2}s>0$ in the limit $\sigma^{2}\downarrow0$. 
From the definition of $\mathcal{S}$ in (\ref{extremizer}), we use 
Lemma~\ref{lemma_R_transform_bound} to have the boundedness 
$\sigma^{2}s=R_{\boldsymbol{A}^{\mathrm{T}}\boldsymbol{A}}(-E/\sigma^{2})<\infty$ as 
$\sigma^{2}\downarrow0$. Furthermore, we utilize the upper bound 
$\mathrm{MMSE}(s)\leq s^{-1}$~\cite[Eq.~(6)]{Wu11} to obtain 
$E/\sigma^{2}=\mathrm{MMSE}(s)/\sigma^{2}\leq(\sigma^{2}s)^{-1}<\infty$, 
because of the assumption $\sigma^{2}s>0$.  

Finally, we prove the limit~(\ref{former_limit}). 
Using a general formula~\cite[Theorem~6]{Wu11} between the mutual information 
and R\'enyi information dimension yields 
\begin{equation} \label{Id_relationship}
I(s) = \frac{d_{\mathrm{I}}}{2}\ln s + o(\ln s)
\end{equation}
as $s\to\infty$. Furthermore, we use \cite[Theorem~8]{Wu11} to obtain 
$\mathrm{MMSE}(s) = d_{\mathrm{I}}/s + o(s^{-1})$ as $s\to\infty$, which implies  
\begin{equation} \label{sE}
sE = s\mathrm{MMSE}(s) = d_{\mathrm{I}} + o(1). 
\end{equation}
Substituting the former formula~(\ref{Id_relationship}) into 
the replica-symmetric potential~(\ref{RS_potential}), we have  
\begin{IEEEeqnarray}{l}
f_{\mathrm{RS}}(s,E) + \frac{d_{\mathrm{I}}}{2}\ln\sigma^{2}
= o(\ln(\sigma^{2}s) + \ln\sigma^{-2}) \nonumber \\
+ \frac{d_{\mathrm{I}}}{2}\ln(\sigma^{2}s) 
+ \frac{1}{2}\int_{0}^{E/\sigma^{2}}
R_{\boldsymbol{A}^{\mathrm{T}}\boldsymbol{A}}(-z)dz 
- \frac{sE}{2},
\end{IEEEeqnarray}
where the last three terms are bounded, because of $0<\sigma^{2}s<\infty$,   
$E/\sigma^{2}<\infty$, Lemma~\ref{lemma_R_transform_bound}, 
and (\ref{sE}). Thus, we arrive at the limit~(\ref{former_limit}). 

\subsection{Proof of (\ref{latter_limit})}
\label{proof_latter_limit}
We first prove the three properties $E/\sigma^{2}\to\infty$, $s<\infty$, 
and $E>0$ for the interference-limited extremizer $(s,E)\in\mathcal{S}$ 
satisfying $\sigma^{2}s\downarrow0$ as $\sigma^{2}\downarrow0$. 
From the definition of $\mathcal{S}$ in (\ref{extremizer}), we have  
$R_{\boldsymbol{A}^{\mathrm{T}}\boldsymbol{A}}(-E/\sigma^{2})=\sigma^{2}s\downarrow0$ as 
$\sigma^{2}\downarrow0$. Since $R_{\boldsymbol{A}^{\mathrm{T}}\boldsymbol{A}}(z)$ has 
been assumed to be positive for $z\leq0$, 
we have the first property $E/\sigma^{2}\to\infty$.  

Since the last property $E=\mathrm{MMSE}(s)>0$ follows from 
the second property $s<\infty$, 
we prove the second property $s<\infty$ by contradiction. 
Assume $s\to\infty$. The last assumption in Theorem~\ref{theorem_potential} 
implies $(E/\sigma^{2})R_{\boldsymbol{A}^{\mathrm{T}}\boldsymbol{A}}(-E/\sigma^{2})
=\delta+o(1)$, because of $E/\sigma^{2}\to\infty$. Using 
$R_{\boldsymbol{A}^{\mathrm{T}}\boldsymbol{A}}(-E/\sigma^{2})=\sigma^{2}s$ in 
(\ref{extremizer}) yields $sE=\delta+o(1)$, 
which is a contradiction because of (\ref{sE}) in the limit $s\to\infty$ 
and $\delta\neq d_{\mathrm{I}}$.  
Thus, there are no extremizers $(s,E)\in\mathcal{S}$ 
satisfying $s\to\infty$ and $\sigma^{2}s\downarrow0$ as $\sigma^{2}\downarrow0$. 

We next prove the lower bound~(\ref{latter_limit}). 
Using the non-negativity of mutual information, we lower-bound the 
replica-symmetric potential~(\ref{RS_potential}) as 
\begin{equation} \label{RS_lower_bound} 
f_{\mathrm{RS}}(s,E) 
\geq \frac{1}{2}\int_{0}^{E/\sigma^{2}}
R_{\boldsymbol{A}^{\mathrm{T}}\boldsymbol{A}}(-z)dz - \frac{sE}{2}. 
\end{equation}
Since $s$ is bounded, $sE=s\mathrm{MMSE}(s)<\infty$ is trivial. Thus, we can 
focus on the first term in the lower bound~(\ref{RS_lower_bound}) 
to evaluate (\ref{latter_limit}). 

The last assumption in Theorem~\ref{theorem_potential} implies that,  
for any $\epsilon>0$, there is some $z_{0}>0$ such that 
$|zR_{\boldsymbol{A}^{\mathrm{T}}\boldsymbol{A}}(-z)-\delta|<\epsilon$ holds for all 
$z>z_{0}$. Thus, we use $E/\sigma^{2}\to\infty$ and the positivity assumption 
for $R_{\boldsymbol{A}^{\mathrm{T}}\boldsymbol{A}}(z)$ to have 
\begin{IEEEeqnarray}{rl}
&\liminf_{\sigma^{2}\downarrow0}\frac{1}{2\ln\sigma^{-2}}\int_{0}^{E/\sigma^{2}}
R_{\boldsymbol{A}^{\mathrm{T}}\boldsymbol{A}}(-z)dz
\nonumber \\
>& \liminf_{\sigma^{2}\downarrow0}\frac{1}{2\ln\sigma^{-2}}
\int_{z_{0}}^{E/\sigma^{2}}R_{\boldsymbol{A}^{\mathrm{T}}\boldsymbol{A}}(-z)dz
\nonumber \\
>& \liminf_{\sigma^{2}\downarrow0}\frac{1}{2\ln\sigma^{-2}}
\int_{z_{0}}^{E/\sigma^{2}}\frac{\delta-\epsilon}{z}dz
\nonumber \\
=& \frac{\delta-\epsilon}{2}\liminf_{\sigma^{2}\downarrow0}\frac{
\ln E + \ln\sigma^{-2} - \ln z_{0}}{\ln\sigma^{-2}}
= \frac{\delta-\epsilon}{2},
\end{IEEEeqnarray}
where the last follows from the fact that $z_{0}$ is independent of 
$\sigma^{2}$, as well as $E>0$. Combining these observations, 
we arrive at the lower bound~(\ref{latter_limit}). 

\section{Sufficient Statistic}
\subsection{Memoryless Processing} 
\label{appen_suf}
Consider a virtual AWGN measurement $Y_{t}[w]\in\mathbb{R}$ 
for $w\in\{0,\ldots,W\}$, given by 
\begin{equation} \label{scalar_virtual_model} 
Y_{t}[w] = \gamma[l+w][l]X + Z_{t}[w], \quad
Z_{t}[w]\sim\mathcal{N}(0,v_{t}[w]),
\end{equation}
where $X$ and $\{Z_{t}[w]: w\in\{0,\ldots,W\}\}$ are independent. 
The goal of this appendix is to derive a sufficient statistic for estimation 
of a scalar signal $X$ given $\{Y_{t}[w]\}_{w=0}^{W}$. 

To understand the significance of this problem, we use (\ref{extraction}) to 
find  that $X=x_{n}[l]$ is included only in 
$\{\vec{x}_{n}[l+w][w]: w\in\{0,\ldots,W\}\}$. 
Since $\vec{\boldsymbol{x}}_{\A\to\B,t}[l+w]$ is an estimator of 
$|\mathcal{W}[l+w]|^{-1/2}\vec{\boldsymbol{x}}[l+w]$, 
we associate $Y_{t}[w]$ with the message $\vec{x}_{\A\to\B,n,t}[l+w][w]$ 
to obtain the relationship $v_{t}[w]=v_{\A\to\B,t}[l+w]$.   

We derive a sufficient statistic for estimation of $X$ given 
$\{Y_{t}[w]\}_{w=0}^{W}$. 
Since $\{Z_{t}[w]\}$ are independent, we have the log likelihood 
\begin{equation}
\ln p(\{Y_{t}[w]\} | X) 
= -\sum_{w=0}^{W}\frac{(Y_{t}[w] - \gamma[l+w][l]X)^{2}}{2v_{t}[w]}
+ \mathrm{Const.} 
\end{equation}
Expanding the square implies that the sum 
$\sum_{w=0}^{W}\gamma[l+w][l]Y_{t}[w]/v_{t}[w]$ is a sufficient statistic for 
estimation of $X$ given $\{Y_{t}[w]: w\in\{0,\ldots,W\}\}$.  

Consider the following normalized sufficient statistic 
\begin{equation}
S_{t} = \frac{\sum_{w=0}^{W}\gamma[l+w][l]Y_{t}[w]
v_{t}^{-1}[w]}
{\sum_{w=0}^{W}\gamma^{2}[l+w][l]v_{t}^{-1}[w]}.
\end{equation}
Substituting the virtual measurement~(\ref{scalar_virtual_model}) into this 
expression, we have 
\begin{equation} \label{sufficient_statistic}
S_{t} = X + Z_{t}, \quad
Z_{t} = \frac{\sum_{w=0}^{W}\gamma[l+w][l]
Z_{t}[w]v_{t}^{-1}[w]}
{\sum_{w=0}^{W}\gamma^{2}[l+w][l]v_{t}^{-1}[w]},
\end{equation}
which is a zero-mean Gaussian random variable with variance 
\begin{equation} \label{sufficient_var}
\mathbb{E}[Z_{t}^{2}] 
= \frac{1}{\sum_{w=0}^{W}\gamma^{2}[l+w][l]v_{t}^{-1}[w]}. 
\end{equation}

We derive the mean message $\boldsymbol{x}_{\A\to \B,t}^{\mathrm{suf}}[l]$ and 
variance message $v_{\A\to \B,t}^{\mathrm{suf}}[l]$. 
Applying $v_{t}[w]=v_{\A\to\B,t}[l+w]$ to (\ref{sufficient_var}) yields 
the variance message (\ref{var_suf_B}). Similarly, substituting 
$Y_{t}[w]=\vec{x}_{\A\to\B,n,t}[l+w][w]=x_{\A\to\B,n,t}[l][w]$---obtained from 
(\ref{mean_AB_projection})---and $v_{t}[w]=v_{\A\to\B,t}[l+w]$ into 
(\ref{sufficient_statistic}), we arrive at the mean message~(\ref{mean_suf_B}). 

\subsection{Long Memory Processing}
\label{appen_suf_LM}
Instead of the scalar AWGN measurement~(\ref{scalar_virtual_model}), 
consider the virtual AWGN measurement vector  
$\boldsymbol{Y}_{t}[w]\in\mathbb{R}^{1\times(t+1)}$ for $w\in\{0,\ldots,W\}$, 
given by 
\begin{equation} \label{virtual_model} 
\boldsymbol{Y}_{t}[w] = 
\gamma[l+w][l]X\boldsymbol{1}^{\mathrm{T}} 
+ \boldsymbol{Z}_{t}[w], \quad 
\boldsymbol{Z}_{t}[w]\sim\mathcal{N}(\boldsymbol{0},
\boldsymbol{V}_{t}[w]), 
\end{equation}
where $X$ and $\{\boldsymbol{Z}_{t}[w]: w\in\{0,\ldots,W\}\}$ are independent. 
The goal is to derive a sufficient statistic for estimation 
of a scalar signal $X$ given $\{\boldsymbol{Y}_{t}[w]\}_{w=0}^{W}$. 
The significance of this problem is in the relationship 
$\boldsymbol{V}_{t}[w]=\boldsymbol{V}_{\A\to\B,t}[l+w]$ when 
$[\boldsymbol{Y}_{t}[w]]_{\tau}$ is associated with the 
corresponding element in the message  
$\vec{\boldsymbol{x}}_{\A\to\B,\tau}[l+w]$.   

We derive a sufficient statistic for estimation of $X$ given 
$\{\boldsymbol{Y}_{t}[w]\}$. 
Since $\{\boldsymbol{Z}_{t}[w]\}$ are independent, we have the log likelihood 
\begin{IEEEeqnarray}{rl}
\ln p(\{\boldsymbol{Y}_{t}[w]\} | X) 
&= -\frac{1}{2}\sum_{w=0}^{W}
(\boldsymbol{Y}_{t}[w] - \gamma[l+w][l]X
\boldsymbol{1}^{\mathrm{T}})
\nonumber \\
\cdot\boldsymbol{V}_{t}^{-1}[w](\boldsymbol{Y}_{t}&[w] 
- \gamma[l+w][l]X\boldsymbol{1}^{\mathrm{T}})^{\mathrm{T}}
+ \mathrm{Const.}
\end{IEEEeqnarray}
Expanding the square, we find that the sum $\sum_{w=0}^{W}\gamma[l+w][l]
\boldsymbol{Y}_{t}[w]\boldsymbol{V}_{t}^{-1}[w]\boldsymbol{1}$ 
is a sufficient statistic for estimation of $X$ given 
$\{\boldsymbol{Y}_{t}[w]: w\in\{0,\ldots,W\}\}$. 

Consider the following normalized sufficient statistic 
\begin{equation} \label{sufficient_statistic_LM} 
S_{t} = \frac{\sum_{w=0}^{W}\gamma[l+w][l]
\boldsymbol{Y}_{t}[w]\boldsymbol{V}_{t}^{-1}[w]\boldsymbol{1}}
{\sum_{w=0}^{W}\gamma^{2}[l+w][l]\boldsymbol{1}^{\mathrm{T}} 
\boldsymbol{V}_{t}^{-1}[w]\boldsymbol{1}}.
\end{equation}
Substituting the virtual measurement vector~(\ref{virtual_model}) into this 
expression, we have 
\begin{equation}
S_{t} = X + Z_{t},
\quad Z_{t} = \frac{\sum_{w=0}^{W}\gamma[l+w][l]\boldsymbol{Z}_{t}[w]
\boldsymbol{V}_{t}^{-1}[w]\boldsymbol{1}}
{\sum_{w=0}^{W}\gamma^{2}[l+w][l]\boldsymbol{1}^{\mathrm{T}} 
\boldsymbol{V}_{t}^{-1}[w]\boldsymbol{1}},
\end{equation}
where $\{Z_{t}\}$ are zero-mean Gaussian random variables with 
covariance  
\begin{equation} \label{sufficient_cov}
\mathbb{E}[Z_{t'}Z_{t}] 
= \frac{1}{\sum_{w=0}^{W}\gamma^{2}[l+w][l]
\boldsymbol{1}^{\mathrm{T}} \boldsymbol{V}_{t}^{-1}[w]\boldsymbol{1}}
\end{equation}
for all $t'\leq t$. 

We derive the mean message $\boldsymbol{x}_{\A\to \B,t}^{\mathrm{suf}}[l]$ and 
covariance message $v_{\A\to \B,t',t}^{\mathrm{suf}}[l]$. Substituting 
$\boldsymbol{V}_{t}[w]=\boldsymbol{V}_{\A\to\B,t}[l+w]$ into 
(\ref{sufficient_cov}) yields the covariance message~(\ref{cov_suf_B}).  
Similarly, associating $\boldsymbol{Y}_{t}[w]$ with each row of 
$\boldsymbol{X}_{\A\to\B,t+1}[l][w]$, as well as using 
$\boldsymbol{V}_{t}[w]=\boldsymbol{V}_{\A\to\B,t}[l+w]$, from 
(\ref{sufficient_statistic_LM}) we arrive at 
the mean message~(\ref{mean_suf_B_LM}). 

\section{Proofs of Corollaries}
\subsection{Proof of Corollary~\ref{corollary4}}
\label{proof_corollary4}
It is straightforward to confirm $R_{\boldsymbol{G}[\ell]}(0)=R(0)$ since 
Lemma~\ref{lemma_R_transform_0} and Assumption~\ref{assumption_A} imply 
$R_{\boldsymbol{G}[\ell]}(0)=1$. 

We next prove that the condition~(\ref{R_transform_limit}) in 
Theorem~\ref{theorem_optimality} holds for $z<0$. 
From the representation of the R-transform in (\ref{R_transform_tmp}) 
we obtain 
\begin{equation} \label{R_transform_tmp2}
R_{\boldsymbol{G}[\ell]}\left(
 - \frac{z\eta_{\boldsymbol{G}[\ell]}(z/|\mathcal{W}[\ell]|)}{|\mathcal{W}[\ell]|}
\right)
= \frac{|\mathcal{W}[\ell]|\{1-\eta_{\boldsymbol{G}[\ell]}(z/|\mathcal{W}[\ell]|)\}}
{z\eta_{\boldsymbol{G}[\ell]}(z/|\mathcal{W}[\ell]|)}.  
\end{equation}
Since the normalized rank $r[\ell]/N$ has been assumed to converge toward 
$\delta$ in the large system limit, repeating the derivation of 
the $\eta$-transform~(\ref{eta_transform_tmp}) for 
$\boldsymbol{G}[\ell]$ yields 
\begin{IEEEeqnarray}{l}
\eta_{\boldsymbol{G}[\ell]}\left(
 \frac{z}{|\mathcal{W}[\ell]|} 
\right)
= 1-\frac{\delta}{|\mathcal{W}[\ell]|}
\nonumber \\
+ \lim_{M=\delta N\to\infty}
\frac{1}{|\mathcal{W}[\ell]|N}\sum_{n=1}^{r[\ell]}
\frac{1}{1+\lambda_{n}[\ell]z/|\mathcal{W}[\ell]|}.
\label{eta_transform_G}
\end{IEEEeqnarray}
Applying the assumption~(\ref{eta_transform_limit}) 
in Corollary~\ref{corollary4} to the $\eta$-transform 
$\eta_{\boldsymbol{G}[\ell]}(z/|\mathcal{W}[\ell]|)$ in (\ref{eta_transform_G}) 
yields 
\begin{equation} \label{eta_transform_1}
\eta_{\boldsymbol{G}[\ell]}(z/|\mathcal{W}[\ell]|)
= 1 + {\cal O}(|\mathcal{W}[\ell]|^{-1}) 
\end{equation}
in the continuum limit. Similarly, we have  
\begin{equation}
|\mathcal{W}[\ell]|\left\{
 1 - \eta_{\boldsymbol{G}[\ell]}\left(
 \frac{z}{|\mathcal{W}[\ell]|} 
\right)
\right\}
= 1 - \eta(z) + {\cal O}(a_{W}^{-1}) 
\label{eta_transform_limit_tmp}
\end{equation}
in the continuum limit. 
Using (\ref{eta_transform_1}) and (\ref{eta_transform_limit_tmp}), 
we find that the R-transform~(\ref{R_transform_tmp2}) reduces to 
\begin{equation}
R_{\boldsymbol{G}[\ell]}\left(
 - \frac{z}{|\mathcal{W}[\ell]|}
\right)
= \frac{1-\eta(z)}{z} + {\cal O}(a_{W}^{-1})
\end{equation}
for $z>0$, where we have used Proposition~\ref{proposition_W}.
Thus, the condition~(\ref{R_transform_limit}) holds for 
$R(-z)=\{1-\eta(z)\}/z$ for $z>0$. 

Finally, we prove that $R(z)$ satisfies 
$\lim_{z\to\infty}zR(-z)=\delta$ or equivalently 
$\lim_{z\to\infty}\eta(z)=1-\delta$. From the 
assumption~(\ref{eta_transform_limit}) in Corollary~\ref{corollary4}, 
it is sufficient to prove 
\begin{equation} \label{eta_transform_limit_target}
\lim_{z\to\infty}\lim_{W=\Delta L\to\infty}\lim_{M=\delta N\to\infty}
\frac{1}{N}\sum_{n=1}^{r[\ell]}
\frac{1}{1+\lambda_{n}[\ell]z/|\mathcal{W}[\ell]|}
= 0.  
\end{equation} 
Since the summation is bounded from above by $\delta$, 
we can interchange the limit $z\to\infty$ and the other two limits 
to arrive at (\ref{eta_transform_limit_target}). 

\subsection{Proof of Corollary~\ref{corollary6}}
\label{proof_corollary6}
The representation~(\ref{R_transform_geometric}) implies that 
$R(z)$ is proper, twice continuously differentiable, strictly increasing, 
and positive for all $z\leq0$. Thus, it is sufficient to prove the 
condition~(\ref{eta_transform_limit}) in Corollary~\ref{corollary4}. 

We know that the $k$th moment $\mu_{k}[\ell]$ of the asymptotic eigenvalue 
distribution of $\boldsymbol{G}[\ell]$ is given 
by $\mu_{k}[\ell]=|\mathcal{W}[\ell]|^{k-1}\mu_{k}$ 
for $k>0$~\cite[Eq.~(64)]{Takeuchi211}, with  
\begin{equation}
\mu_{k} 
= \left(
 \frac{C}{1 - \kappa^{-2}} 
\right)^{k}\frac{1 - \kappa^{-2k}}{Ck},
\end{equation}
which is equal to the $k$th moment of the asymptotic eigenvalue distribution 
of $\boldsymbol{G}[\ell]$ for $|\mathcal{W}[\ell]|=1$. 
Thus, we use the series-expansion $(1+z)^{-1}=\sum_{k=0}^{\infty}(-z)^{k}$ for 
all $|z|<1$ to have 
\begin{IEEEeqnarray}{l}
\lim_{M=\delta N\to\infty}\frac{1}{N}\sum_{n=1}^{r[\ell]}
\frac{1}{1+\lambda_{n}[\ell]z/|\mathcal{W}[\ell]|}
\nonumber \\
= \delta + \lim_{M=\delta N\to\infty}
\frac{1}{|\mathcal{W}[\ell]|N}\sum_{n=1}^{r[\ell]}\sum_{k=1}^{\infty}
\frac{\lambda_{n}^{k}[\ell](-z)^{k}}{|\mathcal{W}[\ell]|^{k-1}} 
\nonumber \\
= \delta + \sum_{k=1}^{\infty}\mu_{k}(-z)^{k}
= \delta - \frac{1}{C}\ln\left(
 \frac{\kappa^{2} - 1 + \kappa^{2}Cz}{\kappa^{2} - 1 + Cz}
\right),
\end{IEEEeqnarray}
where the last follows from \cite[Eq.~(65)]{Takeuchi211}. 
This implies that the condition~(\ref{eta_transform_limit}) in 
Corollary~\ref{corollary4} holds for $\eta(z)$, given by 
\begin{equation} \label{geometical_eta_transform} 
\eta(z) = 1 - \frac{1}{C}\ln\left(
 \frac{\kappa^{2} - 1 + \kappa^{2}Cz}{\kappa^{2} - 1 + Cz}
\right).
\end{equation}

We next evaluate $R(z)=\{\eta(-z)-1\}/z$, which satisfies 
the condition~(\ref{R_transform_limit}) in Theorem~\ref{theorem_optimality} 
from Corollary~\ref{corollary4}. By definition, 
\begin{equation}
R(z) 
= - \frac{1}{Cz}\ln\left(
 \frac{\kappa^{2} - 1 - \kappa^{2}Cz}{\kappa^{2} - 1 - Cz}
\right)
\end{equation}
for $z<0$, with $R(0)=1$. It is straightforward to confirm that 
$R(z)$ can be represented as (\ref{R_transform_geometric}) for $z\leq0$.

\section{Proof of Proposition~\ref{proposition_equivalence}}
\label{proof_proposition_equivalence} 
\subsection{Overview}
The proof is by induction. The proof for $t=0$ is omitted since it is the same 
as that for general $t$. The remaining proof consists of two steps. 
In a first step, for some $t$, we assume that 
$\vec{\boldsymbol{x}}_{\B\to\A,\tau}[\ell]$ in (\ref{mean_BA_LM}) 
and $v_{\B\to\A,\tau',\tau}[\ell]$ in (\ref{cov_BA}) or (\ref{cov_BA0}) 
are respectively equal to $\vec{\boldsymbol{x}}_{\B\to\A,\tau}[\ell]$ 
in (\ref{mean_BA}) and $v_{\B\to\A,\tau}[\ell]$ in (\ref{var_BA_opt}) 
for all $\tau\in\{0,\ldots,t\}$ and $\tau'\in\{0,\ldots,\tau\}$. 
We need to prove that $\vec{\boldsymbol{x}}_{\A\to\B,t}[\ell]$ in  
(\ref{mean_AB_LM}) and $v_{\A\to\B,t',t}[\ell]$ in (\ref{cov_AB}) are equal 
to (\ref{mean_AB}) and (\ref{var_AB_LMMSE}) in OAMP for all 
$t'\in\{0,\ldots,t\}$. 

In the second step, we postulate that $\vec{\boldsymbol{x}}_{\A\to\B,\tau}[\ell]$ 
in (\ref{mean_AB_LM}) and $v_{\A\to\B,\tau',\tau}[\ell]$ in (\ref{cov_AB}) 
are respectively equal to $\vec{\boldsymbol{x}}_{\A\to\B,\tau}[\ell]$ 
in (\ref{mean_BA}) and $v_{\A\to\B,\tau}[\ell]$ in (\ref{var_BA_opt}) 
for all $\tau\in\{0,\ldots,t\}$ and $\tau'\in\{0,\ldots,\tau\}$. 
We need to prove that $\vec{\boldsymbol{x}}_{\B\to\A,t+1}[\ell]$ in  
(\ref{mean_BA_LM}) and $v_{\B\to\A,t',t+1}[\ell]$ in (\ref{cov_BA}) or 
(\ref{cov_BA0}) are equal to (\ref{mean_BA}) and 
(\ref{var_BA_opt}) in OAMP for all $t'\in\{0,\ldots,t+1\}$. These proofs by 
induction imply that Proposition~\ref{proposition_equivalence} holds. 

\subsection{Proof for module~A}
For some $t$, assume that $\vec{\boldsymbol{x}}_{\B\to\A,\tau}[\ell]$ 
in (\ref{mean_BA_LM}) and $v_{\B\to\A,\tau',\tau}[\ell]$ in (\ref{cov_BA}) or 
(\ref{cov_BA0}) are respectively equal to 
$\vec{\boldsymbol{x}}_{\B\to\A,\tau}[\ell]$ in (\ref{mean_BA}) 
and $v_{\B\to\A,\tau}[\ell]$ in (\ref{var_BA_opt}) 
for all $\tau\in\{0,\ldots,t\}$ and $\tau'\in\{0,\ldots,\tau\}$. 
We first prove that the posterior messages 
$\vec{\boldsymbol{x}}_{\A,t}^{\mathrm{post}}[\ell]$ and 
$v_{\A,t',t}^{\mathrm{post}}[\ell]$ in LM-OAMP are equivalent to those in 
OAMP when the LMMSE filter~(\ref{LMMSE_LM}) is used. 

Under the induction hypothesis on the equivalence between (\ref{mean_BA}) and 
(\ref{mean_BA_LM}) for $\vec{\boldsymbol{x}}_{\B\to\A,t}[\ell]$, 
as well as $v_{\B\to\A,t,t}[\ell]=v_{\B\to\A,t}[\ell]$, we find that 
$\vec{\boldsymbol{x}}_{\A,t}^{\mathrm{post}}[\ell]$ in (\ref{mean_post_A_LM}) is 
equal to (\ref{mean_post_A}) for OAMP. By substituting the LMMSE 
filter~(\ref{LMMSE_LM}) into the definition of 
$v_{\A,t',t}^{\mathrm{post}}[\ell]$ in (\ref{cov_post_A}), 
it is straightforward to confirm that, under the induction hypothesis 
$v_{\B\to\A,t',t}[\ell]=v_{\B\to\A,t}[\ell]$,  
the posterior covariance $v_{\A,t',t}^{\mathrm{post}}[\ell]$ reduces to 
\begin{equation} \label{cov_post_A_LMMSE} 
v_{\A,t',t}^{\mathrm{post}}[\ell]
= \eta_{\A,t}[\ell]v_{\B\to\A,t}[\ell]
\end{equation}
for all $t'\in\{0,\ldots,t\}$, with $\eta_{\A,t}[\ell]$ defined in 
(\ref{eta_A}). 
This expression implies 
$v_{\A,t',t}^{\mathrm{post}}[\ell]=v_{\A,t}^{\mathrm{post}}[\ell]$ given in 
(\ref{var_post_A_LMMSE}) for all $t'\in\{0,\ldots,t\}$. 

We next prove the equivalence between OAMP and LM-OAMP for the extrinsic 
messages in module~A. Since we have already proved the equivalence between 
(\ref{mean_post_A}) and (\ref{mean_post_A_LM}) for 
$\vec{\boldsymbol{x}}_{\A,t}^{\mathrm{post}}[\ell]$, we find that 
$\vec{\boldsymbol{x}}_{\A\to\B,t}[\ell]$ in (\ref{mean_AB_LM}) is the same as 
(\ref{mean_AB}) in OAMP under the induction hypothesis. For the extrinsic 
covariance~(\ref{cov_AB}), we use the identity~(\ref{cov_post_A_LMMSE}) for 
the LMMSE filter and the induction hypothesis 
$v_{\B\to\A,t',t}[\ell]=v_{\B\to\A,t}[\ell]$ to obtain 
\begin{equation}
v_{\A\to\B,t',t}[\ell] 
= \frac{1}{|\mathcal{W}[\ell]|}\frac{\eta_{\A,t}[\ell]v_{\B\to\A,t}[\ell]}
{1-\eta_{\A,t}[\ell]}, 
\end{equation}
which is equal to (\ref{var_AB_LMMSE}) in OAMP. 

\subsection{Proof for module~B} 
Assume that $\vec{\boldsymbol{x}}_{\A\to\B,\tau}[\ell]$ in (\ref{mean_AB_LM}) 
and $v_{\A\to\B,\tau',\tau}[\ell]$ in (\ref{cov_AB}) 
are respectively equal to $\vec{\boldsymbol{x}}_{\A\to\B,\tau}[\ell]$ 
in (\ref{mean_AB}) and $v_{\A\to\B,\tau}[\ell]$ in (\ref{var_AB_LMMSE}) 
for all $\tau\in\{0,\ldots,t\}$ and $\tau'\in\{0,\ldots,\tau\}$. 
We first prove that $\boldsymbol{x}_{\A\to \B,t}^{\mathrm{suf}}[l]$ 
in (\ref{mean_suf_B_LM}) and $v_{\A\to \B,t',t}^{\mathrm{suf}}[l]$ in 
(\ref{cov_suf_B}) are respectively equal to (\ref{mean_suf_B}) and 
(\ref{var_suf_B}) for OAMP.

As proved in \cite[Lemma~3]{Takeuchi213}, we use the positive-definiteness 
assumption of $\boldsymbol{V}_{\A\to\B,t}$ and the induction hypothesis 
$v_{\A\to\B,\tau',\tau}[\ell]=v_{\A\to\B,\tau}[\ell]$ for all $\tau\in\{0,\ldots,t\}$ 
and $\tau'\in\{0,\ldots,\tau\}$ to obtain 
\begin{equation} \label{reduction}
\boldsymbol{V}_{\A\to\B,t}^{-1}\boldsymbol{1}
=v_{\A\to\B,t}^{-1}\boldsymbol{e}_{t}.
\end{equation} 
Using the identity~(\ref{reduction}), we find that  
$v_{\A\to \B,t',t}^{\mathrm{suf}}[l]$ in (\ref{cov_suf_B}) reduces to 
$v_{\A\to \B,t}^{\mathrm{suf}}[l]$ in (\ref{var_suf_B}) for all 
$t'\in\{0,\ldots,t\}$. Similarly, 
$\boldsymbol{x}_{\A\to \B,t}^{\mathrm{suf}}[l]$ in (\ref{mean_suf_B_LM}) is equal 
to (\ref{mean_suf_B}) for OAMP. 

We have proved that the second step in LM-OAMP is equivalent to that in OAMP.  
In this sense, Bayes-optimal LM-OAMP is regarded as a tool to prove that 
$\boldsymbol{x}_{\A\to \B,t}^{\mathrm{suf}}[l]$ in Bayes-optimal OAMP is a 
sufficient statistic for estimation of $\boldsymbol{x}[l]$ given not only 
$\{\vec{\boldsymbol{x}}_{\A\to\B,t}[\ell]\}$ but also given the preceding 
messages $\{\vec{\boldsymbol{x}}_{\A\to\B,\tau}[\ell]\}$ for all $\tau<t$. 

We next prove that $\boldsymbol{x}_{\B,t+1}^{\mathrm{post}}[l]$ in 
(\ref{mean_post_B_LM}) and $v_{\B,t'+1,t+1}^{\mathrm{post}}[l]$ in (\ref{cov_post_B}) 
are respectively equal to $\boldsymbol{x}_{\B,t+1}^{\mathrm{post}}[l]$ 
in (\ref{mean_post_B}) and $v_{\B,t+1}^{\mathrm{post}}[l]$ in (\ref{var_post_B}) 
for all $t'\in\{0,\ldots,t\}$. Since we have proved the equivalence between 
the second steps in module~B for OAMP and LM-OAMP, it is trivial that 
(\ref{mean_post_B_LM}) is equivalent to (\ref{mean_post_B}). 
From the identity $v_{\A\to \B,t',t}^{\mathrm{suf}}[l]=v_{\A\to \B,t}^{\mathrm{suf}}[l]$ 
for all $t'\in\{0,\ldots,t\}$, on the other hand, we use 
\cite[Lemma~2]{Takeuchi213} for the Bayes-optimal denoiser to obtain 
$v_{\B,t'+1,t+1}^{\mathrm{post}}[l]=v_{\B,t+1}^{\mathrm{post}}[l]$ in (\ref{var_post_B}) 
for all $t'\in\{0,\ldots,t\}$. 

Finally, we prove that $\vec{\boldsymbol{x}}_{\B\to\A,t+1}[\ell]$ 
in (\ref{mean_BA_LM}) and $v_{\B\to\A,t',t+1}[\ell]$ in (\ref{cov_BA}) or 
(\ref{cov_BA0}) are respectively equal to (\ref{mean_BA}) and 
(\ref{var_BA_opt}) in OAMP for all $t'\in\{0,\ldots,t+1\}$. 
We use the identity~(\ref{reduction}) and $v_{\A\to\B,t,t}^{\mathrm{suf}}[l]
=v_{\A\to\B,t}^{\mathrm{suf}}[l]$ to find that $\eta_{\B,t}[\ell][w]$ 
in (\ref{eta_B_w_LM}) is equivalent to (\ref{eta_B_w}) for OAMP. 
For (\ref{zeta_B}) we use the identity~(\ref{reduction}) to 
have $\eta_{\B,t,t}[\ell][w]=\eta_{\B,t}[\ell][w]$ and 
$\eta_{\B,\tau,t}[\ell][w]=0$ for all $\tau\neq t$. These observations 
imply that $\vec{\boldsymbol{x}}_{\B\to\A,t+1}[\ell]$ in (\ref{mean_BA_LM}) is 
equal to (\ref{mean_BA}) for OAMP. 

Consider $v_{\B\to\A,t'+1,t+1}[\ell]$ for 
$t'\in\{0,\ldots,t\}$. Applying the identity 
$\langle f'_{\mathrm{opt}}
(\boldsymbol{x}_{\A\to\B,t}^{\mathrm{suf}}[l];v_{\A\to\B,t}^{\mathrm{suf}}[l])\rangle
=v_{\B,t+1}^{\mathrm{post}}[l]/v_{\A\to\B,t}^{\mathrm{suf}}[l]$ 
for the Bayes-optimal denoiser to the definition of $\eta_{\B,t}[\ell][w]$ 
in (\ref{eta_B_w}) yields 
\begin{equation} \label{eta_B_tmp}
\eta_{\B,t}[\ell][w]v_{\A\to\B,t}[\ell] 
= |\mathcal{W}[\ell]|\gamma^{2}[\ell][\ell-w]
v_{\B,t+1}^{\mathrm{post}}[\ell-w]. 
\end{equation}
Substituting (\ref{eta_B_tmp}) into the definition of $v_{\B\to\A,t'+1,t+1}[\ell]$ 
in (\ref{cov_BA}) with $v_{\B,t'+1,t+1}^{\mathrm{post}}[\ell-w]=
v_{\B,t+1}^{\mathrm{post}}[\ell-w]$ for all $t'\in\{0,\ldots,t\}$ and using 
the definition of $\eta_{\B,t}[\ell]$ in (\ref{eta_B}) and 
the identity~(\ref{reduction}), we arrive at  
\begin{equation}
v_{\B\to\A,t'+1,t+1}[\ell] 
= \frac{\eta_{\B,t}[\ell]v_{\A\to\B,t}[\ell]}
{1 - \eta_{\B,t}[\ell]/|\mathcal{W}[\ell]|},
\end{equation}
which is equivalent to (\ref{var_BA_opt}) for OAMP. 

Consider $v_{\B\to\A,0,t+1}[\ell]$ given in (\ref{cov_BA0}). From the 
definition of $v_{\B\to\A,0,t+1}^{\mathrm{post}}[\ell]$ in (\ref{cov_post_B0}), 
we have $v_{\B\to\A,0,t+1}^{\mathrm{post}}[\ell]=v_{\B\to\A,t+1}^{\mathrm{post}}[\ell]$ 
given in (\ref{var_post_B}). Substituting this identity into 
(\ref{cov_BA0}) and using the identity~(\ref{eta_B_tmp}) and the 
definition of $\eta_{\B,t}[\ell]$ in (\ref{eta_B}), we arrive at  
\begin{equation}
v_{\B\to\A,0,t+1}[\ell] 
= \frac{\eta_{\B,t}[\ell]v_{\A\to\B,t}[\ell]}
{1 - \eta_{\B,t}[\ell]/|\mathcal{W}[\ell]|},
\end{equation}
which is equivalent to (\ref{var_BA_opt}) for OAMP.

\section{Proof of Theorem~\ref{theorem_SE_LM}}
\label{proof_theorem_SE_LM}
\subsection{Overview}
Theorem~\ref{theorem_SE_LM} is a generalization of conventional state 
evolution~\cite{Rangan192,Takeuchi201} to the long-memory and 
spatial coupling cases. We follow \cite{Takeuchi211} to treat the long-memory 
case. The spatial coupling case  was addressed only for zero-mean i.i.d.\ 
Gaussian matrices in existing state evolution~\cite{Javanmard13,Rush21}. 
This paper generalizes the existing state evolution to the spatial coupling 
case for right-orthogonally invariant matrices. 

The proof of Theorem~\ref{theorem_SE_LM} consists of three steps: A first 
step is to propose a general error model for the spatial coupling case and 
prove the inclusion of the error models for both OAMP and LM-OAMP in the 
proposed general error model. The general error model should be defined 
so as to realize the asymptotic Gaussianity of errors. Unless the general 
error model is defined appropriately, the errors are not Gaussian-distributed. 
The design guideline is to define the general error model such that it can 
be analyzed via a natural generalization of conventional state 
evolution~\cite{Takeuchi211}. 

A second step is rigorous state evolution analysis for 
the general error model. This step is the main part in the proof of 
Theorem~\ref{theorem_SE_LM}. As long as the general error model in the first 
step is defined appropriately, the second step can be established via a 
natural generalization of conventional state evolution~\cite{Takeuchi211}. 
In this sense, the significance of the general error model should be 
understood via state evolution analysis in the second step. 

The last step is to prove the state evolution recursions for both OAMP and 
LM-OAMP via the state evolution analysis in the second step. The last step 
itself is elementary since all evaluation tools needed in the last step are 
prepared in the second step. 

\subsection{Pseudo-Lipschitz Function}
Before presenting the proposed general error model, we follow 
\cite[Section~II-A]{Takeuchi211} to define pseudo-Lipschitz functions. 
They are used to regularize separable functions in the proposed 
general error model for the spatial coupling case. 

\begin{definition}
A function $f:\mathbb{R}^{t}\to\mathbb{R}$ is said to be pseudo-Lipschitz 
of order~$k$~\cite{Bayati11} if there are some Lipschitz constant $L>0$ and 
some order $k\in\mathbb{N}$ such that 
for all $\boldsymbol{x}\in\mathbb{R}^{t}$ and $\boldsymbol{y}\in\mathbb{R}^{t}$ 
the following holds: 
\begin{equation}
|f(\boldsymbol{x}) - f(\boldsymbol{y})| 
\leq L(1 + \|\boldsymbol{x}\|^{k-1} + \|\boldsymbol{y}\|^{k-1})
\|\boldsymbol{x} - \boldsymbol{y}\|. 
\end{equation}
\end{definition}

The first-order pseudo-Lipschitz property is equivalent to the 
Lipschitz-continuity. The Lipschitz constant $L$ can depend on the 
dimension~$t$, which is finite throughout state evolution analysis.  

To use \cite[Lemma~3]{Takeuchi211} sequentially, we use the following 
proposition, which is explicitly presented in this paper while it was 
implicitly used in existing state evolution~\cite{Takeuchi201,Takeuchi211}. 

\begin{proposition} \label{proposition_pseudo-Lipschitz}
Suppose that $f:\mathbb{R}^{t}\to\mathbb{R}$ is pseudo-Lipschitz of order~$k$ 
and consider the vector of variables $\boldsymbol{x}
=(\boldsymbol{x}_{1}, \boldsymbol{x}_{2})\in\mathbb{R}^{1\times t}$ with two 
sections $\boldsymbol{x}_{1}$ and $\boldsymbol{x}_{2}$. Then, 
the marginalized function $g(\boldsymbol{x}_{2})
=\mathbb{E}_{\boldsymbol{x}_{1}}[f(\boldsymbol{x}_{1},\boldsymbol{x}_{2})]$ over 
the first section $\boldsymbol{x}_{1}$ is pseudo-Lipschitz of order~$k$ 
if $\mathbb{E}[\|\boldsymbol{x}_{1}\|^{k-1}]$ is bounded.  
\end{proposition}
\begin{IEEEproof}
We first prove 
\begin{equation} \label{inequality_a+b}
(a+b)^{p}\leq\max\{1,2^{p-1}\}(a^{p}+b^{p})
\end{equation}
for all $a, b\geq0$ and $p>0$. For $p\in(0, 1]$, we let $q=1/p\geq1$ to have 
$(a+b)^{p}=\{(a^{1/q})^{q}+(b^{1/q})^{q}\}^{1/q}\leq a^{1/q}+b^{1/q}$, because of 
$\|\cdot\|_{q}\leq\|\cdot\|_{1}$ for the $q$-norm $\|\cdot\|_{q}$. For 
$p>1$, on the other hand, we use H\"older's inequality to obtain 
$a+b\leq 2^{1-1/p}(a^{p}+b^{p})^{1/p}$. Thus, we arrive at (\ref{inequality_a+b}). 

We next prove that $g$ is pseudo-Lipschitz. 
Since $f$ is pseudo-Lipschitz of order~$k$, we have 
\begin{IEEEeqnarray}{rl}
&|g(\boldsymbol{x}_{2}) - g(\boldsymbol{y}_{2})|
= \left|
 \mathbb{E}_{\boldsymbol{x}_{1}}\left[
  f(\boldsymbol{x}_{1},\boldsymbol{x}_{2}) 
  - f(\boldsymbol{x}_{1},\boldsymbol{y}_{2})
 \right] 
\right| 
\nonumber \\
\leq& L\left\{
 1 + \mathbb{E}_{\boldsymbol{x}_{1}}\left[
  (\|\boldsymbol{x}_{1}\|^{2} 
  + \|\boldsymbol{x}_{2}\|^{2})^{(k-1)/2} 
 \right]
\right. \nonumber \\
&\left.
 + \mathbb{E}_{\boldsymbol{x}_{1}}\left[
  (\|\boldsymbol{x}_{1}\|^{2} + \|\boldsymbol{y}_{2}\|^{2})^{(k-1)/2}
 \right]
\right\}\|\boldsymbol{x}_{2} - \boldsymbol{y}_{2}\| 
\end{IEEEeqnarray}
for some Lipschitz constant $L>0$. For $k=1$, the function $g$ is 
obviously Lipschitz-continuous. 

For $k>1$, we use the inequality~(\ref{inequality_a+b}) to obtain 
\begin{IEEEeqnarray}{rl}
&\mathbb{E}_{\boldsymbol{x}_{1}}\left[
 (\|\boldsymbol{x}_{1}\|^{2} 
 + \|\boldsymbol{u}\|^{2})^{(k-1)/2} 
\right]
\nonumber \\
\leq& \max\{1,2^{(k-3)/2}\}\left(
 \mathbb{E}\left[
  \|\boldsymbol{x}_{1}\|^{k-1} 
 \right]  + \|\boldsymbol{u}\|^{k-1} 
\right)
\end{IEEEeqnarray}
for $\boldsymbol{u}=\boldsymbol{x}_{2}$ and $\boldsymbol{u}=\boldsymbol{y}_{2}$. 
Thus, the function $g$ is pseudo-Lipschitz of order~$k$. 
\end{IEEEproof}

Separable vector-valued functions are used in state evolution analysis. 
A vector-valued function $\boldsymbol{f}=(f_{1},\ldots,f_{N})^{\mathrm{T}}$ is 
said to be pseudo-Lipschitz if all element functions $\{f_{n}\}$ are 
pseudo-Lipschitz. 

\begin{definition}
A vector-valued function $\boldsymbol{f}:
\mathbb{R}^{N\times t}\to\mathbb{R}^{N}$ is said to be 
separable if $[\boldsymbol{f}(\boldsymbol{x}_{1},\ldots,\boldsymbol{x}_{t})]_{n}
=f_{n}(x_{n,1},\ldots,x_{n,t})$ holds for all $n$. 
\end{definition}  
\begin{definition}
A separable pseudo-Lipschitz function 
$\boldsymbol{f}:\mathbb{R}^{N\times t}\to\mathbb{R}^{N}$ is said to be proper 
if the Lipschitz constant $L_{n}>0$ for the $n$th function $f_{n}$ satisfies 
\begin{equation}
\limsup_{N\to\infty}\frac{1}{N}\sum_{n=1}^{N}L_{n}^{j}<\infty
\end{equation}
for any $j\in\mathbb{N}$. 
\end{definition}

This paper considers separable, Lipschitz-continuous, and proper denoisers 
while separable, pseudo-Lipschitz, and proper functions are used to treat 
general performance measure. In particular, pseudo-Lipschitz functions of 
order~$k=2$ are used for the error covariance.  

\subsection{General Error Model with Spatial Coupling}
We define a general error model for the spatial coupling case. 
The proposed general error model is 
a discrete-time dynamical system with respect to six vectors 
$\vec{\boldsymbol{b}}_{t}[\ell]$, $\vec{\boldsymbol{m}}_{t}^{\mathrm{post}}[\ell]$, 
$\vec{\boldsymbol{m}}_{t}^{\mathrm{ext}}[\ell]$, 
$\vec{\boldsymbol{h}}_{t}[\ell]$, 
$\vec{\boldsymbol{q}}_{t}^{\mathrm{post}}[\ell]$, and 
$\vec{\boldsymbol{q}}_{t}^{\mathrm{ext}}[\ell]$ 
in the extended signal space $\mathbb{R}^{N_{\mathrm{c}}[\ell]}$ 
for iteration~$t=0, 1,\ldots$ and row section index $\ell\in\mathcal{L}_{W}$.  
 
To present the proposed general error model, we write the SVD of 
$\boldsymbol{A}[\ell]\in\mathbb{R}^{M[\ell]\times N_{\mathrm{c}}[\ell]}$ 
defined in (\ref{A_row}) as 
$\boldsymbol{A}[\ell]=\boldsymbol{U}[\ell]\boldsymbol{\Sigma}[\ell]
\boldsymbol{V}^{\mathrm{T}}[\ell]$. We define $\vec{\boldsymbol{B}}_{t}[\ell]
=(\vec{\boldsymbol{b}}_{0}[\ell],\ldots,\vec{\boldsymbol{b}}_{t-1}[\ell])
\in\mathbb{R}^{N_{\mathrm{c}}[\ell]\times t}$. 
The matrices $\vec{\boldsymbol{M}}_{t}^{\mathrm{ext}}[\ell]$, 
$\vec{\boldsymbol{H}}_{t}[\ell]$, and 
$\vec{\boldsymbol{Q}}_{t}^{\mathrm{ext}}[\ell]$ are defined in the same manner 
as for $\vec{\boldsymbol{B}}_{t}[\ell]$.  

For $N_{\mathrm{c}}=\max_{\ell\in\mathcal{L}_{W}}N_{\mathrm{c}}[\ell]$, we let 
$\mathcal{B}_{t}=\{
[\vec{\boldsymbol{B}}_{t}^{\mathrm{T}}[\ell], \boldsymbol{O}]^{\mathrm{T}}
\in\mathbb{R}^{N_{\mathrm{c}}\times t}: \ell\in\mathcal{L}_{W}\}$, 
$\Omega=\{[(\boldsymbol{U}^{\mathrm{T}}[\ell]\boldsymbol{n}[\ell])^{\mathrm{T}},
\boldsymbol{0}^{\mathrm{T}}]^{\mathrm{T}}\in\mathbb{R}^{N_{\mathrm{c}}}: 
\ell\in\mathcal{L}_{W}\}$, and $\Lambda=\{
[\vec{\boldsymbol{\lambda}}^{\mathrm{T}}[\ell], \boldsymbol{0}]^{\mathrm{T}}
\in\mathbb{R}^{N_{\mathrm{c}}}: \ell\in\mathcal{L}_{W}\}$, in which 
$\vec{\boldsymbol{\lambda}}[\ell]
\in\mathbb{R}^{N_{\mathrm{c}}[\ell]}$ is a vector that consists of all 
eigenvalues of $\boldsymbol{A}^{\mathrm{T}}[\ell]\boldsymbol{A}[\ell]$. 
Similarly, we define 
$\mathcal{H}_{t}=\{[\vec{\boldsymbol{H}}_{t}^{\mathrm{T}}[\ell], 
\boldsymbol{O}]^{\mathrm{T}}\in\mathbb{R}^{N_{\mathrm{c}}\times t}: 
\ell\in\mathcal{L}_{W}\}$ and $\mathcal{X}
=\{(\mathrm{diag}\{\boldsymbol{1}_{N_{\mathrm{c}}[\ell]},\boldsymbol{0}\},
\boldsymbol{O})\boldsymbol{P}_{\mathrm{x}}[\ell]\boldsymbol{x}
\in\mathbb{R}^{N_{\mathrm{c}}}: \ell\in\mathcal{L}_{W}\}$ for deterministic 
$N_{\mathrm{all}}\times N_{\mathrm{all}}$ permutation matrices 
$\{\boldsymbol{P}_{\mathrm{x}}[\ell]\}$. The permutation matrix 
$\boldsymbol{P}_{\mathrm{x}}[\ell]$ are used to extract $N_{\mathrm{c}}[\ell]$ 
desired elements from the signal vector $\boldsymbol{x}$. 
 
For vector-valued functions 
$\boldsymbol{\phi}_{t}[\ell]:\mathbb{R}^{N_{\mathrm{c}}\times (t+3)|\mathcal{L}_{W}|}
\to\mathbb{R}^{N_{\mathrm{c}}}$ and $\boldsymbol{\psi}_{t}[\ell]:
\mathbb{R}^{N_{\mathrm{c}}\times (t+2)|\mathcal{L}_{W}|}\to\mathbb{R}^{N_{\mathrm{c}}}$, 
the general error model for the spatial coupling case is given by 
\begin{equation} \label{b}
\vec{\boldsymbol{b}}_{t}[\ell]
=\boldsymbol{V}^{\mathrm{T}}[\ell]\vec{\boldsymbol{q}}_{t}^{\mathrm{ext}}[\ell],
\end{equation}
\begin{equation}
\vec{\boldsymbol{m}}_{t}^{\mathrm{post}}[\ell]
= (\boldsymbol{I}_{N_{\mathrm{c}}[\ell]}, \boldsymbol{O})
\boldsymbol{\phi}_{t}[\ell](\mathcal{B}_{t+1}, \Omega, \Lambda),  
\end{equation}
\begin{equation} \label{m_ext}
\vec{\boldsymbol{m}}_{t}^{\mathrm{ext}}[\ell] 
= \vec{\boldsymbol{m}}_{t}^{\mathrm{post}}[\ell]
- \sum_{\tau=0}^{t}\xi_{\A,\tau,t}[\ell]\vec{\boldsymbol{b}}_{\tau}[\ell], 
\end{equation}
\begin{equation} \label{h}
\vec{\boldsymbol{h}}_{t}[\ell] 
= \boldsymbol{V}[\ell]\vec{\boldsymbol{m}}_{t}^{\mathrm{ext}}[\ell],
\end{equation}
\begin{equation}
\vec{\boldsymbol{q}}_{t+1}^{\mathrm{post}}[\ell]
= (\boldsymbol{I}_{N_{\mathrm{c}}[\ell]}, \boldsymbol{O})
\boldsymbol{\psi}_{t}[\ell](\mathcal{H}_{t+1}, \mathcal{X}),  
\end{equation}
\begin{equation} \label{q_ext}
\vec{\boldsymbol{q}}_{t+1}^{\mathrm{ext}}[\ell] 
= \vec{\boldsymbol{q}}_{t+1}^{\mathrm{post}}[\ell]
- \sum_{\tau=0}^{t}\xi_{\B,\tau,t}[\ell]\vec{\boldsymbol{h}}_{\tau}[\ell], 
\end{equation}
with the initial condition $\vec{\boldsymbol{q}}_{0}^{\mathrm{ext}}[\ell]
=(\boldsymbol{I}_{N_{\mathrm{c}}[\ell]}, \boldsymbol{O})
\boldsymbol{\psi}_{-1}[\ell](\mathcal{X})$. 
In (\ref{m_ext}) and (\ref{q_ext}), 
$\xi_{\A,\tau,t}[\ell]\in\mathbb{R}$ and $\xi_{\B,\tau,t}[\ell]\in\mathbb{R}$ 
are given by  
\begin{equation}
\xi_{\A,\tau,t}[\ell] 
= \frac{1}{N_{\mathrm{c}}[\ell]}
\sum_{n=1}^{N_{\mathrm{c}}[\ell]}\partial_{(t+1)\ell+\tau}
[\vec{\boldsymbol{m}}_{t}^{\mathrm{post}}[\ell]]_{n}, 
\end{equation}
\begin{equation} \label{xi_B}
\xi_{\B,\tau,t}[\ell] 
= \frac{1}{N_{\mathrm{c}}[\ell]}
\sum_{n=1}^{N_{\mathrm{c}}[\ell]}\partial_{(t+1)\ell+\tau}
\left[
 \vec{\boldsymbol{q}}_{t+1}^{\mathrm{post}}[\ell]
\right]_{n}.
\end{equation}
The Onsager correction of $\vec{\boldsymbol{m}}_{t}^{\mathrm{post}}[\ell]$ and 
$\vec{\boldsymbol{q}}_{t+1}^{\mathrm{post}}[\ell]$ in (\ref{m_ext}) and 
(\ref{q_ext}) has been defined so as to realize the asymptotic Gaussianity 
for the vectors $\vec{\boldsymbol{b}}_{t}[\ell]$ and 
$\vec{\boldsymbol{h}}_{t}[\ell]$ in the large system limit. 
The significance of the Onsager correction should be understood via state 
evolution in the second step. 

\begin{assumption} \label{assumption_function}
The function $\boldsymbol{\phi}_{t}[\ell]$ is separable with respect to 
all variables and proper Lipschitz-continuous with 
respect to $\mathcal{B}_{t+1}$ and $\Omega$ while $\boldsymbol{\psi}_{t}[\ell]$ 
is separable and proper Lipschitz-continuous with respect to all variables.  
Furthermore, $\|\vec{\boldsymbol{m}}_{t}^{\mathrm{ext}}[\ell]\|\neq0$ and 
$\|\vec{\boldsymbol{q}}_{t+1}^{\mathrm{ext}}[\ell]\|\neq0$ hold for all $t$. 
\end{assumption}

The properties $\|\vec{\boldsymbol{m}}_{t}^{\mathrm{ext}}[\ell]\|\neq0$ and 
$\|\vec{\boldsymbol{q}}_{t+1}^{\mathrm{ext}}[\ell]\|\neq0$ are required to 
guarantee the positive definiteness of the covariance matrix  
$\boldsymbol{V}_{\A\to \B,t}[\ell]$ in the large system limit. Intuitively, 
$\|\vec{\boldsymbol{m}}_{t}^{\mathrm{ext}}[\ell]\|=0$ or  
$\|\vec{\boldsymbol{q}}_{t+1}^{\mathrm{ext}}[\ell]\|=0$ implies that asymptotically 
zero MSE is achieved in iteration~$t$. Thus, additional iteration cannot 
improve the MSE anymore. 

A general error model proposed in \cite{Takeuchi211} was used to conduct 
state evolution of LM-MP in the conventional system~(\ref{uncoupled_model}) 
without spatial coupling, such as CAMP~\cite{Takeuchi211}, 
MAMP~\cite{Liu222}, WS-CG VAMP~\cite{Skuratovs221,Skuratovs222}, 
and LM-OAMP~\cite{Takeuchi213}. Furthermore, it was utilized in 
\cite{Takeuchi19} to reproduce state evolution for AMP~\cite{Bayati11}. 
The proposed general error model~(\ref{b})--(\ref{q_ext}) can be applied 
to state evolution of such LM-MP algorithms in the spatially coupled 
system~(\ref{vector_model}). Furthermore, it may be utilized to reproduce 
state evolution of AMP for the spatial coupling case~\cite{Javanmard13}. 
Since such applications are outside the scope of 
this paper, however, we focus on state evolution analysis for LM-OAMP 
in the spatially coupled system. 

The following lemma implies that the general error model contains error 
models for both OAMP and LM-OAMP in the spatially coupled 
system~(\ref{vector_model}): 

\begin{lemma} \label{lemma_inclusion}
Suppose that Assumption~\ref{assumption_A}--\ref{assumption_denoiser} hold. 
Let $\vec{\boldsymbol{q}}_{0}^{\mathrm{ext}}[\ell]
=-\vec{\boldsymbol{x}}[\ell]$,  
\begin{equation} \label{m_post_OAMP} 
\vec{\boldsymbol{m}}_{t}^{\mathrm{post}}[\ell] 
= |\mathcal{W}[\ell]|^{-1/2}\frac{\boldsymbol{V}^{\mathrm{T}}[\ell]
(\vec{\boldsymbol{x}}_{\A,t}^{\mathrm{post}}[\ell] 
- \vec{\boldsymbol{x}}[\ell])}
{1 - \bar{\eta}_{\A,t}[\ell]},
\end{equation}
and 
\begin{equation} \label{q_post_OAMP}
\vec{\boldsymbol{q}}_{t+1}^{\mathrm{post}}[\ell] 
= \frac{\sqrt{|W[\ell]|}\boldsymbol{\Gamma}[\ell]
\boldsymbol{x}_{\B,t+1}^{\mathrm{post}} 
- \vec{\boldsymbol{x}}[\ell]}{1-\bar{\eta}_{\B,t}[\ell]/|\mathcal{W}[\ell]|}. 
\end{equation}
The messages $\vec{\boldsymbol{x}}_{\A,t}^{\mathrm{post}}[\ell]$, 
$\bar{\eta}_{\A,t}[\ell]$, $\boldsymbol{x}_{\B,t+1}^{\mathrm{post}}$, and 
$\bar{\eta}_{\B,t}[\ell]$ are given in (\ref{mean_post_A_LM}), (\ref{eta_A_SE}), 
(\ref{mean_post_B_LM}), and (\ref{eta_B_SE_LM}) for LM-OAMP, respectively, 
while they are given in (\ref{mean_post_A}), (\ref{eta_A_SE}), 
(\ref{mean_post_B}), and (\ref{eta_B_SE}) for OAMP.  
\begin{itemize}
\item If $\vec{\boldsymbol{q}}_{t}^{\mathrm{ext}}[\ell]\aeq
\vec{\boldsymbol{x}}_{\B\to\A,t}[\ell] - \vec{\boldsymbol{x}}[\ell]+o(1)$ holds 
in the large system limit and if $\eta_{\A,t}[\ell]$ in (\ref{eta_A}) 
converges almost surely to $\bar{\eta}_{\A,t}[\ell]$, then 
$\vec{\boldsymbol{h}}_{t}[\ell]
\aeq\vec{\boldsymbol{x}}_{\A\to\B,t}[\ell] - |\mathcal{W}[\ell]|^{-1/2}
\vec{\boldsymbol{x}}[\ell]+o(1)$ and 
the properties for $\boldsymbol{\phi}_{t}[\ell]$ 
in Assumption~\ref{assumption_function} hold. 

\item If $\vec{\boldsymbol{h}}_{\tau}[\ell]
\aeq\vec{\boldsymbol{x}}_{\A\to\B,\tau}[\ell] - |\mathcal{W}[\ell]|^{-1/2}
\vec{\boldsymbol{x}}[\ell]+o(1)$ holds 
in the large system limit for all $\tau\in\{0,\ldots,t\}$, 
if $\boldsymbol{V}_{\A\to\B,t}[\ell]$ given via (\ref{cov_AB}) converges 
almost surely to positive definite $\bar{\boldsymbol{V}}_{\A\to\B,t}[\ell]$ 
for LM-OAMP, and if $\eta_{\B,t}[\ell]$ in (\ref{eta_B}) converges almost 
surely to $\bar{\eta}_{\B,t}[\ell]$, 
then $\vec{\boldsymbol{q}}_{t+1}^{\mathrm{ext}}[\ell]\aeq
\vec{\boldsymbol{x}}_{\B\to\A,t+1}[\ell] - \vec{\boldsymbol{x}}[\ell]+o(1)$ and 
the properties for $\boldsymbol{\psi}_{t}[\ell]$
in Assumption~\ref{assumption_function} hold. 
\end{itemize}
\end{lemma}
\begin{IEEEproof}
We only prove Lemma~\ref{lemma_inclusion} for LM-OAMP since the lemma for 
OAMP can be proved in the same manner as for LM-OAMP. 
Prove the former statement. By proving the following identity: 
\begin{equation} \label{m_ext_OAMP} 
\vec{\boldsymbol{m}}_{t}^{\mathrm{ext}}[\ell]\aeq \boldsymbol{V}^{\mathrm{T}}[\ell]
(\vec{\boldsymbol{x}}_{\A\to\B,t}[\ell] 
- |\mathcal{W}[\ell]|^{-1/2}\vec{\boldsymbol{x}}[\ell]) + o(1),
\end{equation} 
we use the definition of $\vec{\boldsymbol{h}}_{t}[\ell]$ in (\ref{h}) to 
arrive at $\vec{\boldsymbol{h}}_{t}[\ell]
\aeq\vec{\boldsymbol{x}}_{\A\to\B,t}[\ell] 
- |\mathcal{W}[\ell]|^{-1/2}\vec{\boldsymbol{x}}[\ell]+o(1)$.  

We first evaluate the LHS in (\ref{m_ext_OAMP}). 
Substituting the definition of $\vec{\boldsymbol{x}}_{\A,t}^{\mathrm{post}}[\ell]$ 
in (\ref{mean_post_A_LM}) into (\ref{m_post_OAMP}) and using the spatially 
coupled system~(\ref{vector_model}), 
the SVD $\boldsymbol{A}[\ell]=\boldsymbol{U}[\ell]\boldsymbol{\Sigma}[\ell]
\boldsymbol{V}^{\mathrm{T}}[\ell]$, and $\boldsymbol{F}_{t}[\ell]
=\boldsymbol{U}[\ell]\boldsymbol{\Sigma}_{\boldsymbol{F}_{t}[\ell]}
\boldsymbol{V}^{\mathrm{T}}[\ell]$ in Assumption~\ref{assumption_filter}, 
we have  
\begin{IEEEeqnarray}{r} 
\sqrt{|\mathcal{W}[\ell]|}(1 - \bar{\eta}_{\A,t}[\ell])
\vec{\boldsymbol{m}}_{t}^{\mathrm{post}}[\ell] 
= \boldsymbol{\Sigma}_{\boldsymbol{F}_{t}[\ell]}^{\mathrm{T}}
\boldsymbol{U}^{\mathrm{T}}[\ell]\boldsymbol{n}[\ell]
\nonumber \\
+ (\boldsymbol{I} - \boldsymbol{\Sigma}_{\boldsymbol{F}_{t}[\ell]}^{\mathrm{T}}
\boldsymbol{\Sigma}[\ell])
\boldsymbol{V}^{\mathrm{T}}[\ell]
(\vec{\boldsymbol{x}}_{\B\to\A,t}[\ell] - \vec{\boldsymbol{x}}[\ell]).
\label{m_post_OAMP_tmp}
\end{IEEEeqnarray}
Under the assumption $\vec{\boldsymbol{q}}_{t}^{\mathrm{ext}}[\ell]
\aeq\vec{\boldsymbol{x}}_{\B\to\A,t}[\ell] - \vec{\boldsymbol{x}}[\ell]+o(1)$, 
we use the definition of $\vec{\boldsymbol{b}}_{t}[\ell]$ in (\ref{b}) to 
find 
\begin{IEEEeqnarray}{rl} 
\sqrt{|\mathcal{W}[\ell]|}(1 - \bar{\eta}_{\A,t}[\ell])&
\vec{\boldsymbol{m}}_{t}^{\mathrm{post}}[\ell] 
\aeq \boldsymbol{\Sigma}_{\boldsymbol{F}_{t}[\ell]}^{\mathrm{T}}
\boldsymbol{U}^{\mathrm{T}}[\ell]\boldsymbol{n}[\ell]
\nonumber \\
+& (\boldsymbol{I} - \boldsymbol{\Sigma}_{\boldsymbol{F}_{t}[\ell]}^{\mathrm{T}}
\boldsymbol{\Sigma}[\ell])
\vec{\boldsymbol{b}}_{t}[\ell] + o(1).
\label{m_post_OAMP_tmp2}
\end{IEEEeqnarray}
Since $\vec{\boldsymbol{m}}_{t}^{\mathrm{post}}[\ell]$ is a function of 
$\vec{\boldsymbol{b}}_{t}[\ell]$, applying this expression to the definition of 
$\vec{\boldsymbol{m}}_{t}^{\mathrm{ext}}[\ell]$ in (\ref{m_ext}) yields 
\begin{equation} \label{m_ext_OAMP_tmp}
\vec{\boldsymbol{m}}_{t}^{\mathrm{ext}}[\ell]
\aeq \vec{\boldsymbol{m}}_{t}^{\mathrm{post}}[\ell] 
- |\mathcal{W}[\ell]|^{-1/2}\frac{\eta_{\A,t}[\ell]}{1 - \bar{\eta}_{\A,t}[\ell]}
\vec{\boldsymbol{b}}_{t}[\ell] + o(1),
\end{equation}
with $\eta_{\A,t}[\ell]$ defined in (\ref{eta_A}), in which we have used 
the identity $\mathrm{Tr}(\boldsymbol{I} 
- \boldsymbol{\Sigma}_{\boldsymbol{F}_{t}[\ell]}^{\mathrm{T}}
\boldsymbol{\Sigma}[\ell])=\mathrm{Tr}(\boldsymbol{I} 
- \boldsymbol{F}_{t}^{\mathrm{T}}[\ell]\boldsymbol{A}[\ell])$.  

We next evaluate the RHS in (\ref{m_ext_OAMP}). 
Using the definitions of $\vec{\boldsymbol{x}}_{\A\to\B,t}[\ell]$ 
and $\vec{\boldsymbol{m}}_{t}^{\mathrm{post}}[\ell]$ in (\ref{mean_AB_LM}) 
and (\ref{m_post_OAMP}) yields 
\begin{IEEEeqnarray}{rl}
&\boldsymbol{V}^{\mathrm{T}}[\ell]
(\vec{\boldsymbol{x}}_{\A\to\B,t}[\ell] 
- |\mathcal{W}[\ell]|^{-1/2}\vec{\boldsymbol{x}}[\ell])
\nonumber \\
\aeq& \vec{\boldsymbol{m}}_{t}^{\mathrm{post}}[\ell] 
- |\mathcal{W}[\ell]|^{-1/2}\frac{\eta_{\A,t}[\ell]}{1 - \eta_{\A,t}[\ell]}
\vec{\boldsymbol{b}}_{t}[\ell] + o(1)
\end{IEEEeqnarray}
under the assumptions $\vec{\boldsymbol{q}}_{t}^{\mathrm{ext}}[\ell]
\aeq\vec{\boldsymbol{x}}_{\B\to\A,t}[\ell] - \vec{\boldsymbol{x}}[\ell]+o(1)$ 
and $\eta_{\A,t}[\ell]\ato\bar{\eta}_{\A,t}[\ell]$. 
Comparing the obtained two results under the assumption 
$\eta_{\A,t}[\ell]\ato\bar{\eta}_{\A,t}[\ell]$, 
we arrive at the identity~(\ref{m_ext_OAMP}). 

Let us prove the properties for $\vec{\boldsymbol{m}}_{t}^{\mathrm{post}}[\ell]
=\boldsymbol{\phi}_{t}[\ell]$ in Assumption~\ref{assumption_function}. 
The separability and proper Lipschitz-continuity of 
$\vec{\boldsymbol{m}}_{t}^{\mathrm{post}}[\ell]$ 
follow from (\ref{m_post_OAMP_tmp2}) and Assumption~\ref{assumption_A}.   
The property $\|\vec{\boldsymbol{m}}_{t}^{\mathrm{ext}}[\ell]\|\neq0$ is 
satisfied, because of (\ref{m_post_OAMP_tmp2}), (\ref{m_ext_OAMP_tmp}), 
and Assumption~\ref{assumption_w}. Thus, the former statement in 
Lemma~\ref{lemma_inclusion} holds. 

We prove $\vec{\boldsymbol{q}}_{t+1}^{\mathrm{ext}}[\ell]
\aeq\vec{\boldsymbol{x}}_{\B\to\A,t+1}[\ell] - \vec{\boldsymbol{x}}[\ell]+o(1)$ 
in the latter statement.  
We first evaluate the LHS $\vec{\boldsymbol{q}}_{t+1}^{\mathrm{ext}}[\ell]$. 
Define 
\begin{equation}
\vec{\boldsymbol{q}}_{t+1}^{\mathrm{post}}[\ell]
=\begin{pmatrix}
\vec{\boldsymbol{q}}_{t+1}^{\mathrm{post}}[\ell][\min\{W,\ell\}] \\
\vdots \\
\vec{\boldsymbol{q}}_{t+1}^{\mathrm{post}}[\ell][\max\{\ell-(L-1),0\}]
\end{pmatrix}. 
\end{equation}
Applying the definition of $\vec{\boldsymbol{q}}_{t+1}^{\mathrm{post}}[\ell]$ 
in (\ref{q_post_OAMP}) and $\vec{\boldsymbol{x}}[\ell]
=|\mathcal{W}[\ell]|^{1/2}\boldsymbol{\Gamma}[\ell]\boldsymbol{x}$ yields  
\begin{equation} \label{q_post_OAMP_tmp}
\vec{\boldsymbol{q}}_{t+1}^{\mathrm{post}}[\ell][w] 
= \frac{\sqrt{|\mathcal{W}[\ell]|}
\gamma[\ell][\ell-w]}{1-\bar{\eta}_{\B,t}[\ell]/|\mathcal{W}[\ell]|}
(\boldsymbol{x}_{\B,t+1}^{\mathrm{post}}[\ell-w] - \boldsymbol{x}[\ell-w])  
\end{equation}
for $w\in\mathcal{W}[\ell]$, 
where we have utilized the relationship~(\ref{extraction}) between 
the signal vectors in the original and extended spaces. 
The difference $\boldsymbol{x}_{\B,t+1}^{\mathrm{post}}[\ell-w]
-\boldsymbol{x}[\ell-w]$ defined in (\ref{mean_post_B_LM}) depends on 
$\{\vec{\boldsymbol{h}}_{\tau}[\ell]\}_{\tau=0}^{t}$ through 
$\boldsymbol{x}_{\A\to\B,t}^{\mathrm{suf}}[\ell-w]$ in (\ref{mean_suf_B_LM}): 
We use the definition of $\boldsymbol{x}_{\A\to\B,\tau}[l][w]$ 
in (\ref{mean_AB_projection}) and the assumption  
$\vec{\boldsymbol{h}}_{\tau}[\ell]
\aeq\vec{\boldsymbol{x}}_{\A\to\B,\tau}[\ell]
-|\mathcal{W}[\ell]|^{-1/2}\vec{\boldsymbol{x}}[\ell]+o(1)$
to find that $\boldsymbol{x}_{\A\to\B,\tau}[l][w]$ contained in 
$\boldsymbol{x}_{\A\to\B,t}^{\mathrm{suf}}[l]$ is given by  
\begin{IEEEeqnarray}{rl}
\boldsymbol{x}_{\A\to\B,\tau}[l][w] 
\aeq& |\mathcal{W}[l+w]|^{-1/2}\vec{\boldsymbol{x}}[l+w][w] 
+ \vec{\boldsymbol{h}}_{\tau}[l+w][w] 
\nonumber \\
&+ o(1) 
\label{mean_AB_projection_tmp} 
\end{IEEEeqnarray}
for $\tau\in\{0,\ldots,t\}$, 
with $\vec{\boldsymbol{h}}_{\tau}[\ell][w]$ defined in the same manner 
as for $\vec{\boldsymbol{q}}_{t+1}^{\mathrm{post}}[\ell][w]$.  

We evaluate $\xi_{\B,\tau,t}[\ell]$ given in (\ref{xi_B}). 
Using the definition of $\boldsymbol{x}_{\B,t+1}^{\mathrm{post}}[\ell-w]$ 
in (\ref{mean_post_B_LM}) and the representation~(\ref{q_post_OAMP_tmp}) yields 
\begin{IEEEeqnarray}{l}
\frac{\partial \vec{q}_{n,t+1}^{\mathrm{post}}[\ell][w]}
{\partial \vec{h}_{n,\tau}[\ell][w]}
= \frac{\sqrt{|\mathcal{W}[\ell]|}\gamma[\ell][\ell-w]}
{1-\bar{\eta}_{\B,t}[\ell]/|\mathcal{W}[\ell]|}
\nonumber \\
\cdot f_{t}'[l](x_{\A\to\B,n,t}^{\mathrm{suf}}[\ell-w])
\frac{\partial x_{\A\to\B,n,t}^{\mathrm{suf}}[l]}
{\partial \vec{h}_{n,\tau}[l+w][w]}, 
\end{IEEEeqnarray}
with $l=\ell-w$. Under the assumption $\boldsymbol{V}_{\A\to\B,t}[\ell]
\ato\bar{\boldsymbol{V}}_{\A\to\B,t}[\ell]$, the covariance  
$v_{\A\to\B,t',t}^{\mathrm{suf}}[l]$ in (\ref{cov_suf_B}) converges almost surely 
to $\bar{v}_{\A\to\B,t',t}^{\mathrm{suf}}[l]$. Thus, the last factor reduces to 
\begin{IEEEeqnarray}{r}
\frac{\partial x_{\A\to\B,n,t}^{\mathrm{suf}}[l]}
{\partial \vec{h}_{n,\tau}[l+w][w]}
\aeq \bar{v}_{\A\to \B,t,t}^{\mathrm{suf}}[l]
\sum_{w'=0}^{W}\gamma[l+w'][l]
\nonumber \\
\cdot\boldsymbol{e}_{\tau}^{\mathrm{T}}\bar{\boldsymbol{V}}_{\A\to \B,t}^{-1}[l+w']
\boldsymbol{1}
\frac{\partial x_{\A\to\B,n,\tau}[l][w']}
{\partial \vec{h}_{n,\tau}[l+w][w]} + o(1)
\end{IEEEeqnarray}
because of the definition of $\boldsymbol{x}_{\A\to\B,t}^{\mathrm{suf}}[l]$ 
in (\ref{mean_suf_B_LM}), with   
\begin{equation}
\frac{\partial x_{\A\to\B,n,\tau}[l][w']}
{\partial \vec{h}_{n,\tau}[l+w][w]}
\aeq \delta_{w,w'} + o(1),
\end{equation}
obtained from the representation~(\ref{mean_AB_projection_tmp}). 
Applying these results to the definition of $\xi_{\B,\tau,t}[\ell]$ 
in (\ref{xi_B}), we arrive at  
\begin{IEEEeqnarray}{rl} 
\xi_{\B,\tau,t}[\ell]
=& \frac{1}{N_{\mathrm{c}}[\ell]}\sum_{w\in\mathcal{W}[\ell]} 
\sum_{n=1}^{N[\ell-w]}\frac{\partial \vec{q}_{n,t+1}^{\mathrm{post}}[\ell][w]}
{\partial \vec{h}_{n,\tau}[\ell][w]}
\nonumber \\
\aeq& \frac{|\mathcal{W}[\ell]|^{-1/2}}
{1-\bar{\eta}_{\B,t}[\ell]/|\mathcal{W}[\ell]|}
\eta_{\B,\tau,t}[\ell]
+ o(1), 
\label{xi_B_tmp}
\end{IEEEeqnarray}
with $\eta_{\B,\tau,t}[\ell]$ given by (\ref{eta_B_ell_LM}), where we have 
used $\boldsymbol{V}_{\A\to\B,t}[\ell]
\ato\bar{\boldsymbol{V}}_{\A\to\B,t}[\ell]$ and 
$v_{\A\to\B,t',t}^{\mathrm{suf}}[l]\ato\bar{v}_{\A\to\B,t',t}^{\mathrm{suf}}[l]$ again.  

We next evaluate the RHS $\vec{\boldsymbol{x}}_{\B\to\A,t+1}[\ell] 
- \vec{\boldsymbol{x}}[\ell]$. Using the definitions of 
$\vec{\boldsymbol{x}}_{\B\to\A,t+1}[\ell]$ and 
$\vec{\boldsymbol{q}}_{t+1}^{\mathrm{post}}[\ell]$ in (\ref{mean_BA_LM}) and 
(\ref{q_post_OAMP}), as well as the assumptions 
$\vec{\boldsymbol{h}}_{\tau}[\ell]\aeq\vec{\boldsymbol{x}}_{\A\to\B,\tau}[\ell] 
- |\mathcal{W}[\ell]|^{-1/2}\vec{\boldsymbol{x}}[\ell]+o(1)$ for all 
$\tau\in\{0,\ldots,t\}$ and $\eta_{\B,t}[\ell]\ato\bar{\eta}_{\B,t}[\ell]$, 
yields 
\begin{IEEEeqnarray}{l}
\vec{\boldsymbol{x}}_{\B\to\A,t+1}[\ell] 
- \vec{\boldsymbol{x}}[\ell]
\aeq \vec{\boldsymbol{q}}_{t+1}^{\mathrm{post}}[\ell]
\nonumber \\
- \frac{|\mathcal{W}[\ell]|^{-1/2}}
{1 - \eta_{\B,t}[\ell]/|\mathcal{W}[\ell]|}
\sum_{\tau=0}^{t}
\eta_{\B,\tau,t}[\ell]
\vec{\boldsymbol{h}}_{\tau}[\ell]
+ o(1),  
\label{mean_BA_tmp}
\end{IEEEeqnarray}
where we have used the definitions of $\eta_{\B,t}[\ell]$ and 
$\eta_{\B,\tau,t}[\ell]$ in (\ref{eta_B}) and (\ref{eta_B_ell_LM}), respectively, 
and the identity $\sum_{\tau=0}^{t}\eta_{\B,\tau,t}[\ell][w] 
= \eta_{\B,t}[\ell][w]$ obtained from (\ref{zeta_B}). 
Comparing this expression to the definition of 
$\vec{\boldsymbol{q}}_{t+1}^{\mathrm{ext}}[\ell]$ in (\ref{q_ext}) 
with $\xi_{\B,\tau,t}[\ell]$ given in (\ref{xi_B_tmp}), we arrive at 
$\vec{\boldsymbol{q}}_{t+1}^{\mathrm{ext}}[\ell]
\aeq\vec{\boldsymbol{x}}_{\B\to\A,t+1}[\ell] - \vec{\boldsymbol{x}}[\ell]+o(1)$.  

The separability and proper Lipschitz-continuity of 
$\vec{\boldsymbol{q}}_{t+1}^{\mathrm{post}}[\ell]$  
follow from (\ref{mean_post_B_LM}), (\ref{q_post_OAMP_tmp}), and 
Assumption~\ref{assumption_denoiser}. Furthermore, 
the nonlinearity of the denoiser in Assumption~\ref{assumption_denoiser} 
implies $\|\vec{\boldsymbol{q}}_{t+1}^{\mathrm{ext}}[\ell]\|\neq0$, so that  
we have the properties for $\vec{\boldsymbol{q}}_{t+1}^{\mathrm{post}}[\ell]
=\boldsymbol{\psi}_{t}[\ell]$ in Assumption~\ref{assumption_function}. 
Thus, the latter statement holds. 
\end{IEEEproof}

Lemma~\ref{lemma_inclusion} implies that the asymptotic dynamics of both OAMP 
and LM-OAMP can be analyzed via state evolution of the general error model 
in the spatial coupling case. 

\subsection{State Evolution} 
As the second step in the proof of Theorem~\ref{theorem_SE_LM}, we present 
state evolution analysis for the general error model in the spatial coupling 
case. The proof is based on Bolthausen's conditioning 
technique~\cite{Bolthausen14}. 
The set $\mathfrak{F}=\{\Omega, \Lambda, \mathcal{X}\}$ 
is always conditioned. Define the set 
$\mathfrak{E}_{t,t'}=\{\mathcal{B}_{t'}, \mathcal{M}_{t'}^{\mathrm{ext}}, 
\mathcal{H}_{t}, \mathcal{Q}_{t+1}^{\mathrm{ext}}\}$ with 
$\mathcal{M}_{t}^{\mathrm{ext}}=\{\vec{\boldsymbol{M}}_{t}^{\mathrm{ext}}[\ell]: 
\ell\in\mathcal{L}_{W}\}$ and $\mathcal{Q}_{t}^{\mathrm{ext}}
=\{\vec{\boldsymbol{Q}}_{t}^{\mathrm{ext}}[\ell]: \ell\in\mathcal{L}_{W}\}$. 
The set $\mathfrak{E}_{t,t}$ contains the messages that have been already 
computed just before updating $\vec{\boldsymbol{b}}_{t}[\ell]$ in (\ref{b}) 
while $\mathfrak{E}_{t,t+1}$ includes the messages just before updating 
$\vec{\boldsymbol{h}}_{t}[\ell]$ in (\ref{h}). The asymptotic dynamics of 
the general error model is analyzed via the conditional distributions of 
$\{\boldsymbol{V}[\ell]\}$ given $\mathfrak{F}$ and $\mathfrak{E}_{t,t'}$. 

The following theorem presents five asymptotic properties in each module. 
A first property in each module is obtained via Bolthausen's 
conditioning technique. Second and third properties are the asymptotic 
Gaussianity for the general error model. The last two properties are 
technical results for establishing a proof by induction. 

\begin{theorem} \label{theorem_SE_tech} 
Suppose that Assumptions~\ref{assumption_x}, \ref{assumption_A}, 
\ref{assumption_w}, and~\ref{assumption_function} hold. 
Then, the following properties in module A hold  
for all $\tau=0,1,\ldots$ in the large system limit.  
\begin{enumerate}[label=(A\arabic*)]
\item \label{property_A1}
Suppose that $\{\tilde{\boldsymbol{V}}[\ell]
\in\mathcal{O}_{N_{\mathrm{c}}[\ell]-2\tau}\}$ are independent Haar-distributed 
orthogonal matrices. 
Let 
$\vec{\boldsymbol{q}}_{\tau}^{\mathrm{ext},\perp}[\ell]
=\boldsymbol{P}_{\vec{\boldsymbol{Q}}_{\tau}^{\mathrm{ext}}[\ell]}^{\perp}
\vec{\boldsymbol{q}}_{\tau}^{\mathrm{ext}}[\ell]$, 
$\vec{\boldsymbol{\beta}}_{\tau}[\ell]=(
\vec{\boldsymbol{Q}}_{\tau}^{\mathrm{ext}}[\ell])^{\dagger}
\vec{\boldsymbol{q}}_{\tau}^{\mathrm{ext}}[\ell]$, 
and 
\begin{equation} 
\vec{\boldsymbol{\omega}}_{\A,\tau}[\ell] 
= \tilde{\boldsymbol{V}}[\ell]
(\boldsymbol{\Phi}_{(\vec{\boldsymbol{Q}}_{\tau}^{\mathrm{ext}}[\ell], 
\vec{\boldsymbol{H}}_{\tau}[\ell])}^{\perp})^{\mathrm{T}}
\vec{\boldsymbol{q}}_{\tau}^{\mathrm{ext}}[\ell]. 
\end{equation}
Then, for $\tau>0$ we have 
\begin{IEEEeqnarray}{rl} 
\vec{\boldsymbol{b}}_{\tau}[\ell]
\sim& \vec{\boldsymbol{B}}_{\tau}[\ell]\vec{\boldsymbol{\beta}}_{\tau}[\ell] 
+ \boldsymbol{\Phi}_{(\vec{\boldsymbol{B}}_{\tau}[\ell], 
\vec{\boldsymbol{M}}_{\tau}^{\mathrm{ext}}[\ell])}^{\perp}
\vec{\boldsymbol{\omega}}_{\A,\tau}[\ell]
\nonumber \\
& + \vec{\boldsymbol{M}}_{\tau}^{\mathrm{ext}}[\ell]\boldsymbol{o}(1) 
+ \vec{\boldsymbol{B}}_{\tau}[\ell]\boldsymbol{o}(1)  
\label{b_tau}
\end{IEEEeqnarray}
conditioned on $\mathfrak{F}$ and $\mathfrak{E}_{\tau,\tau}$  
in the large system limit, with 
\begin{equation}  
\frac{1}{N_{\mathrm{c}}[\ell]}\left\{
 \|\vec{\boldsymbol{\omega}}_{\A,\tau}[\ell]\|^{2} 
 - \|\vec{\boldsymbol{q}}_{\tau}^{\mathrm{ext},\perp}[\ell]\|^{2} 
\right\}
\ato 0.
\end{equation}

\item \label{property_A2}
Suppose that  
$\tilde{\boldsymbol{\phi}}_{\tau}(\mathcal{B}_{\tau+1},\Omega, \Lambda):
\mathbb{R}^{N_{\mathrm{c}}\times(\tau+3)|\mathcal{L}_{W}|}\to\mathbb{R}^{N_{\mathrm{c}}}$ 
is separable, pseudo-Lipschitz of order~$2$ with respect to 
$\mathcal{B}_{\tau+1}$ and $\Omega$, and proper. 
If $N_{\mathrm{c}}^{-1}[\ell](\vec{\boldsymbol{q}}_{t'}^{\mathrm{ext}}[\ell])^{\mathrm{T}}
\vec{\boldsymbol{q}}_{t}^{\mathrm{ext}}[\ell]$ 
converges almost surely to some constant $\kappa_{t',t}[\ell]\in\mathbb{R}$ 
in the large system limit for all $t', t=0,\ldots,\tau$, then   
\begin{equation} 
\langle\tilde{\boldsymbol{\phi}}_{\tau}(\mathcal{B}_{\tau+1}, \Omega, \Lambda)
\rangle
- \mathbb{E}\left[
 \langle\tilde{\boldsymbol{\phi}}_{\tau}(\mathcal{Z}_{\A,\tau+1}, 
 \tilde{\Omega},\Lambda)\rangle 
\right]\ato 0, 
\label{phi_SLLN}
\end{equation}
with the sets of independent random matrices 
$\mathcal{Z}_{\A,\tau+1}=\{
[\vec{\boldsymbol{Z}}_{\A,\tau+1}^{\mathrm{T}}[\ell], \boldsymbol{O}]^{\mathrm{T}}
\in\mathbb{R}^{N_{\mathrm{c}}\times(\tau+1)}: \ell\in\mathcal{L}_{W}\}$ and vectors 
$\tilde{\Omega}=\{
[\tilde{\boldsymbol{n}}^{\mathrm{T}}[\ell], \boldsymbol{0}]^{\mathrm{T}}
\in\mathbb{R}^{N_{\mathrm{c}}}: \ell\in\mathcal{L}_{W}\}$. 
Here, each $\vec{\boldsymbol{Z}}_{\A,\tau+1}[\ell]
=(\vec{\boldsymbol{z}}_{\A,0}[\ell],\ldots,
\vec{\boldsymbol{z}}_{\A,\tau}[\ell])$ has zero-mean Gaussian random 
vectors with covariance $\mathbb{E}[\vec{\boldsymbol{z}}_{\A,t}[\ell]
\vec{\boldsymbol{z}}_{\A,t'}^{\mathrm{T}}[\ell]]
=\kappa_{t',t}[\ell]\boldsymbol{I}_{N_{\mathrm{c}}[\ell]}$ for all 
$t', t\in\{0,\ldots,\tau\}$. 
Furthermore, each $\tilde{\boldsymbol{n}}[\ell]\in\mathbb{R}^{M[\ell]}$ has 
independent zero-mean Gaussian random elements with variance $\sigma^{2}$. 

\item \label{property_A3}  
Suppose that $\tilde{\boldsymbol{\phi}}_{\tau}(\mathcal{B}_{\tau+1},\Omega, 
\Lambda):\mathbb{R}^{N_{\mathrm{c}}\times(\tau+3)|\mathcal{L}_{W}|}
\to\mathbb{R}^{N_{\mathrm{c}}}$ is separable, Lipschitz-continuous 
with respect to $\mathcal{B}_{\tau+1}$ and $\Omega$, and proper. Then, 
\begin{equation} 
\langle\partial_{\tau'}\tilde{\boldsymbol{\phi}}_{\tau}
(\mathcal{B}_{\tau+1},\Omega,\Lambda) \rangle
- \mathbb{E}\left[
 \langle\partial_{\tau'}\tilde{\boldsymbol{\phi}}_{\tau}
 (\mathcal{Z}_{\A,\tau+1},\tilde{\Omega},\Lambda)\rangle 
\right]\ato 0
\label{phi_SLLN_1}
\end{equation}
for all $\tau'\in\{0,\ldots,(\tau+1)|\mathcal{L}_{W}|-1\}$ and 
\begin{equation} \label{orthogonality_A}
\frac{1}{N_{\mathrm{c}}[\ell]}
\vec{\boldsymbol{b}}_{\tau'}^{\mathrm{T}}[\ell] 
\left(
 (\boldsymbol{I}_{N_{\mathrm{c}}[\ell]},\boldsymbol{O})
 \tilde{\boldsymbol{\phi}}_{\tau}
 - \sum_{t'=0}^{\tau}\tilde{\xi}_{\A,t',\tau}[\ell] 
 \vec{\boldsymbol{b}}_{t'}[\ell] 
\right) 
\ato0 
\end{equation}
for all $\tau'\in\{0,\ldots,\tau\}$ hold, with 
\begin{equation} \label{tilde_xi_A}
\tilde{\xi}_{\A,t',\tau}[\ell] 
= \frac{1}{N_{\mathrm{c}}[\ell]}
\sum_{n=1}^{N_{\mathrm{c}}[\ell]}\partial_{(\tau+1)\ell+t'}
[\tilde{\boldsymbol{\phi}}_{\tau}]_{n}.
\end{equation}

\item \label{property_A4}
The inner product 
$N_{\mathrm{c}}^{-1}[\ell](\vec{\boldsymbol{m}}_{\tau'}^{\mathrm{ext}}[\ell])^{\mathrm{T}}
\vec{\boldsymbol{m}}_{\tau}^{\mathrm{ext}}[\ell]$ 
converges almost surely to some constant $\pi_{\tau',\tau}[\ell]\in\mathbb{R}$ 
for all $\tau'\in\{0,\ldots,\tau\}$. 

\item \label{property_A5}
For some $\epsilon>0$ and $C>0$, 
\begin{equation} 
\mathbb{E}\left[
 |\vec{m}_{n,\tau}^{\mathrm{ext}}[\ell]|^{2+\epsilon}
\right]<\infty, 
\end{equation} 
\begin{equation} 
\lambda_{\mathrm{min}}\left(
 \frac{1}{N_{\mathrm{c}}[\ell]}
 (\vec{\boldsymbol{M}}_{\tau+1}^{\mathrm{ext}}[\ell])^{\mathrm{T}}
 \vec{\boldsymbol{M}}_{\tau+1}^{\mathrm{ext}}[\ell]
\right)
\ag C 
\end{equation}
in the large system limit. 
\end{enumerate}

The following properties in module~B hold for all $\tau=0,1,\ldots$ 
in the large system limit. 
\begin{enumerate}[label=(B\arabic*)]
\item \label{property_B1}
Suppose that $\{\tilde{\boldsymbol{V}}[\ell]
\in\mathcal{O}_{N_{\mathrm{c}}[\ell]-(2\tau+1)}\}$ 
are independent Haar-distributed orthogonal matrices. 
Let $\vec{\boldsymbol{\alpha}}_{\tau}[\ell]
=(\vec{\boldsymbol{M}}_{\tau}^{\mathrm{ext}}[\ell])^{\dagger}
\vec{\boldsymbol{m}}_{\tau}^{\mathrm{ext}}[\ell]$, 
$\vec{\boldsymbol{m}}_{0}^{\mathrm{ext},\perp}[\ell]
=\vec{\boldsymbol{m}}_{0}^{\mathrm{ext}}[\ell]$, 
$\vec{\boldsymbol{\omega}}_{\B,0}[\ell]=\tilde{\boldsymbol{V}}[\ell]
(\boldsymbol{\Phi}_{\vec{\boldsymbol{b}}_{0}[\ell]}^{\perp})^{\mathrm{T}}
\vec{\boldsymbol{m}}_{0}^{\mathrm{ext}}[\ell]$, 
$\vec{\boldsymbol{m}}_{\tau}^{\mathrm{ext},\perp}[\ell]
=\boldsymbol{P}_{\vec{\boldsymbol{M}}_{\tau}^{\mathrm{ext}}[\ell]}^{\perp}
\vec{\boldsymbol{m}}_{\tau}^{\mathrm{ext}}[\ell]$, and 
$\vec{\boldsymbol{\omega}}_{\B,\tau}[\ell] 
= \tilde{\boldsymbol{V}}[\ell]
(\boldsymbol{\Phi}_{(\vec{\boldsymbol{M}}_{\tau}^{\mathrm{ext}}[\ell], 
\vec{\boldsymbol{B}}_{\tau+1}[\ell])}^{\perp})^{\mathrm{T}}
\vec{\boldsymbol{m}}_{\tau}^{\mathrm{ext}}[\ell]$ 
for $\tau>0$. Then, we have 
\begin{equation}  \label{h_0}
\vec{\boldsymbol{h}}_{0}[\ell]
\sim o(1)\vec{\boldsymbol{q}}_{0}^{\mathrm{ext}}[\ell] 
+ \boldsymbol{\Phi}_{\vec{\boldsymbol{q}}_{0}^{\mathrm{ext}}[\ell]}^{\perp}
\vec{\boldsymbol{\omega}}_{\B,0}[\ell] 
\end{equation}
conditioned on $\mathfrak{F}$ and $\mathfrak{E}_{0,1}
=\{\vec{\boldsymbol{b}}_{0}, \vec{\boldsymbol{m}}_{0}^{\mathrm{ext}}, 
\vec{\boldsymbol{q}}_{0}^{\mathrm{ext}}\}$ in the large system limit. For $\tau>0$ 
\begin{IEEEeqnarray}{rl} 
\vec{\boldsymbol{h}}_{\tau}[\ell]
\sim& \vec{\boldsymbol{H}}_{\tau}[\ell]\vec{\boldsymbol{\alpha}}_{\tau}[\ell]
+ \boldsymbol{\Phi}_{(\vec{\boldsymbol{H}}_{\tau}[\ell], 
\vec{\boldsymbol{Q}}_{\tau+1}^{\mathrm{ext}}[\ell])}^{\perp}
\vec{\boldsymbol{\omega}}_{\B,\tau}[\ell] 
\nonumber \\
&+ \vec{\boldsymbol{Q}}_{\tau+1}^{\mathrm{ext}}[\ell]\boldsymbol{o}(1) 
+ \vec{\boldsymbol{H}}_{\tau}[\ell]\boldsymbol{o}(1)  
\end{IEEEeqnarray}
conditioned on $\mathfrak{F}$ and $\mathfrak{E}_{\tau,\tau+1}$ 
in the large system limit, with 
\begin{equation}  \label{omega_B_norm}
\frac{1}{N_{\mathrm{c}}[\ell]}\left\{
 \|\vec{\boldsymbol{\omega}}_{\B,\tau}[\ell]\|^{2}  
 - \|\vec{\boldsymbol{m}}_{\tau}^{\mathrm{ext},\perp}[\ell]\|^{2}
\right\} \ato 0.  
\end{equation}

\item \label{property_B2}
Suppose that $\tilde{\boldsymbol{\psi}}_{\tau}(\mathcal{H}_{\tau+1}, 
\mathcal{X}):\mathbb{R}^{N_{\mathrm{c}}\times(\tau+2)|\mathcal{L}_{W}|}
\to\mathbb{R}^{N_{\mathrm{c}}}$ 
is a separable and proper pseudo-Lipschitz function of order~$2$.  
If $N_{\mathrm{c}}^{-1}[\ell](\vec{\boldsymbol{m}}_{t'}^{\mathrm{ext}}[\ell])^{\mathrm{T}}
\vec{\boldsymbol{m}}_{t}^{\mathrm{ext}}[\ell]$ 
converges almost surely to some constant $\pi_{t',t}[\ell]\in\mathbb{R}$ 
in the large system limit for all $t', t\in\{0,\ldots,\tau\}$, then   
\begin{equation} \label{psi_SLLN}
\langle\tilde{\boldsymbol{\psi}}_{\tau}(\mathcal{H}_{\tau+1}, \mathcal{X})
\rangle
- \mathbb{E}\left[
 \langle\tilde{\boldsymbol{\psi}}_{\tau}(\mathcal{Z}_{\B,\tau+1},\mathcal{X})
 \rangle 
\right]\ato 0, 
\end{equation}
with the set of independent random matrices 
$\mathcal{Z}_{\B,\tau+1}=\{
[\vec{\boldsymbol{Z}}_{\B,\tau+1}^{\mathrm{T}}[\ell],\boldsymbol{O}]^{\mathrm{T}}
\in\mathbb{R}^{N_{\mathrm{c}}\times(\tau+1)}: \ell\in\mathcal{L}_{W}\}$. Here, each 
$\vec{\boldsymbol{Z}}_{\B,\tau+1}[\ell]
=(\vec{\boldsymbol{z}}_{\B,0}[\ell],\ldots,
\vec{\boldsymbol{z}}_{\B,\tau}[\ell])$ has zero-mean Gaussian random vectors 
with covariance $\mathbb{E}[\vec{\boldsymbol{z}}_{\B,t}[\ell]
\vec{\boldsymbol{z}}_{\B,t'}^{\mathrm{T}}[\ell]]=\pi_{t',t}[\ell]
\boldsymbol{I}_{N_{\mathrm{c}}[\ell]}$ for all $t', t\in\{0,\ldots,\tau\}$. 

\item \label{property_B3}
Suppose that $\tilde{\boldsymbol{\psi}}_{\tau}(\mathcal{H}_{\tau+1},
\mathcal{X}):\mathbb{R}^{N_{\mathrm{c}}\times(\tau+2)|\mathcal{L}_{W}|}
\to\mathbb{R}^{N_{\mathrm{c}}}$ is separable and proper 
Lipschitz-continuous. Then, 
\begin{equation} 
\langle\partial_{\tau'}\tilde{\boldsymbol{\psi}}_{\tau}
(\mathcal{H}_{\tau+1}, \mathcal{X})\rangle
- \mathbb{E}\left[
 \langle\partial_{\tau'}\tilde{\boldsymbol{\psi}}_{\tau}
 (\mathcal{Z}_{\B,\tau+1}, \mathcal{X})\rangle 
\right]\ato 0 \label{psi_SLLN_1}
\end{equation}
for all $\tau'\in\{0,\ldots,(\tau+1)|\mathcal{L}_{W}|-1\}$ and 
\begin{equation}  
\frac{1}{N_{\mathrm{c}}[\ell]}
\vec{\boldsymbol{h}}_{\tau'}^{\mathrm{T}}[\ell]\left(
 (\boldsymbol{I}_{N_{\mathrm{c}}[\ell]},\boldsymbol{O})
 \tilde{\boldsymbol{\psi}}_{\tau}
 - \sum_{t'=0}^{\tau}\tilde{\xi}_{\B,t',\tau}[\ell]
 \vec{\boldsymbol{h}}_{t'}[\ell] 
\right)
\ato0 \label{orthogonality_B}
\end{equation}
for all $\tau'\in\{0,\ldots,\tau\}$ hold, with 
\begin{equation}
\tilde{\xi}_{\B,t',\tau}[\ell] 
= \frac{1}{N_{\mathrm{c}}[\ell]}
\sum_{n=1}^{N_{\mathrm{c}}[\ell]}\partial_{(\tau+1)\ell+t'}
[\tilde{\boldsymbol{\psi}}_{\tau}]_{n}.
\end{equation}

\item  \label{property_B4}
The inner product 
$N_{\mathrm{c}}^{-1}[\ell](\vec{\boldsymbol{q}}_{\tau'}^{\mathrm{ext}}[\ell])^{\mathrm{T}}
\vec{\boldsymbol{q}}_{\tau+1}^{\mathrm{ext}}[\ell]$ 
converges almost surely to some constant 
$\kappa_{\tau',\tau+1}[\ell]\in\mathbb{R}$ for all $\tau'\in\{0,\ldots,\tau+1\}$. 

\item \label{property_B5}
For some $\epsilon>0$ and $C>0$, 
\begin{equation} 
\mathbb{E}\left[
 |\vec{q}_{n,\tau+1}^{\mathrm{ext}}[\ell]|^{2+\epsilon}
\right]<\infty,  
\end{equation} 
\begin{equation} 
\lambda_{\mathrm{min}}\left(
 \frac{1}{N_{\mathrm{c}}[\ell]}(
 \vec{\boldsymbol{Q}}_{\tau+2}^{\mathrm{ext}}[\ell])^{\mathrm{T}}
 \vec{\boldsymbol{Q}}_{\tau+2}^{\mathrm{ext}}[\ell]
\right)
\ag C
\end{equation}
in the large system limit. 
\end{enumerate}
\end{theorem}
\begin{IEEEproof}
The proof is by induction. In a first step, we prove the properties of 
module~A for $\tau=0$ in Appendix~\ref{appen_CE}. In a second step, 
the properties of module~B are proved for $\tau=0$ in Appendix~\ref{appen_CF}. 
In a third step, for some $t$ we assume that Theorem~\ref{theorem_SE_tech} 
is correct for all $\tau<t$ and prove the properties of module~A for $\tau=t$ 
in Appendix~\ref{appen_CG}. The last step is a proof for the properties of 
module~B for $\tau=t$ under induction hypotheses where the properties of 
modules~A and B are correct for all $\tau\leq t$ and $\tau<t$, respectively.  
The proof in the last step is omitted since it is the same as that in the 
third step. By induction, we arrive at Theorem~\ref{theorem_SE_tech} for 
all $\tau$. 
\end{IEEEproof}

The second and third properties in Theorem~\ref{theorem_SE_tech} are 
regarded as evaluation tools. The property~\ref{property_A2} allows us to 
replace the non-tractable vector $\vec{\boldsymbol{b}}_{t}[\ell]$ with 
a tractable zero-mean Gaussian vector. The property~\ref{property_A3} 
is useful for reducing evaluation for the inner product of 
$\vec{\boldsymbol{b}}_{t}[\ell]$ and its nonlinear mapping 
to that for the squared norm of $\vec{\boldsymbol{b}}_{t}[\ell]$. 
The properties \ref{property_B2} and \ref{property_B3} play the same roles 
as for $\vec{\boldsymbol{h}}_{t}[\ell]$. 

As the last step, we prove Theorem~\ref{theorem_SE_LM}. 

\begin{IEEEproof}[Proof of Theorem~\ref{theorem_SE_LM}]
We only prove Theorem~\ref{theorem_SE_LM} for LM-OAMP since the theorem 
for OAMP can be proved in the same manner as for LM-OAMP. 
For any $t$, we prove the identity 
\begin{equation} \label{target_identity} 
\frac{1}{N_{\mathrm{c}}[\ell]}
(\vec{\boldsymbol{q}}_{\tau'}^{\mathrm{ext}}[\ell])^{\mathrm{T}}
\vec{\boldsymbol{q}}_{\tau+1}^{\mathrm{ext}}[\ell]\aeq
\bar{v}_{\B\to\A,\tau',\tau+1}[\ell]+o(1)
\end{equation}
for all $\tau\in\{0,\ldots,t\}$ and $\tau'\in\{0,\ldots,\tau+1\}$, 
with (\ref{cov_BA_SE}). 
To use Lemma~\ref{lemma_inclusion}, we need 
Properties~\ref{property_A3} and \ref{property_B3} 
in Theorem~\ref{theorem_SE_tech} to prove the almost sure 
convergence of $\eta_{\A,t}[\ell]$ and $\eta_{\B,t}[\ell]$. On the other 
hand, we need Lemma~\ref{lemma_inclusion} to use 
Theorem~\ref{theorem_SE_tech}. To resolve this dilemma, 
we prove $\vec{\boldsymbol{h}}_{\tau}[\ell]
\aeq\vec{\boldsymbol{x}}_{\A\to\B,\tau}[\ell] - |\mathcal{W}[\ell]|^{-1/2}
\vec{\boldsymbol{x}}[\ell]+o(1)$ and 
$\vec{\boldsymbol{q}}_{\tau+1}^{\mathrm{ext}}[\ell]\aeq
\vec{\boldsymbol{x}}_{\B\to\A,\tau+1}[\ell] - \vec{\boldsymbol{x}}[\ell]+o(1)$ 
for all $\tau\in\{0,\ldots,t\}$ by induction, 
as well as the identity~(\ref{target_identity}). 

The proof for $t=0$ is omitted since it is the same as that 
for general $t$. For some $t>0$, assume $\vec{\boldsymbol{h}}_{\tau}[\ell]
\aeq\vec{\boldsymbol{x}}_{\A\to\B,\tau}[\ell] - |\mathcal{W}[\ell]|^{-1/2}
\vec{\boldsymbol{x}}[\ell]+o(1)$ and 
$\vec{\boldsymbol{q}}_{\tau+1}^{\mathrm{ext}}[\ell]\aeq
\vec{\boldsymbol{x}}_{\B\to\A,\tau+1}[\ell] - \vec{\boldsymbol{x}}[\ell]+o(1)$, 
and (\ref{target_identity}) for all $\tau\in\{0,\ldots,t-1\}$ and 
$\tau'\in\{0,\ldots,\tau+1\}$. 
We need to prove $\vec{\boldsymbol{h}}_{t}[\ell]
\aeq\vec{\boldsymbol{x}}_{\A\to\B,t}[\ell] - |\mathcal{W}[\ell]|^{-1/2}
\vec{\boldsymbol{x}}[\ell]+o(1)$ and 
$\vec{\boldsymbol{q}}_{t+1}^{\mathrm{ext}}[\ell]\aeq
\vec{\boldsymbol{x}}_{\B\to\A,t+1}[\ell] - \vec{\boldsymbol{x}}[\ell]+o(1)$, 
and (\ref{target_identity}) for $\tau=t$ and all $\tau'\in\{0,\ldots,t+1\}$. 

The induction hypotheses imply that the error model for 
LM-OAMP is included in the general error model as long as at most $t-1$ 
iterations are considered. Thus, we can use the properties of module~A 
in Theorem~\ref{theorem_SE_tech} for $\tau=t$. 

We first evaluate $N_{\mathrm{c}}^{-1}[\ell] 
(\vec{\boldsymbol{x}}_{\A,t'}^{\mathrm{post}}[\ell] 
- \vec{\boldsymbol{x}}[\ell])^{\mathrm{T}} 
(\vec{\boldsymbol{x}}_{\A,t}^{\mathrm{post}}[\ell] 
- \vec{\boldsymbol{x}}[\ell])\ato\bar{v}_{\A,t',t}^{\mathrm{post}}[\ell]$ 
for all $t'\in\{0,\ldots,t\}$.  
From the induction hypothesis $\vec{\boldsymbol{q}}_{t}^{\mathrm{ext}}[\ell]\aeq
\vec{\boldsymbol{x}}_{\B\to\A,t}[\ell] - \vec{\boldsymbol{x}}[\ell]+o(1)$, 
we use the definitions of 
$\vec{\boldsymbol{m}}_{t}^{\mathrm{post}}[\ell]$ in (\ref{m_post_OAMP}) and 
(\ref{m_post_OAMP_tmp2}) to obtain 
\begin{IEEEeqnarray}{r}
\boldsymbol{V}^{\mathrm{T}}[\ell]
(\vec{\boldsymbol{x}}_{\A,t}^{\mathrm{post}}[\ell] 
- \vec{\boldsymbol{x}}[\ell])
\aeq (\boldsymbol{I} - \boldsymbol{\Sigma}_{\boldsymbol{F}_{t}[\ell]}^{\mathrm{T}}
\boldsymbol{\Sigma}[\ell])\vec{\boldsymbol{b}}_{t}[\ell] 
\nonumber \\
+ \boldsymbol{\Sigma}_{\boldsymbol{F}_{t}[\ell]}^{\mathrm{T}}
\boldsymbol{U}^{\mathrm{T}}[\ell]\boldsymbol{n}[\ell] + o(1). 
\end{IEEEeqnarray}
Applying Property~\ref{property_A2} in Theorem~\ref{theorem_SE_tech} yields  
\begin{IEEEeqnarray}{l}
\bar{v}_{\A,t',t}^{\mathrm{post}}[\ell] 
\aeq o(1) + \frac{\sigma^{2}}{N_{\mathrm{c}}[\ell]}
\mathbb{E}\left[
 \mathrm{Tr}\left(
  \boldsymbol{\Sigma}_{\boldsymbol{F}_{t'}[\ell]}
  \boldsymbol{\Sigma}_{\boldsymbol{F}_{t}[\ell]}^{\mathrm{T}}
 \right)
\right]
\nonumber \\
+ \frac{\bar{v}_{\B\to\A,t',t}[\ell]}{N_{\mathrm{c}}[\ell]}\mathbb{E}\left[
 \mathrm{Tr}
 (\boldsymbol{I} - \boldsymbol{\Sigma}_{\boldsymbol{F}_{t}[\ell]}^{\mathrm{T}}
 \boldsymbol{\Sigma}[\ell])
 (\boldsymbol{I} - \boldsymbol{\Sigma}_{\boldsymbol{F}_{t'}[\ell]}^{\mathrm{T}}
 \boldsymbol{\Sigma}[\ell])^{\mathrm{T}}
\right]
\nonumber \\
= o(1) + \frac{\sigma^{2}}{N_{\mathrm{c}}[\ell]}
\mathbb{E}\left[
 \mathrm{Tr}\left(
  \boldsymbol{F}_{t}^{\mathrm{T}}[\ell]\boldsymbol{F}_{t'}[\ell]
 \right)
\right]
\nonumber \\
+ \frac{\bar{v}_{\B\to\A,t',t}[\ell]}{N_{\mathrm{c}}[\ell]}\mathbb{E}\left[
 \mathrm{Tr}
 (\boldsymbol{I} - \boldsymbol{F}_{t}^{\mathrm{T}}[\ell]\boldsymbol{A}[\ell])
 (\boldsymbol{I} - \boldsymbol{F}_{t'}^{\mathrm{T}}[\ell]
 \boldsymbol{A}[\ell])^{\mathrm{T}}
\right]. 
\nonumber \\
\end{IEEEeqnarray}
In the derivation of the first equality, we have used the induction hypothesis 
(\ref{target_identity}) for all $\tau\in\{0,\ldots,t-1\}$ and 
$\tau'\in\{0,\ldots,\tau+1\}$. 
The last expression is equivalent to the state evolution 
recursion~(\ref{cov_post_A_SE}). 

Let us prove $\vec{\boldsymbol{h}}_{t}[\ell]\aeq
\vec{\boldsymbol{x}}_{\A\to\B,t}[\ell] 
- |\mathcal{W}[\ell]|^{-1/2}\vec{\boldsymbol{x}}[\ell]+o(1)$. 
Since $\vec{\boldsymbol{m}}_{t}^{\mathrm{post}}[\ell]$ in (\ref{m_post_OAMP_tmp2}) 
is a separable and proper-Lipschitz function of $\boldsymbol{b}_{t}[\ell]$, 
we can use Property~\ref{property_A3} in Theorem~\ref{theorem_SE_tech} 
for $\tau=t$ to obtain $\eta_{\A,t}[\ell]\ato\bar{\eta}_{\A,t}[\ell]$. 
Under the induction hypothesis $\vec{\boldsymbol{q}}_{t}^{\mathrm{ext}}[\ell]\aeq
\vec{\boldsymbol{x}}_{\B\to\A,t}[\ell] - \vec{\boldsymbol{x}}[\ell]+o(1)$, 
thus, Lemma~\ref{lemma_inclusion} implies $\vec{\boldsymbol{h}}_{t}[\ell]\aeq
\vec{\boldsymbol{x}}_{\A\to\B,t}[\ell] 
- |\mathcal{W}[\ell]|^{-1/2}\vec{\boldsymbol{x}}[\ell]+o(1)$. 

We next evaluate $N_{\mathrm{c}}^{-1}[\ell] 
(\vec{\boldsymbol{x}}_{\A\to\B,t'}[\ell] 
- |\mathcal{W}[\ell]|^{-1/2}\vec{\boldsymbol{x}}[\ell])^{\mathrm{T}}$ 
$ (\vec{\boldsymbol{x}}_{\A\to\B,t}[\ell] 
- |\mathcal{W}[\ell]|^{-1/2}\vec{\boldsymbol{x}}[\ell])
\ato\bar{v}_{\A\to\B,t',t}[\ell]$. 
Using $\vec{\boldsymbol{h}}_{t}[\ell]\aeq
\vec{\boldsymbol{x}}_{\A\to\B,t}[\ell] 
- |\mathcal{W}[\ell]|^{-1/2}\vec{\boldsymbol{x}}[\ell]+o(1)$ and  
the definition of $\vec{\boldsymbol{m}}_{t}^{\mathrm{ext}}[\ell]$ in 
(\ref{h}) yields 
\begin{equation} \label{cov_AB_SE_tmp} 
\bar{v}_{\A\to\B,t',t}[\ell] 
\aeq \frac{1}{N_{\mathrm{c}}[\ell]}
(\vec{\boldsymbol{m}}_{t'}^{\mathrm{ext}}[\ell])^{\mathrm{T}}
\vec{\boldsymbol{m}}_{t}^{\mathrm{ext}}[\ell].  
\end{equation}
Applying Property~\ref{property_A2} in Theorem~\ref{theorem_SE_tech} to 
the expression of $\vec{\boldsymbol{m}}_{t}^{\mathrm{ext}}[\ell]$ 
in (\ref{m_ext_OAMP_tmp}) yields 
\begin{IEEEeqnarray}{l}
\bar{v}_{\A\to\B,t',t}[\ell]
\aeq \frac{1}{N_{\mathrm{c}}[\ell]}
(\vec{\boldsymbol{m}}_{t'}^{\mathrm{post}}[\ell])^{\mathrm{T}}
\vec{\boldsymbol{m}}_{t}^{\mathrm{ext}}[\ell] + o(1) 
\nonumber \\
\aeq \frac{\bar{v}_{\A,t',t}^{\mathrm{post}}[\ell] 
- \eta_{\A,t}[\ell]
N_{\mathrm{c}}^{-1}[\ell]\vec{\boldsymbol{b}}_{t'}^{\mathrm{T}}[\ell]
(\boldsymbol{I} - \boldsymbol{\Sigma}_{\boldsymbol{F}_{t'}[\ell]}^{\mathrm{T}}
\boldsymbol{\Sigma}[\ell])^{\mathrm{T}}\vec{\boldsymbol{b}}_{t}[\ell]}
{|\mathcal{W}[\ell]|(1 - \bar{\eta}_{\A,t'}[\ell])(1-\bar{\eta}_{\A,t}[\ell])} 
\nonumber \\
+ o(1)
\ato \frac{\bar{v}_{\A,t',t}^{\mathrm{post}}[\ell] 
- \bar{\eta}_{\A,t'}[\ell]\bar{\eta}_{\A,t}[\ell]
\bar{v}_{\B\to\A,t',t}[\ell]}
{|\mathcal{W}[\ell]|(1 - \bar{\eta}_{\A,t'}[\ell])(1-\bar{\eta}_{\A,t}[\ell])},
\end{IEEEeqnarray}
which is equivalent to the state evolution recursion~(\ref{cov_AB_SE}). 
Here, the first equality follows from the asymptotic orthogonality 
$N_{\mathrm{c}}^{-1}[\ell]\vec{\boldsymbol{b}}_{t'}^{\mathrm{T}}[\ell]
\vec{\boldsymbol{m}}_{t}^{\mathrm{ext}}[\ell]\aeq o(1)$. The second equality 
is due to the definitions of $\vec{\boldsymbol{m}}_{t}^{\mathrm{post}}[\ell]$ 
and $\vec{\boldsymbol{m}}_{t}^{\mathrm{ext}}[\ell]$ in (\ref{m_post_OAMP_tmp2}) 
and (\ref{m_ext_OAMP_tmp}), as well as 
the expression of $\vec{\boldsymbol{m}}_{t}^{\mathrm{post}}[\ell]$ in 
(\ref{m_post_OAMP}). The last follows from Property~\ref{property_A2} 
in Theorem~\ref{theorem_SE_tech}.  

Let us prove $\vec{\boldsymbol{q}}_{t+1}^{\mathrm{ext}}[\ell]\aeq
\vec{\boldsymbol{x}}_{\B\to\A,t+1}[\ell] - \vec{\boldsymbol{x}}[\ell]+o(1)$.  
We have already proved that the covariance message $v_{\A\to\B,t',t}[\ell]$ 
in (\ref{cov_AB}) is a consistent estimator of the error covariance 
$\bar{v}_{\A\to\B,t',t}[\ell]$ in the large system limit. Furthermore, 
we use Property~\ref{property_A5} in Theorem~\ref{theorem_SE_tech} to 
confirm that the covariance matrix $\bar{\boldsymbol{V}}_{\A\to\B,t}[\ell]$ 
is positive definite. Using the definition of 
$\vec{\boldsymbol{q}}_{t+1}^{\mathrm{post}}[\ell]$ in (\ref{q_post_OAMP_tmp}) 
and $\vec{\boldsymbol{h}}_{\tau}[\ell]
\aeq\vec{\boldsymbol{x}}_{\A\to\B,\tau}[\ell] - |\mathcal{W}[\ell]|^{-1/2}
\vec{\boldsymbol{x}}[\ell]+o(1)$ for all $\tau\in\{0,\ldots,t\}$, we find that 
$\vec{\boldsymbol{q}}_{t+1}^{\mathrm{post}}[\ell]$ is a separable and 
proper-Lipschitz function of 
$\{\vec{\boldsymbol{h}}_{\tau}[\ell]: \tau\in\{0,\ldots,t\}\}$. 
Thus, we can utilize Property~\ref{property_B3} in 
Theorem~\ref{theorem_SE_tech} for $\tau=t$ to obtain 
$\eta_{\B,t}[\ell]\ato\bar{\eta}_{\B,t}[\ell]$, which implies that 
$\vec{\boldsymbol{q}}_{t+1}^{\mathrm{ext}}[\ell]\aeq
\vec{\boldsymbol{x}}_{\B\to\A,t+1}[\ell] - \vec{\boldsymbol{x}}[\ell]+o(1)$ 
holds from Lemma~\ref{lemma_inclusion}.  

We derive the state evolution recursions~(\ref{cov_suf_B_SE}) and 
(\ref{cov_post_B_SE}) with respect to $\bar{v}_{\A\to\B,t',t}^{\mathrm{suf}}[l]$ 
and $\bar{v}_{\B,t'+1,t+1}^{\mathrm{post}}[l]$, respectively. The almost sure 
convergence $\boldsymbol{V}_{\A\to\B,t}[\ell]
\ato\bar{\boldsymbol{V}}_{\A\to\B,t}[\ell]$ 
implies that the covariance message $v_{\A\to\B,t',t}^{\mathrm{suf}}[l]$ in 
(\ref{cov_suf_B}) converges almost surely to 
$\bar{v}_{\A\to\B,t',t}^{\mathrm{suf}}[l]$ in 
(\ref{cov_suf_B_SE}) in the large system limit. Since 
$v_{\A\to\B,t',t}^{\mathrm{suf}}[l]$ is a consistent estimator of the error 
covariance $N^{-1}[l]\mathbb{E}[(\boldsymbol{x}[l]
-\boldsymbol{x}_{\A\to\B,t'}^{\mathrm{suf}}[l])^{\mathrm{T}}
(\boldsymbol{x}[l]-\boldsymbol{x}_{\A\to\B,t}^{\mathrm{suf}}[l])]$ in the large 
system limit, we use Property~\ref{property_B2} in 
Theorem~\ref{theorem_SE_tech} to arrive at the almost sure convergence 
$N^{-1}[l]\mathbb{E}[(\boldsymbol{x}[l]
-\boldsymbol{x}_{\B,t'+1}^{\mathrm{post}}[l])^{\mathrm{T}}
(\boldsymbol{x}[l]-\boldsymbol{x}_{\B,t+1}^{\mathrm{post}}[l])]
\ato\bar{v}_{\B,t'+1,t+1}^{\mathrm{post}}[l]$ given in (\ref{cov_post_B_SE}). 

Finally, we evaluate $\bar{v}_{\B\to\A,t'+1,t+1}[\ell]$ in 
(\ref{target_identity}) for all $t'\in\{-1,\ldots,t\}$ 
Consider the case $t'\geq0$. 
Using $\vec{\boldsymbol{q}}_{t+1}^{\mathrm{ext}}[\ell]
=\vec{\boldsymbol{x}}_{\B\to\A,t+1}[\ell] - \vec{\boldsymbol{x}}[\ell]$, 
the definition of $\vec{\boldsymbol{x}}_{\B\to\A,t+1}[\ell]$ in 
(\ref{mean_BA_tmp}), and the asymptotic orthogonality 
$N_{\mathrm{c}}^{-1}[\ell]\vec{\boldsymbol{h}}_{t'}^{\mathrm{T}}[\ell]
\vec{\boldsymbol{q}}_{t+1}^{\mathrm{next}}[\ell]\aeq o(1)$ yields 
\begin{IEEEeqnarray}{rl}
&\bar{v}_{\B\to\A,t'+1,t+1}[\ell] 
\aeq \frac{(\vec{\boldsymbol{q}}_{t'+1}^{\mathrm{post}}[\ell])^{\mathrm{T}}
\vec{\boldsymbol{q}}_{t+1}^{\mathrm{ext}}[\ell]}{N_{\mathrm{c}}[\ell]}
+ o(1) \nonumber \\
=& \frac{(\vec{\boldsymbol{q}}_{t'+1}^{\mathrm{post}}[\ell])^{\mathrm{T}}
\vec{\boldsymbol{q}}_{t+1}^{\mathrm{post}}[\ell]}{N_{\mathrm{c}}[\ell]}
- \frac{|\mathcal{W}[\ell]|^{-1/2}}
{1 - \bar{\eta}_{\B,t}[\ell]/|\mathcal{W}[\ell]|}
\nonumber \\
&\cdot\sum_{\tau=0}^{t}\eta_{\B,\tau,t}[\ell]
\frac{(\vec{\boldsymbol{q}}_{t'+1}^{\mathrm{post}}[\ell])^{\mathrm{T}}
\vec{\boldsymbol{h}}_{\tau}[\ell]}{N_{\mathrm{c}}[\ell]}
+ o(1), \label{cov_BA_SE_tmp}
\end{IEEEeqnarray} 
where we have used $\eta_{\B,t}[\ell]\ato\bar{\eta}_{\B,t}[\ell]$. 
From the definition of $\vec{\boldsymbol{q}}_{t+1}^{\mathrm{post}}[\ell]$ 
in (\ref{q_post_OAMP_tmp}), we find that the first term reduces to 
\begin{IEEEeqnarray}{l}
\frac{(\vec{\boldsymbol{q}}_{t'+1}^{\mathrm{post}}[\ell])^{\mathrm{T}}
\vec{\boldsymbol{q}}_{t+1}^{\mathrm{post}}[\ell]}{N_{\mathrm{c}}[\ell]}
= \frac{1}{N_{\mathrm{c}}[\ell]}\sum_{w\in\mathcal{W}[\ell]}
(\vec{\boldsymbol{q}}_{t'+1}^{\mathrm{post}}[\ell][w])^{\mathrm{T}}
\vec{\boldsymbol{q}}_{t+1}^{\mathrm{post}}[\ell][w]
\nonumber \\
\aeq \sum_{w\in\mathcal{W}[\ell]}\frac{N[\ell-w]}{N_{\mathrm{c}}[\ell]}
\frac{|\mathcal{W}[\ell]|\gamma^{2}[\ell][\ell-w]
\bar{v}_{\B,t'+1,t+1}^{\mathrm{post}}[\ell-w]}
{(1-\bar{\eta}_{\B,t'}[\ell]/|\mathcal{W}[\ell]|)
(1-\bar{\eta}_{\B,t}[\ell]/|\mathcal{W}[\ell]|)}
\nonumber \\
+o(1). \label{cov_BA_SE_1}
\end{IEEEeqnarray}

For the second term, on the other hand, we use the expression of 
$\vec{\boldsymbol{q}}_{t+1}^{\mathrm{ext}}[\ell]
=\vec{\boldsymbol{x}}_{\B\to\A,t+1}[\ell] 
- \vec{\boldsymbol{x}}[\ell]$ in (\ref{mean_BA_tmp}) and 
the asymptotic orthogonality $N_{\mathrm{c}}^{-1}[\ell]
(\vec{\boldsymbol{q}}_{t'+1}^{\mathrm{ext}}[\ell])^{\mathrm{T}}
\vec{\boldsymbol{h}}_{\tau}[\ell]\aeq o(1)$ to obtain 
\begin{IEEEeqnarray}{rl}
&\frac{|\mathcal{W}[\ell]|^{-1/2}}
{1 - \bar{\eta}_{\B,t}[\ell]/|\mathcal{W}[\ell]|}
\sum_{\tau=0}^{t}\eta_{\B,\tau,t}[\ell]
\frac{(\vec{\boldsymbol{q}}_{t'+1}^{\mathrm{post}}[\ell])^{\mathrm{T}}
\vec{\boldsymbol{h}}_{\tau}[\ell]}{N_{\mathrm{c}}[\ell]}
\nonumber \\
\aeq& \frac{|\mathcal{W}[\ell]|^{-1}}
{(1 - \bar{\eta}_{\B,t}[\ell]/|\mathcal{W}[\ell]|)
(1 - \bar{\eta}_{\B,t'}[\ell]/|\mathcal{W}[\ell]|)}
\nonumber \\
&\cdot\sum_{\tau'=0}^{t'}\sum_{\tau=0}^{t}
\eta_{\B,\tau',t'}[\ell]\eta_{\B,\tau,t}[\ell]
\bar{v}_{\A\to\B,\tau',\tau}[\ell]
+ o(1). 
\end{IEEEeqnarray}
Using the following identity obtained from (\ref{eta_B}), (\ref{zeta_B}), 
and (\ref{eta_B_ell_LM}):  
\begin{equation}
\eta_{\B,\tau,t}[\ell] 
= \eta_{\B,t}[\ell]
\frac{\boldsymbol{e}_{\tau}^{\mathrm{T}}\boldsymbol{V}_{\A\to\B,t}^{-1}[\ell]
\boldsymbol{1}}
{\boldsymbol{1}^{\mathrm{T}}\boldsymbol{V}_{\A\to\B,t}^{-1}[\ell]
\boldsymbol{1}}, 
\end{equation}
as well as the expression $\sum_{\tau'=0}^{t'}\sum_{\tau=0}^{t}
\bar{v}_{\A\to\B,\tau',\tau}[\ell]
\boldsymbol{e}_{\tau'}\boldsymbol{e}_{\tau}^{\mathrm{T}}
=(\boldsymbol{I}_{t'+1}, \boldsymbol{O})
\bar{\boldsymbol{V}}_{\A\to\B,t}[\ell]$ for $t'\leq t$ and 
$\boldsymbol{V}_{\A\to\B,t}[\ell]\ato\bar{\boldsymbol{V}}_{\A\to\B,t}[\ell]$, 
we find that the last factor reduces to
\begin{IEEEeqnarray}{rl}
&\sum_{\tau'=0}^{t'}\sum_{\tau=0}^{t}\eta_{\B,\tau',t'}[\ell]
\eta_{\B,\tau,t}[\ell]\bar{v}_{\A\to\B,\tau',\tau}[\ell]
\nonumber \\
\aeq& \frac{\bar{\eta}_{\B,t'}[\ell]\bar{\eta}_{\B,t}[\ell]}
{\boldsymbol{1}^{\mathrm{T}}\bar{\boldsymbol{V}}_{\A\to\B,t}^{-1}[\ell]
\boldsymbol{1}} + o(1),
\end{IEEEeqnarray}
with 
\begin{equation}
\bar{\eta}_{\B,t}[\ell]=\sum_{w\in\mathcal{W}[\ell]}
\frac{N[\ell-w]}{N_{\mathrm{c}}[\ell]}\bar{\eta}_{\B,t}[\ell][w],
\end{equation}
where we have used the fact that $\eta_{\B,t}[\ell][w]$ in (\ref{eta_B_w_LM}) 
converges almost surely to $\bar{\eta}_{\B,t}[\ell][w]$ in (\ref{zeta_B_SE}), 
because of $\boldsymbol{V}_{\A\to\B,t}[\ell]\ato
\bar{\boldsymbol{V}}_{\A\to\B,t}[\ell]$, $v_{\A\to\B,t,t}^{\mathrm{suf}}[l]
\ato\bar{v}_{\A\to\B,t,t}^{\mathrm{suf}}[l]$, and Property~\ref{property_B3} 
in Theorem~\ref{theorem_SE_tech} for $\tau=t$.
Combining these results, we arrive at 
\begin{IEEEeqnarray}{rl}
&\frac{|\mathcal{W}[\ell]|^{-1/2}}
{1 - \bar{\eta}_{\B,t}[\ell]/|\mathcal{W}[\ell]|}
\sum_{\tau=0}^{t}\eta_{\B,\tau,t}[\ell]
\frac{(\vec{\boldsymbol{q}}_{t'+1}^{\mathrm{post}}[\ell])^{\mathrm{T}}
\vec{\boldsymbol{h}}_{\tau}[\ell]}{N_{\mathrm{c}}[\ell]}
\nonumber \\
\aeq& \frac{|\mathcal{W}[\ell]|^{-1}}
{(1 - \bar{\eta}_{\B,t'}[\ell]/|\mathcal{W}[\ell]|)
(1 - \bar{\eta}_{\B,t}[\ell]/|\mathcal{W}[\ell]|)}
\frac{\bar{\eta}_{\B,t'}[\ell]\bar{\eta}_{\B,t}[\ell]}
{\boldsymbol{1}^{\mathrm{T}}\bar{\boldsymbol{V}}_{\A\to\B,t}^{-1}[\ell]
\boldsymbol{1}} 
\nonumber \\
&+ o(1).  \label{cov_BA_SE_2}
\end{IEEEeqnarray}

We are ready to prove (\ref{target_identity}) for $\tau'>0$. 
Substituting (\ref{cov_BA_SE_1}) and (\ref{cov_BA_SE_2}) into 
(\ref{cov_BA_SE_tmp}) yields 
\begin{IEEEeqnarray}{rl}
&(1 - \bar{\eta}_{\B,t'}[\ell]/|\mathcal{W}[\ell]|)
(1 - \bar{\eta}_{\B,t}[\ell]/|\mathcal{W}[\ell]|)\bar{v}_{\B\to\A,t'+1,t+1}[\ell] 
\nonumber \\
\aeq& \sum_{w\in\mathcal{W}[\ell]}\frac{N[\ell-w]}{N_{\mathrm{c}}[\ell]}
|\mathcal{W}[\ell]|\gamma^{2}[\ell][\ell-w]
\bar{v}_{\B,t'+1,t+1}^{\mathrm{post}}[\ell-w]
\nonumber \\
&- \frac{1}{|\mathcal{W}[\ell]|}
\frac{\bar{\eta}_{\B,t'}[\ell]\bar{\eta}_{\B,t}[\ell]}
{\boldsymbol{1}^{\mathrm{T}}\bar{\boldsymbol{V}}_{\A\to\B,t}^{-1}[\ell]
\boldsymbol{1}} + o(1), 
\end{IEEEeqnarray}
which is equivalent to the state evolution recursion~(\ref{cov_BA_SE}). 

In the case of $t'=-1$, for $f_{-1}[l]=0$ we repeat the same proof as that 
for the case $t'\geq0$ to find that $\bar{v}_{\B\to\A,0,t+1}[\ell]$ is 
equivalent to (\ref{cov_BA_SE_0}).   
Thus, Theorem~\ref{theorem_SE_LM} holds. 
\end{IEEEproof}

\subsection{Module~A for $\tau=0$} \label{appen_CE}
\begin{IEEEproof}[Proof of \ref{property_A2}]
We first prove the strong law of large numbers with respect to 
$\vec{\boldsymbol{b}}_{0}[0]$. Consider
\begin{equation}
\langle\tilde{\boldsymbol{\phi}}_{0}(\mathcal{B}_{1},\Omega,\Lambda)\rangle
= \frac{1}{N_{\mathrm{c}}}\sum_{n=1}^{N_{\mathrm{c}}[0]}\tilde{\phi}_{n,0}
+ \frac{1}{N_{\mathrm{c}}}\sum_{n=N_{\mathrm{c}}[0]+1}^{N_{\mathrm{c}}}\tilde{\phi}_{n,0}.  
\end{equation}
By definition, the second term is independent of $\vec{\boldsymbol{b}}_{0}[0]$. 

Let us evaluate the first term. Under Assumptions~\ref{assumption_x} and 
\ref{assumption_function}, for some $\kappa_{0,0}[\ell]>0$ we find 
$N_{\mathrm{c}}^{-1}[\ell]\|\vec{\boldsymbol{q}}_{0}^{\mathrm{ext}}[\ell]\|^{2}
\ato\kappa_{0,0}[\ell]$ for the initial condition 
$\vec{\boldsymbol{q}}_{0}^{\mathrm{ext}}[\ell]
=(\boldsymbol{I}_{N_{\mathrm{c}}[\ell]}, \boldsymbol{O})
\boldsymbol{\psi}_{-1}[\ell](\mathcal{X})$. Consider 
$\boldsymbol{A}_{3|\mathcal{L}_{W}|-1}=\{\{[\vec{\boldsymbol{b}}_{0}^{\mathrm{T}}[\ell],
\boldsymbol{0}]^{\mathrm{T}}\}_{\ell=1}^{|\mathcal{L}_{W}|-1},
\Omega,\Lambda\}$, $\boldsymbol{a}_{3|\mathcal{L}_{W}|-1}=\boldsymbol{\epsilon}
=\boldsymbol{0}$, $\boldsymbol{\Phi}_{\boldsymbol{E}}^{\perp}=\boldsymbol{I}$, 
and $\boldsymbol{\omega}=\vec{\boldsymbol{b}}_{0}[0]$ in 
\cite[Lemma 3]{Takeuchi211}. Since $\vec{\boldsymbol{b}}_{0}[0]=
\boldsymbol{V}^{\mathrm{T}}[0]\vec{\boldsymbol{q}}_{0}^{\mathrm{ext}}[0]$ 
is orthogonally invariant and $\boldsymbol{V}[0]$ is independent of 
the other matrices $\{\boldsymbol{V}[\ell]: \ell\neq0\}$, 
we can use \cite[Lemma 3]{Takeuchi211} to obtain  
\begin{IEEEeqnarray}{rl} 
&\langle\tilde{\boldsymbol{\phi}}_{0}(\mathcal{B}_{1},\Omega,\Lambda)\rangle
\nonumber \\
\aeq& \mathbb{E}_{\vec{\boldsymbol{z}}_{\A,0}[0]}\left[
 \langle\tilde{\boldsymbol{\phi}}_{0}(
 [\vec{\boldsymbol{z}}_{\A,0}^{\mathrm{T}}[0], \boldsymbol{0}]^{\mathrm{T}},
\boldsymbol{A}_{3|\mathcal{L}_{W}|-1})\rangle
\right] + o(1),
\end{IEEEeqnarray}
with $\vec{\boldsymbol{z}}_{\A,0}[\ell]\sim\mathcal{N}(\boldsymbol{0},
\kappa_{0,0}[\ell]\boldsymbol{I}_{N_{\mathrm{c}}[\ell]})$. 

We next prove the strong law of large numbers with respect to the remaining 
vectors in the same manner. 
Since $\phi_{0}$ is separable and proper pseudo-Lipschitz, 
$\mathbb{E}_{\vec{\boldsymbol{z}}_{\A,0}[0]}[
\langle\tilde{\boldsymbol{\phi}}_{0}(
[\vec{\boldsymbol{z}}_{\A,0}^{\mathrm{T}}[0], \boldsymbol{0}]^{\mathrm{T}},
\boldsymbol{A}_{3|\mathcal{L}_{W}|-1})]$ is also separable and proper 
pseudo-Lipschitz from Proposition~\ref{proposition_pseudo-Lipschitz}. 
Thus, we can repeat the same derivation for 
$\{\vec{\boldsymbol{b}}_{0}[\ell]\}_{\ell=1}^{|\mathcal{L}_{W}|-1}$ and 
$\{\boldsymbol{n}[\ell]\}_{\ell=1}^{|\mathcal{L}_{W}|-1}$ to arrive at
\begin{equation} 
\langle\tilde{\boldsymbol{\phi}}_{0}(\mathcal{B}_{1},\Omega,\Lambda)\rangle
\aeq \mathbb{E}\left[
 \langle\tilde{\boldsymbol{\phi}}_{0}(\mathcal{Z}_{\A,1},
 \tilde{\Omega}, \Lambda)\rangle 
\right]
+ o(1),
\end{equation}
where the strong law of large  numbers with respect to 
$\{\vec{\boldsymbol{\lambda}}[\ell]\}$ follows from 
Assumption~\ref{assumption_A}. 
Thus, the former property~(\ref{phi_SLLN}) holds for $\tau=0$. 
\end{IEEEproof}

\begin{IEEEproof}[Proof of \ref{property_A3}]
The former property~(\ref{phi_SLLN_1}) for $\tau=0$ follows from 
Property~\ref{property_A2} and a technical result in 
\cite[Lemma 5]{Bayati11}. Thus, we only prove the latter property for $\tau=0$. 

Without loss of generality, we focus on $\ell=0$. 
We use Property~\ref{property_A2} and (\ref{phi_SLLN_1}) for $\tau=0$ to find 
that the LHS of (\ref{orthogonality_A}) reduces to its expectation 
\begin{equation}
\frac{1}{N_{\mathrm{c}}[0]}\mathbb{E}\left[
 \vec{\boldsymbol{z}}_{\A,0}^{\mathrm{T}}[0]\left(
  (\boldsymbol{I}_{N_{\mathrm{c}}[0]},\boldsymbol{O})\tilde{\boldsymbol{\phi}}_{0}
  - \mathbb{E}[\tilde{\xi}_{\A,0,0}[0]]
  \vec{\boldsymbol{z}}_{\A,0}[0]
 \right)
\right]
\end{equation}
in the large system limit, with $\tilde{\boldsymbol{\phi}}_{0}
=\tilde{\boldsymbol{\phi}}_{0}(\mathcal{Z}_{\A,0},\tilde{\Omega},\Lambda)$.   
Using \cite[Lemma 2]{Takeuchi211} yields 
\begin{IEEEeqnarray}{rl}
&\hbox{LHS of (\ref{orthogonality_A})}
\nonumber \\
\aeq& \frac{1}{N_{\mathrm{c}}[0]}\sum_{n=1}^{N_{\mathrm{c}}[0]}
\mathbb{E}[\vec{z}_{\A,n,0}^{2}[0]]\mathbb{E}[\langle \partial_{0}\tilde{\phi}_{n,0}
\rangle] \nonumber \\
&- \mathbb{E}[\tilde{\xi}_{\A,0,0}[0]]
\frac{\mathbb{E}[\|\vec{\boldsymbol{z}}_{\A,0}[0]\|^{2}]}
{N_{\mathrm{c}}[0]} + o(1) = o(1), 
\end{IEEEeqnarray}
where the last equality follows from the definition of 
$\tilde{\xi}_{\A,0,0}[0]$ in (\ref{tilde_xi_A}). 
Thus, Property~\ref{property_A3} holds for $\tau=0$.
\end{IEEEproof}

\begin{IEEEproof}[Proofs of \ref{property_A4} and \ref{property_A5}]
The proofs are omitted since they are the same as in 
\cite[p.~4419]{Takeuchi211}. 
\end{IEEEproof}

\subsection{Module~B for $\tau=0$} \label{appen_CF}
\begin{IEEEproof}[Proof of \ref{property_B1}]
The proof is omitted since it is the same as in 
\cite[pp.~4419--4420]{Takeuchi211}. 
\end{IEEEproof}

\begin{IEEEproof}[Proof of \ref{property_B2}]
We prove the strong law of large numbers with respect to 
$\vec{\boldsymbol{h}}_{0}[0]$. Consider 
\begin{equation}
\langle\tilde{\boldsymbol{\psi}}_{0}(\mathcal{H}_{1},\mathcal{X})\rangle
= \frac{1}{N_{\mathrm{c}}}\sum_{n=1}^{N_{\mathrm{c}}[0]}\tilde{\psi}_{n,0}
+ \frac{1}{N_{\mathrm{c}}}\sum_{n=N_{\mathrm{c}}[0]+1}^{N_{\mathrm{c}}}\tilde{\psi}_{n,0}.  
\end{equation}
By definition, the second term is independent of $\vec{\boldsymbol{h}}_{0}[0]$. 

To evaluate the first term, we utilize \cite[Lemma~3]{Takeuchi211} with 
$\boldsymbol{a}_{2|\mathcal{L}_{W}|-1}=\boldsymbol{0}$, 
$\boldsymbol{A}_{2|\mathcal{L}_{W}|-1}=\{\{
[\vec{\boldsymbol{h}}_{0}^{\mathrm{T}}[\ell], \boldsymbol{O}]^{\mathrm{T}}
\}_{\ell=1}^{|\mathcal{L}_{W}|-1}, \mathcal{X}\}$, 
$\boldsymbol{\epsilon}=o(1)\vec{\boldsymbol{q}}_{0}^{\mathrm{ext}}[0]$, 
$\boldsymbol{E}=\vec{\boldsymbol{q}}_{0}^{\mathrm{ext}}[0]$, and 
$\boldsymbol{\omega}=\vec{\boldsymbol{\omega}}_{\B,0}[0]$. 
Using Property~\ref{property_B1} for $\tau=0$ and 
\cite[Lemma~3]{Takeuchi211} yields 
\begin{IEEEeqnarray}{rl}
&\langle\tilde{\boldsymbol{\psi}}_{0}(\mathcal{H}_{1}, \mathcal{X})\rangle
- \mathbb{E}_{\vec{\boldsymbol{z}}_{\B,0}[0]}\left[
 \langle\tilde{\boldsymbol{\psi}}_{0}([\vec{\boldsymbol{z}}_{\B,0}^{\mathrm{T}}[0],
 \boldsymbol{0}]^{\mathrm{T}}, \boldsymbol{A}_{2|\mathcal{L}_{W}|-1})
 \rangle 
\right] 
\nonumber \\
&\ato 0, 
\end{IEEEeqnarray}
with $\vec{\boldsymbol{z}}_{\B,0}[0]\sim\mathcal{N}(\boldsymbol{0},
\pi_{0,0}[0]\boldsymbol{I}_{N_{\mathrm{c}}[0]})$, where Property~\ref{property_A4} 
for $\tau=0$ implies the existence of $\pi_{0,0}[0]$. 

The strong law of large numbers with respect to the remaining vectors 
can be proved in the same manner. 
Repeating the application of \cite[Lemma~3]{Takeuchi211} for 
$\ell=1,\ldots,|\mathcal{L}_{W}|-1$, we arrive at
\begin{equation}
\langle\tilde{\boldsymbol{\psi}}_{0}(\mathcal{H}_{1}, \mathcal{X})\rangle
- \mathbb{E}\left[
 \langle\tilde{\boldsymbol{\psi}}_{0}(\mathcal{Z}_{\B,0},\mathcal{X})
 \rangle 
\right]\ato 0, 
\end{equation}
where the strong law of large numbers for $\mathcal{X}$ follows from 
Assumption~\ref{assumption_x}. 
Thus, Property~\ref{property_B2} holds for $\tau=0$.  
\end{IEEEproof}

\begin{IEEEproof}[Proofs of \ref{property_B3} and \ref{property_B4}]
The proofs are omitted since they are the same as in 
Properties~\ref{property_A3} and \ref{property_A4} for $\tau=0$, 
respectively.  
\end{IEEEproof}

\begin{IEEEproof}[Proof of \ref{property_B5}]
The proof is omitted since it is the same as in \cite[p.~4420]{Takeuchi211}. 
\end{IEEEproof}

\subsection{Module~A by Induction} \label{appen_CG}
Assume that Theorem~\ref{theorem_SE_tech} is correct for all $\tau<t$. 
We prove Theorem~\ref{theorem_SE_tech} for $\tau=t$. Since the properties 
for module~B can be proved in the same manner as for module~A, we only 
prove the properties for module~A. 

\begin{IEEEproof}[Proof of \ref{property_A1} for $\tau=t$]
The proof is omitted since it is the same as in \cite[p.~4420]{Takeuchi211}. 
\end{IEEEproof}

\begin{IEEEproof}[Proof of \ref{property_A2} for $\tau=t$]
The proof is essentially the same as in the proof of 
Property~\ref{property_A2} for $\tau=0$. We prove the strong law of large 
numbers with respect to $\{\vec{\boldsymbol{b}}_{\tau}[0]\}$ in the order 
$\tau=t,\ldots, 0$. Consider 
\begin{equation}
\langle\tilde{\boldsymbol{\phi}}_{t}(\mathcal{B}_{t+1},\Omega,\Lambda)\rangle
= \frac{1}{N_{\mathrm{c}}}\sum_{n=1}^{N_{\mathrm{c}}[0]}\tilde{\phi}_{n,t}
+ \frac{1}{N_{\mathrm{c}}}\sum_{n=N_{\mathrm{c}}[0]+1}^{N_{\mathrm{c}}}\tilde{\phi}_{n,t}.  
\end{equation}
By definition, the second term is independent of $\vec{\boldsymbol{b}}_{t}[0]$. 

To evaluate the first term, we utilize \cite[Lemma~3]{Takeuchi211} under 
Property~\ref{property_A1} for $\tau\leq t$. Let 
$\boldsymbol{a}_{(t+3)|\mathcal{L}_{W}|-1} 
=\vec{\boldsymbol{B}}_{t}[0]\vec{\boldsymbol{\beta}}_{t}[0]$, 
$\boldsymbol{A}_{(t+3)|\mathcal{L}_{W}|-1}=\{[\vec{\boldsymbol{B}}_{t}^{\mathrm{T}}[0], 
\boldsymbol{O}]^{\mathrm{T}}, 
\{[\vec{\boldsymbol{B}}_{t+1}^{\mathrm{T}}[\ell], \boldsymbol{O}]^{\mathrm{T}}
\}_{\ell=1}^{|\mathcal{L}_{W}|-1}, \Omega$, $\Lambda\}$, 
$\boldsymbol{\epsilon}=\vec{\boldsymbol{M}}_{t}^{\mathrm{ext}}[0]\boldsymbol{o}(1)$
$+ \vec{\boldsymbol{B}}_{t}[0]\boldsymbol{o}(1)$, 
$\boldsymbol{E}=(\vec{\boldsymbol{B}}_{t}[0], 
\vec{\boldsymbol{M}}_{t}^{\mathrm{ext}}[0])$, 
and $\boldsymbol{\omega}=\vec{\boldsymbol{\omega}}_{\A,t}[0]$. 
Using Property~\ref{property_A1} for $\tau=t$ and 
\cite[Lemma~3]{Takeuchi211} yields 
\begin{IEEEeqnarray}{rl}
\langle\tilde{\boldsymbol{\phi}}_{t}(\mathcal{B}_{t+1}, \Omega,\Lambda)\rangle
- &\mathbb{E}_{\vec{\boldsymbol{z}}_{\A,t}[0]}\left[
 \langle\tilde{\boldsymbol{\phi}}_{t}(
 [(\vec{\boldsymbol{B}}_{t}[0]
 \vec{\boldsymbol{\beta}}_{t}[0]+\vec{\boldsymbol{z}}_{\A,t}[0])^{\mathrm{T}},
\right. \nonumber \\
&\left.
 \boldsymbol{0}]^{\mathrm{T}}, 
 \boldsymbol{A}_{(t+3)|\mathcal{L}_{W}|-1}) \rangle 
\right] 
\ato 0.
\end{IEEEeqnarray}
Repeating this argument, we follow \cite[p.~4420]{Takeuchi211} to arrive at 
\begin{IEEEeqnarray}{rl}
\langle\tilde{\boldsymbol{\phi}}_{t}(\mathcal{B}_{t+1}, \Omega,\Lambda)\rangle
- &\mathbb{E}_{\vec{\boldsymbol{Z}}_{\A,t+1}[0]}\left[
 \langle\tilde{\boldsymbol{\phi}}_{t}(
 [\vec{\boldsymbol{Z}}_{\A,t+1}^{\mathrm{T}}[0],
 \boldsymbol{0}]^{\mathrm{T}}, 
\right. \nonumber \\
&\left.
 \boldsymbol{A}_{(t+3)|\mathcal{L}_{W}|-t-1}) \rangle 
\right] 
\ato 0, 
\end{IEEEeqnarray}
with $\boldsymbol{A}_{(t+3)|\mathcal{L}_{W}|-(t+1)} 
= \{\{[\vec{\boldsymbol{B}}_{t+1}^{\mathrm{T}}[\ell], \boldsymbol{O}]^{\mathrm{T}}
\}_{\ell=1}^{|\mathcal{L}_{W}|-1}, \Omega,$ $\Lambda\}$. 

The strong law of large numbers with respect to the remaining vectors 
can be proved in the same manner. 
Repeating the application of \cite[Lemma~3]{Takeuchi211} for 
$\ell=1,\ldots,|\mathcal{L}_{W}|-1$, we obtain 
\begin{equation}
\langle\tilde{\boldsymbol{\phi}}_{t}(\mathcal{B}_{t+1}, \Omega,\Lambda)\rangle
- \mathbb{E}\left[
 \langle\tilde{\boldsymbol{\phi}}_{t}(\mathcal{Z}_{\A,t+1},\tilde{\Omega}, 
\Lambda)\rangle 
\right]\ato 0, 
\end{equation}
where the strong law of large numbers for $\Omega$ and $\Lambda$ follows 
from Assumptions~\ref{assumption_A} and \ref{assumption_w}. 
Thus, Property~\ref{property_A2} holds for $\tau=t$.  
\end{IEEEproof}

\begin{IEEEproof}[Proof of \ref{property_A3} for $\tau=t$]
The former property~(\ref{phi_SLLN_1}) for $\tau=t$ follows from 
Property~\ref{property_A2} for $\tau=t$ and a technical result in 
\cite[Lemma 5]{Bayati11}. Thus, we only prove the latter property for $\tau=t$. 

Without loss of generality, we focus on $\ell=0$. 
We use Property~\ref{property_A2} and (\ref{phi_SLLN_1}) for $\tau=t$ to find 
that the LHS of (\ref{orthogonality_A}) reduces to 
its expectation 
\begin{equation}
\mathbb{E}\left[
 \frac{\vec{\boldsymbol{z}}_{\A,\tau'}^{\mathrm{T}}[0]}{N_{\mathrm{c}}[0]}\left(
  (\boldsymbol{I}_{N_{\mathrm{c}}[0]},\boldsymbol{O})\tilde{\boldsymbol{\phi}}_{t}
  - \sum_{t'=0}^{t}\mathbb{E}[\tilde{\xi}_{\A,t',t}[0]]
  \vec{\boldsymbol{z}}_{\A,t'}[0]
 \right)
\right]
\end{equation}
in the large system limit, with $\tilde{\boldsymbol{\phi}}_{t}
=\tilde{\boldsymbol{\phi}}_{t}(\mathcal{Z}_{\A,t+1},\tilde{\Omega},\Lambda)$.   
Using \cite[Lemma 2]{Takeuchi211} yields 
\begin{IEEEeqnarray}{rl}
&\hbox{LHS of (\ref{orthogonality_A})}
\nonumber \\
\aeq& o(1) + \frac{1}{N_{\mathrm{c}}[0]}\sum_{n=1}^{N_{\mathrm{c}}[0]}
\sum_{t'=0}^{t}\mathbb{E}[\vec{z}_{\A,n,\tau'}[0]\vec{z}_{\A,n,t'}[0]]
\mathbb{E}[\langle \partial_{t'}\tilde{\phi}_{n,t}
\rangle] \nonumber \\
&- \sum_{t'=0}^{t}\mathbb{E}[\tilde{\xi}_{\A,t',t}[0]]
\frac{\mathbb{E}[\vec{\boldsymbol{z}}_{\A,\tau'}^{\mathrm{T}}[0]
\vec{\boldsymbol{z}}_{\A,t'}[0]]}{N_{\mathrm{c}}[0]}  = o(1) 
\end{IEEEeqnarray}
where the last equality follows from the definition of 
$\tilde{\xi}_{\A,t',t}[0]$ in (\ref{tilde_xi_A}). 
Thus, Property~\ref{property_A3} holds for $\tau=t$.
\end{IEEEproof}

\begin{IEEEproof}[Proofs of \ref{property_A4} and \ref{property_A5} 
for $\tau=t$]
The proofs are omitted since they are the same as in 
\cite[p.~4420]{Takeuchi211}. 
\end{IEEEproof}

\section{Proof of Lemma~\ref{lemma_approximation}}
\label{proof_lemma_approximation} 
\subsection{Proof of (\ref{E_t_dif})} 
\label{proof_lemma_approximation_former} 
The former property~(\ref{E_t_dif}) follows from the latter~(\ref{s_t_dif}). 
We first evaluate an upper bound on $\tilde{s}_{\tau}[l]$. 
From Lemma~\ref{lemma_R_transform_0} and the 
assumption~(\ref{R_transform_limit}) in Theorem~\ref{theorem_optimality}, 
we have $R(0)=1$. Using this result, the definition of $g(z)$ in 
(\ref{g_func}), and the non-decreasing assumption for 
$R(z)$ on $(-\infty, 0]$, 
we have the upper bound $g(z)\leq 1/\sigma^{2}$ for all $z\geq0$. 
Thus, from the definition of $\tilde{s}_{\tau}[l]$ in (\ref{s_t_tilde}) 
we arrive at $\tilde{s}_{\tau}[l]\leq 1/\sigma^{2}$. 

We next evaluate $s_{\tau}[l]$. 
From the latter property~(\ref{s_t_dif}), the difference 
$|\tilde{s}_{\tau}[l] - s_{\tau}[l]|$ is bounded for fixed $\tau$ and 
all $l\in\mathcal{L}_{0}$, i.e.\ $|\tilde{s}_{\tau}[l] - s_{\tau}[l]|<d$ 
for some $d>0$, which implies $s_{\tau}[l]<\tilde{s}_{\tau}[l]+d\leq
1/\sigma^{2}+d$. 
Thus, we can restrict the domain of $\mathrm{MMSE}(s)$ to 
the interval $[0, 1/\sigma^{2}+d]$. Since Proposition~\ref{proposition_MSE} 
implies the continuous differentiability of $\mathrm{MMSE}(s)$ for all 
$s\geq0$, $\mathrm{MMSE}(s)$ is Lipschitz-continuous 
for all $s\in[0, 1/\sigma^{2}+d]$. 

Let us prove the former property~(\ref{E_t_dif}). 
From the definitions of $E_{\tau+1}[\ell]$ and $\tilde{E}_{\tau+1}$ 
in (\ref{E_t}) and (\ref{E_t_tilde}), we use the triangle inequality and 
the Lipschitz-continuity of $\mathrm{MMSE}(s)$ to obtain  
\begin{IEEEeqnarray}{rl}
&|\tilde{E}_{\tau+1}[\ell] - E_{\tau+1}[\ell]|
\nonumber \\
\leq& \frac{1}{W+1}\sum_{w\in\mathcal{W}[\ell]}
|\mathrm{MMSE}(\tilde{s}_{\tau}[\ell-w])
- \mathrm{MMSE}(s_{\tau}[\ell-w])|
\nonumber \\  
<& \frac{C}{W+1}\sum_{w\in\mathcal{W}[\ell]}
|\tilde{s}_{\tau}[\ell-w] - s_{\tau}[\ell-w]| 
= {\cal O}(a_{W}^{-1}),
\end{IEEEeqnarray}
with some constant $C>0$, 
where the last follows from the latter property~(\ref{s_t_dif}) and 
$|\mathcal{W}[\ell]|\leq W+1$ for all $\ell\in\mathcal{L}_{W}$. 
Thus, the former property~(\ref{E_t_dif}) holds 
if the latter property~(\ref{s_t_dif}) is correct. 

\subsection{Proof of (\ref{s_t_dif})}
We prove the latter property~(\ref{s_t_dif}) by induction. 
The proof of (\ref{s_t_dif}) for $\tau=0$ is trivial from the initial 
condition $\tilde{s}_{0}[l]=s_{0}[l]$. For some $t\in\mathbb{N}$, suppose that 
(\ref{s_t_dif}) is correct for $\tau=t-1$. 
We need to prove (\ref{s_t_dif}) for $\tau=t$.  

From the definitions of $s_{t}[l]$ and $\tilde{s}_{t}[l]$ in (\ref{s_t}) and 
(\ref{s_t_tilde}), respectively, we use the triangle inequality to obtain 
\begin{equation}
|\tilde{s}_{t}[l] - s_{t}[l]|
< \frac{1}{W+1}\sum_{\ell=l}^{l+W}\left(
 T_{t}^{(1)}[\ell] + T_{t}^{(2)}[\ell] + T_{t}^{(3)}[\ell]
\right),
\end{equation}
where $T_{t}^{(i)}[\ell]$ for $i\in\{1, 2, 3\}$ is given by 
\begin{equation}
T_{t}^{(1)}[\ell] 
= |g(\tilde{E}_{t}[\ell]) - g(E_{t}[\ell])|,
\end{equation}
\begin{equation}
T_{t}^{(2)}[\ell] 
=\left|
 g(\nu_{t}[\ell]E_{t}[\ell])
 - g[\ell](\nu_{t}[\ell]E_{t}[\ell])
\right|, 
\end{equation}
\begin{equation}
T_{t}^{(3)}[\ell] 
= \left|
 g(E_{t}[\ell]) - g(\nu_{t}[\ell]E_{t}[\ell])
\right|,
\end{equation}
with 
\begin{equation} \label{nu}
\nu_{t}[\ell] = \frac{\bar{\eta}_{\A,t}[\ell]}
{1 - \bar{\eta}_{\B,t-1}[\ell]/|\mathcal{W}[\ell]|}>0.  
\end{equation}

We evaluate the first term. 
The induction hypothesis~(\ref{s_t_dif}) for $\tau=t-1$ implies the former 
property~(\ref{E_t_dif}) for $\tau=t-1$, 
\begin{equation} \label{E_t_hypothesis}
|\tilde{E}_{t}[\ell] - E_{t}[\ell]|={\cal O}(a_{W}^{-1}).
\end{equation}
The continuous-differentiability assumption of $R(z)$ in 
Lemma~\ref{lemma_approximation} implies that $g(z)$ in (\ref{g_func}) is also 
continuously differentiable for all $z\geq0$. Furthermore, from the upper 
bound $\mathrm{MMSE}(s)\leq1$ in (\ref{MMSE_upper_bound}) and 
the induction hypothesis~(\ref{E_t_hypothesis}) we find that both 
$\tilde{E}_{t}[\ell]$ and $E_{t}[\ell]$ are bounded. 
Repeating the same proof as that for the former property~(\ref{E_t_dif}) 
in Appendix~\ref{proof_lemma_approximation_former}, 
we obtain 
\begin{equation} 
\frac{1}{W+1}\sum_{\ell=l}^{l+W}T_{t}^{(1)}[\ell] = {\cal O}(a_{W}^{-1}). 
\end{equation}

In evaluating the remaining terms, we use the following lemma on 
${\cal O}(W/a_{W})$ sections in both ends: 

\begin{lemma} \label{lemma_boundary}
Suppose that $\mathbb{E}[x_{1}^{2}[l]]=1$ holds. 
Let $\{a_{W}>0\}_{W=1}^{\infty}$ denote a positive and diverging sequence 
at a sublinear speed in $W$: $\lim_{W\to\infty}a_{W}=\infty$ and 
$\lim_{W\to\infty}a_{W}/W=0$. 
For all $t$ and $\ell\in\{0,\ldots,\lceil W/a_{W}\rceil\}\cup
\{L+W-1-\lceil W/a_{W}\rceil,\ldots,L+W-1\}$, $E_{t}[\ell]={\cal O}(a_{W}^{-1})$ 
and $\tilde{E}_{t}[\ell]={\cal O}(a_{W}^{-1})$ hold 
in the continuum limit $L, W\to\infty$ with $\Delta=W/L$ kept constant. 
\end{lemma}
\begin{IEEEproof}
Without loss of generality, we focus on $\tilde{E}_{t}[\ell]$ given in 
(\ref{E_t_tilde}) and only consider the case of 
$\ell\in\{0,\ldots,\lceil W/a_{W}\rceil\}$ since the case of 
$\ell\in\{L+W-1-\lceil W/a_{W}\rceil,\ldots,L+W-1\}$ can be proved in the 
same manner. 
  
For all $\ell\in\{0,\ldots,\lceil W/a_{W}\rceil\}$, 
we use the definition of $\mathcal{W}[\ell]$ in (\ref{set_W}) and 
the upper bound $\mathrm{MMSE}(\tilde{s}_{t}[l])\leq\mathbb{E}[x_{1}^{2}]=1$ 
in (\ref{MMSE_upper_bound}) to have
\begin{IEEEeqnarray}{rl}
&\tilde{E}_{t}[\ell] 
= \frac{1}{W+1}\sum_{w=0}^{\ell}\mathrm{MMSE}(\tilde{s}_{t-1}[\ell-w])
\nonumber \\
\leq& \frac{\ell+1}{W+1}
< \frac{2+W/a_{W}}{W+1} = {\cal O}(a_{W}^{-1})   
\end{IEEEeqnarray}
in the continuum limit. 
\end{IEEEproof}

We arrive at the latter property~(\ref{s_t_dif}) for $\tau=t$, by proving 
\begin{equation} \label{term2} 
\frac{1}{W+1}\sum_{\ell=l}^{l+W}T_{t}^{(2)}[\ell] = {\cal O}(a_{W}^{-1}), 
\end{equation}
\begin{equation} \label{term3} 
\frac{1}{W+1}\sum_{\ell=l}^{l+W}T_{t}^{(3)}[\ell] = {\cal O}(a_{W}^{-1}) 
\end{equation}
for all $l\in\mathcal{L}_{0}$. 
They are proved in Appendices~\ref{proof_term2}, and \ref{proof_term3}, 
respectively. 
 
\subsection{Proof of (\ref{term2})}
\label{proof_term2}
\subsubsection{Case 1}
We evaluate the summation~(\ref{term2}) in the case of  
$l\in\{\lceil W/a_{W}\rceil,\ldots,L-1-\lceil W/a_{W}\rceil\}$. 
In this case, we can use the assumption~(\ref{R_transform_limit}) in 
Theorem~\ref{theorem_optimality} for all $\ell\in\{l,\ldots,l+W\}$.  
Applying $g[\ell](z)$ in (\ref{g_l_func}) and $g(z)$ in (\ref{g_func}) 
to (\ref{R_transform_limit}), we have $|g[\ell](z)-g(z)|={\cal O}(a_{W}^{-1})$ 
in the continuum limit for all $\ell\in\{l,\ldots,l+W\}$. Thus, 
\begin{equation} \label{T2_bound}
\frac{1}{W+1}\sum_{\ell=l}^{l+W}T_{t}^{(2)}[\ell] 
= {\cal O}(a_{W}^{-1}).  
\end{equation}

\subsubsection{Case 2}
In the case of $l\in\{0,\ldots,\lceil W/a_{W}\rceil-1\}$, we decompose the 
summation~(\ref{term2}) into two terms, 
\begin{IEEEeqnarray}{r}
\frac{1}{W+1}\sum_{\ell=l}^{l+W}T_{t}^{(2)}[\ell] 
= \frac{1}{W+1}\sum_{\ell=l}^{\lceil W/a_{W}\rceil-1}T_{t}^{(2)}[\ell]
\nonumber \\
+ \frac{1}{W+1}\sum_{\ell=\lceil W/a_{W}\rceil}^{l+W}T_{t}^{(2)}[\ell].  
\end{IEEEeqnarray}
Repeating the derivation of (\ref{T2_bound}) for the second term yields 
\begin{equation}
\frac{1}{W+1}\sum_{\ell=\lceil W/a_{W}\rceil}^{l+W}T_{t}^{(2)}[\ell]
= {\cal O}(a_{W}^{-1})
\end{equation}
in the continuum limit. 

For the first term, we use the triangle inequality to obtain 
\begin{equation}
T_{t}^{(2)}[\ell]
< \left|
  g(\nu_{t}[\ell]E_{t}[\ell]) - \frac{1}{\sigma^{2}}
\right|
+ \left|
 \frac{1}{\sigma^{2}}
 -  g[\ell](\nu_{t}[\ell]E_{t}[\ell])
\right|. 
\end{equation}
From $g[\ell](z)$ in (\ref{g_l_func}) and Lemma~\ref{lemma_R_transform_0}, 
we have $|g[\ell](z)-\sigma^{-2}|=o(1)$ for all $\ell\in\mathcal{L}_{W}$  
as $z\to0$. Similarly, we use the definition of $g(z)$ in (\ref{g_func}), 
the assumption~(\ref{R_transform_limit}) in Theorem~\ref{theorem_optimality}, 
and the continuity assumption of $R(z)$ to obtain  
$|g(z)-\sigma^{-2}|=o(1)$.
Furthermore, Lemma~\ref{lemma_boundary} implies $E_{t}[\ell]\to0$ 
for $\ell\in\{l,\ldots,\lceil W/a_{W}\rceil-1\}$. Combining these results,  
we arrive at
\begin{equation} \label{T2_bound_case2}
\frac{1}{W+1}\sum_{\ell=l}^{\lceil W/a_{W}\rceil-1}T_{t}^{(2)}[\ell]
= {\cal O}(a_{W}^{-1})
\end{equation}
in the continuum limit. Thus, (\ref{term2}) holds.  

\subsubsection{Case 3}
The proof in the case of $l\in\{L-\lceil W/a_{W}\rceil,\ldots,L-1\}$ is 
omitted since it is the same as that in the case of 
$l\in\{0,\ldots,\lceil W/a_{W}\rceil-1\}$. 

\subsection{Proof of (\ref{term3})}
\label{proof_term3}
\subsubsection{Case 1}
We evaluate the summation~(\ref{term3}) in the case of  
$l\in\{\lceil W/a_{W}\rceil,\ldots,L-1-\lceil W/a_{W}\rceil\}$. 
Since we have already proved the boundedness of $E_{t}[\ell]$, we use 
Lemma~\ref{lemma_eta} for all $\ell\in\{l,\ldots,l+W\}$ to find the 
boundedness of $\nu_{t}[\ell]E_{t}[\ell]$. 
We repeat the proof in Appendix~\ref{proof_lemma_approximation_former} 
to arrive at 
\begin{IEEEeqnarray}{rl} 
\frac{1}{W+1}\sum_{\ell=l}^{l+W}T_{t}^{(3)}[\ell] 
<& \frac{C}{W+1}\sum_{\ell=l}^{l+W}|1-\nu_{t}[\ell]|E_{t}[\ell]
\nonumber \\
=& {\cal O}(a_{W}/W)  
\label{T3_bound}
\end{IEEEeqnarray}
for some constant $C>0$, where the last equality follows from 
Lemma~\ref{lemma_eta}. 

\subsubsection{Case 2}
In the case of $l\in\{0,\ldots,\lceil W/a_{W}\rceil-1\}$, we decompose the 
summation~(\ref{term3}) into two terms, 
\begin{IEEEeqnarray}{r}
\frac{1}{W+1}\sum_{\ell=l}^{l+W}T_{t}^{(3)}[\ell] 
= \frac{1}{W+1}\sum_{\ell=l}^{\lceil W/a_{W}\rceil-1}T_{t}^{(3)}[\ell]
\nonumber \\
+ \frac{1}{W+1}\sum_{\ell=\lceil W/a_{W}\rceil}^{l+W}T_{t}^{(3)}[\ell].  
\end{IEEEeqnarray}
Since we have already proved $|g(z)-\sigma^{-2}|=o(1)$, for the first term 
we use Lemma~\ref{lemma_boundary} to obtain   
\begin{equation}
\frac{1}{W+1}\sum_{\ell=l}^{\lceil W/a_{W}\rceil-1}T_{t}^{(3)}[\ell]
= {\cal O}(a_{W}^{-1}).  
\end{equation}
Repeating the derivation of (\ref{T3_bound}) for the second term, we have 
\begin{equation}
\frac{1}{W+1}\sum_{\ell=\lceil W/a_{W}\rceil}^{l+W}T_{t}^{(3)}[\ell]
= {\cal O}(a_{W}/W). 
\end{equation}
Thus, (\ref{term3}) holds. 

\subsubsection{Case 3}
The proof in the case of $l\in\{L-\lceil W/a_{W}\rceil,\ldots,L-1\}$ is 
omitted since it is the same as that in the case of 
$l\in\{0,\ldots,\lceil W/a_{W}\rceil-1\}$.

\section*{Acknowledgment}
The author thanks the anonymous reviewers for their suggestions that have 
improved the quality of the manuscript greatly.

\ifCLASSOPTIONcaptionsoff
  \newpage
\fi



\bibliographystyle{IEEEtran}
\bibliography{IEEEabrv,kt-it2022_1}

\begin{thebibliography}{10}
\providecommand{\url}[1]{#1}
\csname url@samestyle\endcsname
\providecommand{\newblock}{\relax}
\providecommand{\bibinfo}[2]{#2}
\providecommand{\BIBentrySTDinterwordspacing}{\spaceskip=0pt\relax}
\providecommand{\BIBentryALTinterwordstretchfactor}{4}
\providecommand{\BIBentryALTinterwordspacing}{\spaceskip=\fontdimen2\font plus
\BIBentryALTinterwordstretchfactor\fontdimen3\font minus
  \fontdimen4\font\relax}
\providecommand{\BIBforeignlanguage}[2]{{%
\expandafter\ifx\csname l@#1\endcsname\relax
\typeout{** WARNING: IEEEtran.bst: No hyphenation pattern has been}%
\typeout{** loaded for the language `#1'. Using the pattern for}%
\typeout{** the default language instead.}%
\else
\language=\csname l@#1\endcsname
\fi
#2}}
\providecommand{\BIBdecl}{\relax}
\BIBdecl

\bibitem{Donoho06}
D.~L. Donoho, ``Compressed sensing,'' \emph{{IEEE} Trans. Inf. Theory},
  vol.~52, no.~4, pp. 1289--1306, Apr. 2006.

\bibitem{Candes061}
E.~J. Cand\'es, J.~Romberg, and T.~Tao, ``Robust uncertainty principles: Exact
  signal reconstruction from highly incomplete frequency information,''
  \emph{{IEEE} Trans. Inf. Theory}, vol.~52, no.~2, pp. 489--509, Feb. 2006.

\bibitem{Renyi59}
A.~R\'enyi, ``On the dimension and entropy of probability distributions,''
  \emph{Acta Math. Acad. Sci. Hung.}, vol.~10, no. 1--2, pp. 193--215, Mar.
  1959.

\bibitem{Wu10}
Y.~Wu and S.~Verd\'u, ``R\'enyi information dimension: Fundamental limits of
  almost lossless analog compression,'' \emph{{IEEE} Trans. Inf. Theory},
  vol.~56, no.~8, pp. 3721--3748, Aug. 2010.

\bibitem{Donoho09}
D.~L. Donoho, A.~Maleki, and A.~Montanari, ``Message-passing algorithms for
  compressed sensing,'' \emph{Proc. Nat. Acad. Sci.}, vol. 106, no.~45, pp.
  18\,914--18\,919, Nov. 2009.

\bibitem{Rangan11}
S.~Rangan, ``Generalized approximate message passing for estimation with random
  linear mixing,'' in \emph{Proc. 2011 IEEE Int. Symp. Inf. Theory}, Saint
  Petersburg, Russia, Aug. 2011, pp. 2168--2172.

\bibitem{Kabashima03}
Y.~Kabashima, ``A {CDMA} multiuser detection algorithm on the basis of belief
  propagation,'' \emph{J. Phys. A: Math. Gen.}, vol.~36, no.~43, pp.
  11\,111--11\,121, Oct. 2003.

\bibitem{Bayati11}
M.~Bayati and A.~Montanari, ``The dynamics of message passing on dense graphs,
  with applications to compressed sensing,'' \emph{{IEEE} Trans. Inf. Theory},
  vol.~57, no.~2, pp. 764--785, Feb. 2011.

\bibitem{Bayati15}
M.~Bayati, M.~Lelarge, and A.~Montanari, ``Universality in polytope phase
  transitions and message passing algorithms,'' \emph{Ann. Appl. Probab.},
  vol.~25, no.~2, pp. 753--822, Apr. 2015.

\bibitem{Bolthausen14}
E.~Bolthausen, ``An iterative construction of solutions of the {TAP} equations
  for the {Sherrington}-{Kirkpatrick} model,'' \emph{Commun. Math. Phys.}, vol.
  325, no.~1, pp. 333--366, Jan. 2014.

\bibitem{Takeuchi15}
K.~Takeuchi, T.~Tanaka, and T.~Kawabata, ``Performance improvement of iterative
  multiuser detection for large sparsely-spread {CDMA} systems by spatial
  coupling,'' \emph{{IEEE} Trans. Inf. Theory}, vol.~61, no.~4, pp. 1768--1794,
  Apr. 2015.

\bibitem{Kudekar11}
S.~Kudekar, T.~Richardson, and R.~Urbanke, ``Threshold saturation via spatial
  coupling: Why convolutional {LDPC} ensembles perform so well over the
  {BEC},'' \emph{{IEEE} Trans. Inf. Theory}, vol.~57, no.~2, pp. 803--834, Feb.
  2011.

\bibitem{Hassani12}
S.~H. Hassani, N.~Macris, and R.~Urbanke, ``Chains of mean field models,''
  \emph{J. Stat. Mech.}, no.~2, p. P02011, Feb. 2012.

\bibitem{Takeuchi12}
K.~Takeuchi, T.~Tanaka, and T.~Kawabata, ``A phenomenological study on
  threshold improvement via spatial coupling,'' \emph{IEICE Trans.
  Fundamentals}, vol. E95-A, no.~5, pp. 974--977, May 2012.

\bibitem{Krzakala12}
F.~Krzakala, M.~M\'ezard, F.~Sausset, Y.~F. Sun, and L.~Zdeborov\'a,
  ``Statistical-physics-based reconstruction in compressed sensing,''
  \emph{Phys. Rev. X}, vol.~2, pp. 021\,005--1--18, May 2012.

\bibitem{Donoho13}
D.~L. Donoho, A.~Javanmard, and A.~Montanari, ``Information-theoretically
  optimal compressed sensing via spatial coupling and approximate message
  passing,'' \emph{{IEEE} Trans. Inf. Theory}, vol.~59, no.~11, pp. 7434--7464,
  Nov. 2013.

\bibitem{Javanmard13}
A.~Javanmard and A.~Montanari, ``State evolution for general approximate
  message passing algorithms, with applications to spatial coupling,''
  \emph{Inf. Inference: A Journal of the IMA}, vol.~2, no.~2, pp. 115--144,
  Dec. 2013.

\bibitem{Barbier17}
J.~Barbier and F.~Krzakala, ``Approximate message-passing decoder and capacity
  achieving sparse superposition codes,'' \emph{{IEEE} Trans. Inf. Theory},
  vol.~63, no.~8, pp. 4894--4927, Aug. 2017.

\bibitem{Rush21}
C.~Rush, K.~Hsieh, and R.~Venkataramanan, ``Capacity-achieving spatially
  coupled sparse superposition codes with {AMP} decoding,'' \emph{{IEEE} Trans.
  Inf. Theory}, vol.~67, no.~7, pp. 4446--4484, Jul. 2021.

\bibitem{Joseph12}
A.~Joseph and A.~R. Barron, ``Least squares superposition codes of moderate
  dictionary size are reliable at rates up to capacity,'' \emph{{IEEE} Trans.
  Inf. Theory}, vol.~58, no.~5, pp. 2541--2557, May 2012.

\bibitem{Yedla14}
A.~Yedla, Y.~Jian, P.~S. Nguyen, and H.~D. Pfister, ``A simple proof of
  {Maxwell} saturation for coupled scalar recursions,'' \emph{{IEEE} Trans.
  Inf. Theory}, vol.~60, no.~11, pp. 6943--6965, Nov. 2014.

\bibitem{Tanaka02}
T.~Tanaka, ``A statistical-mechanics approach to large-system analysis of
  {CDMA} multiuser detectors,'' \emph{{IEEE} Trans. Inf. Theory}, vol.~48,
  no.~11, pp. 2888--2910, Nov. 2002.

\bibitem{Guo051}
D.~Guo and S.~Verd\'u, ``Randomly spread {CDMA}: Asymptotics via statistical
  physics,'' \emph{{IEEE} Trans. Inf. Theory}, vol.~51, no.~6, pp. 1983--2010,
  Jun. 2005.

\bibitem{Mezard87}
M\'ezard, G.~Parisi, and M.~A. Virasoro, \emph{Spin Glass Theory and
  Beyond}.\hskip 1em plus 0.5em minus 0.4em\relax Singapore: World Scientific,
  1987.

\bibitem{Nishimori01}
H.~Nishimori, \emph{Statistical Physics of Spin Glasses and Information
  Processing}.\hskip 1em plus 0.5em minus 0.4em\relax New York: Oxford
  University Press, 2001.

\bibitem{Reeves19}
G.~Reeves and H.~D. Pfister, ``The replica-symmetric prediction for random
  linear estimation with {Gaussian} matrices is exact,'' \emph{{IEEE} Trans.
  Inf. Theory}, vol.~65, no.~4, pp. 2252--2283, Apr. 2019.

\bibitem{Barbier20}
J.~Barbier, N.~Macris, M.~Dia, and F.~Krzakala, ``Mutual information and
  optimality of approximate message-passing in random linear estimation,''
  \emph{{IEEE} Trans. Inf. Theory}, vol.~66, no.~7, pp. 4270--4303, Jul. 2020.

\bibitem{Wu11}
Y.~Wu and S.~Verd\'u, ``{MMSE} dimension,'' \emph{{IEEE} Trans. Inf. Theory},
  vol.~57, no.~8, pp. 4857--4879, Aug. 2011.

\bibitem{Takeda06}
K.~Takeda, S.~Uda, and Y.~Kabashima, ``Analysis of {CDMA} systems that are
  characterized by eigenvalue spectrum,'' \emph{Europhys. Lett.}, vol.~76,
  no.~6, pp. 1193--1199, 2006.

\bibitem{Tulino13}
A.~M. Tulino, G.~Caire, S.~Verd\'u, and S.~{Shamai (Shitz)}, ``Support recovery
  with sparsely sampled free random matrices,'' \emph{{IEEE} Trans. Inf.
  Theory}, vol.~59, no.~7, pp. 4243--4271, Jul. 2013.

\bibitem{Hiai00}
F.~Hiai and D.~Petz, \emph{The Semicircle Law, Free Random Variables and
  Entropy}.\hskip 1em plus 0.5em minus 0.4em\relax Providence, RI, USA: Amer.
  Math. Soc., 2000.

\bibitem{Tulino04}
A.~M. Tulino and S.~Verd\'{u}, \emph{Random Matrix Theory and Wireless
  Communications}.\hskip 1em plus 0.5em minus 0.4em\relax Hanover, MA USA: Now
  Publishers Inc., 2004.

\bibitem{Barbier18}
J.~Barbier, N.~Macris, A.~Maillard, and F.~Krzakala, ``The mutual information
  in random linear estimation beyond i.i.d. matrices,'' in \emph{Proc. 2018
  IEEE Int. Symp. Inf. Theory}, Vail, CO, USA, Jun. 2018, pp. 1390--1394.

\bibitem{Li22}
Y.~Li, Z.~Fan, S.~Sen, and Y.~Wu, ``Random linear estimation with
  rotationally-invariant designs: Asymptotics at high temperature,'' Dec. 2022,
  [Online] Available: https://arxiv.org/abs/2212.10624.

\bibitem{Caltagirone14}
F.~Caltagirone, L.~Zdeborov\'a, and F.~Krzakala, ``On convergence of
  approximate message passing,'' in \emph{Proc. 2014 IEEE Int. Symp. Inf.
  Theory}, Honolulu, HI, USA, Jul. 2014, pp. 1812--1816.

\bibitem{Rangan191}
S.~Rangan, P.~Schniter, A.~Fletcher, and S.~Sarkar, ``On the convergence of
  approximate message passing with arbitrary matrices,'' \emph{{IEEE} Trans.
  Inf. Theory}, vol.~65, no.~9, pp. 5339--5351, Sep. 2019.

\bibitem{Kabashima14}
Y.~Kabashima and M.~Vehkaper\"a, ``Signal recovery using expectation consistent
  approximation for linear observations,'' in \emph{Proc. 2014 IEEE Int. Symp.
  Inf. Theory}, Honolulu, HI, USA, Jul. 2014, pp. 226--230.

\bibitem{Vila15}
J.~Vila, P.~Schniter, S.~Rangan, F.~Krzakala, and L.~Zdeborov\'a, ``Adaptive
  damping and mean removal for the generalized approximate message passing
  algorithm,'' in \emph{Proc. 2015 IEEE Int. Conf. Acoust. Speech Signal
  Process.}, South Brisbane, Australia, Apr. 2015, pp. 2021--2025.

\bibitem{Manoel15}
A.~Manoel, F.~Krzakala, E.~W. Tramel, and L.~Zdeborov\'a, ``Swept approximate
  message passing for sparse estimation,'' in \emph{Proc. 32nd Int. Conf. Mach.
  Learn.}, Lille, France, Jul. 2015, pp. 1123--1132.

\bibitem{Rangan17}
S.~Rangan, A.~K. Fletcher, P.~Schniter, and U.~S. Kamilov, ``Inference for
  generalized linear models via alternating directions and {Bethe} free energy
  minimization,'' \emph{{IEEE} Trans. Inf. Theory}, vol.~63, no.~1, pp.
  676--697, Jan. 2017.

\bibitem{Ma17}
J.~Ma and L.~Ping, ``Orthogonal {AMP},'' \emph{IEEE Access}, vol.~5, pp.
  2020--2033, Jan. 2017.

\bibitem{Rangan192}
S.~Rangan, P.~Schniter, and A.~K. Fletcher, ``Vector approximate message
  passing,'' \emph{{IEEE} Trans. Inf. Theory}, vol.~65, no.~10, pp. 6664--6684,
  Oct. 2019.

\bibitem{Yuan21}
Z.~Yuan, Q.~Guo, and M.~Luo, ``Approximate message passing with unitary
  transformation for robust bilinear recovery,'' \emph{{IEEE} Trans. Signal
  Process.}, vol.~69, pp. 617--630, 2021.

\bibitem{Opper05}
M.~Opper and O.~Winther, ``Expectation consistent approximate inference,''
  \emph{J. Mach. Learn. Res.}, vol.~6, pp. 2177--2204, Dec. 2005.

\bibitem{Cespedes14}
J.~C\'espedes, P.~M. Olmos, M.~S\'anchez-Fern\'andez, and F.~Perez-Cruz,
  ``Expectation propagation detection for high-order high-dimensional {MIMO}
  systems,'' \emph{{IEEE} Trans. Commun.}, vol.~62, no.~8, pp. 2840--2849, Aug.
  2014.

\bibitem{Takeuchi201}
K.~Takeuchi, ``Rigorous dynamics of expectation-propagation-based signal
  recovery from unitarily invariant measurements,'' \emph{{IEEE} Trans. Inf.
  Theory}, vol.~66, no.~1, pp. 368--386, Jan. 2020.

\bibitem{Minka01}
T.~P. Minka, ``Expectation propagation for approximate {Bayesian} inference,''
  in \emph{Proc. 17th Conf. Uncertainty Artif. Intell.}, Seattle, WA, USA, Aug.
  2001, pp. 362--369.

\bibitem{Takeuchi22}
K.~Takeuchi, ``On the convergence of orthogonal/vector {AMP}: Long-memory
  message-passing strategy,'' in \emph{Proc. 2022 IEEE Int. Symp. Inf. Theory},
  Espoo, Finland, Jun.--Jul. 2022, pp. 1366--1371.

\bibitem{Takeuchi213}
------, ``On the convergence of orthogonal/vector {AMP}: Long-memory
  message-passing strategy,'' \emph{{IEEE} Trans. Inf. Theory}, vol.~68,
  no.~12, pp. 8121--8138, Dec. 2022.

\bibitem{Liu221}
L.~Liu, S.~Huang, and B.~M. Kurkoski, ``Sufficient statistic memory approximate
  message passing,'' in \emph{Proc. 2022 IEEE Int. Symp. Inf. Theory}, Espoo,
  Finland, Jun.--Jul. 2022, pp. 1378--1383.

\bibitem{Opper16}
M.~Opper, B.~\c{C}akmak, and O.~Winther, ``A theory of solving {TAP} equations
  for {Ising} models with general invariant random matrices,'' \emph{J. Phys.
  A: Math. Theor.}, vol.~49, no.~11, p. 114002, Feb. 2016.

\bibitem{Fan22}
Z.~Fan, ``Approximate message passing algorithms for rotationally invariant
  matrices,'' \emph{Ann. Statist.}, vol.~50, no.~1, pp. 197--224, Feb. 2022.

\bibitem{Venkataramanan21}
R.~Venkataramanan, K.~K\"ogler, and M.~Mondelli, ``Estimation in rotationally
  invariant generalized linear models via approximate message passing,'' in
  \emph{Proc. 39th Int. Conf. Mach. Learn.}, Baltimore, MD, USA, Jul. 2022.

\bibitem{Takeuchi211}
K.~Takeuchi, ``Bayes-optimal convolutional {AMP},'' \emph{{IEEE} Trans. Inf.
  Theory}, vol.~67, no.~7, pp. 4405--4428, Jul. 2021.

\bibitem{Takeuchi202}
------, ``Convolutional approximate message-passing,'' \emph{{IEEE} Signal
  Process. Lett.}, vol.~27, pp. 416--420, 2020.

\bibitem{Takeuchi221}
------, ``On the convergence of convolutional approximate message-passing for
  {Gaussian} signaling,'' \emph{IEICE Trans. Fundamentals.}, vol. E105-A,
  no.~2, pp. 100--108, Feb. 2022.

\bibitem{Liu222}
L.~Liu, S.~Huang, and B.~M. Kurkoski, ``Memory {AMP},'' \emph{{IEEE} Trans.
  Inf. Theory}, vol.~68, no.~12, pp. 8015--8039, Dec. 2022.

\bibitem{Skuratovs221}
N.~Skuratovs and M.~E. Davies, ``Compressed sensing with upscaled vector
  approximate message passing,'' \emph{{IEEE} Trans. Inf. Theory}, vol.~68,
  no.~7, pp. 4818--4836, Jul. 2022.

\bibitem{Skuratovs222}
------, ``Warm-starting in message passing algorithms,'' in \emph{Proc. 2022
  IEEE Int. Symp. Inf. Theory}, Espoo, Finland, Jun.--Jul. 2022, pp.
  1187--1192.

\bibitem{Takeuchi172}
K.~Takeuchi and C.-K. Wen, ``Rigorous dynamics of expectation-propagation
  signal detection via the conjugate gradient method,'' in \emph{Proc. 18th
  IEEE Int. Workshop Sig. Process. Advances Wirel. Commun.}, Sapporo, Japan,
  Jul. 2017, pp. 88--92.

\bibitem{Takeuchi23}
K.~Takeuchi, ``Long-memory message-passing for spatially coupled systems,'' in
  \emph{Proc. 2023 IEEE Int. Conf. Acoust. Speech Signal Process.}, Rhodes
  Island, Greece, Jun. 2023.

\bibitem{Candes062}
E.~J. Cand\'es and T.~Tao, ``Near-optimal signal recovery from random
  projections: Universal encoding strategies?'' \emph{{IEEE} Trans. Inf.
  Theory}, vol.~52, no.~12, pp. 5406--5425, Dec. 2006.

\bibitem{Anderson14}
G.~W. Anderson and B.~Farrell, ``Asymptotically liberating sequences of random
  unitary matrices,'' \emph{Adv. Math.}, vol. 255, pp. 381--413, Apr. 2014.

\bibitem{Male20}
C.~Male, \emph{Traffic Distributions and Independence: Permutation Invariant
  Random Matrices and the Three Notions of Independence}.\hskip 1em plus 0.5em
  minus 0.4em\relax Providence, RI, USA: Amer. Math. Soc., 2020.

\bibitem{Dudeja22}
R.~Dudeja, Y.~M. Lu, and S.~Sen, ``Universality of approximate message passing
  with semi-random matrices,'' Apr. 2022, [Online] Available:
  https://arxiv.org/abs/2204.04281.

\bibitem{Barbier191}
J.~Barbier, F.~Krzakala, N.~Macris, L.~Miolane, and L.~Zdeborov\'a, ``Optimal
  errors and phase transitions in high-dimensional generalized linear models,''
  \emph{Proc. Nat. Acad. Sci.}, vol. 116, no.~12, pp. 5451--5460, Mar. 2019.

\bibitem{Mingo17}
J.~A. Mingo and R.~Speicher, \emph{Free Probability and Random Matrices}.\hskip
  1em plus 0.5em minus 0.4em\relax New York, NY, USA: Springer, 2017.

\bibitem{Guo052}
D.~Guo, S.~{Shamai (Shitz)}, and S.~Verd\'u, ``Mutual information and minimum
  mean-square error in {Gaussian} channels,'' \emph{{IEEE} Trans. Inf. Theory},
  vol.~51, no.~4, pp. 1261--1282, Apr. 2005.

\bibitem{Guo11}
D.~Guo, Y.~Wu, S.~{Shamai (Shitz)}, and S.~Verd\'u, ``Estimation in {Gaussian}
  noise: Properties of the minimum mean-square error,'' \emph{{IEEE} Trans.
  Inf. Theory}, vol.~57, no.~4, pp. 2371--2385, Apr. 2011.

\bibitem{Berthier19}
R.~Berthier, A.~Montanari, and P.-M. Nguyen, ``State evolution for approximate
  message passing with non-separable functions,'' \emph{Inf. Inference: A
  Journal of the IMA}, 2019, doi:10.1093/imaiai/iay021.

\bibitem{Ma19}
Y.~Ma, C.~Rush, and D.~Baron, ``Analysis of approximate message passing with
  non-separable denoisers and {Markov} random field priors,'' \emph{{IEEE}
  Trans. Inf. Theory}, vol.~65, no.~11, pp. 7367--7389, Nov. 2019.

\bibitem{Fletcher19}
A.~K. Fletcher, P.~Pandit, S.~Rangan, S.~Sarkar, and P.~Schniter, ``Plug-in
  estimation in high-dimensional linear inverse problems a rigorous analysis,''
  \emph{J. Stat. Mech.: Theory Exp.}, vol. 2019, pp. 124\,021--1--15, Dec.
  2019.

\bibitem{Ellis06}
R.~S. Ellis, \emph{Entropy, Large Deviations, and Statistical Mechanics
  (Reprint of the 1985 Edition)}.\hskip 1em plus 0.5em minus 0.4em\relax
  Berlin: Springer-Verlag, 2006.

\bibitem{Takeuchi19}
K.~Takeuchi, ``A unified framework of state evolution for message-passing
  algorithms,'' in \emph{Proc. 2019 IEEE Int. Symp. Inf. Theory}, Paris,
  France, Jul. 2019, pp. 151--155.

\end{thebibliography}
%

%







\end{document}